\documentclass[12pt]{article}
\usepackage{graphicx,amssymb,amsmath,empheq}
\usepackage{authblk}
\usepackage{amsthm}
\usepackage{empheq}
\usepackage{subfig} 

\usepackage{hyperref}
\hypersetup{colorlinks = true, linkcolor=black, citecolor=black, urlcolor=blue}
\usepackage{url}

\theoremstyle{plain}

\usepackage{tikz}
\usepackage{tikz-cd}
\usetikzlibrary{arrows}
\usetikzlibrary{intersections}
\usetikzlibrary{shapes.geometric}
\usetikzlibrary{decorations.pathmorphing, patterns,shapes}
\usetikzlibrary{decorations.markings}

\tikzset{
  mid arrow/.style={postaction={decorate,decoration={
        markings,
        mark=at position .575 with {\arrow[#1]{stealth}}
      }}},
  near arrow/.style={postaction={decorate,decoration={
        markings,
        mark=at position .275 with {\arrow[#1]{stealth}}
      }}},
   far arrow/.style={postaction={decorate,decoration={
        markings,
        mark=at position .800 with {\arrow[#1]{stealth}}
      }}},
}

\pgfdeclarepatternformonly{south west lines}{\pgfqpoint{-0pt}{-0pt}}{\pgfqpoint{3pt}{3pt}}{\pgfqpoint{3pt}{3pt}}{
        \pgfsetlinewidth{0.4pt}
        \pgfpathmoveto{\pgfqpoint{0pt}{0pt}}
        \pgfpathlineto{\pgfqpoint{3pt}{3pt}}
        \pgfpathmoveto{\pgfqpoint{2.8pt}{-.2pt}}
        \pgfpathlineto{\pgfqpoint{3.2pt}{.2pt}}
        \pgfpathmoveto{\pgfqpoint{-.2pt}{2.8pt}}
        \pgfpathlineto{\pgfqpoint{.2pt}{3.2pt}}
        \pgfusepath{stroke}}

\usepackage[nosort]{cite}

\usepackage[nosort]{cite}

\newenvironment{defn}[1][Definition]{\begin{trivlist}
\item[\hskip \labelsep {\bfseries #1}]}{\end{trivlist}}

\renewcommand{\bar}{\overline}
\renewcommand{\tilde}{\widetilde}

\renewcommand{\leq}{\leqslant}
\renewcommand{\geq}{\geqslant}

\newcommand{\Tr}{\operatorname{Tr}}

\newcommand{\dkap}{\delta\kern-1.25pt\varkappa}

\newcommand{\UU}{\operatorname{U}}

\newcommand{\SU}{\operatorname{SU}}

\newcommand{\SL}{\operatorname{SL}}

\newcommand{\CC}{\mathbb{C}}
\newcommand{\RR}{\mathbb{R}}

\newcommand{\ZZ}{\mathbb{Z}}

\newcommand{\bbD}{\mathbb{D}}
\newcommand{\bbZ}{\mathbb{Z}}

\newcommand{\calC}{\mathcal{C}}

\newcommand{\calM}{\mathcal{M}}

\newcommand{\calO}{\mathcal{O}}

\makeatletter

\newcommand*{\wideboxed}[1]{\setlength{\fboxsep}{1ex}%
  \fbox{\m@th$\displaystyle#1$}}
\makeatother

\newcommand{\eqnref}[1]{Eq.\,\eqref{#1}}
\newcommand{\figref}[1]{Fig.~\,\ref{#1}}
\newcommand{\secref}[1]{Sec.\,\ref{#1}}
\newcommand{\appref}[1]{Appendix.\,\ref{#1}}

\newcommand{\XW}[1]{\textcolor{blue}{#1}}

\def\be{\begin{equation}}
\def\ee{\end{equation}}

\title{
Periodically, Quasi-periodically, and Randomly
Driven \\
Conformal Field Theories: Part I
}

\author[1]{Xueda Wen}
\author[2]{Ruihua Fan}

\author[2]{Ashvin Vishwanath}
\author[2]{Yingfei Gu}
\affil[1]{\normalsize\it Department of Physics, Massachusetts Institute of Technology, Cambridge, MA 02139, USA}
\affil[2]{\normalsize\it Department of Physics, Harvard University, Cambridge MA 02138, USA}

\begin{document}
\maketitle

\begin{abstract}

In this paper and its sequel, we study non-equilibrium dynamics in driven 1+1D conformal field theories (CFTs) with periodic, quasi-periodic, and random driving. We study a soluble family of drives in which the Hamiltonian only involves the energy-momentum density spatially modulated at a single wavelength. The resulting time evolution is then captured by a M\"obius coordinate transformation.
In this Part I, we establish the general framework and focus on the first two classes. 
In  periodically driven CFTs, we generalize earlier work and 
study the generic features of entanglement/energy evolution
in different phases, i.e. the heating, non-heating phases and the phase transition between them. 
In quasi-periodically driven CFTs, we mainly focus on the case of driving with a Fibonacci sequence. We find that (i) the non-heating phases form a Cantor set of measure zero; 
(ii) in the heating phase, the Lyapunov exponents 
(which characterize the growth 
rate of the entanglement entropy and energy) 
exhibit self-similarity, and can be arbitrarily small;  (iii) the heating phase exhibits  periodicity in the location of spatial structures at the Fibonacci times;   (iv) one can find exactly the non-heating fixed point, where the entanglement entropy/energy oscillate at the Fibonacci numbers, but grow logarithmically/polynomially 
at the non-Fibonacci numbers; 
(v) for certain choices of driving Hamiltonians, the non-heating  phases of the Fibonacci driving CFT can be mapped to the energy spectrum of electrons propagating in a Fibonacci quasi-crystal. In addition, another quasi-periodically driven CFT with an Aubry-Andr\'e like sequence is also studied. We compare the CFT results to lattice calculations and find remarkable agreement.  

\end{abstract}

\newpage

\tableofcontents

\section{Introduction}
\label{Sec: Introduction}

Non-equilibrium dynamics in time-dependent driven 
quantum many-body systems has received extensive recent attention.
A time-dependent drive, such as a periodic drive, 
creates a new stage in the search for novel systems that 
may not have an equilibrium analog, e.g., Floquet topological phases\cite{Jiang:2011xv,Oka2009,demler2010prb,rudner2013anomalous,else2016prb,ashvin2016prx,roy2016prb,Po:2016qlt,roy2017prb,roy2017prl,po2017radical,yao2017floquetspt,po2017timeglide,ashvin2019prb} and time crystals\cite{khemani2016phase, else2016timecrystal, sondhi2016phaseI, sonhdi2016phaseII, Else:2017ghz, normal2017timecrystal, lukin2017exp, zhang2017observation, normal2018classical}. 
It is also one of the basic protocols to study non-equilibrium phenomena, such as localization-thermalization transitions, prethermalization,
dynamical localization,
dynamical Casimir effect, etc\cite{rigol2014prx, abanin2014mbl, abanin2014theory, abanin2015prl, abanin2015rigorous, abanin2015effectiveh,Oka2016,
Dunlap1986,Hanggi1991,
Law1994,Dodonov1996,martin2019floquet,RMP2017}.

Despite the rich phenomena and applications in 
the time-dependent driving physics, 
exactly solvable setups are, in general, very rare. Usually, we have to resort to numerical methods limited to small system size. 
In this work, we are interested in a quantum $(1+1)$ dimensional conformal field theory (CFT), which may be
viewed as the low energy effective field theory of a many-body system at the critical point. 
The property of conformal invariance at the critical point can be exploited to  constrain the operator content
of the critical theory\cite{belavin1984infinite,francesco2012conformal}.
In particular, for $(1+1)$D CFTs, the conformal symmetry is enlarged to the full Virasoro symmetry, which makes tractable the study of non-equilibrium dynamics, such as the quantum quench problems
\cite{Calabrese_2005,CC2006}.
For a time-dependent driven CFT, however, relatively little is known.

\begin{figure}[t]
\small
\centering
\begin{tikzpicture}
\draw [>=stealth,->] (0pt, -10pt)--(110pt,-10pt);
\draw [>=stealth,->] (0pt, -10pt)--(0pt,80pt);
\node at (120pt, -10pt){$t$};
\node at (-15pt, 80pt){$S_A(t)$};

\node at (115pt, 30pt){\color{red}$S_A(t)\simeq \frac{c}{3}\log t$, phase transition};

\node at (110pt, 8pt){$S_A(t)=a+b\cos(\omega t)$, non-heating phase};


\draw [thick,red](0pt,0pt)..controls (4.5pt,11.787*0.55pt) and (9pt,17.047*0.55pt)..(13.5pt,20.477*0.55pt)..controls (18pt,23.026*0.55pt) and (22.5pt,25.055*0.55pt)..(27pt,26.741*0.55pt)..controls (31.5pt,28.184*0.55pt) and (36pt,29.444*0.55pt)..(40.5pt,30.564*0.55pt)..controls (45pt,31.570*0.55pt) and (49.5pt,32.484*0.55pt)..(54pt,33.322*0.55pt)..controls
(58.5pt,34.095*0.55pt) and (63pt,34.812*0.55pt)..(67.5pt,35.482*0.55pt)..controls (72pt,36.109*0.55pt) and (76.5pt,36.7*0.55pt)..(81pt,37.257*0.55pt)..
controls (85.5pt,37.785*0.55pt) and (90pt,38.286*0.55pt)..(94.5pt,38.764*0.55pt);


\draw[thick] (0pt,0pt) sin (5pt,2pt);
\draw[thick] (5pt,2pt) cos (10pt,0pt);
\draw[thick] (10pt,0pt) sin (15pt,-2pt);
\draw[thick] (15pt,-2pt) cos (20pt,0pt);
\draw[thick] (20pt,0pt) sin (25pt,2pt);
\draw[thick] (25pt,2pt) cos (30pt,0pt);
\draw[thick] (30pt,0pt) sin (35pt,-2pt);
\draw[thick] (35pt,-2pt) cos (40pt,0pt);
\draw[thick] (40pt, 0pt) sin (45pt,2pt);
\draw[thick] (45pt,2pt) cos (50pt,0pt);
\draw[thick] (50pt,0pt) sin (55pt,-2pt);
\draw[thick] (55pt,-2pt) cos (60pt,0pt);
\draw[thick] (60pt,0pt) sin (65pt,2pt);
\draw[thick] (65pt,2pt) cos (70pt,0pt);
\draw[thick] (70pt,0pt) sin (75pt,-2pt);
\draw[thick] (75pt,-2pt) cos (80pt,0pt);
\draw[thick] (80pt,0pt) sin (85pt,2pt);
\draw[thick] (85pt,2pt) cos (90pt,0pt);
\draw[thick] (90pt,0pt) sin (95pt,-2pt);

\draw [thick,blue] (0pt, 0pt)--(70pt,72pt);
\node at (90pt, 79pt){\color{blue}$S_A(t)\propto t$, heating phase};

\end{tikzpicture}
    \caption{Typical features of the time evolution of entanglement entropy
    in different phases of a periodically driven CFT.
    The entanglement entropy grows linearly in time in the heating phase, 
    grows logarithmically at the phase transition, and simply oscillates in
    the non-heating phase.
    }
    \label{FloquetCFT_EE}
\end{figure}
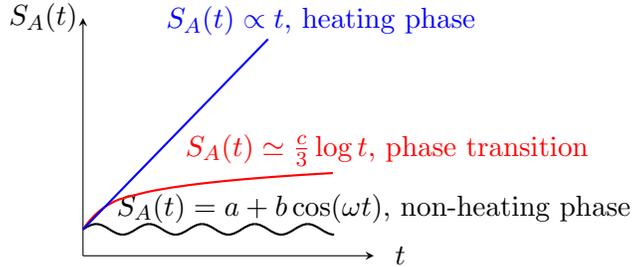

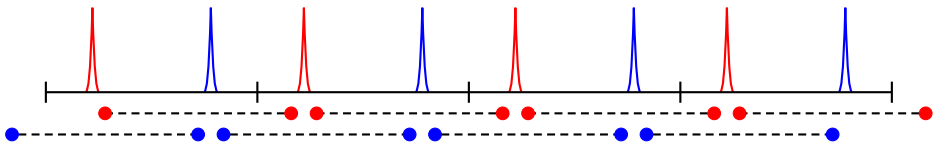
\begin{figure}[t]
\center
    \begin{tikzpicture}[scale=0.8, baseline={([yshift=-6pt]current bounding box.center)}]
    \draw [red][thick](-31pt,0pt)..controls (-30pt,2pt) and (-29pt,5pt)..(-28pt,40pt)..controls (-27pt,5pt) and (-26pt,2pt)..(-25pt,0pt);
    \draw [blue][thick](31pt,0pt)..controls (30pt,2pt) and (29pt,5pt)..(28pt,40pt)..controls (27pt,5pt) and (26pt,2pt)..(25pt,0pt);
    \draw [thick] (-50pt,0pt)--(50pt,0pt);
    \draw [thick] (-50pt,-5pt) -- (-50pt,5pt);
    \draw[densely dashed, thick] (-22pt,-10pt) -- (66pt,-10pt);
    \filldraw[red][fill=red] (-22pt,-10pt) circle (3pt);
    \filldraw[red][fill=red] (66pt,-10pt) circle (3pt);
    \draw[densely dashed, thick] (22pt,-20pt) -- (-66pt,-20pt);
    \filldraw[blue][fill=blue] (22pt,-20pt) circle (3pt);
    \filldraw[blue][fill=blue] (-66pt,-20pt) circle (3pt);
    
     \draw [red][xshift=100pt,thick](-31pt,0pt)..controls (-30pt,2pt) and (-29pt,5pt)..(-28pt,40pt)..controls (-27pt,5pt) and (-26pt,2pt)..(-25pt,0pt);
    \draw [blue][xshift=100pt,thick](31pt,0pt)..controls (30pt,2pt) and (29pt,5pt)..(28pt,40pt)..controls (27pt,5pt) and (26pt,2pt)..(25pt,0pt);
    \draw [xshift=100pt,thick] (-50pt,0pt)--(50pt,0pt);
    \draw [xshift=100pt,thick] (-50pt,-5pt) -- (-50pt,5pt);
    \draw[xshift=100pt,densely dashed, thick] (-22pt,-10pt) -- (66pt,-10pt);
    \filldraw[red][xshift=100pt,fill=red] (-22pt,-10pt) circle (3pt);
    \filldraw[red][xshift=100pt,fill=red] (66pt,-10pt) circle (3pt);
    \draw[xshift=100pt,densely dashed, thick] (22pt,-20pt) -- (-66pt,-20pt);
    \filldraw[blue][xshift=100pt,fill=blue] (22pt,-20pt) circle (3pt);
    \filldraw[blue][xshift=100pt,fill=blue] (-66pt,-20pt) circle (3pt);
    
    \draw [red][xshift=200pt,thick](-31pt,0pt)..controls (-30pt,2pt) and (-29pt,5pt)..(-28pt,40pt)..controls (-27pt,5pt) and (-26pt,2pt)..(-25pt,0pt);
    \draw [blue][xshift=200pt,thick](31pt,0pt)..controls (30pt,2pt) and (29pt,5pt)..(28pt,40pt)..controls (27pt,5pt) and (26pt,2pt)..(25pt,0pt);
    \draw [xshift=200pt,thick] (-50pt,0pt)--(50pt,0pt);
    \draw [xshift=200pt,thick] (-50pt,-5pt) -- (-50pt,5pt);
    \draw[xshift=200pt,densely dashed, thick] (-22pt,-10pt) -- (66pt,-10pt);
    \filldraw[red][xshift=200pt,fill=red] (-22pt,-10pt) circle (3pt);
    \filldraw[red][xshift=200pt,fill=red] (66pt,-10pt) circle (3pt);
    \draw[xshift=200pt,densely dashed, thick] (22pt,-20pt) -- (-66pt,-20pt);
    \filldraw[blue][xshift=200pt,fill=blue] (22pt,-20pt) circle (3pt);
    \filldraw[blue][xshift=200pt,fill=blue] (-66pt,-20pt) circle (3pt);
    
    \draw [red][xshift=300pt,thick](-31pt,0pt)..controls (-30pt,2pt) and (-29pt,5pt)..(-28pt,40pt)..controls (-27pt,5pt) and (-26pt,2pt)..(-25pt,0pt);
    \draw [blue][xshift=300pt,thick](31pt,0pt)..controls (30pt,2pt) and (29pt,5pt)..(28pt,40pt)..controls (27pt,5pt) and (26pt,2pt)..(25pt,0pt);
    \draw [xshift=300pt,thick] (-50pt,0pt)--(50pt,0pt);
    \draw [xshift=300pt,thick] (-50pt,-5pt) -- (-50pt,5pt);
    \draw[xshift=300pt,densely dashed, thick] (-22pt,-10pt) -- (66pt,-10pt);
    \filldraw[red][xshift=300pt,fill=red] (-22pt,-10pt) circle (3pt);
    \filldraw[red][xshift=300pt,fill=red] (66pt,-10pt) circle (3pt);
    \draw[xshift=300pt,densely dashed, thick] (22pt,-20pt) -- (-66pt,-20pt);
    \filldraw[blue][xshift=300pt,fill=blue] (22pt,-20pt) circle (3pt);
    \filldraw[blue][xshift=300pt,fill=blue] (-66pt,-20pt) circle (3pt);
    
     \draw [thick] (350pt,-5pt) -- (350pt,5pt);
       \end{tikzpicture}
    \caption{A cartoon of the entanglement pattern and the energy-momentum density distribution in real space
    in the heating phase of a periodically driven CFT,
    where we drive the system with $H_0(x)=\int_0^L h(x) dx$ and 
    $H_1(x) =\int_0^L 2\sin^2\frac{q\pi x}{L} \,h(x)\, dx$ with $q=4$ here.
    Red and blue color stand for two different chiralities. 
    Each peak is entangled with its nearest neighbor with the same chirality/color. Periodic boundary conditions are assumed here.
    }
    \label{fig:entanglement pattern generalized}    
\end{figure}

Most recently, an analytically solvable setup on the periodically driven CFT
was proposed in Ref.\cite{wen2018floquet}. The authors 
implement the periodic 
driving with two non-commuting Hamiltonians $H_0$ and $H_1$
for time durations $T_0$ and $T_1$ respectively,
where $H_0=\int_0^L h(x)\, dx$ is the uniform CFT Hamiltonian 
on a line of length $L$, and 
$H_1$ is obtained from $H_0$ by deforming the Hamiltonian density $h(x)$ 
as $H_1(x) =\int_0^L 2\sin^2\frac{\pi x}{L}\, h(x) dx$, which 
is also called sine-square deformation (SSD) in literature
\cite{Nishino2011prb,2011freefermionssd,katsura2012sine,ishibashi2015infinite,ishibashi2016dipolar, Okunishi:2016zat, WenWu2018quench,Ryu1604,
Tamura:2017vbx, Tada:2017wul,
MacCormack_2019,
tada2019time,caputa2020geometry,liu2020analysis}.
Interestingly, it was found that different phases can emerge during the driving, depending on duration of the two time evolutions.
As depicted in Fig.~\ref{FloquetCFT_EE}, there exits a heating phase
with the entanglement entropy growing linearly in time, and a non-heating 
phase with the entanglement entropy simply oscillating in time.
At the phase transition, the entanglement entropy grows logarithmically
in time. 
Later in Ref.\cite{fan2019emergent}, these emergent phases and 
the phase diagram were further confirmed by studying how
the system absorbs energy.
More explicitly, the total energy of the system grows
exponentially in time in the heating phase, oscillates 
in the non-heating phase, and grows polynomially  
at the phase transition. 
Furthermore, the system develops interesting spatial structures in the heating phase. The energy density forms an array of peaks
\footnote{See also Ref.\cite{Zurich1} for a related study on the emergent
spatial structure of the energy-momentum density.}
with simple patterns of entanglement as shown in Fig.~\ref{fig:entanglement pattern generalized}.  


\bigskip


In this work and its sequel, 
we introduce and study a general class of soluble models of driven CFTs with a 
variety of driving protocols. We determine their dynamical phase diagrams 
of heating versus non-heating behavior, particularly when the periodicity 
of the drive is absent. 
We extend the previous study on periodic driving to quasi-periodic\footnote{
See also Ref.~\cite{dumitrescu2018logarithmically,zhao2019floquet,else2020long} for studies on quasi-periodically driven quantum systems.
}
and random drivings, and make a connection to the familiar 
concepts of crystal, quasi-crystal, and disordered systems. 
The connection is based on the coincidence of the group structures  underlying the two problems:
\begin{enumerate}
    \item The driving protocol we considered for the CFTs involves $\SL_2$ deformed Hamiltonians. These are generalizations of the SSD Hamiltonian protocols, where the deformed Hamiltonians $H_q$ are chosen as
    $H_q=\int_0^L [f_q(x)\, h(x)+g_q(x) \,p(x)]dx$. Here $h(x)$ and $p(x)$
    are the energy and momentum densities, and $f_q(x)$ ($g_q(x)$)
    are real functions of the form $a+b\,\cos\frac{2\pi q x}{L}+c\,\sin\frac{2\pi q x}{L}$, with $q\in\mathbb Z$.
    The remarkable aspect of these protocols, which is the key to their solubility, is that the time evolution of many physical quantities after a prescribed time is captured simply by a $2\times 2$ matrix transformation, i.e. a $\SL_2$ or M\"obius transformation. This simplification occurs despite the  fact that we are discussing a spatially extended system. In this case, the operator evolution can be recast into a sequence of M\"obius transformations on a suitable Riemann surface  (See Fig.~\ref{OperatorEvolutionIntro}),
        \begin{equation}
        \label{OP_evolv_intro}
        z_n=(M_1\cdot M_2\cdots M_n)\, z,
        \quad M_j\in \SU(1,1)
        \end{equation}

\begin{figure}[t]
\center
\begin{tikzpicture}[x=0.75pt,y=0.75pt,yscale=-1,xscale=1]

\draw    [thick](371.5,132) .. controls (393.5,90) and (393.5,92) .. (415.5,49) ;
\draw    [thick](214.5,65) .. controls (236.5,23) and (365.5,72) .. (415.5,49) ;
\draw    [thick](170.5,148) .. controls (192.5,106) and (321.5,155) .. (371.5,132) ;
\draw  [dash pattern={on 4.5pt off 4.5pt}]  (247.5,85) .. controls (246.5,62) and (337.5,53) .. (345,92) ;

\draw (345,92) node    {$\bullet$};
\draw (348,102) node    {$z$};
\draw (355,82) node    {$\mathcal{O}$};

\draw (325,71) node    {$\bullet$};
\draw (325,81) node    {$z_1$};

\draw (295,65) node    {$\bullet$};
\draw (295,76) node    {$z_2$};

\draw (246,93) node    {$\bullet$};
\draw (246,103) node    {$z_n$};

\end{tikzpicture}
    \caption{A local view of the operator evolution on a 
    Riemann surface. By choosing a suitable coordinates, each step of the driving can be characterized by a M\"obius transformation that is determined by the $\SL_2$ deformed Hamiltonian. 
    }
    \label{OperatorEvolutionIntro}    
\end{figure}
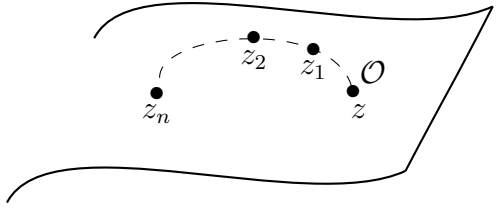

    \item The hopping problem in tight binding model can be solved using transfer matrix method, namely reformulating the discrete Sch\"ordinger equation $E \psi_j= [H \psi]_j=\psi_{j+1}+\psi_{j-1}+V_j \psi_j$ by product of the transfer matrices
    \be\label{TransferMatrix_WF}
    \Psi_n=\left(T_n\cdots T_2\cdot T_1\right) \Psi_0,\quad T_j
    = \begin{pmatrix}
    E-V_j & -1 \\
    1 & 0
    \end{pmatrix}
    \in \SL(2,\mathbb R).
    \ee
    where $\Psi_n=(\psi_{n+1},\psi_{n})^T$ represents the corresponding wave-function.
\end{enumerate}
Both problems are now solved by analyzing products of $\SU(1,1)\simeq \SL(2,\RR)$ matrices, creating intriguing analogies. In fact, the main part of the paper is to dive into the analogies and exam whether the rich phenomenons in solids can reassemble in the time domain.

\subsection{Outline and main results of this paper}


\begin{itemize}

   \item 
   In Sec.~\ref{Sec: Time-dependent CFT} we explain the details of the general setup of our study, which is a time-dependent driven $(1+1)$ CFT with arbitrary $\SL_2$ deformations. As mentioned in the introduction, the physical consequence of such driving is encoded in the product 
   \begin{equation}
   \Pi_n = M_1 \cdot M_2 \cdots M_n\,, \quad \text{where} \qquad M_j \in \SU(1,1)
   \end{equation}
   of a sequence of $\SU(1,1)$ matrices that correspond to the driving steps. 
   
   \item In Sec.~\ref{sec: diag}, we introduce the main 
   diagnostics of our driven CFT: the Lyapunov exponent and group walking. The former is a useful characterization to quantify the growth of $\Pi_n$ w.r.t. the number of driving step $n$, i.e. 
        \begin{equation}
            \lambda_L := \lim_{n\rightarrow \infty} \frac{\log \lVert M_1 \cdot M_2 \cdots M_n \rVert }{n}
        \end{equation}
    where $\lVert \cdot \rVert$ is a matrix norm. Applying to our driven system, the Lyapunov exponent has the meaning of the heating rate and serves as a good ``order parameter'' in carving the phase diagram. For example, $\lambda_L>0$ represents a heating phase, and we show that total energy of the system grows exponentially as $E(n)\propto e^{2 \lambda_L \cdot n}$ and the entanglement entropy of the subsystem that includes the energy-momentum density peaks grows linearly in time as $S(n)\propto \lambda_L\cdot n$. 
    One interesting universal phenomenon here is that the total energy and the entanglement are not distributed evenly in the system, instead the driven state will develop an array of peaks of energy-momentum density in the real space. This phenomenon has been reported in Ref.~\cite{fan2019emergent} for special setups, now we verify the universality in a larger class. 
    
    For $\lambda_L=0$, the system is either in the non-heating phase where total energy and the entanglement entropy oscillate or at the phase transition where the total energy grows polynomially and the entanglement grows logarithmically. 
    
    The second diagnostic we introduce is the notion of group walking, which is particularly useful in analyzing and visualizing the details of the spatial structures. This tool is necessary in the cases such as quasi-periodic and random driving when we need to resort to the numerics to identify the universal features.


   \item 
   In Sec.~\ref{Sec: PeriodicDriving}, we study the properties of the periodic driving, providing criteria of the heating phase, non-heating phase, and the phase transition. We discuss the generic features in each phase. This section generalizes the minimal setup in Ref.~\cite{wen2018floquet,fan2019emergent}, and also provides the necessary tools for the discussions in quasi-periodic driving where technically we approach the quasi-periodic limit via a family of periodic driving.


    
    \item In  Sec.~\ref{Sec: QuasiPeridic}, we consider the quasi-periodic driving using two examples: Fibonacci type and Aubry-Andr\'e type.  The Fibonacci driving is the main focus. 
    In the Fibonacci driving, we use the Fibonacci bitstring/word (see Appendix.~\ref{appendix Fib}) 
$
X_{j=1,2,3\ldots} = 10110101 \ldots
$
and two distinct unitary operators $U_A=e^{-iH_AT_A}$, $U_B=e^{-iH_BT_B}$ 
to generate a quasi-periodic driving sequence $U_j= X_j U_A + (1-X_j) U_B $.
The simplest way to generate the Fibonacci
bitstring is through the following substitution rule:
Begin with a single bit $1$, and apply the  substitution rule $1\to 10$, $0\to1$
at each step, then we will generate the following sequence $1\to 10\to 101\to 10110\to 10110101\to...$,
which approaches the Fibonacci bitstring in the infinite step limit.
Denoting the $n$-th Fibonacci number as $F_n$, 
 namely  $F_n=F_{n-1}+F_{n-2}$ with $F_0=F_1=1$, the 
 Fibonacci bitstring/word satisfies: 
$X_{j+F_n}=X_j$,
where $n\geq 2$ and $1\leq j<F_n$.
In the Fibonacci driving, we find the following features:

    \begin{enumerate}

    \item In the heating phase, the distribution of Lyapunov exponents 
    (heating rates) exhibits self-similarity in the parameter space (See Fig.\ref{LyapunovSelfSimilarity}). 
    This also implies  
    there exist heating phases with arbitrarily small positive Lyapunov exponents.
    At these points, the growth of entanglement entropy/energy can be arbitrarily slow. 
    In addition, there are very rich patterns in the time evolution of entanglement/energy in the heating phase. In particular, the locations of the energy-momentum density peaks exhibit even/odd effects at 
    those driving steps that correspond to the Fibonacci numbers.

    \item Exact non-heating fixed points. We find that there always exist \textit{exact} non-heating fixed 
    points in the phase diagram, as long as both of the two driving Hamiltonians are elliptic [See the definition in Eq.\eqref{HamiltonianType}].
    At the non-heating fixed point, 
    the time evolution of the entanglement entropy and the total 
    energy can be analytically obtained
    at the Fibonacci numbers $F_n$. They exhibit an oscillating feature of period $6$, i.e., 
    $S_A(F_n)=S_A(F_{n+6})$ and $E(F_n)=E(F_{n+6})$. 
    At the driving steps that are not Fibonacci numbers, the envelope of the entanglement entropy grows logarithmically in time, and the total energy grows in a power law. 
    
        \item We find an exact mapping between the phase diagram of a Fibonacci driving CFT and the energy spectrum of a Fibonacci quasi-crystal. More precisely, the non-heating phase in the parameter space of a 
    Fibonacci driving CFT corresponds to the energy spectrum of a Fibonacci quasi-crystal. Both form a Cantor set of measure zero. 
    \end{enumerate}
    
   As a complement, we also investigate the quasi-periodic driving with an Aubry-Andr\'e like sequence, where the phase diagram 
   has a nested structure that resembles the famous Hofstadter butterfly found in the Landau level problem \cite{hofstadter1976}. We also exam the measure of the non-heating phase and show it vanishes similar to the Fibonacci driving.

  
      
      
    \item In Sec.~\ref{Sec: Discussion} we conclude with discussions.
    We also provide several appendices with details of calculations and  examples. 
\end{itemize}

\section{Time-dependent driven CFT with SL$_2$ deformations}
\label{Sec: Time-dependent CFT}

In this section, we introduce the general setup and basic properties of a time-dependent driven
CFT with SL$_2$ deformations. The formalism in this section is general, i.e. suitable for arbitrary driving sequence. In the end of this section, we will explain the three classes of driving that we will focus on in this paper and its sequel \cite{RandomCFT}:  the 
periodically, quasi-periodically, and randomly driven CFTs as advertised in the introduction. 
More technical details can be found in Appendix \ref{Sec: TechnicalDetail}
(See also Refs.\cite{WenWu2018quench,wen2018floquet,fan2019emergent}).

We are mainly interested in the time-dependent driven CFT
with discrete time steps. That is, we drive the CFT with $H_1$ for a time
interval $T_1$, then with $H_2$ for a time interval $T_2$, and so on, 
where $H_{1,2,\ldots}$  are $\SL_2$ deformed CFT Hamiltonians that we will explain momentarily. 
Starting from an initial state $|\Psi_0\rangle$, the wavefunction after $n$
steps of driving has the form:
\be\label{Psi_n}
|\Psi_n\rangle= U_n\cdots U_2\cdot U_1|\Psi_0\rangle,\quad \text{with}\quad U_j=e^{-iH_j T_j}.
\ee
The initial state here is not limited to a ground state. 
For instance, it can be chosen as  a highly excited pure state or 
a thermal ensemble at finite temperature, as will be studied in detail in Ref.~\cite{Thermal}.
It is found that the emergent phase diagram of the time-dependent driven CFT is independent of the choices of the initial state, and only depends on the concrete protocols of driving, namely the driving sequences $\{U_j\}$ here.
For simplicity, throughout this work we will choose the initial state $|\Psi_0\rangle$ as the ground state 
of a ``uniform CFT'', i.e. with uniform Hamiltonian density
\begin{equation}
\label{eqn: uniform}
    H_0=\frac{1}{2\pi}\int_0^L \big[T(x)+\bar{T}(x)\big]dx \,.
\end{equation}
where $T(x)$ ($\bar{T}(x)$) are the chiral (anti-chrial) energy-momentum tensor with translation symmetry, $L$ is the total length of the system.

Now let us 
specify the choices of the Hamiltonians $\{H_j\}$ in Eq.~\eqref{Psi_n}, we require them to be generated by
a deformation $\{(f_j,g_j)\}$
 as follows:
\begin{equation}
\label{H_general}
H_j=\frac{1}{2\pi}\int_0^L \Big[f_j(x) \, T(x)+g_j(x) \bar{T}(x)\Big] dx,
\end{equation}
where 
 $f_j(x)$ and $g_j(x)$ are two independent real functions with 
periodic boundary conditions.\footnote{One can of course choose open boundary conditions at the two ends. Then 
$f_j(x)$ and $g_j(x)$ should satisfy the following constrain
$f_j(x)T(x)=g_j(x)\bar{T}(x)$ at $x=0$, $L$, which 
implies that there is no momentum flow across the boundary.
Since we already have $T(x)=\bar{T}(x)$ at $x=0,\, L$ in the uniform case, this indicates 
$f_j(x)=g_j(x)$ at $x=0,\, L$ in the case of open boundary conditions.
}
That is to say, in general we can deform the chiral and anti-chiral modes independently in a system with periodic boundary conditions. 

An alternative way to view the deformation in \eqref{H_general} is to rewrite \eqref{H_general} using energy density $T_{00}(x)=\frac{1}{2\pi}(T(x)+\bar{T}(x))$ and the momentum density $T_{01}(x)=\frac{1}{2\pi}(T(x)-\bar{T}(x))$ as follows
\begin{equation}
\label{H_general2}
H_j=\int_0^L\Big[
\frac{f_j(x)+g_j(x)}{2}T_{00}(x)+\frac{f_j(x)-g_j(x)}{2}T_{01}(x)
\Big]dx.
\end{equation}
Although the formulas and results we obtain in the this work hold for the general case, in many places of this paper we will choose $f_j(x)=g_j(x)$ such that the deformed Hamiltonian takes the following simple form
\begin{equation}
\label{Ht}
H_j=\int_0^L f_j(x) \, T_{00}(x) dx \,.
\end{equation}
The study of the energy spectrum of such Hamiltonian 
can be found in \cite{Ryu1604}. 
In particular, the so-called sine-square deformation (SSD) with
$f_j(x)=\sin^2(\frac{\pi x}{L})$ in Eq.~\eqref{Ht} has received  
extensive study in both condensed matter physics and 
string theory recently\cite{Nishino2011prb,2011freefermionssd,katsura2012sine,ishibashi2015infinite,ishibashi2016dipolar, Okunishi:2016zat, 
WenWu2018quench,Ryu1604,
Tamura:2017vbx, Tada:2017wul,
MacCormack_2019,
tada2019time,caputa2020geometry,liu2020analysis}.
In fact, the initial study of the Floquet CFT in 
Refs.\cite{wen2018floquet,fan2019emergent} is also based on SSD.

\subsection{$\SL_2$ deformation}
\label{Sec:SL2deformation}
A convenient parametrization of the deformed Hamiltonian $H_j$ in Eq.~\eqref{H_general} is to use the Fourier components of $T(x)$ and $\bar{T}(x)$ denoted as $L_n$ and $\bar{L}_n$
\begin{equation}
\label{VirasoroGenerator}
    L_n:=\frac{c}{24}\delta_{n,0}+\frac{L}{2\pi}\int_0^L \frac{dx}{2\pi}\, e^{i\frac{2\pi n}{L}x} \, T(x)  \,, \quad 
    \bar{L}_n:=\frac{c}{24}\delta_{n,0}+\frac{L}{2\pi}\int_0^L \frac{dx}{2\pi}\, e^{-i\frac{2\pi n}{L}x}\, \bar{T}(x)\,,\quad n\in \ZZ\,.
\end{equation}
The operators  $L_n$($\overline{L}_n$) form a Virasoro algebra
\begin{equation}
\label{Virasoro}
[L_m,\, L_n]=(m-n)L_{m+n}+\frac{c}{12}(m^3-m)\delta_{m+n,0},\quad n, m\in \mathbb Z,
\end{equation}
with $c$ being the central charge of the underlying CFT. For example, the uniform Hamiltonian  $H_0$  defined in \eqref{eqn: uniform} can be expressed as 
\begin{eqnarray}
H_0 = \frac{2\pi}{L}(L_0 + \bar{L}_0)-\frac{\pi c}{6L} \,, 
\end{eqnarray}
and what we 
will call a ``$\SL_2$ deformed'' Hamiltonian corresponds to the following enveloping function
\be\label{Envelope_F}
f_j(x)=\sigma_{j}^0+\sigma_{j}^+\cos \frac{2\pi q x}{L}+\sigma_{j}^-\sin\frac{2\pi q x}{L}, 
\quad \sigma_{j}^0,\sigma_{j}^+, \sigma_{j}^-\in\mathbb R, \quad q\in\mathbb Z,
\ee
and similarly for $g_j(x)$. In this case, the corresponding $H_j$ is a linear superposition of $\{L_0,L_{\pm q}\}$ and $\{\bar{L}_0,\bar{L}_{\pm q}\}$, which are the generators of the $\SL^{(q)}(2,\mathbb R)$ subgroup.\footnote{More precisely, $\SL^{(q)}(2,\mathbb R)$.
is isomorphic to an $q$-fold cover of $\SL(2,\mathbb R)$. See, e.g., Ref.~\cite{witten1988}.}
To be concrete, we have $H_j=H_{j,\text{chiral}}+H_{j,\text{anti-chrial}}$,
with
\be\label{Hdeform_Virasoro}
H_{j,\text{chiral}}=\frac{2\pi}{L}\left(\sigma_j^0 L_0+\sigma_j^+ L_{q,+}
+\sigma_j^- L_{q,-}
\right)-\frac{\pi c}{12 L}, 
\ee
where we have defined
$
L_{q,+}:=\frac{1}{2}(L_q+L_{-q})$ and $L_{q,-}:=\frac{1}{2i}(L_q-L_{-q}).
$
Note the SSD deformation mentioned above corresponds to the special case when
\begin{equation}
\label{SSD_Hamiltonian}
    q=1 \,, \quad \sigma^0 = \frac{1}{2} \,, \quad \sigma^+ = -\frac{1}{2} \,, \quad \sigma^- =0 \qquad \text{(SSD)} \,.
\end{equation}
Therefore, the $\SL_2$ deformation can be thought as a generalization of the SSD deformation, while retaining the analytic tractability.\footnote{Solving the non-equilibrium dynamics with the most general deformations that correspond to the infinite dimensional Virasoro algebra is more challenging and will not be discussed in this paper. }

In general, by defining the quadratic Casimir 
\be
c^{(2)}:=-(\sigma^0)^2+(\sigma^+)^2+(\sigma^-)^2,
\ee
the (chiral and anti-chiral) SL$_2$ deformed Hamiltonians can be classified into three types as follows\cite{ishibashi2015infinite,ishibashi2016dipolar,
HanEffectiveHamiltonian,HanWenClassify}
\be\label{HamiltonianType}
\left\{
\begin{split}
&c^{(2)}<0:\quad \text{Elliptic Hamiltonian},\\
&c^{(2)}=0:\quad \text{Parabolic Hamiltonian},\\
&c^{(2)}>0:\quad \text{Hyperbolic Hamiltonian}.\\
\end{split}
\right.
\ee
Different types of Hamiltonians will determine the operator evolution
in different ways (See Appendix.\ref{Sec:OperatorEvoAppendix}).

With the SL$_2$ deformation, many physical properties of the driven system including the phase diagram, the time dependence of the entanglement entropy \cite{WenWu2018quench,wen2018floquet} and the energy-momentum density\cite{fan2019emergent,Zurich1} have been obtained in a \textit{periodically} driven CFT system. 
Here we generalize the driving to an arbitrary sequence $\{U_j\}$ as shown in 
Eq.~\eqref{Psi_n}. In the following, we will derive general formulas based on the $\SL_2$ deformation sequence, and later apply to the 
periodically, quasi-periodically, and randomly driven CFTs. 


\subsection{Operator evolution}
\label{Sec:OperatorEvolution}

For $\SL_2$ driven quantum states, it is convenient to compute observables via Heisenberg picture, namely the correlation functions are given by  $\langle  \Psi_0 | \mathcal{O}_1(x_1,t_1)\cdots \mathcal{O}_n(x_n,t_n)|\Psi_0\rangle$, 
where the Heisenberg operators $\calO(x,t)$ are defined by discrete time evolution 
\begin{equation}
    \calO(x,t) :=U^\dagger(t) O(x) U(t)\,, \quad \text{with} \quad U(t)=U_m\cdots U_2\cdot U_1 \,.
\end{equation}
For each step, 
 $U_j=e^{-iH_j T_j}$ is generated by the $\SL_2$ deformed Hamiltonian and $t=\sum_{j=1}^m T_j$ is only defined for a discrete set of times in our setting. 
 
 The virtue of the driving Hamiltonian in \eqref{H_general} is that the operator evolution can be represented by a conformal mapping $(z,\bar{z})\rightarrow (z',\bar{z}')$, under which the primary operator $O(z,\bar{z})$ transform as 
 \begin{equation}
\label{OperatorTransf}
U_j^{\dag}\, \mathcal O(z,\bar{z})\, U_j=\left(\frac{\partial z'}{\partial z}\right)^{h}
\left(\frac{\partial \bar{z}'}{\partial \bar{z}}\right)^{\bar{h}} \mathcal{O}\big(z',\bar{z}'\big),
\end{equation}
where $h$ ($\bar{h}$) are conformal dimensions of $\mathcal{O}$.
 Then the full unitary $U(t)$ is a composition of a sequence of conformal mappings. For the special type of enveloping function \eqref{Envelope_F}, a convenient coordinate is given as follows (see Fig.~\ref{ConformalMapQ} for an illustration.)
 \begin{eqnarray}
 z = \exp \frac{2\pi q w}{L }\,, \quad w = \tau+ ix\,,
 \end{eqnarray}
 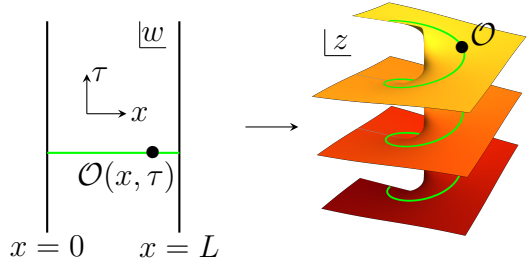
\begin{figure}[t]
\centering
\begin{tikzpicture}
\draw  [xshift=-40pt][thick](20pt,-40pt)--(20pt,40pt);
\draw  [xshift=-40pt][thick](70pt,-40pt)--(70pt,40pt);

\draw [xshift=-40pt][>=stealth,->] (35pt, 5pt)--(35pt,20pt);
\draw [xshift=-40pt][>=stealth,->] (35pt, 5pt)--(50pt,5pt);

\draw [xshift=-40pt](55pt, 30pt)--(55pt,40pt);
\draw [xshift=-40pt](55pt, 30pt)--(65pt,30pt);

\node at (55-40pt,5pt){$x$};
\node at (40-40pt,20pt){$\tau$};
\node at (60-40pt,35pt){$w$};

\draw [green,thick](-20pt, -9.8pt)--(30pt,-9.8pt);
\node at (60-40pt,-10pt){$\bullet$};
\node at (50-40pt,-19pt){$\mathcal{O}(x,\tau)$};

\node at (20-40pt,-45pt){$x=0$};
\node at (70-40pt,-45pt){$x=L$};
\draw [>=stealth,->] (55pt, 0pt)--(75pt,0pt);
\node[inner sep=0pt] (russell) at (120pt,0pt)
    {\includegraphics[width=.18\textwidth]{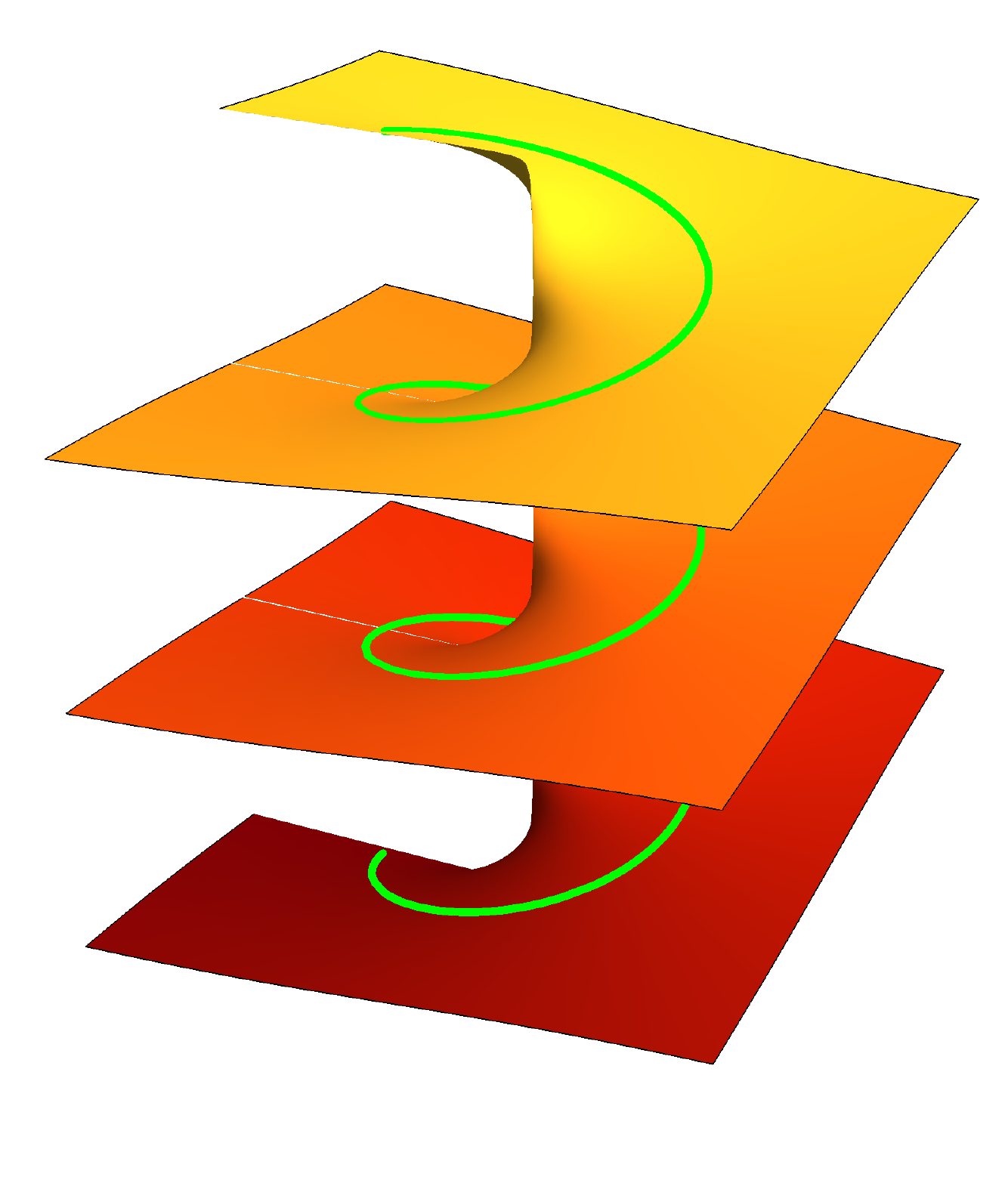}};
    \draw (85pt, 27pt)--(85pt,37pt);
\draw (85pt, 27pt)--(95pt,27pt);
\node at (91pt,32pt){$z$};	
\node at (145pt,35pt){$\mathcal{O}$};
\node at (137pt,30pt){$\bullet$};
\end{tikzpicture}
\caption{Conformal map $z=e^{\frac{2\pi q w}{L}}$ from the $w$-cylinder (where
$w=\tau+ix$, and $x=L$ and 
$x=0$ are identified)
to the $q$-sheet Riemann surface $z$.} 
\label{ConformalMapQ}
\end{figure}
 under which the evolution generated by $U_j$ can be expressed as a 
 M\"obius transformation:
\footnote{More details can be found in Appendix.~\ref{Sec: TechnicalDetail}.
See also Refs.\cite{WenWu2018quench,wen2018floquet,fan2019emergent}.}
\be\label{Mobius_Transf}
z'=\frac{a_j z+b_j}{c_j z+d_j}, \qquad
\text{where} 
\underbrace{
\begin{pmatrix}
a_j &b_j\\
c_j &d_j
\end{pmatrix}
}_{\text{denoted as } M_j} \in \SL(2,\CC) \,.
\ee
The explicit form of $M_j$ is determined by the Hamiltonian $H_j$ and the time interval $T_j$. 
An important observation is that the driving protocol \eqref{Envelope_F} we use in fact generates $M_j$ matrix in the following specific form
\begin{equation}
M_j=
\begin{pmatrix}
a_j & b_j\\
b_j^* &a_j^*
\end{pmatrix}
 \quad \text{where }\,\, a_j, \,b_j\in\mathbb C,\quad |a_j|^2-|b_j|^2=1\,,
 \end{equation}
which is a $\SU(1,1)$ matrix. Note $\text{SU}(1,1)\cong \text{SL}(2,\mathbb R)$, both are subgroups of $\text{SL}(2,\mathbb C)$. The isomorphism is expected since we start from a $\SL(2,\RR)$ action on the states. 

Thus, the net effect of the full evolution 
$U=U_n\cdots U_2 \cdot U_1$ is given by the product of 
$n$ $\SU(1,1)$ matrices
\be\label{Pi_n}
\small
\Pi_n=M_1\cdots M_{n-1} M_n=
\left(
\begin{array}{cccc}
a_1 &b_1\\
c_1 &d_1
\end{array}
\right)
\cdots
\left(
\begin{array}{cccc}
a_{n-1} &b_{n-1}\\
c_{n-1} &d_{n-1}
\end{array}
\right)
\left(
\begin{array}{cccc}
a_n &b_n\\
c_n &d_n
\end{array}
\right).
\ee
Note that the later matrix acts on the right 
since we are using the Heisenberg picture of evolution. 
To summarize, the operator evolution under a sequence of driving $\{U_j\}$
is given by the following formula
\be\label{OP_evolution}
U_1^{\dag}\cdot U_2^{\dag}\cdots
U_n^{\dag}\, \mathcal O(z,\bar{z})\, U_n\cdots U_2\cdot U_1=\left(\frac{\partial z_n}{\partial z}\right)^{h}
\left(\frac{\partial \bar{z}_n}{\partial \bar{z}}\right)^{\bar{h}} \mathcal{O}\big(z_n,\bar z_n\big),
\ee
where $z_n$ is related to $z$ through the M\"obius transformation in \eqref{Mobius_Transf} 
with the matrix $\Pi_n \in \SU(1,1)$. 
 \be\label{Accumlate_MobiusMatrix2}
\Pi_n=
\begin{pmatrix}
\alpha_n &\beta_n\\
\beta_n^* &\alpha_n^*
\end{pmatrix}
 \quad \text{where }\,\, \alpha_n, \,\beta_n\in\mathbb C,\quad |\alpha_n|^2-|\beta_n|^2=1\,,
\ee

\subsection{Time evolution of entanglement and energy-momentum density}


To characterize the possible emergent phases, we study the time evolution 
of the entanglement entropy and the energy-momentum density of the system.
In terms of correlation functions, the former is determined by the two point function of twist operator, while the latter is determined by the one point function of energy-momentum tensor. One can also consider two point functions of general operators, which are discussed in Appendix.\ref{app:two-point functions}.


For example, the $m$-th Renyi entropy of the subsystem $A=[x_1,\, x_2]$ can be obtained by the following formula
\be
S^{(m)}_A(n)=\frac{1}{1-m}\log\,\langle \Psi_n| \mathcal{T}_m(x_1) \bar{\mathcal{T}}_m(x_2) |\Psi_n\rangle,
\ee
where $|\Psi_n\rangle$ is the time-dependent wavefunction in Eq.~\eqref{Psi_n}, and 
$\mathcal{T}_m$ ($\bar{\mathcal{T}}_m$) are twist (anti-twist) operators that are primary, with conformal dimensions $h=\bar{h}=\frac{c}{24}(m-\frac{1}{m})$.
For initial state $|\Psi_0\rangle$ being the ground state of $H_0$ with  periodic boundary conditions, 
the time evolution of the entanglement entropy
for the subsystem $A=[(k-1/2)l, (k+1/2)l]$ where $k\in\mathbb Z$ and $l=L/q$ is given as 
\footnote{For a general choice of single-interval subsystem $A=[x_1, \, x_2]$, the exact expression of the
entanglement entropy under a time-dependent driving will be quite involved. See, e.g.,
the appendix of Ref.~\cite{wen2018floquet}.
However, if the CFT is in a heating phase, one can obtain an approximated expression
of the entanglement entropy of $A=[x_1, \, x_2]$ by keeping the leading order \cite{fan2019emergent}. 
}
\be\label{EE_general}
S_A(n)-S_A(0)=\frac{c}{3}\Big(
\log\big|\alpha_n-\beta_n\big|+\log\big|\alpha'_n-\beta'_n\big|
\Big).
\ee
Here $\alpha_n$ and $\beta_n$ are the matrix elements appearing in the operator 
evolution in Eq.~\eqref{Accumlate_MobiusMatrix2}.
$\alpha_n'$ and $\beta_n'$ are the corresponding matrix elements for the anti-chiral 
part.

One can also study the time evolution of energy-momentum tensors 
based on the operator evolution as 
discussed in the previous subsection. However, since $T(x)$ is not 
a primary field, the operator evolution in Eq.~\eqref{OperatorTransf} should be 
modified as 
\be\label{OP_evolution_T}
U_j^{\dag}\,T(z)\, U_j=\left(\frac{\partial z'}{\partial z}\right)^{2}
T\big(z'\big)+\frac{c}{12}\text{Sch}(z',z),
\ee
where the last term represents the Schwarzian derivative.
The expectation value of the chiral energy-momentum tensor density is
\cite{fan2019emergent}
\footnote{Hereafter, for convenience of writing, 
we write $T(x,t=\sum_{j=1}^n T_j)$ as
$T(x,n)$.}
\be\label{EnergyMomentumDensity}
\frac{1}{2\pi}\langle T(x,n)\rangle=-\frac{q^2 \pi c}{12 L^2}+
\frac{\pi c}{12 L^2}\cdot(q^2-1)\cdot \frac{1}{|\alpha_n e^{\frac{2\pi i x}{l}}+\beta_n|^4},
\quad \text{where } l=L/q
\ee
For the anti-chiral component $\frac{1}{2\pi}\langle \bar{T}(x,n)\rangle$, the expression is 
the same as above by replacing $\alpha_n$($\beta_n)\to \alpha_n'$($\beta_n'$) and 
$e^{\frac{2\pi i x}{l}}\to e^{-\frac{2\pi i x}{l}}$.
The total energy and momentum of the system are $E(n)=\frac{1}{2\pi}\int_0^L\langle T(x,n)+ \bar{T}(x,n)\rangle dx$, and $P(n)=\frac{1}{2\pi}\int_0^L\langle T(x,n)- \bar{T}(x,n)\rangle dx$,
with the expressions:
\be\label{TotalEnergyMomentum}
\left\{
\begin{split}
&E(n)=-\frac{q^2 \pi c}{6 L}+
\frac{\pi c}{12 L}(q^2-1)\cdot(|\alpha_n|^2+|\beta_n|^2+|\alpha_n'|^2+|\beta_n'|^2)\\
&P(n)=\frac{\pi c}{12 L}(q^2-1)\cdot(|\alpha_n|^2+|\beta_n|^2-|\alpha_n'|^2-|\beta_n'|^2).
\end{split}
\right.
\ee
We would like to make a few remarks here:
\begin{enumerate}
    \item For the periodic boundary conditions we considered here, the time evolution with $q=1$ deformations are trivial as $H_j$ also annihilates the ground state of $H_0$.
    \footnote{This is can be seen by considering $q=1$ in Eq.\eqref{TotalEnergyMomentum}, but may be not obvious by looking at the
    expression of $S_A(n)$ in Eq.\eqref{EE_general}. For $q=1$, the choice 
    of the subsystem $A$ in \eqref{EE_general} fails because 
    $A=[-L/2,\, L/2]$ corresponds to the total system. In our calculation
    of $S_A(n)$ in Eq.\eqref{EE_general}, we have assumed explicitly that 
    the two entanglement cuts do not coincide, or equivalently $A$ is not
    the total system.}
In contrast, if we consider an open boundary condition, the ground state of $H_0$ will no longer be the eigenstate of the deformed Hamiltonian $H_i$. Then one can have a non-trivial time evolution, as studied in Refs.~\cite{wen2018floquet,fan2019emergent}.
    
\item If there is no driving, i.e., $\alpha_n=\alpha'_n=1$ and $\beta_n=\beta'_n=0$,
one can find $E(n)=-\frac{\pi c}{6L}$, which is the Casimir energy with periodic 
boundary conditions.
\item 
If we only deform the Hamiltonian density in Eq.~\eqref{H_general2}, i.e. let $f_j(x)=g_j(x)$, one can find that $\alpha_n=\alpha_n'$ and 
$\beta_n=\beta_n'$\cite{wen2018floquet}, and therefore $P(n)=0$, i.e., 
the total momentum stays zero.
In this case, both the left movers and right movers are excited, but they carry 
opposite momentum and the total momentum are canceled to be zero.

\item 
In later sections, we will compare the CFT calculations with the lattice calculations.
An efficient way to perform numerical calculations on the lattice is to consider $q=1$
with an open boundary condition,
since for larger $q$, the length of the wavelength
of deformation
$l:=L/q$ is effectively suppressed for a fixed $L$.
In this case, by deforming the Hamiltonian in Eq.~\eqref{Ht}, where only
the Hamiltonian density is deformed,
one can find the time evolution of the entanglement entropy as follows:\cite{wen2018floquet}
\be\label{HalfEE_general}
S_A(n)-S_A(0)=\frac{c}{3}\log \big|\alpha_n-\beta_n\big|, \quad \text{where } A=[0, L/2].
\ee
The expectation value of the chiral energy-momentum density is: \cite{fan2019emergent}
\be\label{EnergyMomentumQ1}
\frac{1}{2\pi}\langle T(x,n)\rangle=-\frac{\pi c}{12L^2}+\frac{\pi c}{16L^2}\cdot \frac{1}{|\alpha_n e^{\frac{2\pi x}{L}}+\beta_n|^4}.
\ee
The anti-chiral part has the same expression as above with the replacing $e^{\frac{2\pi ix}{L}}\to
e^{\frac{-2\pi ix}{L}}$.
Then one can find $\frac{1}{2\pi}\int_0^L\langle T(x,n)\rangle
=\frac{1}{2\pi}\int_0^L\langle \bar{T}(x,n)\rangle=-\frac{\pi c}{12L}+\frac{\pi c}{16L}\cdot
(|\alpha_n|^2+|\beta_n|^2)$, based on which one can obtain the total energy as
\be\label{EnergyTotal}
E(n)=\frac{\pi c}{8 L}(|\alpha_n|^2+|\beta_n|^2)-\frac{\pi c}{6L}.
\ee
One can find the similarity and difference by comparing with the case with 
periodic boundary conditions. For example, 
in the ground state with $\alpha_n=1,\,\beta_n=0$ in \eqref{EnergyTotal}, 
one can obtain the Casimir energy
 $E=-\frac{\pi c}{24L}$, which is different from that in periodic boundary conditions.
 Nevertheless, the dependence of the entanglement entropy/energy-momentum denstiy
 on the matrix elements $\alpha_n$($\beta_n$) in the $\Pi_n$ in Eq.~\eqref{Accumlate_MobiusMatrix2}
 are similar.
\end{enumerate}

There is rich information contained in the formula discussed above. 
As will be seen later,  if the CFT is in a heating phase, there will be energy-momentum density peaks
emerging in the real space. The locations of these peaks are determined by $\beta_n/\alpha_n$.
It turns out that both the quantities $S_A(n)$ and $E(n)$
can serve as  `order parameters' to distinguish different emergent phases in the 
time-dependent driven CFTs.
For example, for the periodically driven CFT as studied in Ref.~\cite{wen2018floquet},
it is found there are two different phases with a heating phase and a non-heating phase, 
where the time evolution of entanglement entropy exhibits 
qualitatively different features as shown in Fig.~\ref{FloquetCFT_EE}.
Also, it is found n Ref.~\cite{fan2019emergent} that the total energy grows exponentially fast 
as a function of driving cycles $n$ in 
the heating phase and simply oscillates in the non-heating phase.

As a short summary, once we know the operator evolution in Eq.~\eqref{OP_evolution} 
or equivalently the matrix form in Eq.~\eqref{Accumlate_MobiusMatrix2},
one can study the entanglement/energy-momentum evolution based on 
Eqs.\eqref{EE_general} and \eqref{TotalEnergyMomentum} or Eqs.\eqref{HalfEE_general} and \eqref{EnergyTotal}.

\subsection{Periodic, quasi-periodic, and random driving CFTs}
\label{Sec: 3waysDriving}

In general, the sequence of unitary operators $\{U_j\}$ in Eq.~\eqref{Psi_n} can be chosen
in an arbitrary form. In this work, we are interested in three classes:
periodic, quasi-periodic, and random drivings. 

\begin{enumerate}
\item
\textit{Periodical driving}:
the sequence of unitary operators $\{U_j\}$ in \eqref{Psi_n} are chosen with a `period' $p$
($p\in \mathbb Z_+$) such that $U_j=U_{j+p}$, $\forall j\in\mathbb Z$.
Then the time evolution of wavefunction in \eqref{Psi_n} can be written as
\be
|\Psi_{np}\rangle= \big(
\underbrace{U_p\cdots U_2\cdot U_1}_{\text{one driving period}}\big)^n|\Psi_0\rangle,\quad \text{with}\quad U_j=e^{-iH_j T_j}.
\ee
To obtain the physical properties of the system under periodic driving, we only need to analyze the corresponding transformation matrix $M_1M_2\cdots M_p \in \SU(1,1)$ within a period.  


\item 
\textit{Quasi-periodic driving}: $\{U_j\}$ form a quasi-periodic sequence. 
Quasiperiodicity is the property of a system that displays irregular periodicity, where the sequence
exhibits recurrence with a component of unpredictability 
(For example see the review \cite{damanik_2017} for a 
more rigorous mathematical definition of quasi-periodic sequence).
In this paper, we will focus on the following two protocols of quasi-periodical driving: 
\begin{enumerate}
    \item Fibonacci type. This is the type of quasi-periodic driving we will study in detail in Sec.~\ref{Sec: Fibonacci driving CFT}. 
We use the Fibonacci bitstring/word (see Appendix.~\ref{appendix Fib}) 
\begin{equation}
X_{j=1,2,3\ldots} = 10110101 \ldots
\end{equation}
and two distinct unitaries $U_A$, $U_B$ 
to generate a quasi-periodic driving sequence $U_j= X_j U_A + (1-X_j) U_B $, i.e. we apply $U_A$ ($U_B$) if the bit is $1$ ($0$).  

\item Aubry-Andr\'e type. In this case, 
we generate the quasi-periodic driving sequence as follows
\begin{equation*}
    \begin{tikzpicture}[baseline={(current bounding box.center)}]
	\node at (-5pt, 7pt){$H(t)$:};
	\draw [thick](20pt,0pt)--(40pt,0pt) --(40pt,20pt) -- (60pt,20pt);
	
	\draw [thick](60pt,20pt)--(60pt,0pt) --(100pt,0pt) --(100pt,20pt) -- (120pt,20pt);
	
	\draw [thick](120pt,20pt)--(120pt,0pt)--(180pt,0pt) -- (180pt,20pt) -- (200pt,20pt);
	
	\node at (10+40pt,28pt){$H_1$};
	\node at (30pt, 8pt){$H_0$};
	
	\draw[>=stealth,<->] (20pt,-5pt) --node[below]{$\omega L$} (40pt,-5pt);
	\draw[>=stealth,<->] (60pt,-5pt) --node[below]{$2\omega L$} (100pt,-5pt);
	\draw[>=stealth,<->] (120pt,-5pt) --node[below]{$3\omega L$} (180pt,-5pt);
	
	\draw [>=stealth,->] (210pt, -10pt)--(230pt,-10pt);
	\node at (235pt, -10pt){$t$};
	\end{tikzpicture}
\end{equation*}
That is to say, we consider two Hamiltonians $H_0$ and $H_1$ and fix the driving period $T_1$ for $H_1$ while let the driving period of $H_0$ increase with driving cycle $T_0=n\omega L$ where $\omega$ is an irrational number and $L$ is the total length of the system. Note in terms of the unitary $U=\exp (-i H_0 T_0)$, its action on the operator only depends on $T_0$ mod $L$.

\end{enumerate}


\item
\textit{Random driving}: $\{U_j\}$ form a random sequence. 
More concretely, each $U_j$ is drawn independently from the ensemble $\{ (u_k,p_k)\}_{k=1\ldots m}$, where $u_k=e^{-iH_kT_k}$ is the unitary matrix and $p_k$ is the corresponding probability, with the normalization $\sum_k p_k = 1$.  
\end{enumerate}

In brief, for all the three kinds of time-dependent drivings, our goal is to describe the behavior of the physical properties of the CFT
in the long time driving limit $n\to \infty$, where $n$ is the number of driving cycles.

As a remark, one can find that the types of driving sequence are similar to those in the potentials in crystals, quasi-crystals, and disordered systems.
One can find interesting relations between different phases of time-dependent driven CFTs
and different types of wavefunctions in a lattice, as briefly discussed in the introduction. 
Furthermore, both types of quasi-periodicities we mentioned above have been discussed in the quasi-crystal literature, e.g. see 
Refs.\cite{KKT1983, OPRSS1983,sutHo1989singular}.

\section{Diagnostics}
\label{sec: diag}

The previous section explains how the physical properties of an $\SL_2$ driven CFT state can be extracted from the conformal mapping generated by the driving sequence. The mapping is further encoded in an $\SU(1,1)$ matrix, denoted as $\Pi_n$ in \eqref{Accumlate_MobiusMatrix2}, which itself is a product of $n$ $\SU(1,1)$ matrices. 

Mathematically, the long time asymptotics of the driven state now can be understood by the $n$-dependence of $\Pi_n$. In this section, we will introduce two useful diagnostics to characterize such dependence: (1) Lyapunov exponent; (2) Group walking. The former is a simple scalar quantifying the growth of $\Pi_n$, while the latter is more refined and uses two points on the unit disk to track the trajectory of $\Pi_n$. Although not independent, both of them will be useful and used in the later sections.

\subsection{Lyapunov exponent and heating phase}
\label{Sec:LyapunovHeatingPhase}

For all the three classes of drivings we introduced in the previous subsection, the problem is reduced 
to the study of the product $\Pi_n$ (defined in \eqref{Pi_n}) of a sequence of SU$(1,1)$ matrices that encode the conformal mappings.   
One useful and simple characterization for
the growth rate of this matrix product 
is the so-called Lyapunov exponent (for a review of the subject, see e.g. \cite{viana2014lectures}). 
Generally,  we can consider a product of $n$ matrices $\Pi_n=M_1\cdot M_2\cdots M_n$, where 
$M_j\in \text{SL}(d,\mathbb{R})$. 
Then the (upper) Lyapunov exponent is defined as 
\be\label{Def:Lyapunov}
\lambda_L:=\lim_{n\to \infty} \frac{1}{n} \log \lVert M_1\cdot M_2\cdots M_n\rVert,
\ee
where $\lVert \cdot \Vert$ is a matrix norm. We would like to make a few comments
about the definition here: 
\begin{enumerate}
\item Here, the specific choice of norm $\lVert \cdots \rVert$ is not 
essential. To be explicit, we will choose the Frobenius norm in this paper, i.e. 
\begin{equation}
    \lVert M \rVert_F:=\big(\sum_{j,k} |M_{jk}|^2\big)^\frac{1}{2} \,.
\end{equation}
\item The definition also applies to $\text{SL}(d,\mathbb C)$, as one can always embed 
$\text{SL}(d,\mathbb C)$ in $\text{SL}(2d,\mathbb R)$.
\item 
In general, one can define $d$ Lyapunov exponents for $\text{SL}(d,\mathbb R)$. For example, for $\text{SL}(2,\mathbb R)$, one can define two extremal Lyapunov exponents 
\begin{align}
    &\lambda_+:=\lim_{n\to \infty}\frac{1}{n} \log \lVert M_1\cdot M_2\cdots M_n\rVert, \\
    & \lambda_-:=\lim_{n\to \infty}\frac{1}{n} \log \lVert ( M_1\cdot M_2\cdots M_n)^{-1} \rVert^{-1},
\end{align}
with the property $\lambda_+\geq 0 \geq \lambda_-$, since $\lVert B\rVert \geq 1 \geq \lVert B^{-1}\rVert^{-1}$ for $B\in \SL(2,\RR)$. 
\end{enumerate}
Applying to the $\SU(1,1)$ matrix $\Pi_n$: 
\begin{equation}
    \Pi_n=
\begin{pmatrix}
\alpha_n &\beta_n\\
\beta_n^* &\alpha_n^*
\end{pmatrix}
 \quad \text{where }\,\, \alpha_n, \,\beta_n\in\mathbb C,\quad |\alpha_n|^2-|\beta_n|^2=1\,,
\end{equation}
 a positive Lyapunov exponent $\lambda_L>0$ implies that the matrix elements have the following asymptotics
 \begin{equation}
      |\alpha_n| \sim |\beta_n| \sim \frac{1}{2} e^{\lambda_L n} \qquad \text{at} \quad n \rightarrow \infty.
 \end{equation}
Following Eqs.~\eqref{EE_general} and \eqref{TotalEnergyMomentum}, we find the asymptotics in long time limit 
\be\label{EenergyEE_lyapunov}
S_A(n)-S_A(0)\sim \frac{c}{3}\cdot \lambda_L\cdot n, \quad \quad E(n)\sim
\frac{\pi c}{24L}\cdot(q^2-1)\cdot e^{2\lambda_L\cdot n},
\ee
where we have neglected the contribution from the anti-chiral mode for the moment.\footnote{More precisely, the formula on
$S_A(n)-S_A(0)$ holds when the chiral or anti-chiral 
energy-momentum density peaks are in the interior of $A$ (See Sec.~\ref{Sec: PeriodicDrivingEnergyEE}). 
When the entanglement cuts lie on the centers of the energy-momentum density peaks, 
$S_A(n)$ could even decrease in time (See Appendix.~\ref{Sec: LinearDecreaseEE}).
} 
From this perspective, we may interpret the Lyapunov exponent $\lambda_L$ as the heating rate in the heating phase.
If $\lambda_L>0$, then the time-dependent driven CFT must be in a heating phase, with 
the total energy exponentially growing in time. We will also explain in detail momentarily that  since the norm of the ratio $\beta_n/\alpha_n$  approaches $1$ in the long time limit when $\lambda_L>0$, an array of energy-momentum peaks will emerge in real space whose exact locations will be determined by the phase of the ratio $\beta_n/\alpha_n$. 

On the other hand, if $\lambda_L=0$, the system is either in a non-heating phase or at the phase transition. 
We emphasize here that the vanishing of Lyapunov exponential allows a sub-exponential growth of the matrix norm $|| \Pi_n||$ as a function of $n$, e.g. this could happen at the phase transition/boundary. 

To  summarize, using the Lyapunov exponent $\lambda_L$, we can classify the phases as follows:
\be
\left\{
\begin{split}
&\lambda_L>0: \text{heating phase (with exponentially growing energy)},\\
&\lambda_L=0: \text{non-heating phase or phase transition}.
\end{split}
\right.
\nonumber
\ee%
To further identify the detailed properties of entanglement/energy 
evolution in the non-heating phase and at the phase transition, 
one needs to study the finer structure of the matrix $\Pi_n$, 
which we will pursue in the next subsection.

The Lyapunov exponent works for general matrices. When specialized to $\SU(1,1)$ or $\SL(2,\mathbb C)$, another commonly used classifier is the trace of the matrix. Namely, $|\Tr M|>2$, $|\Tr M|=2$ and $|\Tr M|<2$ correspond to the hyperbolic, parabolic and elliptic types of matrix, respectively. This criterion was used to identity different phases for the Floquet driving CFT studied in previous works\cite{wen2018floquet,fan2019emergent}.

For periodic driving, it follows from the definition of matrix norm that this trace classifier is equivalent to the Lyapunov exponent. We have $|\Tr M|>2$ 
if and only if $\lambda_L>0$, $|\Tr M|\le 2$ if and only if $\lambda_L=0$. 
One can also extend it to the quasi-periodic driving as follows. As will be detailed discussed later, any quasi-periodic driving corresponding to an irrational number $w$ can be considered as the limit of a sequence of periodic driving, which is generated by the continued fractions of $\omega$. For each element in the sequence, we can apply the trace classifier to obtain a sequence of phase diagram, with its limit being the true phase diagram for the quasi-periodic driving system.

For the random driving, the Lyapunov exponent will become a more appropriate definition, which we use exclusively in the corresponding discussion.

\subsection{Group walking: fine structures of the time-dependent driving}
\label{Sec: GroupWalking}

The Lyapunov exponent defined in the previous subsection is a single number. 
To view the `internal structure' in the matrix product $\Pi_n$ in Eq.~\eqref{Accumlate_MobiusMatrix2}, 
it is helpful to study how the matrix elements evolve in time, which  determines 
the time evolution of the entanglement entropy and the energy-momentum density.


A convenient parametrization of the $\SU(1,1)$ matrices such as $M_j$ and $\Pi_n$ in \eqref{Pi_n} is given as follows
\be\label{M_alpha_z}
\Pi(\rho, \zeta)=\frac{1}{N_{\rho}}
\begin{pmatrix}
\sqrt{\zeta} &-\rho^*\frac{1}{\sqrt{\zeta}}\\
-\rho \sqrt{\zeta} &\frac{1}{\sqrt{\zeta}}
\end{pmatrix} \quad \text{where} \quad \rho\in \mathbb D, \,\, \zeta\in \partial \mathbb D,
\ee
and $N_{\rho}=\sqrt{1-|\rho|^2}$ is the normalization factor.\footnote{More precisely, the above parametrization $(\rho\in \mathbb{D},\zeta\in \partial \mathbb{D})$ of matrix $\Pi$ only covers the $\SU(1,1)/\ZZ_2$, to obtain the full $\SU(1,1)$ group, one need to let $\zeta$ live on the double cover of the boundary circle. However, our physical quantities are obtained from the M\"obius transformation rather than the $\SU(1,1)$ matrix directly, the former is indeed isomorphic to the $\ZZ_2$ quotient of the latter, namely $\SU(1,1)/\ZZ_2$ and agrees with our parametrization.}
The unit disk $\mathbb D:=\{z\in \mathbb C, |z|<1\}$, the boundary(or edge) of the disk $\partial \mathbb D:=\{z\in\mathbb C, |z|=1\}$,
and the complex numbers $\rho$ and $\zeta$ are depicted as follows:
\be
\begin{tikzpicture}[x=0.75pt,y=0.75pt,yscale=-1,xscale=1]

\draw  [fill={rgb, 255:red, 126; green, 211; blue, 33 }  ,fill opacity=0.21 ][line width=1.5]  (120,130) .. controls (120,99.62) and (144.62,75) .. (175,75) .. controls (205.38,75) and (230,99.62) .. (230,130) .. controls (230,160.38) and (205.38,185) .. (175,185) .. controls (144.62,185) and (120,160.38) .. (120,130) -- cycle ;
\node at (220pt, 89pt){\color{black}$\rho\in\mathbb D$};
\node at (220pt, 105pt){\color{black}$\zeta\in \partial \mathbb D$};
\node at (168pt, 79pt){\color{black}$\bullet$};
\node at (174pt, 72pt){\color{black}$\zeta$};
\node at (94.7+20pt, 79+20pt){\color{black}$\bullet$};
\node at (89+20pt, 75+20pt){\color{black}$\rho$};
\end{tikzpicture}
\ee
Thus, the evolution of matrix $\Pi_n$ as a  function of step $n$ can be captured by the evolution of a pair of points $(\rho_n, \zeta_n)$ on the unit disk. 
We will call this process `group walking' for brevity.
An equivalent but more convenient parameterization of the trajectory $(\rho_n,\zeta_n)$ is to use $(\rho_n, \rho_n \zeta_n)$.
For example, the total energy \eqref{TotalEnergyMomentum}, locations of the energy-momentum density peaks \eqref{EnergyMomentumDensity} and entanglement entropy \eqref{EE_general} are expressible using $(\rho_n, \rho_n \zeta_n)$: 
\begin{enumerate}

\item The energy formula in Eq.~\eqref{TotalEnergyMomentum} only depends on $\rho_n$ 
\be\label{Etotal_rho}
E(n)=
-\frac{q^2 \pi c}{12 L}+
\frac{\pi c}{12 L}(q^2-1)\cdot \frac{1+|\rho_n|^2}{1-|\rho_n|^2}+\text{anti-chiral part}\,,
\ee
and increase monotonically w.r.t. $|\rho_n|$. 

In the heating phase, the exponential growth of $E(n)$ as a function of $n$ is tied to the phenomenon that $|\rho_n|$ approaches exponentially close to the boundary $\partial \mathbb D$. 
On the other hand, if the total energy simply oscillates in $n$, e.g., 
in the non-heating phase of a periodically driven CFT, then $|\rho_n|$ should follow the same 
oscillation pattern. In between, as a we approach the phase transition, the orbit of $\rho$
should be closer and closer to the boundary. As a summary, the behaviors of $\rho$ described above can be visualized by the following cartoon:
\begin{equation}
\label{Disk_Phase}
\begin{tikzpicture}[x=0.75pt,y=0.75pt,yscale=-1,xscale=1]

\draw  [fill={rgb, 255:red, 126; green, 211; blue, 33 }  ,fill opacity=0.21 ][line width=1.5]  (120,130) .. controls (120,99.62) and (144.62,75) .. (175,75) .. controls (205.38,75) and (230,99.62) .. (230,130) .. controls (230,160.38) and (205.38,185) .. (175,185) .. controls (144.62,185) and (120,160.38) .. (120,130) -- cycle ;


\node at (172.5pt, 99pt){\color{black}$\bullet$};
\node at (185pt, 99pt){\color{black}$\rho_b$};

\draw [thick,red][>=stealth,->](114.7pt, 114pt)..controls (128pt, 126pt) and (150pt,123pt)..(160pt,114pt)..controls(165pt,110pt) and 
(168pt,105pt)..(172pt,99pt);

\node at (94.7+20pt, 79+35pt){\color{black}$\bullet$};
\node at (91+20pt, 70+35pt){\color{black}$\rho$};

\node at (91+40pt, 70+85pt){\color{black}Heating phase};
\node at (91+40pt, 70+98pt){\color{white}$Heating phase$};
\end{tikzpicture}
\quad
\begin{tikzpicture}[x=0.75pt,y=0.75pt,yscale=-1,xscale=1]

\draw  [fill={rgb, 255:red, 126; green, 211; blue, 33 }  ,fill opacity=0.21 ][line width=1.5]  (120,130) .. controls (120,99.62) and (144.62,75) .. (175,75) .. controls (205.38,75) and (230,99.62) .. (230,130) .. controls (230,160.38) and (205.38,185) .. (175,185) .. controls (144.62,185) and (120,160.38) .. (120,130) -- cycle ;
\draw [thick,red][>=stealth,->](114.7pt, 114pt)..controls (118pt, 129pt) and (150pt,129pt)..(143pt,100pt)
..controls (135pt,65pt) and (100pt,83pt)..(114pt,113pt);

\node at (94.7+20pt, 79+35pt){\color{black}$\bullet$};
\node at (95+10pt, 70+42pt){\color{black}$\rho$};

\node at (91+40pt, 70+85pt){\color{black}Non-heating phase};
\node at (91+40pt, 70+98pt){\color{white}$non-heating phase$};
\end{tikzpicture}
\quad
\begin{tikzpicture}[x=0.75pt,y=0.75pt,yscale=-1,xscale=1]
\draw  [fill={rgb, 255:red, 126; green, 211; blue, 33 }  ,fill opacity=0.21 ][line width=1.5]  (120,130) .. controls (120,99.62) and (144.62,75) .. (175,75) .. controls (205.38,75) and (230,99.62) .. (230,130) .. controls (230,160.38) and (205.38,185) .. (175,185) .. controls (144.62,185) and (120,160.38) .. (120,130) -- cycle ;
\draw [thick,red][>=stealth,->](114.7pt, 114pt)..controls (118pt, 129pt) and (180pt,129pt)..(170pt,90pt)
..controls (155pt,41pt) and (100pt,60pt)..(114pt,113pt);
\node at (94.7+20pt, 79+35pt){\color{black}$\bullet$};
\node at (94+10pt, 70+42pt){\color{black}$\rho$};
\node at (91+40pt, 70+85pt){\color{black}Non-heating phase};
\node at (91+40pt, 70+98pt){\color{black}near phase transition};
\end{tikzpicture}
\end{equation}
Here we sketch the rough features of the group walking of $\rho_n$, this cartoon are not meant to be exact.
As we will see in Sec.~\ref{Sec: QuasiPeridic} on the quasi-periodically driven CFT,
in general there are many rich fine structures on the orbit of $\rho$.
    
\item  The locations of the energy-momentum density peaks are determined by $(\rho_n\zeta_n)$ only. 
Recall that the poles of the (chiral) energy-momentum density  \eqref{EnergyMomentumDensity}
\begin{equation}
    \frac{1}{2\pi}\langle T(x,n)\rangle=-\frac{q^2 \pi c}{12 L^2}+
\frac{\pi c}{12 L^2}\cdot(q^2-1)\cdot \frac{1}{|\alpha_n e^{\frac{2\pi i x}{l}}+\beta_n|^4},
\quad \text{where } l=L/q
\end{equation}
locate at $-\frac{\beta_n}{\alpha_n}= (\rho_n\zeta_n)^*$, 
which determines the locations of peaks if $|\frac{\beta_n}{\alpha_n}|=|\rho_n\zeta_n|=|\rho_n|\to 1$, with:
\be\label{PeakLocation}
e^{\frac{2\pi i x_{\text{peak}}}{l}}=-(\rho_n \zeta_n)^*\,,\quad \text{if }
|\rho_n|\to 1,
\ee
i.e. $x_{\text{peak}}=\frac{l}{2\pi i}\log\big(
-(\rho_n\zeta_n)^*
\big)+kl$ with $k=0,\ldots ,q-1$. 
For $x$ away from $x_{\text{peak}}$, 
the energy-momentum density will be greatly suppressed. The same conclusion also holds for the anti-chiral component $\langle \bar{T}(x,n)\rangle$.


    \item The entanglement entropy in Eq.~\eqref{EE_general} depends on both $\rho_n$ and $\rho_n \zeta_n$
\be\label{HalfEE_general_GroupWalking}
S_A(n)-S_A(0)=\frac{c}{3}\log \frac{\big| 1+(\rho_n\zeta_n)^*\big|}{\sqrt{1-|\rho_n|^2}},
\ee
where we have neglected the contribution of the anti-chiral mode.
\end{enumerate}




In summary, the group walking of $\rho_n$ and $(\rho_n\zeta_n)$ on a unit disk $\mathbb D$
determine the behaviors of the energy-momentum/entanglement evolution as follows:
\begin{enumerate}

\item $|\rho_n|$ determines the growth of total energy.

\item In the heating phase, $(\rho_n\zeta_n)$ in the long driving limit ($n\gg 1$) determines the 
location of peaks of the energy-momentum density.

\item $|\rho_n|$ and $(\rho_n\zeta_n)$ together determine the time evolution of the entanglement entropy
$S_A(n)$.

\end{enumerate}

\section{Periodic driving}
\label{Sec: PeriodicDriving}

This section is a generalization 
of previous works in Refs.\cite{wen2018floquet,fan2019emergent} by considering a more general setup of periodic drivings.
Apart from its own interesting features, 
this generalized setup can be used to analyze the quasi-periodically driven CFTs in Sec.~\ref{Sec: QuasiPeridic}.

In Refs.~\cite{wen2018floquet,fan2019emergent}, 
a minimal setup of a periodic driving with two driving steps
within one driving period was considered. The two different driving Hamiltonians are chosen as $H_0$ 
and $H_1=H_{\text{SSD}}$ being the sine-square deformed Hamiltonian.
Here we generalize this minimal setup in two aspects: one is to consider arbitrary 
SL$_2$ deformed Hamiltonians, and the other is to consider more general periodic sequences.
In general, as the number of driving steps within a driving period increases, 
the phase diagram will become 
quite rich.\footnote{See, e.g., Fig.~\ref{FibonacciPhase0} in the next section where we use increasingly long periodic drivings to approach the quasi-periodic driving.}

\subsection{General protocol for periodic driving}
\label{Sec: PeriodcDrivingGP}

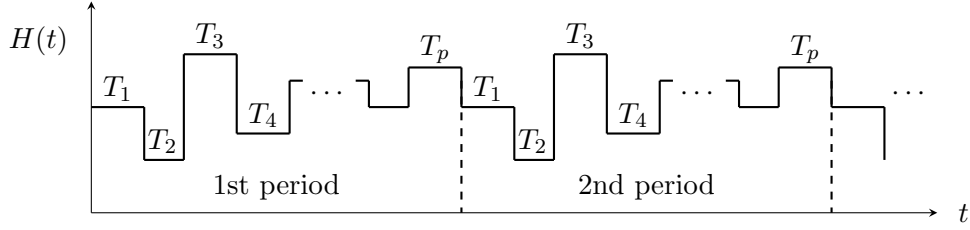
\begin{figure}[t]
\label{fig:FloquetSetup}
\begin{center}
\begin{tikzpicture}[baseline={(current bounding box.center)}]
\small
\node at (-20pt, 45pt){$H(t)$};
\draw [thick](0pt,20pt)--(20pt,20pt);
\draw [thick](20pt,20pt)--(20pt,0pt);
\draw [thick](20pt,0pt)--(35pt,0pt);
\draw [thick](35pt,0pt)--(35pt,40pt);
\draw [thick](35pt,40pt)--(55pt,40pt);
\draw [thick](55pt,40pt)--(55pt,10pt);
\draw [thick](55pt,10pt)--(75pt,10pt);
\draw [thick](75pt,10pt)--(75pt,30pt);
\draw [thick](75pt,30pt)--(80pt,30pt);

\node at (10pt, 27pt){$T_1$};
\node at (27pt, 7pt){$T_2$};
\node at (45pt, 47pt){$T_3$};
\node at (65pt, 17pt){$T_4$};
\node at (130pt, 42pt){$T_p$};


\small
\node at (90pt, 25pt){$\cdots$};

\draw [thick](100pt,30pt)--(105pt,30pt);
\draw [thick](105pt,30pt)--(105pt,20pt);
\draw [thick](105pt,20pt)--(120pt,20pt);
\draw [thick](120pt,20pt)--(120pt,35pt);
\draw [thick](120pt,35pt)--(140pt,35pt);
\draw [thick](140pt,35pt)--(140pt,20pt);

\draw [thick][dashed](140pt,30pt)--(140pt,-20pt);

\node at (70pt, -10pt){1st period};
\node at (70+140pt, -10pt){2nd period};

\draw [thick](0+140pt,20pt)--(20+140pt,20pt);
\draw [thick](20+140pt,20pt)--(20+140pt,0pt);
\draw [thick](20+140pt,0pt)--(35+140pt,0pt);
\draw [thick](35+140pt,0pt)--(35+140pt,40pt);
\draw [thick](35+140pt,40pt)--(55+140pt,40pt);
\draw [thick](55+140pt,40pt)--(55+140pt,10pt);
\draw [thick](55+140pt,10pt)--(75+140pt,10pt);
\draw [thick](75+140pt,10pt)--(75+140pt,30pt);
\draw [thick](75+140pt,30pt)--(75+145pt,30pt);
\node at (90+140pt, 25pt){$\cdots$};

\draw [thick](105+135pt,30pt)--(105+140pt,30pt);
\draw [thick](105+140pt,20pt)--(105+140pt,30pt);
\draw [thick](105+140pt,20pt)--(120+140pt,20pt);
\draw [thick](120+140pt,20pt)--(120+140pt,35pt);
\draw [thick](120+140pt,35pt)--(140+140pt,35pt);
\draw [thick](140+140pt,35pt)--(140+140pt,20pt);

\draw [thick][dashed](280pt,30pt)--(280pt,-20pt);

\draw [thick](0+280pt,20pt)--(20+280pt,20pt);
\draw [thick](20+280pt,20pt)--(20+280pt,0pt);
\node at (310pt, 25pt){$\cdots$};

\node at (10+140pt, 27pt){$T_1$};
\node at (27+140pt, 7pt){$T_2$};
\node at (45+140pt, 47pt){$T_3$};
\node at (65+140pt, 17pt){$T_4$};
\node at (130+140pt, 42pt){$T_p$};

\draw [>=stealth,->](0pt, -20pt)--(320pt,-20pt);
\draw [>=stealth,->](0pt, -20pt)--(0pt,60pt);
\node at (330pt, -20pt){$t$};

\end{tikzpicture}
\end{center}
\caption{A general protocol for a periodically driven CFT. 
There are $p$ steps of driving within each driving period.
In the $i$-th step of driving, we consider the driving with $(H_i, T_i)$, where 
$H_i$ is a SL$_2$ deformed Hamiltonian in Eq.~\eqref{H_general} and $T_i$ is the
corresponding time interval of driving.
}
\end{figure}

For a periodical driving with period $p\in \mathbb Z^+$, we have $U_j=U_{j+p}$ for all $j\in \ZZ_+$.  
Then the time evolution of wavefunction after $n p$ driving steps is determined by the unitary operators
$(U_p\cdots U_2\cdot U_1)$ as follows, 
\be\label{Psi_np}
|\Psi_{np}\rangle= \big(
\underbrace{U_p\cdots U_2\cdot U_1}_{\text{one driving period}}\big)^n|\Psi_0\rangle,\quad \text{with}\quad U_j=e^{-iH_j T_j}.
\ee
In terms of conformal mapping, the operator evolution after $np$ driving steps only depends on the 
the matrix product 
\be\label{Pi_p_sec2}
\Pi_p:=M_1\cdot M_2\cdots M_p \in \SU(1,1)
\ee
Let us denote the matrix elements of $\Pi_p$ and $(\Pi_p)^n$ as follows
\begin{equation}\label{eqn: pi and pin}
\Pi_p = 
\begin{pmatrix}
\alpha_p &\beta_p\\
\beta_p^* &\alpha_p^*
\end{pmatrix}\,, \quad 
(\Pi_p)^n
=\begin{pmatrix}
\alpha_{np} &\beta_{np}\\
\beta_{np}^* &\alpha_{np}^*
\end{pmatrix} : = \begin{pmatrix}
\alpha_p &\beta_p\\
\beta_p^* &\alpha_p^*
\end{pmatrix}^n \,.
\end{equation}
Next, we will determine the phase diagrams 
and relevant physical quantities 
based on the operator evolution given by $\Pi_p$, or equivalently the following M\"obius transformation
\begin{equation}\label{MobiusMap}
    z'=\Pi_p\cdot z=
    \begin{pmatrix}
    \alpha_p &\beta_p\\
    \beta_p^* &\alpha_p^*
    \end{pmatrix} \cdot z 
    = \frac{\alpha_p z+\beta_p}{\beta_p^* z+\alpha_p^*}
\end{equation}
and similarly for $\bar{z}$. 



%


 \subsubsection{Phase diagram and Lyapunov exponents}
 \label{Sec:PhaseDiagram}
 
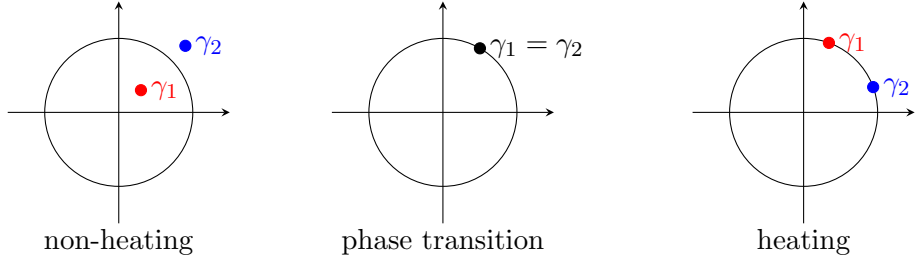
\begin{figure}[t]
\begin{center}
\small
    \begin{tikzpicture}[scale =1.4]
     \draw[->,>=stealth] (-30pt,0pt) -- (30pt,0pt);
     \draw[->,>=stealth] (0pt,-30pt) -- (0pt,30pt);
     \draw (0pt,0pt) circle (20pt);
     \filldraw[blue] (18pt, 18pt) circle (1.5pt) node[right]{$\gamma_2$};
          \filldraw[red] (6pt, 6pt) circle (1.5pt) node[right]{$\gamma_1$};
          \node at (0pt, -35pt){non-heating};
    \end{tikzpicture}
    \hspace{30pt}
        \begin{tikzpicture}[scale =1.4]
     \draw[->,>=stealth] (-30pt,0pt) -- (30pt,0pt);
     \draw[->,>=stealth] (0pt,-30pt) -- (0pt,30pt);
     \draw (0pt,0pt) circle (20pt);
     \filldraw (10pt, 17.32pt) circle (1.5pt) node[right]{$\gamma_1=\gamma_2$};
               \node at (0pt, -35pt){phase transition};
    \end{tikzpicture}
     \hspace{30pt}
        \begin{tikzpicture}[scale =1.4]
     \draw[->,>=stealth] (-30pt,0pt) -- (30pt,0pt);
     \draw[->,>=stealth] (0pt,-30pt) -- (0pt,30pt);
     \draw (0pt,0pt) circle (20pt);
     \filldraw[blue] (18.79pt, 6.84pt) circle (1.5pt) node[right]{$\gamma_2$};
          \filldraw[red] (6.84pt, 18.79pt) circle (1.5pt) node[right]{$\gamma_1$};
                    \node at (0pt, -35pt){heating};
    \end{tikzpicture}
    \end{center}
\caption{Illustration for the locations of fixed points of M\"obius transformation in the three phases. 
In the non-heating phase, the two fixed points are inside and outside the unit circle respectively. 
They will merge at the same point on the unit circle at the phase transition. 
Then the two fixed points will split but still sit on the unit circle in the heating phase.
}
\label{FixedpointFig}
\end{figure}

 The matrix $\Pi_{np}=(\Pi_p)^n$ has three distinct asymptotics depending on the trace of $\Pi_p$, for convenience, let us classify the types of $\SU(1,1)$ matrices in parallel to the classification of M\"obius transformation we used in Ref.~\cite{fan2019emergent}, also see Fig.~\ref{FixedpointFig} for an illustration.  

Let $M\in\text{SU}(1,1)$ not be the central elements $\pm \mathbb I$, then we call the matrix $M$
\begin{enumerate}
\item Elliptic if $|\Tr(M)|<2$. $M$ has two distinct eigenvalues $\lambda_1$, $\lambda_2$
with  $\lambda_2=\lambda_1^*$ and $|\lambda_1|=|\lambda_2|=1$. The corresponding M\"obius transformation has two distinct fixed points, 
one inside the unit circle and the other outside; 

\item Parabolic if $|\Tr(M)|=2$.
$M$ has a single eigenvalue at $+1$ or $-1$. The corresponding fixed points become degenerate (i.e. only one single point) and stay on the circle; 

\item Hyperbolic if $|\Tr(M)|>2$. 
$M$ has two distinct real eigenvalues $\lambda_1$, $\lambda_2$, 
$|\lambda_1|>1>|\lambda_2|$ and $\lambda_2=\lambda_1^{-1}$. The two fixed points are distinct and staying on the circle. 
\end{enumerate}
As a reminder, the fixed points of the M\"obius transformation are  convenient way to characterize the transformation when we repeat it multiple times.\footnote{Therefore, this is the main tool we used in the previous study \cite{fan2019emergent} to visualize the effects of periodic driving.} We rewrite the M\"obius transformation into the following form
 \be\label{FixedPoint}
\frac{z'-\gamma_1}{z'-\gamma_2}=\eta \cdot \frac{z-\gamma_1}{z-\gamma_2}\,,
\ee
where  $\gamma_{1,2}$ are the fixed points we mentioned, and $\eta$ is the multiplier. For $\Pi_p$ parametrized in \eqref{eqn: pi and pin}, we have the following explicit formulas
\be\label{FixedPoint_Eta}
\gamma_{1,2}=\frac{1}{2\beta_p^*}\Big[(\alpha_p-\alpha_p^*\mp \sqrt{(\alpha_p+\alpha_p^*)^2-4})\Big],
\ee
\be\label{Multiplier_eta}
\eta
= \frac{\text{Tr}(\Pi_p)+\sqrt{ [\text{Tr}(\Pi_p)]^2-4} }{\text{Tr}(\Pi_p)-\sqrt{ [\text{Tr}(\Pi_p)]^2-4}}, \qquad \text{where }\text{Tr}(\Pi_p)=\alpha_p+\alpha_p^*.
\ee
Note the sign of the discriminant depends on the trace of $\Pi_p$, which can be used to categorize the M\"obius transform \eqref{FixedPoint} (see Fig.~\ref{FixedpointFig} for an illustration. 
)
For parabolic class when $|\Tr(\Pi_p)|=2$, we have $\gamma_1=\gamma_2$, the transformation \eqref{FixedPoint} becomes trivial and we need to invoke
\begin{equation}
\frac{1}{z'-\gamma}=\frac{1}{z-\gamma}+ \beta_p^*\,, \quad {\text{where}} \quad \gamma=\frac{\alpha_p-\alpha_p^*}{2\beta_p^*} \,.
\end{equation}

When repeating $n$ times, we only need to modify $\eta \rightarrow \eta^n$ for $|\Tr (\Pi_p)| \neq 2 $ case and $\beta_p^* \rightarrow n \beta_p^*$ for $|\Tr (\Pi_p)| = 2 $.  And therefore, we have a simple expression for the matrix elements of $(\Pi_p)^n$ defined in \eqref{eqn: pi and pin}:
\be\label{AlphaBeta}
\alpha_{np}=\frac{\eta^{-\frac{n}{2}}\gamma_1-\eta^{\frac{n}{2}}\gamma_2}{\gamma_1-\gamma_2},\quad 
\beta_{np}=\frac{(\eta^{\frac{n}{2}}-\eta^{-\frac{n}{2}})\cdot \gamma_1\gamma_2}{\gamma_1-\gamma_2} \quad \text{when} \quad |\text{Tr}(\Pi_p)|\neq2
\ee
\be\label{AlphaBeta2}
\alpha_{np}=1+n\gamma\beta_p^*, \quad \beta_{np}=-n\gamma^2\beta_p^*
\quad \text{when} \quad |\text{Tr}(\Pi_p)|=2 \,.
\ee 

Another advantage of the representation using fixed points $\gamma_{1,2}$ and multiplier $\eta$ is that the Lyapunov exponent now only depends on $\eta$ as follows
\be\label{Lyapunov_Eta}
\lambda_L
=\frac{1}{2p}\log\left( \text{max}\left\{|\eta|, \, |\eta|^{-1}\right\}\right)
=\frac{1}{p}\log
\Big|
\frac{|\text{Tr}(\Pi_p)|+ \sqrt{|\text{Tr}(\Pi_p)|^2-4}}{2}
\Big| \,. 
\ee
That is to say, the hyperbolic $\Pi_p$ with $|\Tr (\Pi_p)|>2$ implies a positive Lyapunov exponent and therefore heating phase; while the elliptic and parabolic classes both have $\lambda_L=0$. By analyzing the corresponding group walk in the next subsection we will confirm that $|\Tr (\Pi_p)|<2$ corresponds to non-heating phase while $|\Tr (\Pi_p)|=2$ is the phase transition as expected.


\subsection{Group walking}

The group walking $(\rho_{np}, \rho_{np}\cdot \zeta_{np})$ of $(\Pi_p)^n$ defined in \eqref{M_alpha_z} can be 
straightforwardly obtained by comparing with \eqref{AlphaBeta} for $|\Tr (\Pi_p)|\neq 2$,  
\be\label{rhozeta1}
\rho_{np}=-\frac{1}{\gamma_2}+\frac{1}{\gamma_2}\cdot \frac{\gamma_1-\gamma_2}{\gamma_1-\eta^n\,\gamma_2},
\quad (\rho_{np}\cdot\zeta_{np})=\frac{1}{\gamma_1}+\frac{1}{\gamma_1}\cdot\frac{\gamma_2-\gamma_1}{\eta^n\,\gamma_1-\gamma_2},
\ee
or comparing with 
\eqref{AlphaBeta2} for $|\Tr (\Pi_p)|= 2$
\be\label{GroupWalking_prabolic}
\rho_{np}=-\frac{n\beta^*}{1+n\gamma\beta^*}, \quad (\rho_{np} \cdot \zeta_{np})=-\frac{n\beta^*}{1-n\gamma\beta^*}.
\ee
Now we are ready to discuss the trajectories of $(\rho_{np}, \rho_{np}\cdot \zeta_{np})$ with increasing $n$:
\begin{enumerate}
    \item For $|\Tr (\Pi_p) |<2$, the multiplier $\eta \in \UU(1)$ is a pure phase and implies that both $\rho_{np}$ and $(\rho_{np}\cdot \zeta_{np})$ will form a closed loop in the unit disk. 
    
    \item For $|\Tr (\Pi_p) |>2$, the multiplier $|\eta|\neq 1 $ and we have the following limit at $n\to \infty$, 
    \be\label{GroupWalkPeriodLimit}
    \lim_{n\to \infty}\rho_{np}=
    \left\{
    \begin{split}
    -\gamma_2^*,\quad &\eta>1\\
    -\gamma_1^*,\quad &\eta<1
    \end{split}
    \right.
    \quad\quad
    \lim_{n\to \infty}(\rho_{np}\cdot \zeta_{np})=
    \left\{
    \begin{split}
    \gamma_1^*,\quad &\eta>1\\
    \gamma_2^*,\quad &\eta<1
    \end{split}
    \right.
    \ee
Recall that both $\gamma_1$ and $\gamma_2$ live on $\partial \mathbb D$ as shown in Fig.~\ref{FixedpointFig}. 
Therefore, in this case, both $\rho_n$ and $(\rho_{np}\zeta_{np})$ will approach exponentially close to the boundary of the unit disk $\partial \mathbb D$.

    \item For $|\Tr (\Pi_p)|=2$, we have the following limit 
    \be
    \lim_{n\to \infty}\rho_{np}=-\gamma^*,\quad
    \lim_{n\to \infty}(\rho\zeta)_{np}=\gamma^*,
    \ee
    where $\gamma^*\in \mathbb D$ as shown in Fig.~\ref{FixedpointFig}.
    From Eq.~\eqref{GroupWalking_prabolic}, we notice that in this case, both $\rho_{np}$ and $(\rho_{np}\zeta_{np})$ will 
    approach $\partial \mathbb D$ polynomially (in $n$) close. 
\end{enumerate}
The above behavior confirms that $|\Tr(\Pi_p)|<2$, $|\Tr(\Pi_p)|>2$ and  $|\Tr(\Pi_p)|=2$ correspond to non-heating, heating and phase transition respectively. 





\subsection{Entanglement/energy evolution}
\label{Sec: PeriodicDrivingEnergyEE}

Given the explicit expressions of the matrix elements of $\Pi_{np}=(\Pi_p)^n$ in \eqref{AlphaBeta} and \eqref{AlphaBeta2}, 
we can further obtain the time evolution of the entanglement entropy $S_A(N=np)$ and the total energy $E(N=np)$ based on 
formulas \eqref{EE_general} and \eqref{TotalEnergyMomentum}, respectively.

For the total energy, it grows exponentially in the heating phase
\be\label{Eenergy_lyapunov_p}
 E(N)\simeq
\frac{\pi c}{24L}\cdot(q^2-1)\cdot e^{2\lambda_L\cdot N}, \quad \text{where }N=n\cdot p,
\ee
and the exponent is exactly twice the Lyapunov exponent, the latter is given in \eqref{Lyapunov_Eta}. 
In the non-heating phase and the phase transition, the total energy oscillates and grows polynomially. Here we only consider the contribution of the chiral modes, the anti-chiral modes follow parallel discussions. 


As noted in Ref.~\cite{fan2019emergent}, the energy-momentum density has interesting spatial structures. In fact, 
as mentioned in Sec.~\ref{Sec:LyapunovHeatingPhase},
a positive Lyapunov exponent $\lambda_L$ indicates there is an array 
 of peaks in the energy-momentum density $\langle T(x,n)\rangle$ in real space. The same spatial structure, namely the array of peaks, is also present at the phase transition with $\lambda_L=0$, although the growth is polynomial in $n$, significantly slower than the heating phase. 
 
Following Eqs.\eqref{EnergyMomentumDensity}, \eqref{AlphaBeta} and \eqref{AlphaBeta2}, we find the locations of the (chiral) energy-momentum peaks are given as follows
 \be
 \label{PeakLocationHP}
 \small
 \left\{
 \begin{split}
& \exp\left(\frac{2\pi i x_{\text{peak}}}{l}\right)=-\lim_{n\to \infty}\frac{\beta_{np}}{\alpha_{np}}=\gamma_2, \quad &\text{in the heating phase,}\\
&\exp\left(\frac{2\pi i x_{\text{peak}}}{l}\right)=-\lim_{n\to \infty}\frac{\beta_{np}}{\alpha_{np}}=\gamma, \quad &\text{at the phase transition},
\end{split}
 \right.
 \ee
 where we have assumed $0<\eta<1$ in the above formula, for $\eta>1$, we need to replace $\gamma_2$ by $\gamma_1$. 
 Here $\gamma_2$ corresponds to the unstable fixed point in
 the M\"obius transformation 
 in the heating phase, and $\gamma$ is the unique fixed point 
 at the phase transition. 
 A cartoon plot of the energy-momentum density distribution in real space is shown as follows:
\be\label{EE_growth_Cartoon}
    \begin{tikzpicture}[scale=0.8, baseline={([yshift=-6pt]current bounding box.center)}]
    \draw [dotted][red][thick](-31pt,0pt)..controls (-30pt,2pt) and (-29pt,5pt)..(-28pt,40pt)..controls (-27pt,5pt) and (-26pt,2pt)..(-25pt,0pt);
    \draw [blue][thick](31pt,0pt)..controls (30pt,2pt) and (29pt,5pt)..(28pt,40pt)..controls (27pt,5pt) and (26pt,2pt)..(25pt,0pt);
    \draw [thick] (-50pt,0pt)--(50pt,0pt);
    \draw [thick] (-50pt,-5pt) -- (-50pt,5pt);

     \draw [dotted][red][xshift=100pt,thick](-31pt,0pt)..controls (-30pt,2pt) and (-29pt,5pt)..(-28pt,40pt)..controls (-27pt,5pt) and (-26pt,2pt)..(-25pt,0pt);
    \draw [blue][xshift=100pt,thick](31pt,0pt)..controls (30pt,2pt) and (29pt,5pt)..(28pt,40pt)..controls (27pt,5pt) and (26pt,2pt)..(25pt,0pt);
    \draw [xshift=100pt,thick] (-50pt,0pt)--(50pt,0pt);
    \draw [xshift=100pt,thick] (-50pt,-5pt) -- (-50pt,5pt);

    \draw [dotted][red][xshift=200pt,thick](-31pt,0pt)..controls (-30pt,2pt) and (-29pt,5pt)..(-28pt,40pt)..controls (-27pt,5pt) and (-26pt,2pt)..(-25pt,0pt);
    \draw [blue][xshift=200pt,thick](31pt,0pt)..controls (30pt,2pt) and (29pt,5pt)..(28pt,40pt)..controls (27pt,5pt) and (26pt,2pt)..(25pt,0pt);
    \draw [xshift=200pt,thick] (-50pt,0pt)--(50pt,0pt);
    \draw [xshift=200pt,thick] (-50pt,-5pt) -- (-50pt,5pt);

    \draw [dotted][red][xshift=300pt,thick](-31pt,0pt)..controls (-30pt,2pt) and (-29pt,5pt)..(-28pt,40pt)..controls (-27pt,5pt) and (-26pt,2pt)..(-25pt,0pt);
    \draw [blue][xshift=300pt,thick](31pt,0pt)..controls (30pt,2pt) and (29pt,5pt)..(28pt,40pt)..controls (27pt,5pt) and (26pt,2pt)..(25pt,0pt);
    \draw [xshift=300pt,thick] (-50pt,0pt)--(50pt,0pt);
    \draw [xshift=300pt,thick] (-50pt,-5pt) -- (-50pt,5pt);

    \draw[dashed] (28-40pt,0pt)--(28-40pt,65pt);
    \draw[dashed] (228-40pt,0pt)--(228-40pt,65pt);
     \node at (30-40pt, -10pt){$x_1$};
    \node at (230-40pt, -10pt){$x_2$};
    
        \draw[->,>=stealth] (128-40pt,50pt)--(228-40pt,50pt);
         \draw[->,>=stealth] (128-40pt,50pt)--(28-40pt,50pt);

 \node at (130-40pt, 63pt){subsystem $A$};
    
     \draw [thick] (350pt,-5pt) -- (350pt,5pt);
       \end{tikzpicture}
\ee
where different colors represent different chiralities.
For simplicity, let us keep the anti-chiral part (red) undeformed, and only deform the chiral part (blue).
Then the entanglement entropy in the heating phase depends on the choice of subsystem $A$ as follows \cite{fan2019emergent}:
\begin{equation}
\label{EEpeak1}
\small
    S_A(N=np) - S_A(0) \simeq \left\{ \begin{array}{ll} \vspace{5pt}
        \calO(1) & [x_1,x_2]\, \text{does not include peaks}, \\ \vspace{5pt}
        \frac{c}{3}\cdot \lambda_L\cdot N
        \quad& [x_1,x_2]\,\text{includes peak(s)}, \\
    \end{array}
    \right.
\end{equation}
If one also deforms the anti-chiral part and let it live in heating phase with Layapunov exponent $\lambda_L'$, then we need to add up two contributions when $A$ also includes any anti-chiral peaks. Note in general $\lambda_L \neq \lambda_L'$ as they can be deformed independently in the CFT with periodic boundary condition. 
One can further check the entanglement pattern by looking into the mutual information as studied in \cite{fan2019emergent}, and find each peak is mainly entangled with 
the two peaks of its nearest neighbor with the same chirality, 
as schematically shown in Fig.~\ref{fig:entanglement pattern generalized}.



At the phase transition, similar to the energy, the spatial structure of the entanglement persists, while the growth is slower 
\begin{equation}
\small
    S_A(N=np) - S_A(0) \simeq \left\{ \begin{array}{ll} \vspace{5pt}
        \calO(1) & [x_1,x_2]\, \text{does not include peaks}, \\ \vspace{5pt}
        \dfrac{c}{3} \log n \quad& [x_1,x_2]\,\text{includes peak(s)}, \\
    \end{array}
    \right.
\end{equation}

One final remark is that in the above discussions, the entanglement cuts are chosen to avoid 
the centers of the energy-momentum density peaks.
In Appendix.~\ref{Sec: LinearDecreaseEE}, we also consider the 
cases when the entanglement cuts are 
located at the center(s) of the energy-momentum density peaks. 
Then some interesting features in the entanglement 
entropy could arise.

To summarize, we put the phase diagrams and related quantities in the periodically driven CFT in Table.~\ref{PhaseDiagramPeriodic}. 
\begin{table}[t]
\small
\begin{center}
\begin{tabular}{cccccccccccc}
& \shortstack{Phases}&\vline& \shortstack{M$\ddot{\text{o}}$bius transf. }  &\vline& $\left|\text{tr}\, \Pi_p \right|$  &\vline&$\lambda_L$  &\vline&EE growth &\vline&Energy growth \\ \hline
& Heating     &\vline& Hyperbolic   &\vline&  $>2$&\vline& $\lambda_L>0$ &\vline&linear &\vline&exponential \\ \hline
&Non-heating  &\vline& Elliptic &\vline&$<2$&\vline&  $\lambda_L=0$ &\vline&logarithmic &\vline&power law \\ \hline
&Phase transition &\vline& Parabolic    &\vline&$=2$&\vline& $\lambda_L=0$&\vline&oscillating &\vline&oscillalting\\ \hline
\end{tabular}
\end{center}
\caption{Correspondence of the phase diagram in a \textit{periodically} driven CFT and other quantities.}
\label{PhaseDiagramPeriodic}
\end{table}

\subsection{A minimal setup}
\label{Sec:MinimalSetup}

Now we consider a minimal setup of the periodically driven CFT to 
demonstrate the main features in the previous discussions.
In this setup, we consider only $p=2$ driving steps within one period
\begin{eqnarray}\label{FloquetSetup}
\small
\begin{tikzpicture}[baseline={(current bounding box.center)}]
\node at (-5pt, 7pt){$H(t)$:};
\draw [thick](20pt,0pt)--(40pt,0pt);

\draw [thick](40pt,0pt)--(40pt,20pt);

\draw [thick](40pt,20pt)--(60pt,20pt);
\draw [thick](60pt,20pt)--(60pt,0pt);
\draw [thick](60pt,0pt)--(80pt,0pt);

\draw [thick](80pt,0pt)--(80pt,20pt);

\draw [thick](80pt,20pt)--(100pt,20pt);
\draw [thick](100pt,20pt)--(100pt,0pt);
\draw [thick](100pt,0pt)--(120pt,0pt);

\draw [thick](120pt,20pt)--(120pt,0pt);

\draw [thick](120pt,20pt)--(140pt,20pt);
\draw [thick](140pt,20pt)--(140pt,0pt);
\draw [thick](140pt,0pt)--(160pt,0pt);

\node at (10+40pt,27pt){$H_1$};
\node at (30pt, 7pt){$H_0$};

\draw [>=stealth,->] (120pt, -10pt)--(145pt,-10pt);
\node at (150pt, -10pt){$t$};
\node at (85pt, -20pt){periodical driving};


\end{tikzpicture}
\end{eqnarray}
That is, we drive the CFT with $(H_0, T_0)$ and $(H_1, T_1)$, where $T_0$ and $T_1$ are the time intervals.
We consider a SL$_2$ deformed Hamiltonian with $q=1$
with open boundary conditions\footnote{We choose open boundary condition here for the purpose of providing a comparison with the lattice simulation that will be shown momentarily, where it is natural to take open boundary condition.}:
\be\label{H_theta_A}
H_{\theta}=
\int_0^L \left(1-\tanh(2\theta)\cdot \cos\frac{2\pi q x}{L}\right) T_{00}(x) dx, \quad q=1, \, \theta>0.
\ee
We choose $H_0$ and $H_1$ as $H_{\theta=0}$ and $H_{\theta\neq 0}$, 
and $T_0$ and $T_1$ as $T_{\theta=0}$ and $T_{\theta\neq 0}$, respectively.
Note that $H_{\theta=0}$ corresponds 
to the uniform Hamiltonian, and $H_{\theta=\infty}$
corresponds to the SSD Hamlitonian in Eq.\eqref{SSD_Hamiltonian}
up to an overall factor $2$.
Denoting the time interval of driving as $T_{\theta}$, then the corresponding M\"obius transformation 
$M(H_{\theta}, T_{\theta})$ has the following form
\begin{equation}\label{MobiusTheta}
\wideboxed{
M(H_{\theta}, T_{\theta}) = \begin{pmatrix}
\alpha & \beta \\
\beta^* & \alpha^*
\end{pmatrix}
\quad \text{with} \quad \left\{
\begin{split}
&\alpha=\cos{\left( \frac{\pi T_{\theta}}{L_{\text{eff}}} \right)} + i\cosh(2\theta)\cdot \sin{\left( \frac{\pi T_{\theta} }{L_{\text{eff}}} \right)},\\
&\beta=- i\sinh(2\theta)\cdot\sin{\left( \frac{\pi T_{\theta}}{L_{\text{eff}}} \right)}.
\end{split}
\right.
}
\end{equation}
Here $L_{\text{eff}}=L\cosh(2\theta)$ denotes the effective length of the total system. 
Physically, it characterizes the effective distance that the quasiparticle needs to
travel to return to its original location\cite{WenWu2018quench}.

\subsubsection{Phase diagram and Lyapunov exponent}

\begin{figure}[t]
\center
\centering
\includegraphics[width=6.5in]{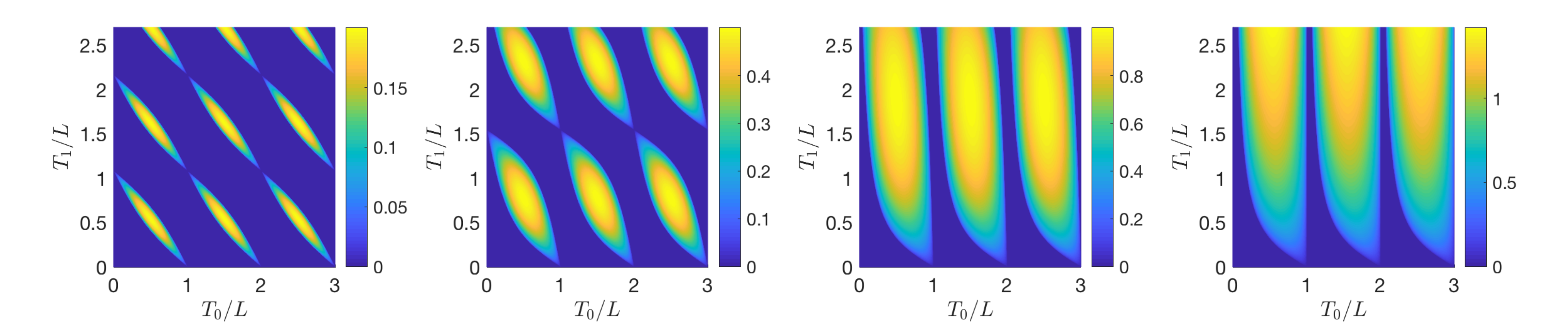}
\caption{
Lyapunov exponent with (from left to right)
$\theta=0.2$, $\theta=0.5$, $\theta=1$, and $\theta=\infty$. The regime with $\lambda_L>0$ corresponds to the heating phase, while the dark blue regime with $\lambda_L=0$ corresponds to non-heating phase, and boundary between them is the phase transition. 
}
\label{Lyapunov_PeriodicDriving}
\end{figure}
\begin{figure}[t]
\begin{center} 
\includegraphics[width=5.30in]{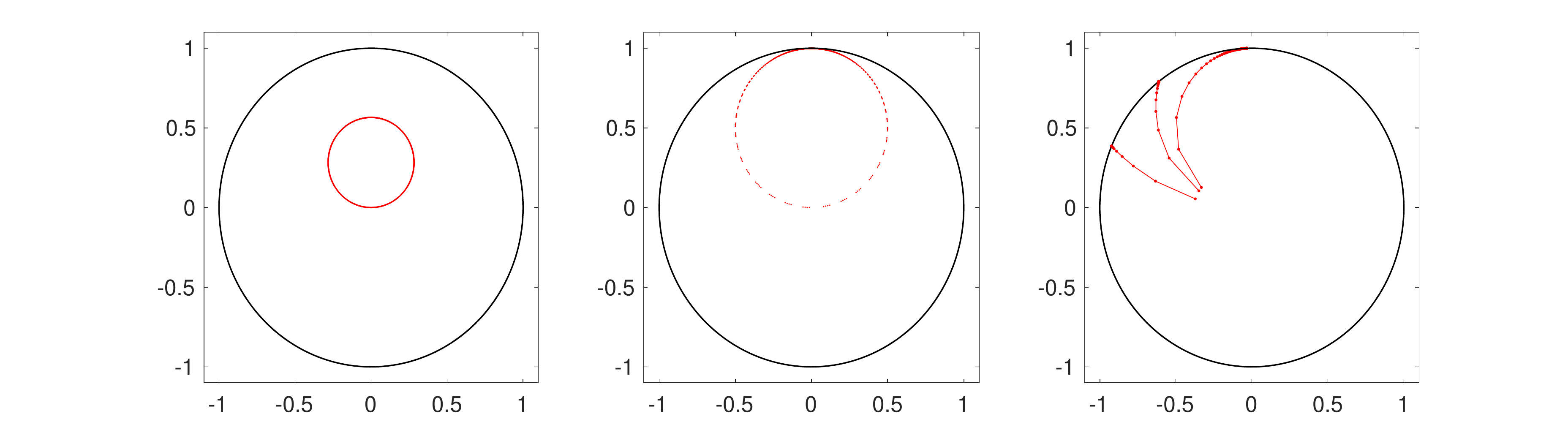}
\end{center}
\caption{
Trajectories of $(\rho_{np}\zeta_{np})$ on the unit disk $\mathbb D$ in a periodically driven CFT 
driven with $H_0$ and $H_1=H_{\theta}$ in Eq.~\eqref{H_theta_A}. 
Here we choose $\theta=0.2$, $T_0=L/2$, and 
$T_1/(L\cosh(2\theta))=0.3$ in the non-heating phase (left),
$T_1/(L\cosh(2\theta))=0.3758$ in the non-heating phase near the phase transition (middle, the phase transition happens at
$\frac{1}{\pi}\arcsin[1/\cosh(2\theta)]\simeq 0.3759$),
and $T_1/(L\cosh(2\theta))=0.376, \, 0.4$, and $0.45$ (in the counterclockwise order) in the heating phase (right). 
}
\label{GroupWalkingPeriodic}
\end{figure}

The Lyapunov exponent $\lambda_L$ is determined by the trace of the transformation matrix $\Pi_p$ as shown in \eqref{Lyapunov_Eta}. In our setting $p=2$ and 
\be
\label{TrPi2}
\big|\text{Tr}(\Pi_{p=2})\big|=
\big|\text{Tr}(M_0M_1)\big|=2\cdot \big|\cosh(2\theta)\cdot \sin x_1\cdot \sin x_0-\cos x_1\cdot \cos x_0\big|,
\ee
where $x_0=\frac{\pi T_0}{L}$ and $x_1=\frac{\pi T_1}{L_{\text{eff}}}$, with $L_{\text{eff}}=L\cosh(2\theta)$. 
Therefore, inserting into 
\begin{equation}
    \lambda_L
=\frac{1}{p}\log
\Big|
\frac{|\text{Tr}(\Pi_p)|+ \sqrt{|\text{Tr}(\Pi_p)|^2-4}}{2}
\Big| \,, 
\end{equation}
we obtain the result shown in Fig.~\ref{Lyapunov_PeriodicDriving}. 
From the figure, we can also read out the phase diagram straightforwardly, namely the regime with $\lambda_L>0$ corresponds to the heating phase, while the dark blue regime with $\lambda_L=0$ corresponds to non-heating phase, and the boundary between them is the phase transition. We also show in Fig.~\ref{GroupWalkingPeriodic} the group walking pictures for  $\theta=0.2$ with different choices of $T_1/L$.


To gain some analytical understanding of the formula for Lyapunov exponent, let us consider a simple example when 
$\theta=\infty$, namely $H_1=H_{\text{SSD}}$. 
Along the line $T_0=L/2$, \eqref{TrPi2} simplifies to  $\big|\text{Tr}(\Pi_{p=2})\big|=\frac{2\pi T_1}{L}$. And therefore, the Lyapunov exponent is a function of $T_1$ given as follows
\be
\lambda_L(T_1)=\frac{1}{2}\log\left(\frac{\pi T_1+\sqrt{(\pi T_1)^2-L^2}}{L} \right), \quad \text{where }\theta=\infty, \, T_0=L/2.
\ee
In particular, in the limit $T_1\gg L$, we have 
\be\label{LyapunovLogGrowth}
\lambda_L(T_1)\simeq \frac{1}{2}\log\left(\frac{\pi T_1}{L}\right).
\ee
That is to say, along the line $T_0=L/2$, the Lyapunov exponent grows logarithmically with $T_1/L$ in the large $T_1$ limit ($T_1/L\gg 1$). 
In fact, for $\theta=\infty$, 
the result in \eqref{LyapunovLogGrowth} holds for arbitrary $T_0/L\neq n\pi$ ($n\in\mathbb Z$) in the large $T_1/L$ limit.

Now we would like to make a few comments on  Fig.~\ref{Lyapunov_PeriodicDriving}
\begin{enumerate}
\item The area of regime with larger Lyapunov exponent grows when we increase the parameter $\theta$ in Hamiltonian. The heuristic argument is that when the `difference' between $H_0$ and $H_{\theta}$ is greater, the driving protocol is easier to heat the system.


\item When we push $\theta\to 0$, the area of heating phase, namely the regime with $\lambda_L>0$ decrease to zero. 
However, there is always at least a point $(T_0/L, T_1/L_{\text{eff}})=(1/2, 1/2)$
staying in the heating phase for arbitrary $\theta\neq 0$. 
This point corresponds to $x_0=x_1=\frac{\pi}{2}$ in Eq.~\eqref{TrPi2} where we find $\big|\text{Tr}(M_0M_1)\big|=2\cosh(2\theta)>2$ for arbitrary $\theta\neq 0$.
\end{enumerate}

Now, let us take a closer look at the special point 
$(T_0/L, T_1/L_{\text{eff}})=(1/2, 1/2)$ in the heating phase. In terms of the transformation matrix $M_0$ and $M_1$ given in \eqref{MobiusTheta}, we have
\be\label{M0M1Exp}
M_0=\left(
\begin{array}{cccc}
i &0\\
0 &-i
\end{array}
\right),
\quad 
M_1=\left(
\begin{array}{cccc}
i\cosh(2\theta) &-i\sinh(2\theta)\\
i\sinh(2\theta) &-i\cosh(2\theta)
\end{array}
\right).
\ee
These two matrices are actually special examples of a larger class of $\SU(1,1)$ matrix defined below
\begin{defn}[Reflection.] Let $M$ be an $\SU(1,1)$ matrix parametrized as \eqref{M_alpha_z}, namely
\begin{equation}
M(\rho,\zeta) =
\frac{1}{\sqrt{1-|\rho|^2}}
\begin{pmatrix}
\sqrt{\zeta} &-\rho^*\frac{1}{\sqrt{\zeta}}\\
-\rho \sqrt{\zeta} &\frac{1}{\sqrt{\zeta}}
\end{pmatrix}
\end{equation}
we call $M$ a reflection if $\zeta=-1$, i.e. 
\begin{equation}
\label{ReflectionMatrix}
M(\rho, \zeta=-1) =
\frac{1}{\sqrt{1-|\rho|^2}}
\begin{pmatrix}
i &i \rho^*\\
-i \rho &-i\\
\end{pmatrix}
\end{equation}
is a reflection. This condition is equivalent to demanding matrix $M$ is traceless $\Tr (M(\rho,\zeta)) = 0 $. 

One important property of the reflection matrix is that it squares to $-1$, i.e. 
\begin{equation}
    M(\rho,-1)^2=-1\,.
\end{equation}

Apparently, a reflection matrix is elliptic since its trace is smaller than 2. But the product of two distinct reflection matrices are hyperbolic, i.e, $|\Tr (M(\rho_1,-1)M(\rho_2,-1))|>2$. 

The reflection matrix will also play an important role in later 
discussions on both the quasi-periodically driven and the randomly driven CFTs\cite{RandomCFT}.
\end{defn}

Applying to our case where we have two distinct reflection matrix $M_0$ and $M_1$, we conclude that $\Pi_2 = M_0 M_1$ is hyperbolic and therefore induces a heating phase for $(T_0, T_1)=(L/2, L_{\text{eff}}(\theta)/2)$.\footnote{Here we comment that for two arbitrary 
elliptic and non-commuting Hamiltonians $H_0$ and $H_1$, 
the corresponding $\SU(1,1)$ matrices can be tuned to
\textit{reflection} matrices by choosing appropriate 
$T_0$ and $T_1$ (See appendix.\ref{Sec: TechnicalDetail}).
At this point, the system will always be in a heating phase.}  
Indeed, we can explicitly check that the Lyapunov exponent has a simple form (assuming $\theta>0$)
\begin{equation}
\label{LyapunovTheta}
\lambda_L=\theta\,.
\end{equation} 
In addition, we have
\be
\Pi_{np}=(\Pi_p)^n=(M_0M_1)^n
=(-1)^n
\left(
\begin{array}{cccc}
\cosh(2n \theta) &-\sinh(2n\theta)\\
-\sinh(2n\theta) &\cosh(2n\theta)
\end{array}
\right)\,,
\ee
which further fixes the location of the energy-momentum peak to be at $x=0$ ($x=L$) for the chiral (anti-chiral) mode (c.f. Eq.~\eqref{EnergyMomentumQ1} for the open boundary condition discussed here). In fact, the chiral and anti-chiral peaks switch positions after each driving period.

The total energy and entanglement entropy are also expressible using $\theta$ (c.f. \eqref{HalfEE_general} and \eqref{EnergyTotal})
\begin{align}\label{EE_energy}
    &E(N=np)=\frac{\pi c}{8L}\cdot \cosh(4n\theta)-\frac{\pi c}{6L}, \\
    &S_A(N=np)-S_A(0)=\frac{c}{3}\cdot 2n\cdot \theta,
\end{align}
where $p=2$ and we consider $A=[0, L/2]$ here. 
We will compare this CFT result with the numerical calculation on a lattice model in the next subsection.


\subsubsection{Numerical simulation on lattice}
\label{Sec:LatticePeriodic}

In Ref.~\cite{wen2018floquet}, the authors compare the CFT and 
lattice calculations on the entanglement entropy
evolution in a periodically driven CFT. 
It was found that the comparison agrees very well in the non-heating phases, but
deviates in the heating phase. The heuristic reason is that the two driving Hamiltonians
in \cite{wen2018floquet} are 
chosen as $H_0$ and $H_{\theta=\infty}$, which result in a large Layapunov exponent
in the heating phase (See Fig.~\ref{Lyapunov_PeriodicDriving}).
Then the system can be easily heated up with only a few driving steps. 
It is noted that the higher energy modes in a lattice system
are no longer well described by the CFT, which results in a deviation between
the lattice and CFT calculations.
Now, by considering the general $H_{\theta}$, we can tune the system to have a small heating rate by choosing a small $\theta$.

\begin{figure}[t]
\centering
\includegraphics[width=6.20in]{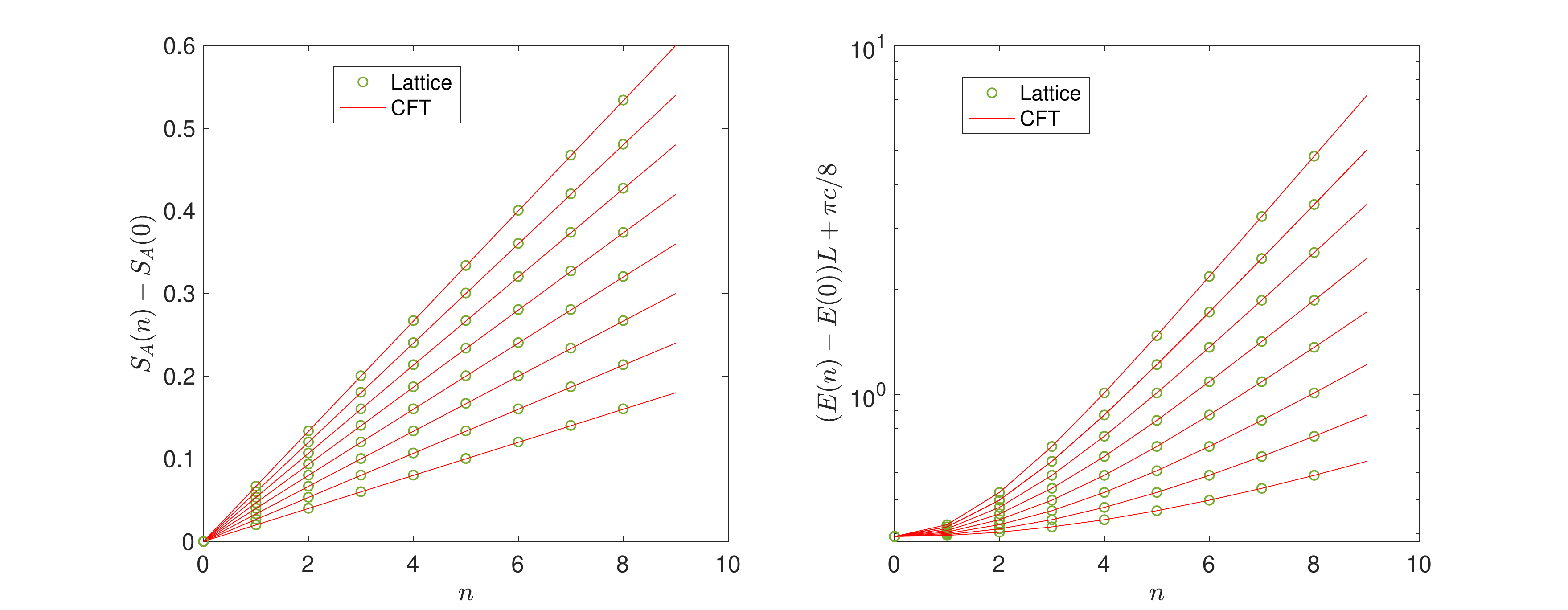}
\caption{
Comparison of the CFT and lattice calculations on the
entanglement entropy (left) and the total energy (right) evolution in 
the heating phase of a periodically driven CFT.
The CFT is periodical driving with $H_0$ and $H_{\theta}$ with time intervals $T_0=L/2$ and $T_1=L_{\text{eff}}(\theta)/2$, respectively.
From bottom to top, we choose $\theta=0.03$, $0.04$, $0.05$, $0.06$, $0.07$,
$0.08$, $0.09$, and $0.1$.
The CFT results are plotted according to Eq.~\eqref{EE_energy}.
}
\label{EELinear1}
\end{figure}


The lattice model we consider is a free fermion lattice,
which has finite sites $L$ with open boundary conditions.
We prepare the initial state as the ground state of
\begin{equation}
    H_0=\frac{1}{2}\sum_{j=1}^{L-1}c_j^{\dag}c_{j+1}+h.c.
\end{equation}
 with half filling.
The SL$_2$ deformed Hamiltonian has the form
\begin{equation}
    H_1=\frac{1}{2}\sum_{j=1}^{L-1}f(j)
c_j^{\dag}c_{j+1}+h.c.
\end{equation}
where $f(j)=1-\tanh(2\theta)\cdot \cos\frac{2\pi  j}{L}$,
$c_j$ are fermionic operator satisfying the
anticommutation relations $\{c_j,c_k\}=\{c_j^{\dag},c_k^{\dag}\}=0$, and
$\{c_j,c_k^{\dag}\}=\delta_{jk}$.
One can refer to the appendix in Ref.~\cite{wen2018floquet} 
for the details of calculation of
the entanglement entropy and correlation functions.
The comparison of the numerical and CFT calculations on both the entanglement and energy time evolution can be found in
Fig.~\ref{EELinear1}. The agreement is remarkable.

One can also refer to Appendix. \ref{app: CompareCFTLattice_Fibonacci} 
for the interesting case that when the entanglement cut lies at the center of 
both the chiral and anti-chiral energy-momentum density peaks, the entanglement
entropy can decrease linearly in time.

\section{Quasi-periodic driving}
\label{Sec: QuasiPeridic}

In this section, we will study the non-equilibrium 
dynamics in a quasi-periodically driven CFT with 
SL$_2$ deformed Hamiltonians. 
We would like to understand the following two questions in this section:
\begin{enumerate}
    \item How does the phase diagram change as we shift the periodic driving protocol to the quasi-periodic driving?
    \item What is the generic feature of the entanglement/energy 
evolution in the quasi-periodically driven CFT?
\end{enumerate}


As an initial effort to answer these questions, 
we will mainly focus on the case of quasi-periodical driving with
a Fibonacci sequence, which is simpler to handle 
compared to a more general quasi-periodic sequence.
Our setup is closely related to the Fibonacci quasi-crystal, 
which was proposed
in the early 1980's by Kohmoto, Kadanoff, 
and Tang\cite{KKT1983},
and Ostlund, Pandit, Rand, Schellnhuber, and Siggia\cite{OPRSS1983}.
It was observed and later proved that
the spectrum of Fibonacci Hamiltonian is a Cantor set of
zero Lebesgue measure\cite{KKT1983,OPRSS1983,sutHo1989singular}.
Since then, the Fibonacci dynamics has been extensively 
studied in both physics and mathematics. 
See, e.g., Ref.~\cite{damanik2016fibonacci} for a recent review.
For simplicity, in the following we may call the 
quasi-periodically driven CFT with a Fibonacci sequence 
as a Fibonacci driven CFT.

In the end of this section, we also discuss another 
kind of quasi-periodic driving with Aubry-Andr\'e-like
sequence by focusing on the properties
of its phase diagram.

\subsection{Fibonacci driving and relation to quasi-crystal}
\label{Sec: Fibonacci driving CFT}

We start with an introduction to the setup and tools we use to analyze the Fibonacci driving, many of which are borrowed from the rich literature of Fibonacci quasi-crystals. 

\subsubsection{Setup and trace map}
\label{Sec:FibonacciSetup}

\begin{figure}[t]
    \centering
    \begin{tikzpicture}[baseline={(current bounding box.center)}]
\draw [>=stealth,->] (0pt, -20pt)--(0pt,40pt);

\draw [thick](0pt,20pt)--(20pt,20pt);
\draw [thick][dashed](20pt, 0pt)--(20pt,-20pt);
\draw [thick](20pt,20pt)--(20pt,0pt);
\draw [thick](20pt,0pt)--(40pt,0pt);

\draw [thick](40pt,0pt)--(40pt,20pt);
\draw [thick][dashed](40pt, 0pt)--(40pt,-20pt);

\draw [thick](40pt,20pt)--(60pt,20pt);
\draw [dashed][thick](60pt,20pt)--(60pt,-20pt);
\draw [thick](60pt,20pt)--(80pt,20pt);

\draw [thick](80pt,0pt)--(80pt,20pt);
\draw [thick][dashed](80pt, 0pt)--(80pt,-20pt);
\draw [thick][dashed](100pt, 0pt)--(100pt,-20pt);

\draw [thick](80pt,0pt)--(100pt,0pt);
\draw [thick](100pt,20pt)--(100pt,0pt);

+100

\draw [thick](0+100pt,20pt)--(20+100pt,20pt);
\draw [thick][dashed](20+100pt, 0pt)--(20+100pt,-20pt);
\draw [thick](20+100pt,20pt)--(20+100pt,0pt);
\draw [thick](20+100pt,0pt)--(40+100pt,0pt);

\draw [thick](40+100pt,0pt)--(40+100pt,20pt);
\draw [thick][dashed](40+100pt, 0pt)--(40+100pt,-20pt);

\draw [thick](40+100pt,20pt)--(60+100pt,20pt);
\draw [dashed][thick](60+100pt,20pt)--(60+100pt,-20pt);
\draw [thick](60+100pt,20pt)--(80+100pt,20pt);

\draw [thick](80+100pt,0pt)--(80+100pt,20pt);
\draw [thick][dashed](80+100pt, 0pt)--(80+100pt,-20pt);
\draw [thick][dashed](100+100pt, 0pt)--(100+100pt,-20pt);

\draw [thick](80+100pt,0pt)--(100+100pt,0pt);
\draw [thick](100+100pt,20pt)--(100+100pt,0pt);

\draw [thick](200pt,20pt)--(240pt,20pt);
\draw [thick][dashed](220pt, 20pt)--(220pt,-20pt);
\draw [thick][dashed](240pt, 0pt)--(240pt,-20pt);
\draw [thick][dashed](260pt, 0pt)--(260pt,-20pt);
\draw [thick](240pt,20pt)--(240pt,0pt);
\draw [thick](240pt,0pt)--(260pt,0pt);

\node at (-10pt,23pt){$H_A$};
\node at (-10pt, 3pt){$H_B$};

\node at (10pt,27pt){$T_A$};
\node at (30pt,7pt){$T_B$};
\node at (50pt,27pt){$T_A$};
\node at (70pt,27pt){$T_A$};
\node at (90pt,7pt){$T_B$};

\node at (10+100pt,27pt){$T_A$};
\node at (30+100pt,7pt){$T_B$};
\node at (50+100pt,27pt){$T_A$};
\node at (70+100pt,27pt){$T_A$};
\node at (90+100pt,7pt){$T_B$};

\node at (210pt,27pt){$T_A$};
\node at (230pt,27pt){$T_A$};
\node at (250pt,7pt){$T_B$};

\draw [>=stealth,->] (0pt, -20pt)--(275pt,-20pt);
\node at (273pt, 0pt){$\cdots$};
\node at (280pt, -20pt){$t$};
\node at (145pt, -40pt){Fibonacci quasi-periodical driving};


\end{tikzpicture}
    \caption{A Fibonacci driving is generated by two unitaries $U_A= e^{-i H_A T_A}$ and $U_B=e^{-i H_B T_B}$ following the pattern of the Fibonacci bitstring $10110101...$}
    \label{fig:fib}
\end{figure}
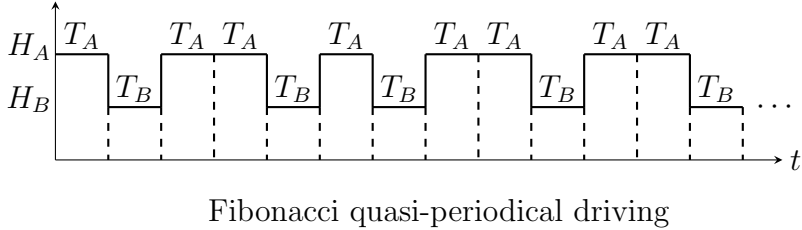

A Fibonacci driving in this paper is generated by two unitaries $U_A= e^{-i H_A T_A}$ and $U_B=e^{-i H_B T_B}$ following the pattern of the Fibonacci bitstring $\{X_j\}$ defined in Appendix.~\ref{appendix Fib}
\begin{equation}
X_{j=1,2,3\ldots} = 10110101 \ldots
\label{eqn: fib string}
\end{equation}
The Hamiltonians $H_A$, $H_B$ are chosen to be the $\SL_2$ deformed Hamiltonian same as the ones used in the previous sections. 
Therefore, each unitary $U_{A(B)}$ corresponds to a conformal map $M_{A(B)}\in \SU(1,1)$ and the final conformal map $\Pi_n$ that determines the operator evolution is given as a product
\begin{equation}
    \Pi_n = \prod_{j=1}^n M_j \,, \quad M_j = X_j M_A + (1-X_j) M_B.
\end{equation}
For example, the first few matrices 
\begin{eqnarray}
\Pi_n = M_A M_B M_A M_A M_B M_A M_B M_A \ldots
\end{eqnarray}
An useful property of the Fibonacci driving $\Pi_n$ is that for $n$ being a Fibonacci number\footnote{Our convention for the Fibonacci number is that $
    F_k = F_{k-1}+F_{k-2}$, $F_1=F_0=1
$.
} $F_k$ with $k\geq 3$
there is a recurrence relation for its trace
\begin{equation}\label{TraceMap}
    x_{F_{k+1}} = 2 x_{F_k} x_{F_{k-1}} -x_{F_{k-2}} \,, \quad \text{where} \quad x_{F_k}=\frac{1}{2} \Tr (\Pi_{F_k}) = \frac{1}{2} \Tr (\Pi_{F_k}^{-1})  \,.
\end{equation}
This relation was used in quasi-crystal literature, e.g. see Ref.~\cite{KKT1983}. Also see Appendix~\ref{appendix Fib} for a derivation following the substitution rule of the Fibonacci bitstring and the property that $\det \Pi_n = 1$. 
The initial conditions for this recurrence relation can be taken as 
\begin{equation}
\label{InitialCondition}
x_{F_1}=\frac{1}{2} \Tr (M_A), \quad x_{F_2}= \frac{1}{2}\Tr (M_A M_B), \quad x_{F_3}= \frac{1}{2} \Tr (M_A M_B M_A)\,.
\end{equation}
It is sometime convenient to define an auxiliary $x_{F_0}=\frac{1}{2} \Tr (M_B)$ regarded as a different element from $x_{F_1}$ although $F_0=F_1=1$. The auxiliary element $x_{F_0}$ is defined such that the recurrence relation \eqref{TraceMap} also holds for $k=2$. 

For SU$(1,1)$ matrices, we have $x_{F_k}\in \mathbb R$. 
To visualize the trace map, let us introduce a three dimensional vector $(x_{F_k}, y_{F_k}, z_{F_k}):=(x_{F_k}, x_{F_{k-1}}, x_{F_{k-2}})$, then the trace map in \eqref{TraceMap} can be expressed as the following mapping between points in three dimensional space 
\be\label{TraceMap2}
\wideboxed{
T: \, \mathbb R^3\to \mathbb R^3,\quad T(x_{F_k}, \, y_{F_k}, \,z_{F_k})=(2x_{F_k}\cdot y_{F_k}-z_{F_k}, \,x_{F_k}, \, y_{F_k}),
}
\ee
with the initial condition $l=(x_{F_3},x_{F_2},x_{F_1})$ given in Eq.~\eqref{InitialCondition}, or alternatively we can use $(x_{F_2},x_{F_1},x_{F_0})$ with the auxiliary element $x_{F_0}$. Remarkably, the trace map 
has a constant of motion \cite{KKT1983}
\be\label{ConstantOfMotion}
\wideboxed{
I=-1+x^2_{F_k}+y^2_{F_k}+z^2_{F_k}-2 x_{F_k} \cdot y_{F_k} \cdot z_{F_k},
}
\ee
see Appendix~\ref{appendix Fib} for an explicit check that $I$ is independent of $k$. 


\subsubsection{Example with $H_\theta$ and fixed point}
\label{Sec:HthetaExample}

\begin{figure}[t]
\begin{center} 
\includegraphics[width=5.00in]{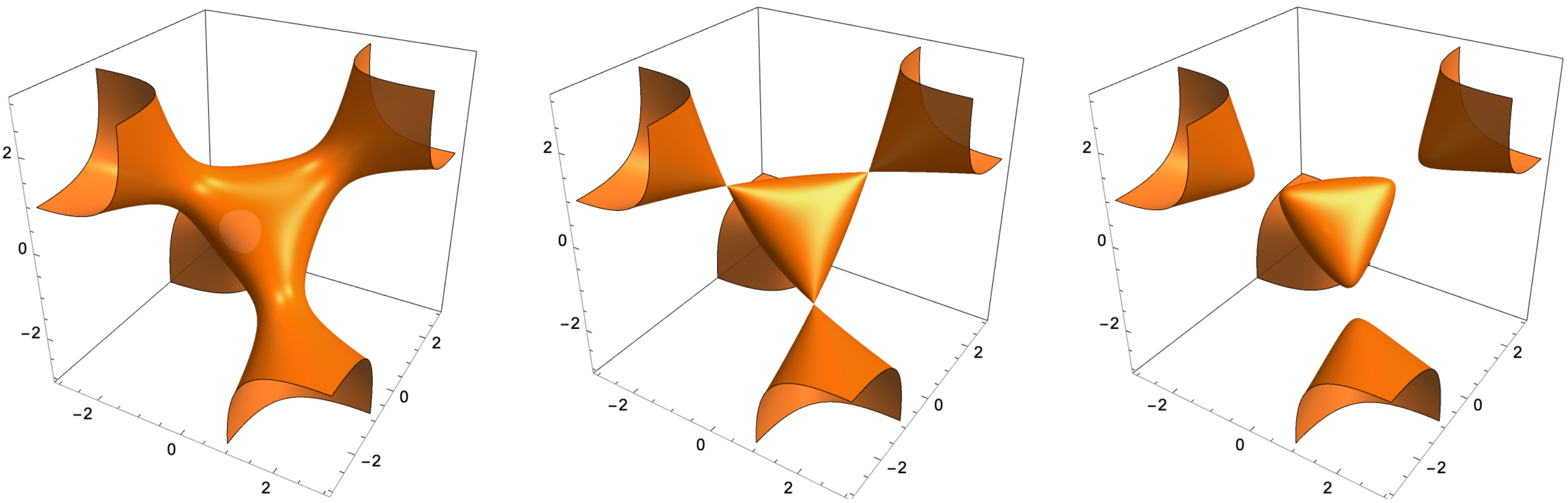}
\end{center}
\caption{
Two dimensional manifolds $\mathcal{M}$ determined by Eq.~\eqref{ConstantOfMotion}, 
with $I=0.5$, $0$, and $-0.1$, respectively.
The manifold with $I=0$ is called the Cayley cubic.
}
\label{Itopology}
\end{figure}

Let us now take explicit example of $\SL_2$ deformed driving Hamiltonians. Consider $(H_A,\, T_A)=(H_\theta,\, T_1)$ and $(H_B, \, T_B)=(H_0, \, T_0)$, where $H_0$ is taken as the CFT Hamiltonian with a uniform Hamiltonian density, and $H_\theta$ is taken as the SL$_2$ deformed one in Eq.~\eqref{H_theta_A} 
\begin{equation}
H_{\theta}=
\int_0^L \left(1-\tanh(2\theta)\cdot \cos\frac{2\pi  x}{L}\right) T_{00}(x) dx, \quad  \theta>0.
\end{equation}
The corresponding conformal transformation $M_A$ and $M_B$ has been computed in \eqref{MobiusTheta} and copied here
\begin{equation}
M(H_{\theta}, T_{\theta}) = \begin{pmatrix}
\alpha & \beta \\
\beta^* & \alpha^*
\end{pmatrix}
\quad \text{with} \quad \left\{
\begin{split}
&\alpha=\cos{\left( \frac{\pi T_{\theta}}{L_{\text{eff}}} \right)} + i\cosh(2\theta)\cdot \sin{\left( \frac{\pi T_{\theta} }{L_{\text{eff}}} \right)},\\
&\beta=- i\sinh(2\theta)\cdot\sin{\left( \frac{\pi T_{\theta}}{L_{\text{eff}}} \right)}.
\end{split}
\right.
\end{equation}
And $L_{\rm eff}=L\cosh (2\theta)$ denotes the effective length of the system under $H_\theta$. 
Therefore, the initial conditon for the trace map is given as
follows
\begin{equation}\label{InitialConditionTheta}
\begin{aligned}
& x_{F_0}=\cos\left(\frac{\pi T_0}{L}\right),\quad x_{F_1}=\cos\left(\frac{\pi T_1}{L_{\text{eff}}}\right), \\
& x_{F_2}=\cos\left(\frac{\pi T_1}{L_{\text{eff}}}\right)\cdot \cos\left(\frac{\pi T_0}{L}\right)-\cosh(2\theta)\cdot \sin\left(\frac{\pi T_1}{L_{\text{eff}}}\right)\cdot \sin\left(\frac{\pi T_0}{L}\right).
\end{aligned}
\end{equation}
And the invariant $I$ defined in \eqref{ConstantOfMotion} is 
\be\label{ConstantGeneral2}
I=\left(\cosh^2(2\theta)-1\right)\cdot \sin^2 \left(\frac{\pi T_0}{L}\right) \cdot \sin^2 \left(\frac{\pi T_1}{L_{\text{eff}}}\right) \in [0,\cosh^2(2\theta)-1]
\ee
Generally speaking, the invariant $I$ constrains the motion of $(x_{F_k},y_{F_k},z_{F_k})$ on a two dimensional manifold $\calM$. 
For $I\in \RR$, there are three topologically distinct scenario as shown in  Fig.~\ref{Itopology}

\begin{enumerate}
\item
$I=0$: The manifold $\mathcal{M}$ can be decomposed into five parts. 
The central part is the curvilinear tetrahedral (`island'), with the vertices/singularities at $A\, (1,1,1)$, $B\,(1,-1,-1)$, $C\,(-1,1,-1)$, and $D\,(-1,-1,1)$.
The tetrahedral is parameterized by $\theta_1$ and $\theta_2$ with $x=\cos(\theta_1+\theta_2)$, $y=\cos \theta_1$, $z=\cos\theta_2$.
The left four parts are funnels. The first funnel is parameterized by $x=\cosh(\theta_1+\theta_2)$, $y=\cosh \theta_1$, and $z=\cosh \theta_2$,
with its vertex at the point $A$. The other three funnels are similar defined with the vertices at $B$, $C$, and $D$.
In the Fibonacci driven CFT,
this case corresponds to $T_0/L\in \ZZ$ or $T_{1}/L_{\text{eff}}\in \ZZ$. 
Physically, this corresponds to a single quantum quench which is not our focus here. 

\item $I>0$: 
The four vertices $A$, $B$, $C$ and $D$ are replaced with four necks, which connect the central part (`island') of the manifold to
the four funnels. The whole manifold is therefore non-compact. 
This case corresponds to all the nontrivial choices of $(T_0, T_1)$ in our setting \eqref{ConstantGeneral2}.
It turns out that for almost all the initial points on the manifold, they will flow
to infinity under the trace map in \eqref{TraceMap}.\cite{damanik_2017}

\item $I<0$: The central part (`island') becomes disconnected to the outside funnels and therefore compacted. 
This case is absent in 
our setting for the Fibonacci driving.
Nevertheless, this case may be related to some non-Hermitian Hamiltonian or non-unitary time evolution and deserves a careful study in future. 
\end{enumerate}

\begin{figure}[t]
\begin{center} 
\includegraphics[width=4.20in]{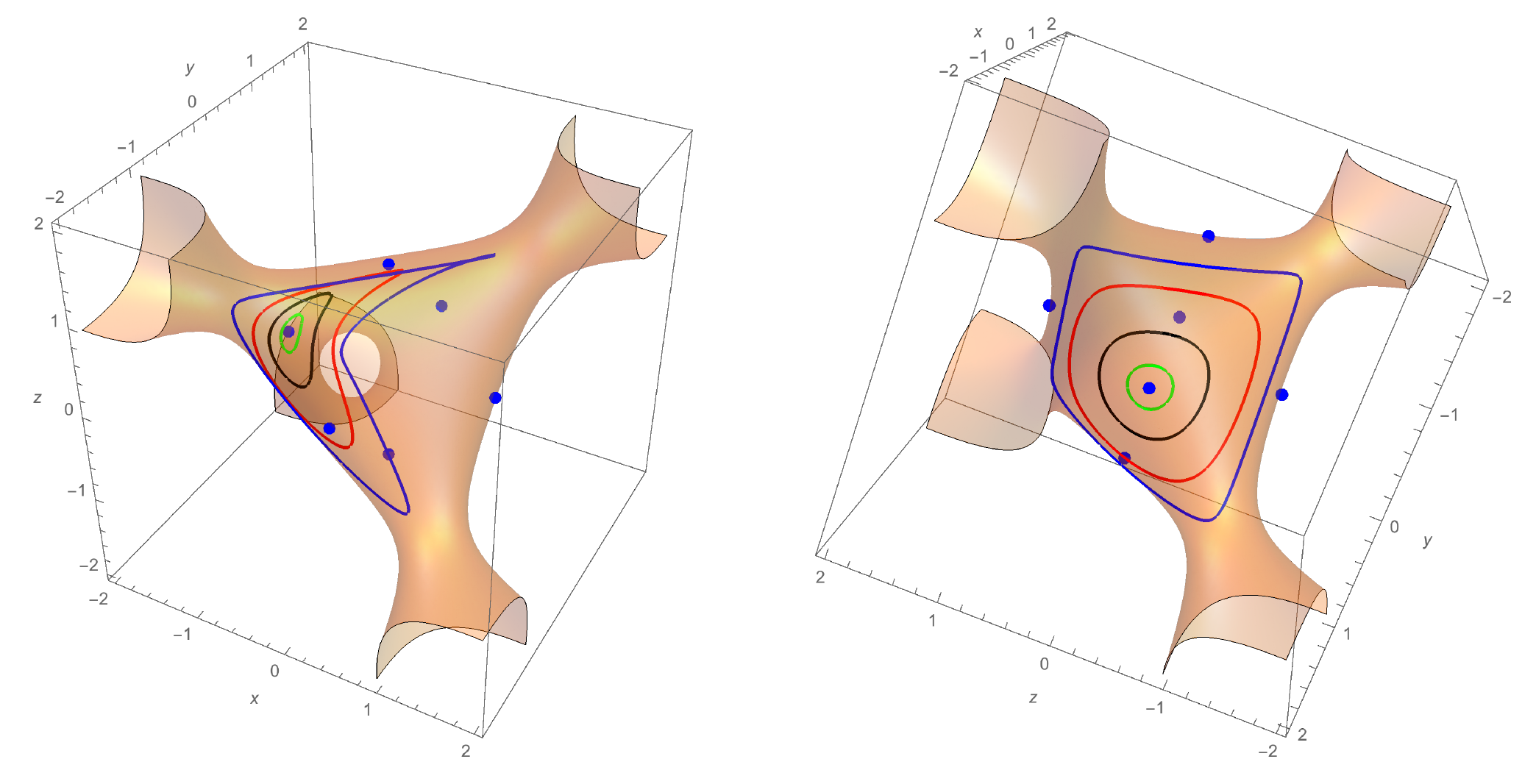}
\end{center}
\caption{Return orbits and initial conditions: here we plot $\calM$ with $I=1/4$ fixed. The right plot is the same as the left one  with a different angle of view. 
The six blue solid dots correspond to the period-6 returning orbit, or equivalently speaking the fixed points of $T^6$ action. This orbit can be viewed as the limit of a family of initial conditions: for each fixed $\theta$, the allowed initial condition forms a line (two parameters $T_0,T_1$ with one constraint \eqref{ConstantGeneral2}.)
In the figure, we set $\theta=1$, $0.4$, $0.27$, and $0.245$ for four loops from big to small. Then, if we further decrease $\theta$ to a critical value $\theta^*$ such that $I=\cosh^2(2\theta^*)-1$, then we have essentially only one possible initial condition $T_0=L/2$ and $T_1=L_{\rm eff}/2$ which will generate the fixed points.
}
\label{InitialLinePlot}
\end{figure}

For a fixed $I>0$, one can tune two of the three parameters $(\theta,T_0,T_1)$ to move the initial point $(x_{F_3},x_{F_2},x_{F_1})$ on the surface $\calM$ then the orbit under the trace map 
\begin{equation}
    T(x_{F_k}, \, y_{F_k}, \,z_{F_k})=(2x_{F_k}\cdot y_{F_k}-z_{F_k}, \,x_{F_k}, \, y_{F_k}),
\end{equation}
is completely determined. As we will show in the following sections, most of the orbits will escape to the infinity and resulting an heating phase. However, there still exists returning orbit, e.g. when we have two zeros in the initial condition $(x_{F_3},x_{F_2},x_{F_1})$, we will end up with a period 6 orbits
\begin{equation}
   (0, 0, a)\to (-a, 0, 0) \to (0, -a, 0)
\to (0,0,-a) \to (a,0, 0)\to (0, a, 0)\to (0, 0, a)\to \ldots
\end{equation}
with $a=(1+I)^{1/2}$, see Fig.~\ref{InitialLinePlot} for an illustration. We will call such initial points that correspond to the non-heating point as ``fixed point'', in the sense that those points are fixed under $T^6$ action.


\subsubsection{Phase diagram: From periodic to quasi-periodical driving}
\label{Sec:PhaseDiagramPeriodicFib}

In this section, we show the shape of the phase diagram of a Fibonacci driven CFT via numerically approaching the Fibonacci bitstring by its finite truncation. 
This strategy has been proven useful 
in the analysis of the energy spectrum of a
Fibonacci quasi-crystal\cite{KKT1983}. 
In the quasi-crystal case, the energy spectrum forms a Cantor set of zero Lebesque measure. 
In this section, we will show numerical evidence of such ``fractal'' structure, while in the next section we will map our phase diagram to the energy spectrum of quasi-crystal and establish the claim. 

Recall that (in Appendix.~\ref{appendix Fib}) we generate the Fibonacci driving using the Fibonacci bitstring 
\begin{equation}
X_j=\chi((j-1)\omega)\,, \quad j=1,2,3\ldots
\end{equation}
where $\chi(t)=\chi(t+1)$ is a
period-1 characteristic function
\begin{equation}
\label{Fibonacci_V}
\chi(t)=\begin{cases}
1 & ~~ -\omega^3\leq t < \omega^2\\
0&   ~~ \omega^2\leq t < 1-\omega^3
\end{cases}
\end{equation}
and $\omega = \frac{\sqrt{5}-1}{2}$ is an irrational number with a simple continued fraction representation
\begin{equation}
\omega = \frac{1}{1+\frac{1}{1+\frac{1}{1+\ldots}}}
\end{equation}
Now to approach the Fibonacci bitstring from a periodic string, we can truncate the continued fraction of $\omega$ at finite order $n$ and obtain a rational number (principal convergent) $\omega_n=F_{n-1}/F_n$, namely the ratio of two nearby Fibonacci number. The corresponding bitstring $\{X_j\}$ now has periodicity $F_n$ and therefore produce a periodic driving. We can now use the tools introduced in Sec.~\ref{Sec: PeriodicDriving} to obtain a phase diagram for each $\omega_n$.


\begin{figure}[t]
\centering
\includegraphics[width=6.00in]{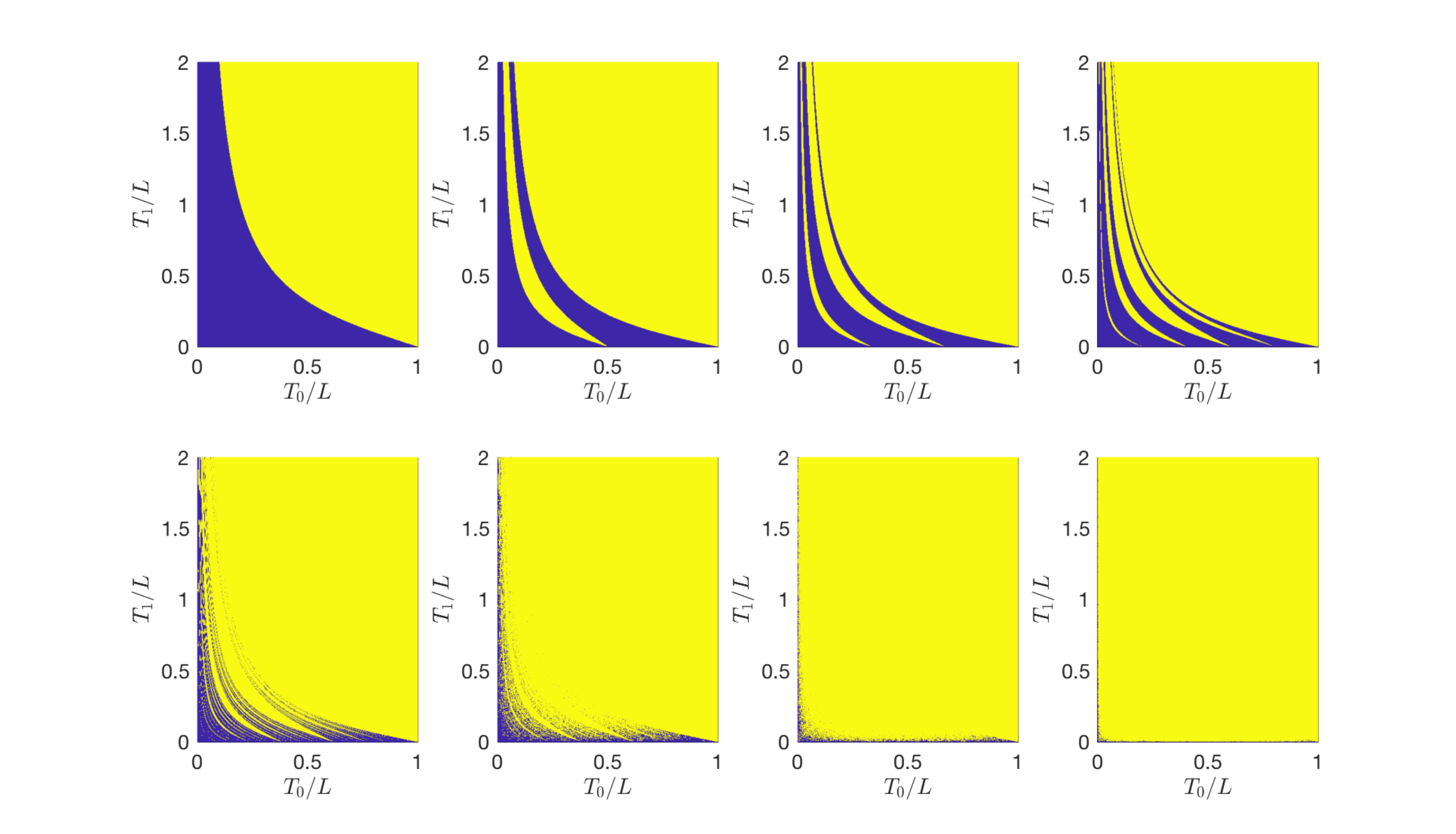}
\caption{
Phase diagrams in a periodically driven CFT with the sequence generated by finitely truncated Fibonacci bitstring, i.e. $\{X_j\}$ with 
 $\omega_n=F_{n-1}/F_n$. Here we choose $n=2$, $4$, $5$, $6$, $10$, $20$, $100$, and $1000$ for 8 plots 
 respectively. The two Hamiltonians we use are $H_0(\theta=0)$ and $H_1(\theta=\infty)$ in \eqref{H_theta_A}.
  The blue (yellow) regions correspond to the heating (non-heating) phases.
}
\label{FibonacciPhase1}
\end{figure}


In Fig.~\ref{FibonacciPhase1}, we show the evolution of phase diagrams
of periodically driven CFTs with protocol $(H_A,\, T_A)=(H_{\theta=\infty},\, T_1)$ and $(H_B, \, T_B)=(H_0, \, T_0)$. 
The phase diagram 
is periodic in $T_0$ direction with period $L$, we only
show the phase diagram within one unit cell $0\leq T_0\leq L$.
As we increase $n$, there are two notable features
\begin{enumerate}
\item The number of regions of the non-heating phases increases with $n$, and tends to 
infinity as $n \rightarrow \infty$.
\item The measure of the non-heating phases decreases with $n$, and tends to zero as 
$n\to \infty$. 
\end{enumerate}

These two features suggest that the non-heating phases in the quasi-periodical driving limit
may form a Cantor set of measure zero, analogous to the feature of the energy spectrum 
in a Fibonacci quasi-crystal. In fact, this is indeed the case, as we will discuss in detail in the 
next subsection.

We also present the evolution of phase diagrams 
 by the Hamiltonians $H_0$ 
and
$H_{\theta}$ with finite $\theta$. 
See Fig.~\ref{FibonacciPhase0} for $\theta=0.5$, and 
 Fig.~\ref{FibonacciPhase3} in the appendix for $\theta=0.2$. 
The two features mentioned above are also observed 
in these cases.
It is noted that for a finite $\theta$ in $H_1(\theta)$, the phase diagram is also periodic in 
$T_1$ direction, with the period $L\cosh(2\theta)$.

\begin{figure}[t]
\centering
\includegraphics[width=6.00in]{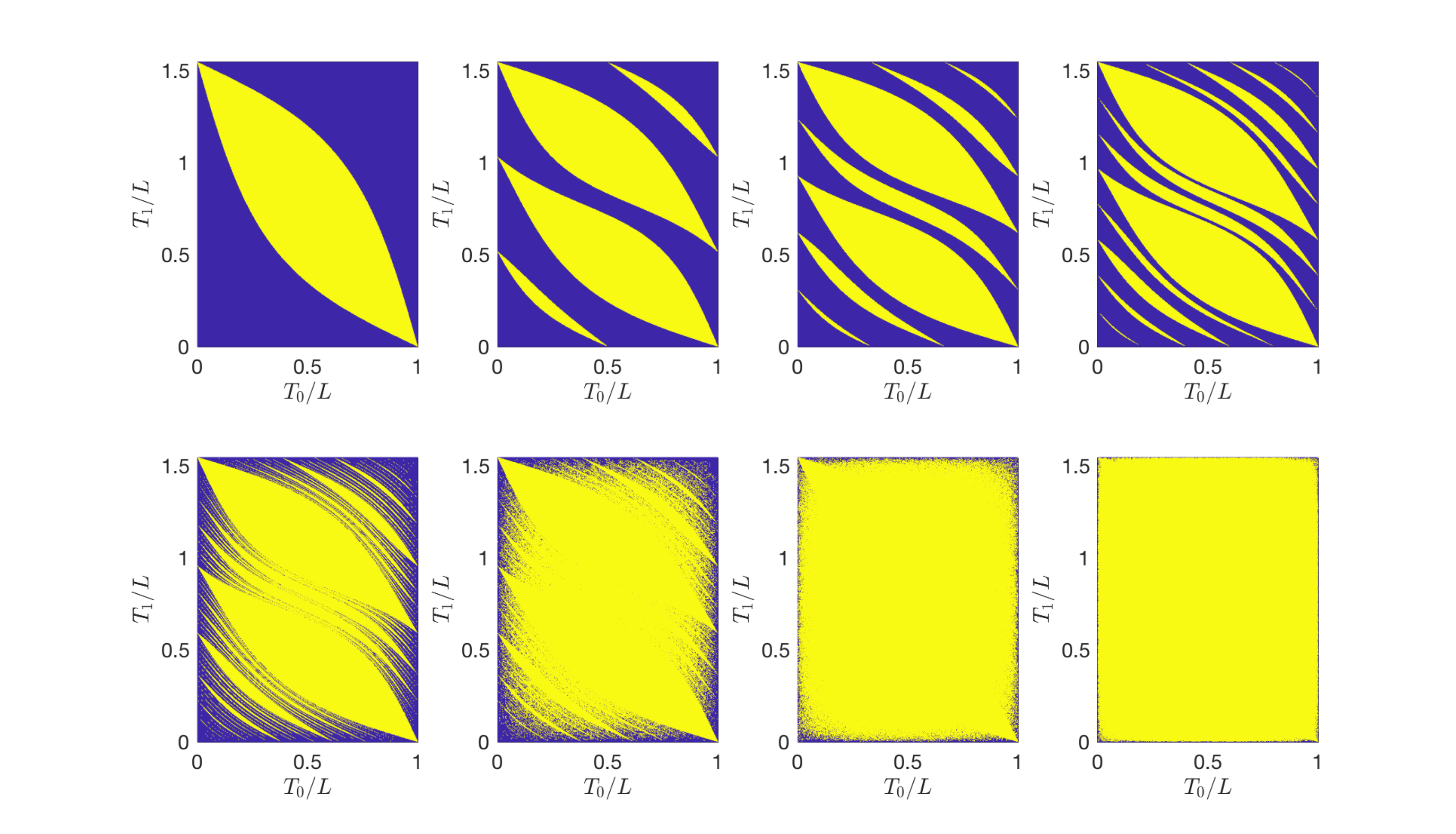}
\caption{
Phase diagrams in a periodically driven CFT with the sequence generated by finitely truncated Fibonacci bitstring, i.e. $\{X_j\}$ with 
 $\omega_n=F_{n-1}/F_n$. Here we choose $n=2$, $4$, $5$, $6$, $10$, $20$, $100$, and $1000$ for 8 plots 
 respectively. The two Hamiltonians we use are $H_0(\theta=0)$ and $H_1(\theta=0.5)$ in \eqref{H_theta_A}.
 The phase diagram is periodic in $T_0$ direction 
 with period $L$ and in $T_1$ direction with
 period $L\cosh(2\theta)\simeq 1.54 L$.
  The blue (yellow) regions correspond to the heating (non-heating) phases.
}
\label{FibonacciPhase0}
\end{figure}

\subsubsection{Exact mapping from a Fibonacci driven CFT to a Fibonacci
quasi-crystal}
\label{Sec: Map_Fibonacci}

The features in the phase diagrams in Fig.~\ref{FibonacciPhase1} and Fig.~\ref{FibonacciPhase0}
suggest that the non-heating phases in the quasi-periodical driving limit may form a Cantor set of 
measure zero. In this subsection, we verify this by performing an exact mapping between the 
phase diagram of a Fibonacci driven CFT and the energy spectrum of a 
Fibonacci quasi-crystal. The latter has been proved mathematically that the energy 
spectrum is indeed a Cantor set \cite{sutHo1989singular} 
(See also Ref.~\cite{damanik_2017} for a review). 

\begin{figure}[h]
\centering
\includegraphics[width=6.0in]{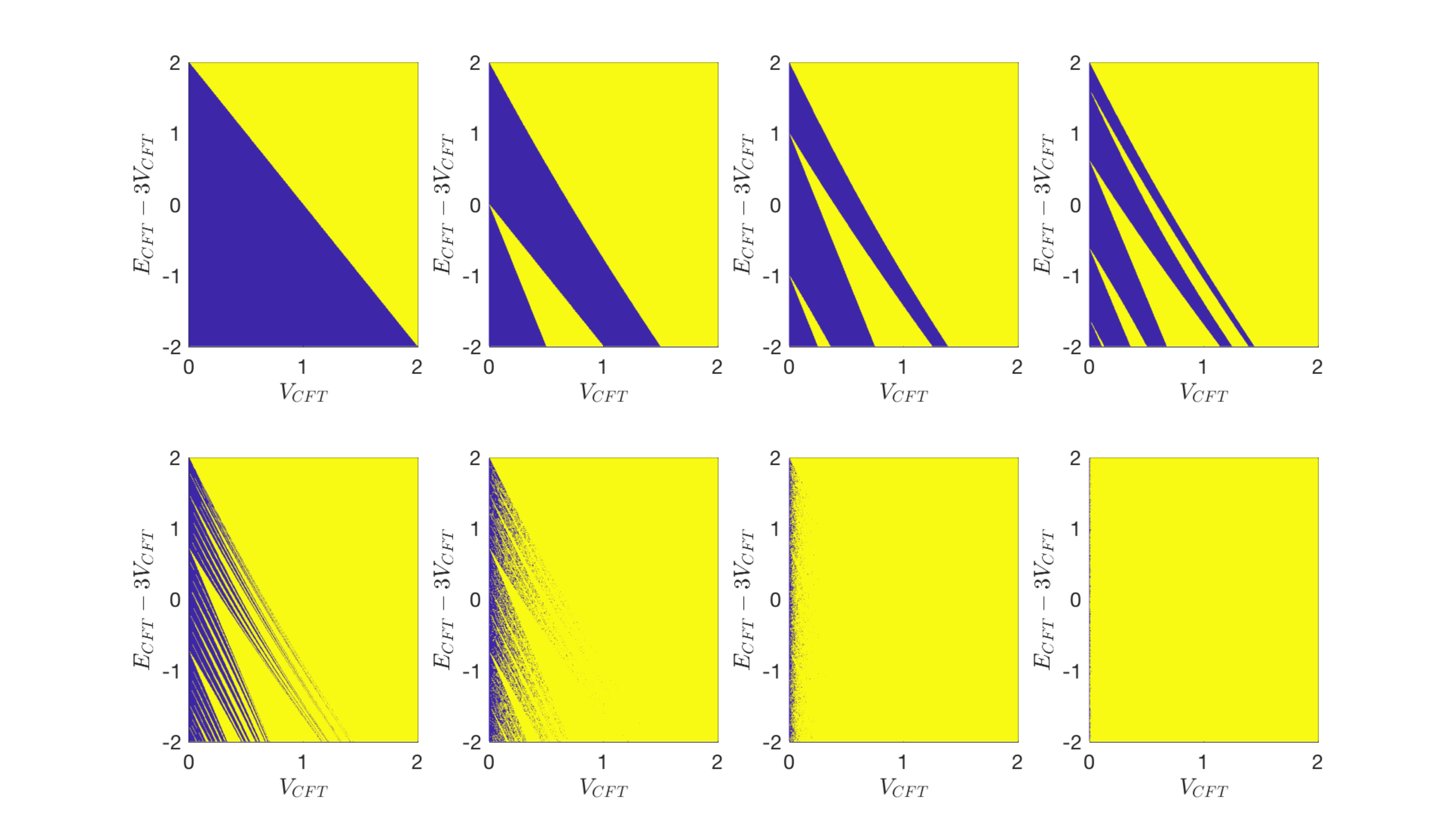}
\caption{
Phase diagrams in a periodically driven CFT with the sequence chosen in \eqref{Fibonacci_V} where
 $\omega_n=F_{n-1}/F_n$.The parameters are the same as those in Fig.~\ref{FibonacciPhase1}
 except that now we change the variables to $V_{CFT}$ and $E_{CFT}$ (See Eq.~\eqref{EV_CFT}).
  The blue (yellow) regions correspond to the heating (non-heating) phases.
}
\label{FibonacciPhaseEV}
\end{figure}

Before introducing the mapping, let us first briefly review the  background of the
Fibonacci quasi-crystal.
We consider the discrete Schr\"odinger operators of the form 
\be\label{SchrodingerOP}
[H \psi]_j=\psi_{j+1}+\psi_{j-1}+V_j \psi_j, \quad j\in \mathbb Z,
\ee
where $\psi_j:=\langle j|\psi\rangle$ is the position-space wavefunction, with $j$ labeling the $j$-th site,
and $V_j$ is the onsite potential.
For eigenvalue problem  
$H \psi=E \psi$, it is useful to consider the transfer matrix
\begin{equation}
    T_j=
\begin{pmatrix}
E-V_j &-1\\
1 &0
\end{pmatrix}
\in \SL(2,\mathbb R).
\end{equation}
Denoting $\Psi_j=(\psi_{j+1},\psi_{j})^T$, we have
\footnote{It is helpful to compare this equation 
with Eq.~\eqref{Psi_n} and Eq.~\eqref{Pi_n} in the time-dependent
driving CFT.
In Eq.~\eqref{Pi_n}, the $\SU(1,1) \simeq \SL(2,\RR)$ matrix $M_j$ may be
considered as a transfer matrix in time direction.
}
\be
\Psi_n=\left(T_n\cdots T_2\cdot T_1\right) \Psi_0.
\ee
In the Fibonacci quasi-crystal, the potential $V_j$ can also be generated by the Fibonacci bitstring $\{X_j \}$ \eqref{eqn: fib potential} as follows
\begin{eqnarray}
V_j = X_j V_A +(1-X_j) V_B.
\end{eqnarray}
The allowed energy spectrum $E$ is determined by requiring that 
\begin{equation}
    \lambda_L:=\lim_{n\to \infty}\frac{1}{n}|| T_n\cdots T_1||=0.
\end{equation}
Defining $\widetilde{T}_{F_n}:=T_{F_n}T_{F_n-1}\cdots T_1$, and 
$x_{F_n}:=\frac{1}{2}\text{Tr}(\widetilde{T}_{F_n})$,
it turns out the traces $\{x_{F_n}\}$ satisfy the same recurrence relation in Eq.~\eqref{TraceMap}.
The only difference between the Fibonacci driving CFTs and the Fibonacci quasi-crystals
is the initial conditions, which we will specify now.
By taking $V_B=-V_A=V$, one can find the initial conditions for the Fibonacci quasi-crystal are\footnote{Here we use the recurrence relation to infer the value of $x_{F_0}$ and $x_{F_{-1}}$ from $(x_{F_{3}}, x_{F_2}, x_{F_{1}})$, the reason we choose to start with $x_{F_{-1}}=1$ for quasi-crystal is that we need a convenient base point to map to the CFT initial point, whose $x_{F_1}$ happens to be $1$ as well. It should be clear later when we construct the mapping.} 
\be\label{InitialCondition_quasi-crystal}
l_{(E; I)}:=(x_{F_{1}}, x_{F_0}, x_{F_{-1}})_{\text{quasi-crystal}}=\left( \frac{E+V}{2}, \frac{E-V}{2},1\right),
\quad \text{where } E\in \mathbb R.
\ee
The invariant $I$ in the constant of motion in Eq.~\eqref{ConstantOfMotion} becomes 
$I=V^2$. In a quasi-crystal, the potential $V$ is fixed, and therefore each $E$ specifies an 
initial condition, which may flow to infinity by iterating the trace map ($E$ is in the gap),
or is bounded ($E$ is in the spectrum).

\begin{figure}[t]
\begin{center} 
\includegraphics[width=6.00in]{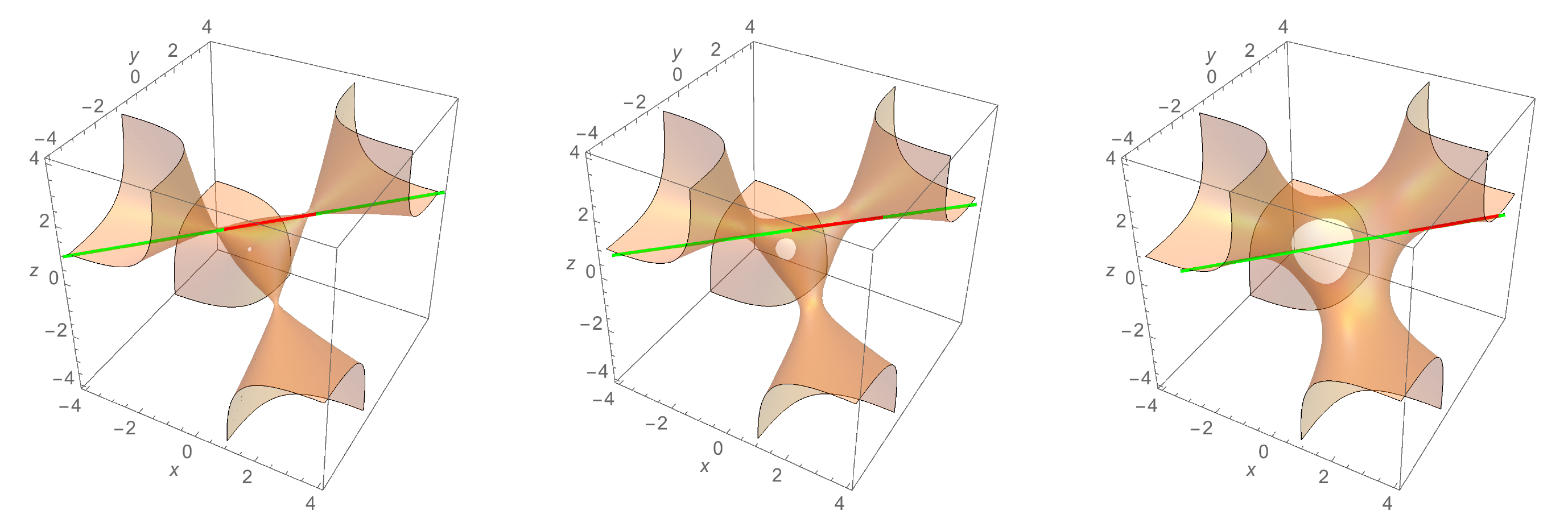}
\end{center}
\caption{
Initial conditions in Eqs.\eqref{InitialCondition_quasi-crystal} and  
\eqref{EV_CFT}
on the two dimensional manifolds $\mathcal{M}$ determined by Eq.~\eqref{ConstantOfMotion}, 
with $I=V^2=0.1^2$, $0.5^2$, and $1.5^2$, respectively.
The green solid lines correspond to the initial condition line $l_{(E,I)}$ in Eq.~\eqref{InitialCondition_quasi-crystal} 
for a quasi-crystal, and the red solid lines correspond to $l_{(E_{\text{CFT}}; I)}$ of a fixed length $2$ in Eq.~\eqref{EV_CFT}
for a quasi-periodically driven CFT. 
For smaller $I$, $l_{(E_{\text{CFT}}; I)}$ overlaps with $l_{(E,I)}$ mainly in the region with 
$|x|$, $|y|\leq 1, \, z=1$.
As $I$ increases, $l_{(E_{\text{CFT}}; I)}$ overlaps with $l_{(E,I)}$ mainly in the region with 
$|x|$, $|y|> 1, \, z=1$. 
That is, as $I$ increases, $l_{(E_{\text{CFT}}; I)}$ moves from the middle `island' 
into the noncompact funnel. This behavior agrees with the feature of the phase diagram in Fig.~\ref{FibonacciPhaseEV}
where the non-heating phases vanishes for larger $V_{\text{CFT}}$. 
}
\label{InitialLineCFT}
\end{figure}

Next, let us compare the initial conditions in the Fibonacci driven CFT. 
We consider the phase diagrams in Fig.~\ref{FibonacciPhase1}, which correspond to $H_0(\theta=0)$ and $H_1(\theta=\infty)$.
By taking the limit $\theta\to \infty$, the initial conditions in Eq.~\eqref{InitialConditionTheta} become
\footnote{Note that by taking the limit $\theta\to \infty$, we always consider finite $T_0$ and $T_1$
such that $T_0, \, T_1 \ll L\cosh(2\theta)$ when $\theta\to \infty$.
In this case, the initial conditions $(x_{F_{2}}, x_{F_1}, x_{F_{0}})_{\text{CFT}}$ form a straight
line with $y=1$, rather than a closed loop in Fig.~\ref{InitialLinePlot}. See Fig.~\ref{InitialLineCFT}
for the initial conditions $(x_{F_{3}}, x_{F_2}, x_{F_{1}})_{\text{CFT}}$ with $\theta=\infty$.
 }
\begin{equation}
(x_{F_{2}}, x_{F_1}, x_{F_{0}})_{\text{CFT}}=\left(
\cos\frac{\pi T_0}{L}-\frac{\pi T_1}{L}\sin\frac{\pi T_0}{L},
\,\, 1, \,\, \cos\frac{\pi T_0}{L}
\right),
\end{equation}
and the invariant $I$ in Eq.~\eqref{ConstantOfMotion} is
\be\label{Invariant_I_02}
I=\left(\frac{\pi T_1}{L}\right)^2 \cdot \sin^2\left(\frac{\pi T_0}{L}\right).
\ee
To compare with the initial conditions of Fibonacci quasi-crystal, 
here we choose $(x_{F_{3}}, x_{F_2}, x_{F_{1}})$ instead of 
$(x_{F_{2}}, x_{F_1}, x_{F_{0}})$ as the initial condition. Based on Eq.~\eqref{TraceMap}, one can obtain
\be
(x_{F_{3}}, x_{F_2}, x_{F_{1}})_{\text{CFT}}=\left(
\cos\frac{\pi T_0}{L}-2\cdot \frac{\pi T_1}{L}\sin\frac{\pi T_0}{L}, \, \cos\frac{\pi T_0}{L}-\frac{\pi T_1}{L}\sin\frac{\pi T_0}{L}, \, 1\right).
\ee
Now by defining
$
E_{\text{CFT}}:=2\cos\frac{\pi T_0}{L}+3V_{\text{CFT}}, \, V_{\text{CFT}}:=-\frac{\pi T_1}{L}\cdot \sin\frac{\pi T_0}{L},
$
then the initial condition line can be written as:
\be\label{EV_CFT}
l_{(E_{\text{CFT}}; I)}:=
(x_{F_{3}}, x_{F_2}, x_{F_{1}})_{\text{CFT}}=\left( \frac{E_{\text{CFT}}+V_{\text{CFT}}}{2}, \frac{E_{\text{CFT}}-V_{\text{CFT}}}{2},1\right), 
\ee
with the invariant $I$ in Eq.~\eqref{Invariant_I_02} expressed as 
\be\label{I_Vcft}
I=V_{\text{CFT}}^2.
\ee
That is, by redefining variables, we can find a map between the initial conditions in 
Eq.~\eqref{InitialCondition_quasi-crystal} and Eq.~\eqref{EV_CFT}.
With this map, the allowed energy $E$ in the spectrum of a Fibonacci quasi-crystal is mapped to 
the non-heating phase in a Fibonacci driven CFT specified by $E_{\text{CFT}}(T_0, T_1)$, and vice versa.

One should note that, however, on the CFT side, $(E_{\text{CFT}}-3 V_{\text{CFT}})\in [-2, 2]$ always lives in a 
window of finite width. On the quasi-crystal side, we have $(E-3V)\in (-\infty, +\infty)$.
This means the non-heating phases in a quasi-periodically driven CFT are only mapped to 
\textit{part} of the energy spectrum in the Fibonacci quasi-crystal.
This can be intuitively seen by considering the initial condition lines in 
Eqs.\eqref{InitialCondition_quasi-crystal} and \eqref{EV_CFT}
on the two dimensional manifold $\mathcal{M}$ determined by Eq.~\eqref{ConstantOfMotion}.
As shown in Fig.~\ref{InitialLineCFT}, the overlap of $l_{(E_{\text{CFT}}; I)}$ and
$l_{(E; I)}$ is always a straight line of finite length $2$.
For smaller $V$ or $V_{\text{CFT}}$, 
$l_{(E_{\text{CFT}}; I)}$ overlaps with $l_{(E; I)}$ mainly in the region with $|x|, \, |y|\leq 1, |z|=1$
in the middle `island'. For the initial conditions in this region, they are much
easier to be bounded as we iterate the trace map\cite{casdagli1986symbolic}.
As $V$ or $V_{\text{CFT}}$ increases, the overlap of $l_{(E_{\text{CFT}}; I)}$ and
$l_{(E; I)}$ moves gradually from the  `island' in the middle to the `funnel' outside.
Then it becomes more difficult for the initial conditions to stay bounded as we iterate the trace map.
This analysis agrees with the fact that in the phase diagrams in Fig.~\ref{FibonacciPhaseEV}, there are
no non-heating phases observed for large $V_{\text{CFT}}$.

This ``inclusion map'' for small $V_{\text{CFT}}$ is totally fine for our goal: 
Since the energy spectrum of a Fibonacci quasi-crystal 
forms a Cantor set of measure zero, then \textit{part} of the energy spectrum 
(which is a connected and finite region in the parameter space)
is also a Cantor set of measure zero. Then with the exact mapping discussed above,
we conclude that the non-heating phases in the quasi-periodically driven CFT form a Cantor set of measure zero.

\begin{figure}[t]
\begin{center} 
\subfloat{\includegraphics[width = 3.2in]{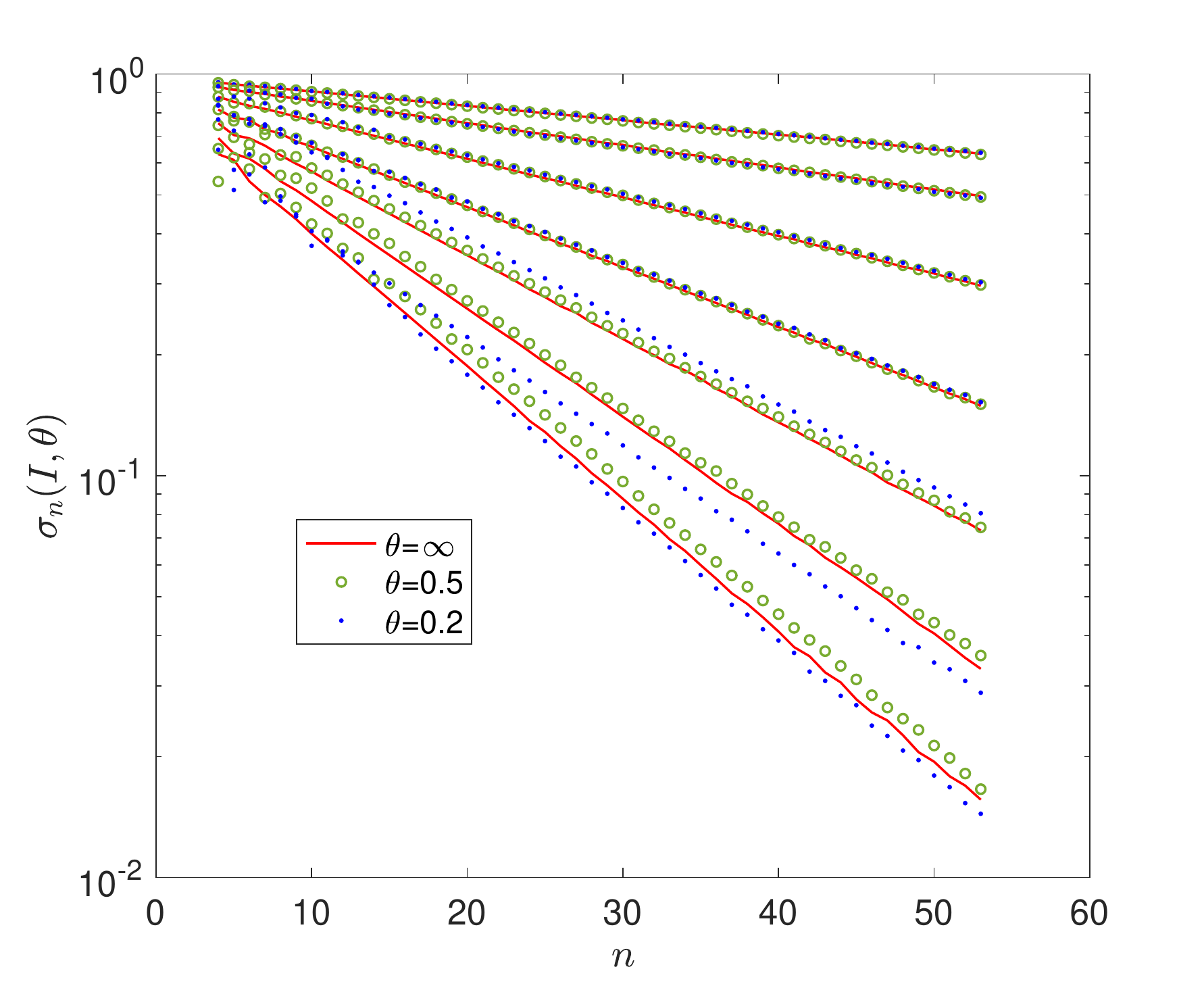}}
\subfloat{\includegraphics[width = 3.2in]{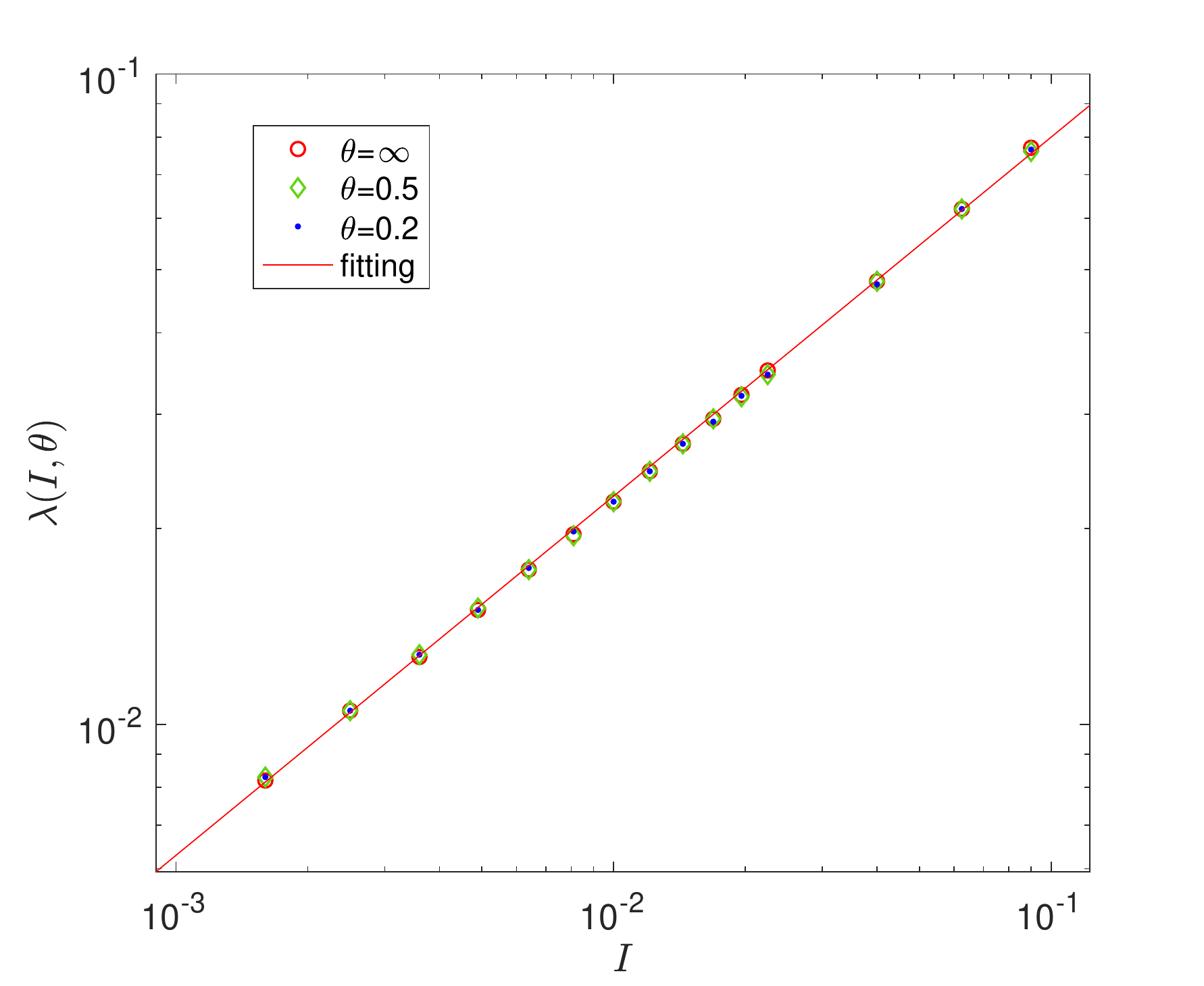}} 
\end{center}
\caption{
Left:
Measure of the non-heating phases $\sigma_n(I,\theta)$ in the phase diagram in Fig.~\ref{FibonacciPhaseEV}
($\theta=\infty$), Fig.~\ref{FibonacciPhaseEV2} ($\theta=0.5$), and Fig.~\ref{FibonacciPhaseEV3} ($\theta=0.2$)
as a function of $n$ for different $I$.
  From top to bottom, we consider $I=0.04^2$, $0.06^2$, $0.1^2$, $0.15^2$, $0.2^2$, $0.25^2$
  and $0.3^2$, respectively.
One can find that the measure is $\sigma_n(I)\propto e^{-\lambda(I,\theta) \cdot n}$,
where $n$ corresponds to the subscript in $F_n$.
Right: The escape rate $\lambda(I,\theta)$ as a function of $I$. It is found that $\lambda(I,\theta)$
with different $\theta$ collapse to the same line described by $y=a\, x^b$, where 
$a\simeq 0.285$ and $b\simeq 0.552$. }
\label{RatioCFTEV}
\end{figure}

Furthermore, in Fig.~\ref{RatioCFTEV}, we also check explicitly the measure of the non-heating phases in the 
phase diagrams in Fig.~\ref{FibonacciPhaseEV} as we approach the quasi-periodic limit.
The procedure of obtaining the measure is as follows:
Fixing a $V_{\text{CFT}}$ (or equivalently the invariant $I$) in Fig.~\ref{FibonacciPhaseEV},
for each $\omega_n=F_{n-1}/F_n$, there are many `energy bands' of non-heating phases. 
Denoting the band width of the $j$-th band as $d_j(n)$, this band width depends on both $I$ and $\theta$ (which is
$\infty$ here).
Then the measure of non-heating phases  with $\omega_n=F_{n-1}/F_n$ is defined as
\be\label{Eq:Measure}
\sigma_n(I,\theta)=\frac{\sum_j d_j(n,I,\theta)}{E_{\text{max}}(n,I,\theta)-E_{\text{min}}(n,I,\theta)}.
\ee
where $E_{\text{max}}(n,I,\theta)-E_{\text{min}}(n,I,\theta)$
is the total width of the energy window, which is $4$ for $\theta=\infty$.
As seen in Fig.~\ref{RatioCFTEV} (left), it is found that $\sigma_n(I,\theta)$ depends on $n$ as
\be
\sigma_n(I,\theta)\propto e^{-\lambda(I,\theta)\cdot n}.
\ee
That is, the measure of the non-heating phases decreases exponentially as a function of $n$,
and tends to become $0$ in the limit $n\to\infty$. This agrees with the fact that the non-heating
phases in the quasi-periodical driving limit form a Cantor set of \textit{measure zero}.
The decaying rate $\lambda(I,\theta)$ may also be interpreted as the escape rate, since it 
describes the rate of initial conditions in Fig.~\ref{InitialLinePlot} escaping into the infinity.
Also, we remind here that the real driving steps are $F_n$ rather than $n$. And  $F_n\sim  \omega^{-n}$ at $n\rightarrow \infty$,  
the measure of non-heating phases depends on $F_n$ as
$\sigma_n(I,\theta)\propto F_n^{\lambda(I,\theta)/\log \omega}$ for large $n$. That is, $\sigma_n(I,\theta)$ 
decays polynomially as a function of the driving steps $F_n$.
In addition, we check how the decaying rate $\lambda(I,\theta)$
depends on the invariant $I$. As shown in Fig.~\ref{RatioCFTEV} (right),
it is found that $\lambda(I,\theta)$ depends on $I$
as $\lambda(I,\theta=\infty)=a\cdot I^b$, with 
$a\simeq 0.285$ and $b\simeq 0.552$.
This monotonic dependence is reasonable in the sense 
that a smaller $I$ corresponds to a narrower neck connecting 
the `island' and `funnel' (See Fig.~\ref{Itopology} and
Fig.~\ref{InitialLineCFT}), which may suppress the escape rate 
from the island to the funnel.

\subsubsection{Cases that cannot be mapped to Fibonacci quasi-crystal}

The exact mapping studied in the previous subsection applies for the case of $\theta\to \infty$ in $H_1(\theta)$. 
For a finite $\theta$, we do not have such an exact mapping.
Here we hope to study the common features among the phase diagrams with different $\theta$ 
(See, e.g., the phase diagrams in Fig.~\ref{FibonacciPhase1}, Fig.~\ref{FibonacciPhase0}, 
and Fig.~\ref{FibonacciPhase3} in the appendix).

\begin{figure}[t]
\centering
\includegraphics[width=6.0in]{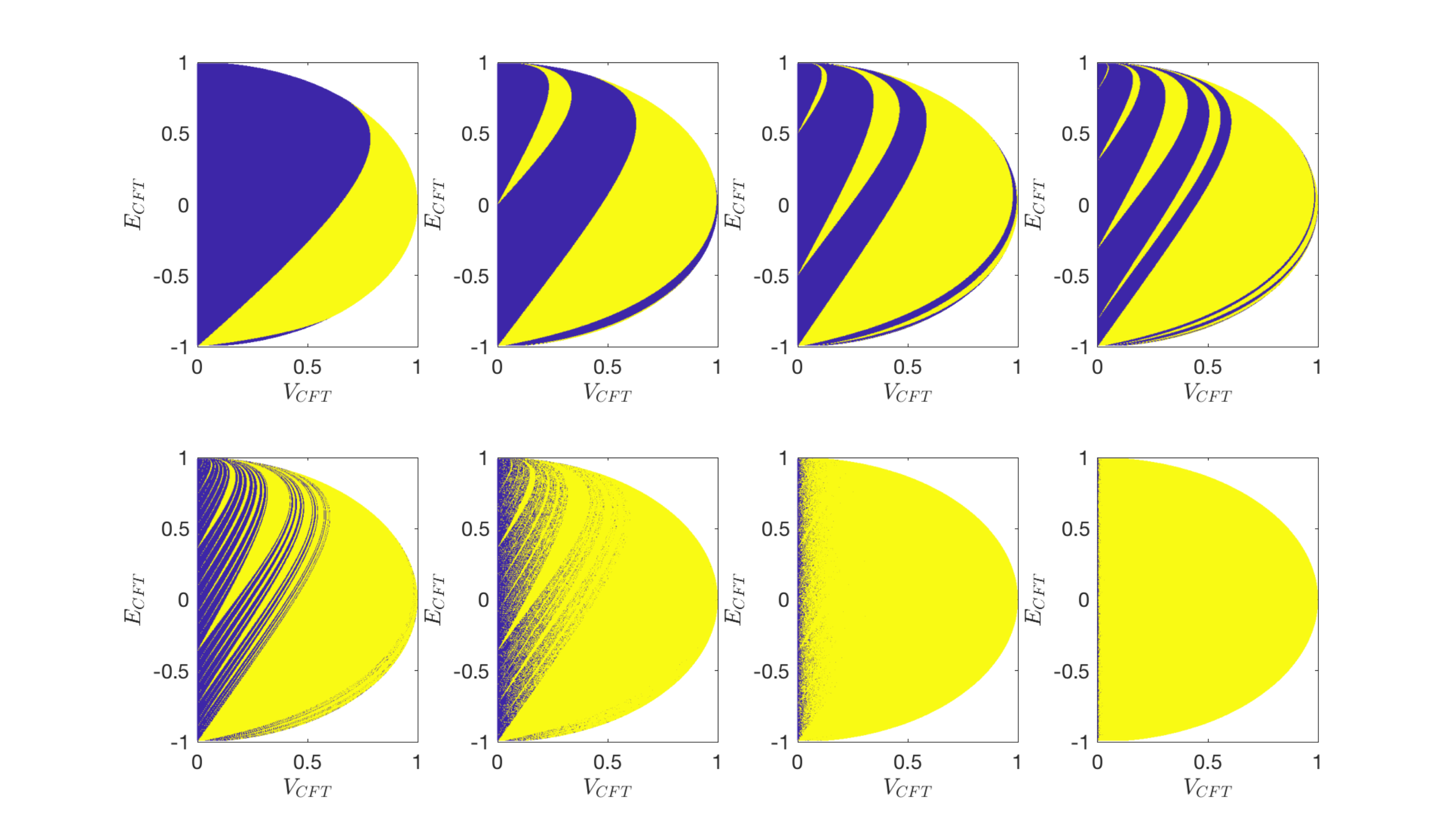}
\caption{
Phase diagrams in a periodically driven CFT with the sequence generated by finitely truncated Fibonacci bitstring, i.e. $\{X_j\}$ with 
 $\omega_n=F_{n-1}/F_n$.
 The parameters are the same as those in Fig.~\ref{FibonacciPhase0}
 except that now we change the variables to $V_{CFT}$ and $E_{CFT}$ (See Eq.~\eqref{EV_CFT}).
  The blue (yellow) regions correspond to the heating (non-heating) phases.
}
\label{FibonacciPhaseEV2}
\end{figure}

As analyzed in the previous subsections, to study the measure of the non-heating phases 
or the escape rate of initial conditions to infinity on the manifold $\mathcal{M}$(See Fig.~\ref{InitialLinePlot}),
it is more appropriate to fix the invariant $I$ in Eq.~\eqref{ConstantOfMotion}.
This is because the trace map in Eqs.\eqref{TraceMap} or \eqref{TraceMap2} 
holds for a fixed invariant $I$. In other words, the points $(x_{F_n}, y_{F_n}, z_{F_n})$ in Eq.~\eqref{TraceMap2}
move on the manifold $\mathcal M$ with a fixed geometry.
For this reason, we can replot the phase diagram in Fig.~\ref{FibonacciPhase0}
by changing variables in the initial conditions in Eq.~\eqref{InitialConditionTheta} 
as follows:
\be\label{EV_finiteTheta}
V_{\text{CFT}}:=\sin\left(\frac{\pi T_1}{L_{\text{eff}}}\right)\cdot \sin\left(\frac{\pi T_0}{L}\right),
\quad
E_{\text{CFT}}=\cos\frac{\pi T_0}{L},
\ee
where $L_{\text{eff}}=L\cosh(2\theta)$
with the invariant 
\be\label{ConstantGeneral3}
I=\left(\cosh^2(2\theta)-1\right)\cdot V_{\text{CFT}}^2.
\ee

With the above procedure, now we map the phase diagram in the 
region $\{(T_0,\, T_1)|0\leq T_0/L\leq 1, 0\leq T_1/L_{\text{eff}}< 1/2\}$ 
in Fig.~\ref{FibonacciPhase0} to Fig.~\ref{FibonacciPhaseEV2}.
The merit of this mapping is that for each $V_{\text{CFT}}$ in Fig.~\ref{FibonacciPhaseEV2},
the invariant $I$ is fixed. 
Then we study the measure of the non-heating phases as defined in Eq.~\eqref{Eq:Measure}, with
the result shown in Fig.~\ref{RatioCFTEV}.
There are two interesting features:

\begin{enumerate}

\item Similar to the case of $\theta=\infty$, the measure of the non-heating phases depends on $n$
as $\sigma_n(I,\theta)\propto e^{-\lambda(I,\theta)\cdot n}$.
That is, the measure of the non-heating phases decays exponentially (power-law) as a function of $n$ ($F_n$),
indicating that the measure will become zero in the quasi-periodical driving limit $n=\infty$.

\item
Interestingly, the decay rate (or escape rate) $\lambda(I,\theta)$ for $\theta=0.5$ and $\theta=\infty$ 
collapse to the same curve with $\lambda_I=a\cdot I^b$, where $b\simeq 0.552$ (See the right plot in Fig.~\ref{RatioCFTEV}). This means $\lambda(I,\theta)$
is only a function of $I$, and is independent of $\theta$.

\end{enumerate}

In addition, in Fig.~\ref{RatioCFTEV}, we also present the results for the measure of non-heating phases
for the case of $\theta=0.2$ (See Fig.~\ref{FibonacciPhase3} and Fig.~\ref{FibonacciPhaseEV3} for the 
corresponding phase diagrams). The decaying rates $\lambda(I,\theta=0.2)$
as a function of $I$
again fall on the same curves as that of $\theta=0.5$ and $\theta=\infty$, 
as seen in Fig.~\ref{RatioCFTEV} (right plot).
This means the decay rate $\lambda(I,\theta)$ is only a function of the invariant $I$, but is independent
of $\theta$ which characterizes the concrete deformation of Hamiltonians.

\subsubsection{Lyapunov exponents in the quasi-periodical driving limit}

\begin{figure}[t]
\centering
\subfloat{\includegraphics[width=3.00in]{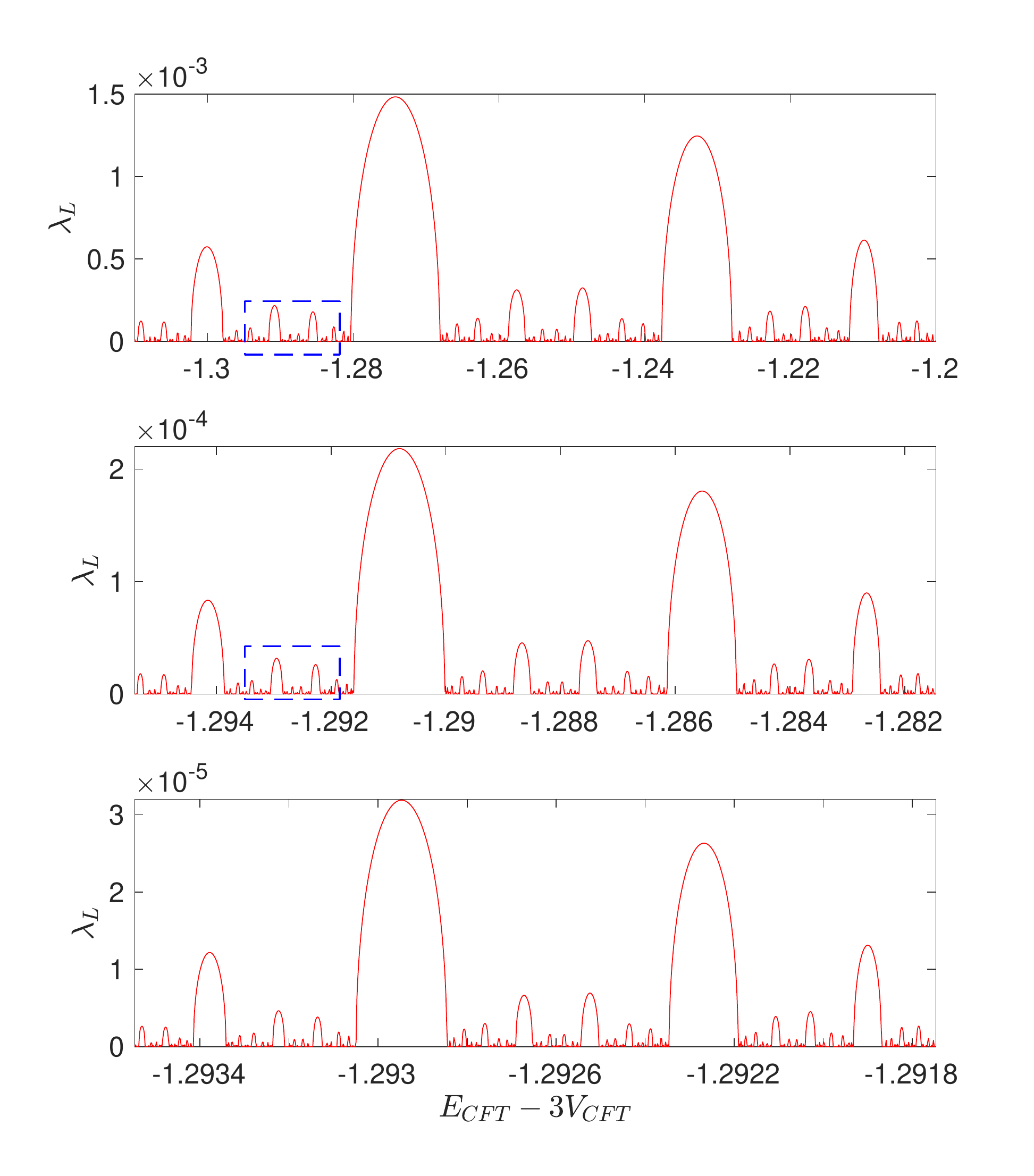}}
\subfloat{\includegraphics[width=3.00in]{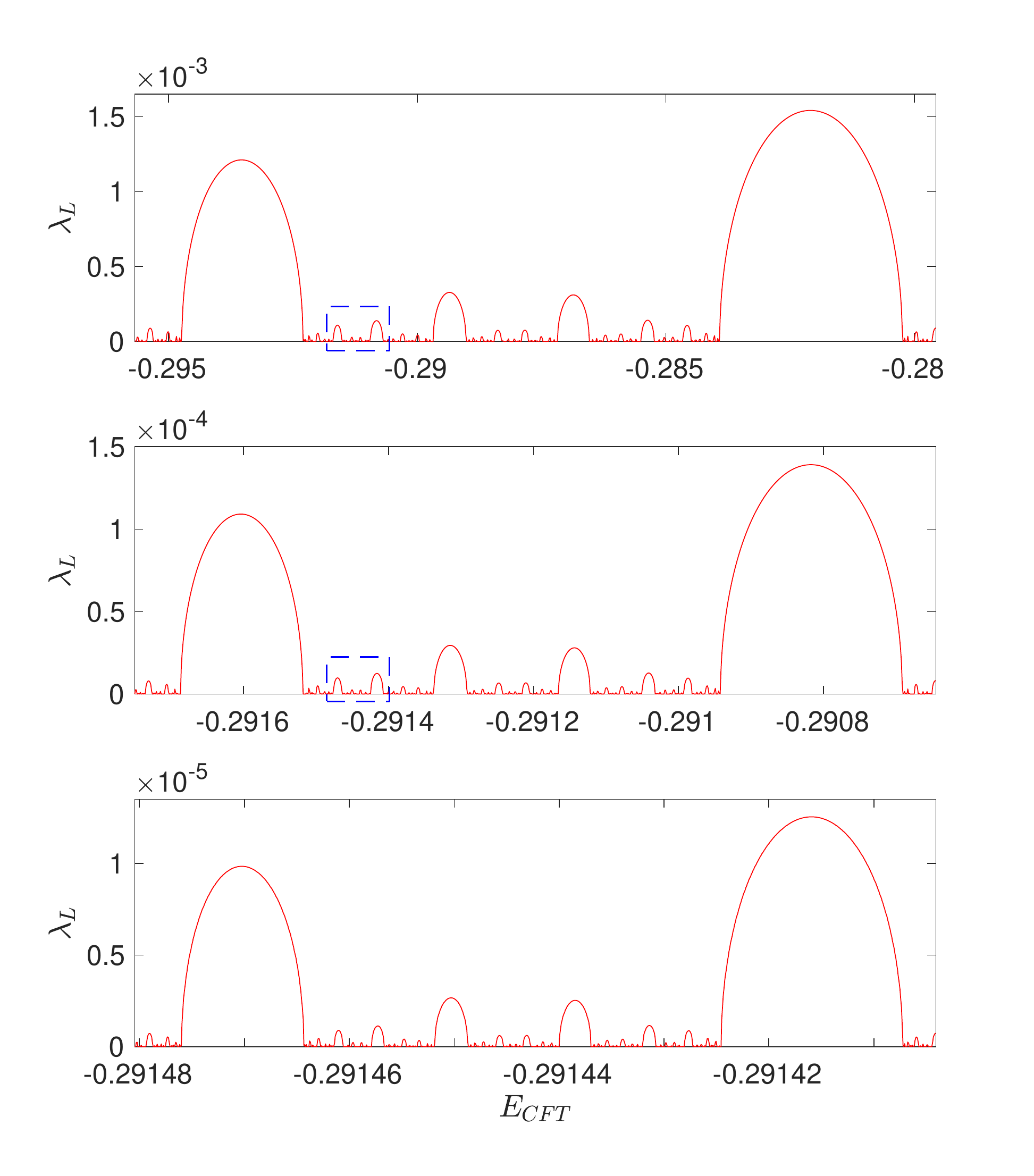}} 
\caption{
Self similarity in the distribution of Lyapunov exponents along
$E_{\text{CFT}}$ in Fig.~\ref{FibonacciPhaseEV} and 
Fig.~\ref{FibonacciPhaseEV2}. 
Left: The CFT is driven by $H_0$ and $H_{\theta=\infty}$.
We fix $I=V^2_{\text{CFT}}=0.3^2$, 
and scan the Lyapunov exponents along $E_{\text{CFT}}-3V_{\text{CFT}}$.
Right: The CFT is driven by $H_0$ and $H_{\theta=0.5}$. We fix $I=\left(\cosh^2(2\theta)-1\right)\cdot V_{\text{CFT}}^2=0.3^2$
and scan the Lyapunov exponents along $E_{\text{CFT}}$.
Each curve in the lower panel is the zoom-in plot of the region in blue dash
in the upper panel. We choose $n=1000$ in $\omega_n=F_{n-1}/F_n$ here. 
}
\label{LyapunovSelfSimilarity}
\end{figure}

The phase diagrams as studied in the previous subsections simply tell us 
whether the CFT is in the heating or non-heating phases. 
As $n\to \infty$, the measure of the non-heating phases becomes zero, and 
one can only ``see" the heating phase in the phase diagram.
In this subsection, we will use Lyapunov exponents 
to further characterize the fine structures in the heating phases in the limit $n\to \infty$.

\begin{figure}[h]
\centering
\subfloat{\includegraphics[width=3.00in]{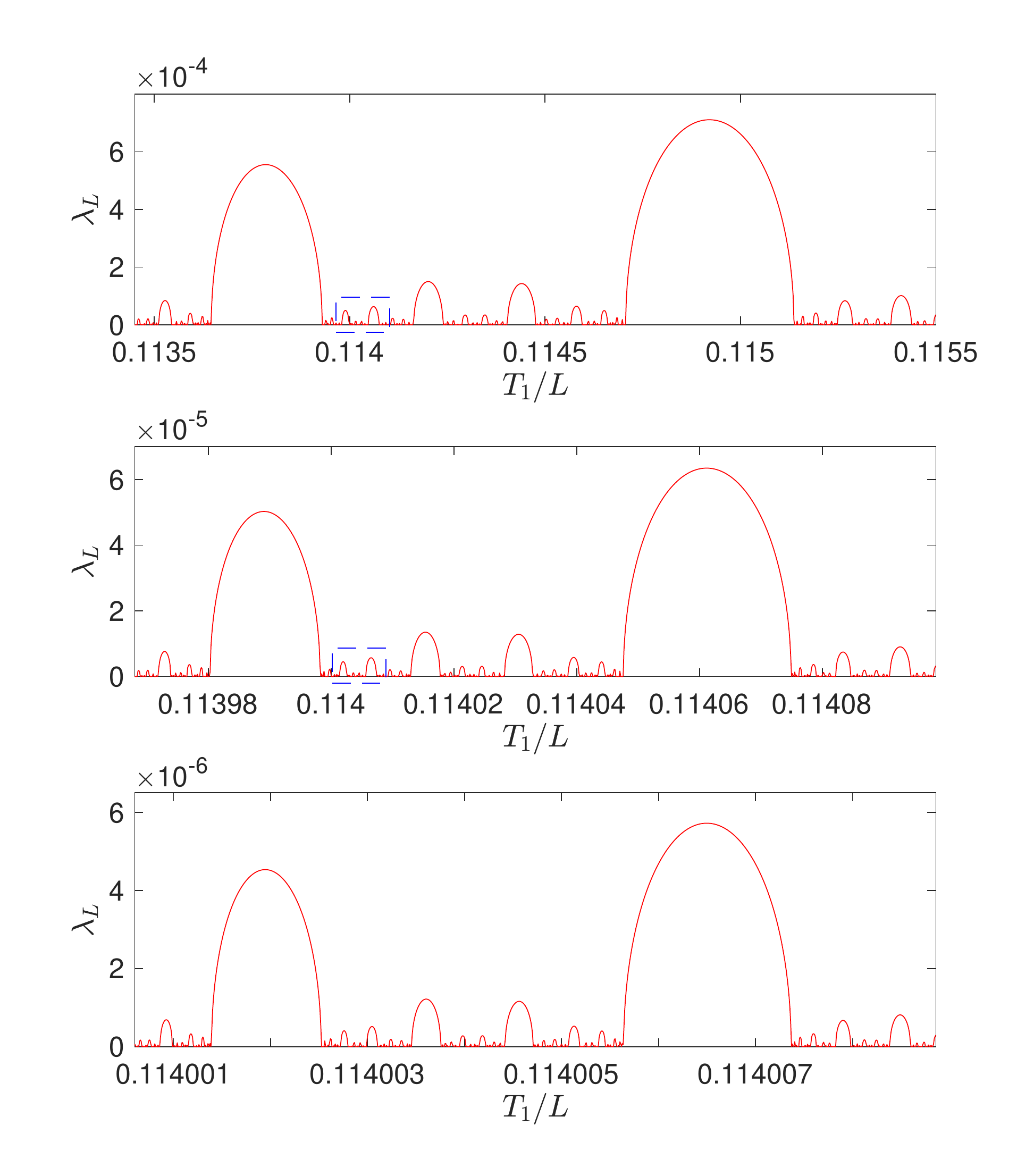}}
\subfloat{\includegraphics[width=3.00in]{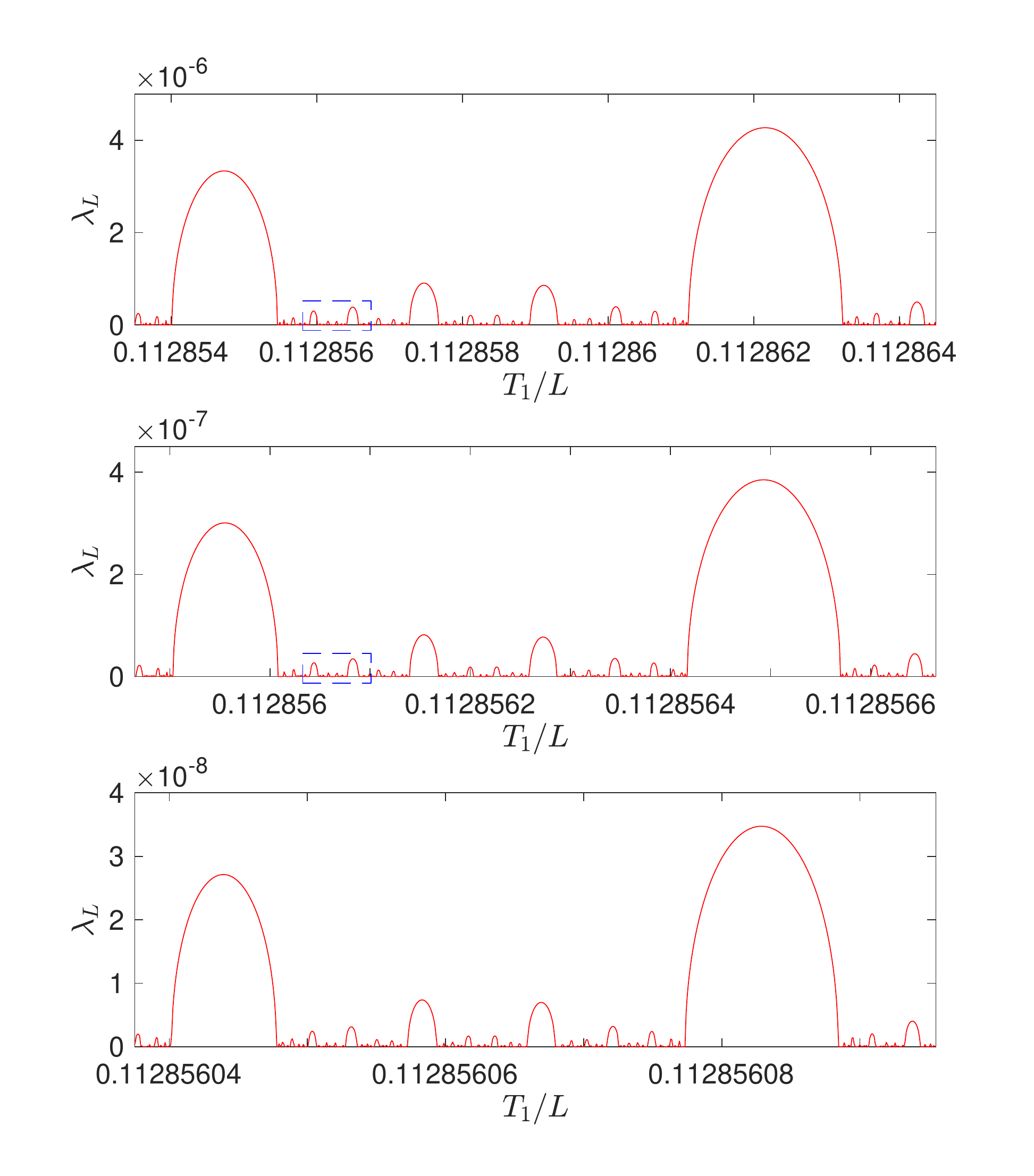}} 
\caption{
Self similarity in the distribution of Lyapunov exponents
along $T_1/L$ in Fig.~\ref{FibonacciPhase1} and Fig.~\ref{FibonacciPhase0}. 
The CFT is driven by $H_0$ and 
$H_{\theta=\infty}$ (left), and $H_0$ and $H_{\theta=0.5}$ (right).
Fixing $T_0/L=1/2$, we scan the Lyapunov exponents along $T_1/L$.
Each curve in the lower panel is the zoom-in plot of the region in blue dash
in the upper panel. We choose $n=1000$ in $\omega_n=F_{n-1}/F_n$ here. 
}
\label{LyapunovSelfSimilarityT0T1}
\end{figure}

Let us first consider a periodical driving with $\omega=\omega_n=F_{n-1}/F_n$ in
Eq.~\eqref{Fibonacci_V}, where the period of driving is $F_n$. 
The Lyapunov exponent in the heating phase can be obtained via  Eq.~\eqref{Lyapunov_Eta} as:
\be\label{Fib_lyapunov_periodic}
\lambda_L(\omega_n)=\frac{1}{F_n}\log\left(
|x_{F_n}|+\sqrt{|x_{F_n}|^2-1}
\right),
\ee
where $x_{F_n}= \frac{1}{2}\Tr (\Pi_{F_n})$ can be efficiently computed using the recurrence relation. 

Now we consider the distribution of Lyapunov exponents in 
Fig.~\ref{FibonacciPhaseEV} in the quasi-periodical driving limit.
To be concrete, we fix $V_{\text{CFT}}$ (or equivalently $I$)
in Fig.~\ref{FibonacciPhaseEV}, and scan
the Lyapunov exponents along $E_{\text{CFT}}$.
As shown in Fig.~\ref{LyapunovSelfSimilarity}, it is found that
the Lyapunov exponents exhibit self-similarity structures.
That is, by zooming in the distribution of Lyapunov exponents, 
one can find the same distributions (in different scales).
One can zoom in the distribution all the way and see 
the self-similarity structure, 
as long as a large enough $n$ is taken.
The self-similarity structure of Lyapunov exponents also 
indicates that the Lyapunov exponents can be arbitrarily small.
In other words, in the heating phases of a Fibonacci quasi-periodic
driving CFT, there exist some regions with arbitrary small 
heating rates for the entanglement/energy growth.

\begin{figure}[h]
\centering
\includegraphics[width=6.50in]{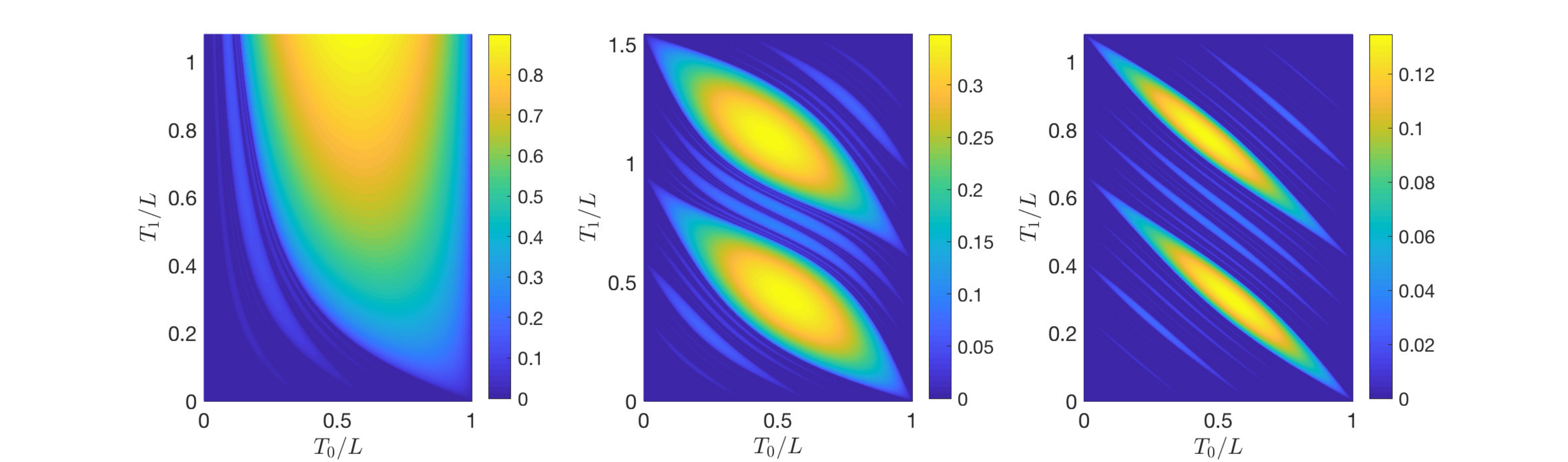}
\caption{
Lyapunov exponents as a function of $T_0/L$ and $T_1/L$ for different choices of drivings
(from left to right): $H_0$ and $H_{\theta=\infty}$, $H_0$ and $H_{\theta=0.5}$,
$H_0$ and $H_{\theta=0.2}$. The Lyapunov exponents are obtained by choosing 
$\omega_n=F_{n-1}/F_n$ in Eq.~\eqref{Fib_lyapunov_periodic}, with $n=1000$.  
}
\label{Lyapunov3d}
\end{figure}

 We also study the distribution of Lypunov exponents in 
 the $T_0/T_1$ parameter space in 
 Fig.~\ref{FibonacciPhase1} and Fig.~\ref{FibonacciPhase0}.
As shown in Fig.~\ref{LyapunovSelfSimilarityT0T1}
are the distributions of $\lambda_L$ along $T_1/L$
with a fixed $T_0/L$. Interestingly, 
although the constant of motion $I$ in Eq.~\eqref{ConstantOfMotion}
varies along $T_1/L$ (with $T_0/L$ fixed), the
self-similarity structures in $\lambda_L$ are still there.

At last, in Fig.~\ref{Lyapunov3d},
we give a color plot of the distribution of
Lyapunov exponents in the parameter space $(T_0/L, T_1/L)$.
One can find the patterns inherit some features of the 
periodically driven CFTs (See Fig.~\ref{Lyapunov_PeriodicDriving}).
It is also helpful to compare these three plots with the phase 
diagrams in Fig.~\ref{FibonacciPhase1}, 
Fig.~\ref{FibonacciPhase0}, and Fig.~\ref{FibonacciPhase3}, respectively.
We emphasis that although there are large areas of regions
with almost zero Lyapunov exponents, they are actually in the heating phase
(See Fig.~\ref{FibonacciPhase1}, 
Fig.~\ref{FibonacciPhase0}, and Fig.~\ref{FibonacciPhase3}).
If we zoom in these regions, one can observe the 
self-similarity structure(See, e.g., Figs.\ref{LyapunovSelfSimilarity} 
and \ref{LyapunovSelfSimilarityT0T1}).

\subsection{Fixed point in the non-heating phase: Entanglement and energy dynamics}
\label{Sec: FixedPoint}

From the previous discussions, we conclude that 
the measure of the non-heating phases shrinks to zero when we approach the quasi-periodical driving limit, i.e. without special guide it is hard to find the exact location of the non-heating point. Indeed, for the case with SSD deformation, namely the driving protocol given by $H_{0}$ and $H_{\theta=\infty}$, we are not able to locate such points. Fortunately, for finite $\theta$, we can use the fixed point discussed in Sec.~\ref{Sec:HthetaExample} to pin down the non-heating point. More explicit, in this section, we will show the followings


\begin{enumerate}
    \item If both the driving Hamiltonians are chosen as elliptic types, 
\footnote{For example, all the Hamiltonians of the form in 
Eq.~\eqref{H_theta_A} with a finite $\theta$ are elliptic.
See Appendix \ref{Sec: TechnicalDetail}.}
one can always find \textit{exact} non-heating phases point.

\item At these non-heating points, both the 
entanglement entropy and the energy evolution are of period $6$ in Fibonacci index, i.e.,
$S_A(F_n)=S_A(F_{n+6})$ and $E(F_n)=E(F_{n+6})$.
It is noted that although the entanglement entropy and energy are 
periodic functions at the Fibonacci numbers, they are not periodic at the
non-Fibonacci numbers. See the following statement.

\item The envelopes of the entanglement entropy and total energy
will grow logarithmically and in a power-law as a function of the 
driving steps $n$ (not the Fibonacci index), respectively.

\end{enumerate}

We will first illustrate the above statements
with simple examples in the following discussions,
and then prove them in Sec.~\ref{Sec:NonheatingFixGeneral}.

\subsubsection{Entanglement and energy evolution at Fibonacci numbers}

In the following, we will study the exact non-heating fixed point 
and its properties with the simple choice of
$H_A=H_{\theta}$ and $H_B=H_{\theta=0}$, where the form of $H_{\theta}$
is given in Eq.~\eqref{H_theta_A}.
The initial conditions for the trace map have been given in Eq.~\eqref{InitialConditionTheta}.
By considering
\be\label{T0T1_fixedpoint}
T_0/L=T_1/L_{\text{eff}}=1/2,\quad \text{where } L_{\text{eff}}=L\cosh(2\theta),
\ee 
the initial condition has the following form
\be\label{InitialConditionFixedPoint}
(x_{F_2}, x_{F_1}, x_{F_0})=(-\cosh(2\theta), 0, 0),
\ee
which will start a fixed point with constant of motion $I=\cosh^2 (2\theta)-1$ under the trace map \eqref{TraceMap2}, i.e. 
\be
\begin{split}
T: \quad &(-a, 0, 0)\,\to\, (0, -a, 0) \,\to\, (0, 0, -a)
\to\, (a,0,0) \to\,  (0, a, 0) \to\,  (0, 0, a) \\
\to\, &(-a,0,0) \,\to\,\cdots.
\end{split}
\ee
where $a:=\cosh(2\theta)$. In fact, for this fixed point, not only the traces (recall $x_{F_k}=\frac{1}{2}\Tr (\Pi_{F_k})$) have periodicity $x_{F_k}=x_{F_{k+6}}$, the corresponding matrices themselves are also returning periodically
\begin{equation}\label{Period6_matrix}
    \begin{aligned}
    &\Pi_{F_{6k+1}} =i \begin{pmatrix}
    \cosh (2\theta) & -\sinh (2\theta) \\
    \sinh (2\theta) & -\cosh (2\theta)
    \end{pmatrix} \qquad 
    \Pi_{F_{6k+2}} = -\begin{pmatrix}
    \cosh (2\theta) & \sinh (2\theta) \\
    \sinh (2\theta) & \cosh (2\theta)
    \end{pmatrix} \\
    &\Pi_{F_{6k+3}} =i \begin{pmatrix}
    -\cosh (4\theta) & \sinh (4\theta) \\
    -\sinh (4\theta) & \cosh (4\theta)
    \end{pmatrix} \qquad 
    \Pi_{F_{6k+4}} = i \begin{pmatrix}
    \cosh (2\theta) & -\sinh (2\theta) \\
    \sinh (2\theta) & -\cosh (2\theta)
    \end{pmatrix} \\
    &\Pi_{F_{6k+5}} =\begin{pmatrix}
    \cosh (2\theta) & -\sinh (2\theta) \\
    -\sinh (2\theta) & \cosh (2\theta)
    \end{pmatrix} \qquad 
    \Pi_{F_{6k+6}} = i \begin{pmatrix}
    1 & 0 \\
    0 & -1
    \end{pmatrix} \\
    \end{aligned}
\end{equation}



\begin{figure}[t]
\centering
\subfloat[$\theta=0.5$]{\includegraphics[width=3.20in]{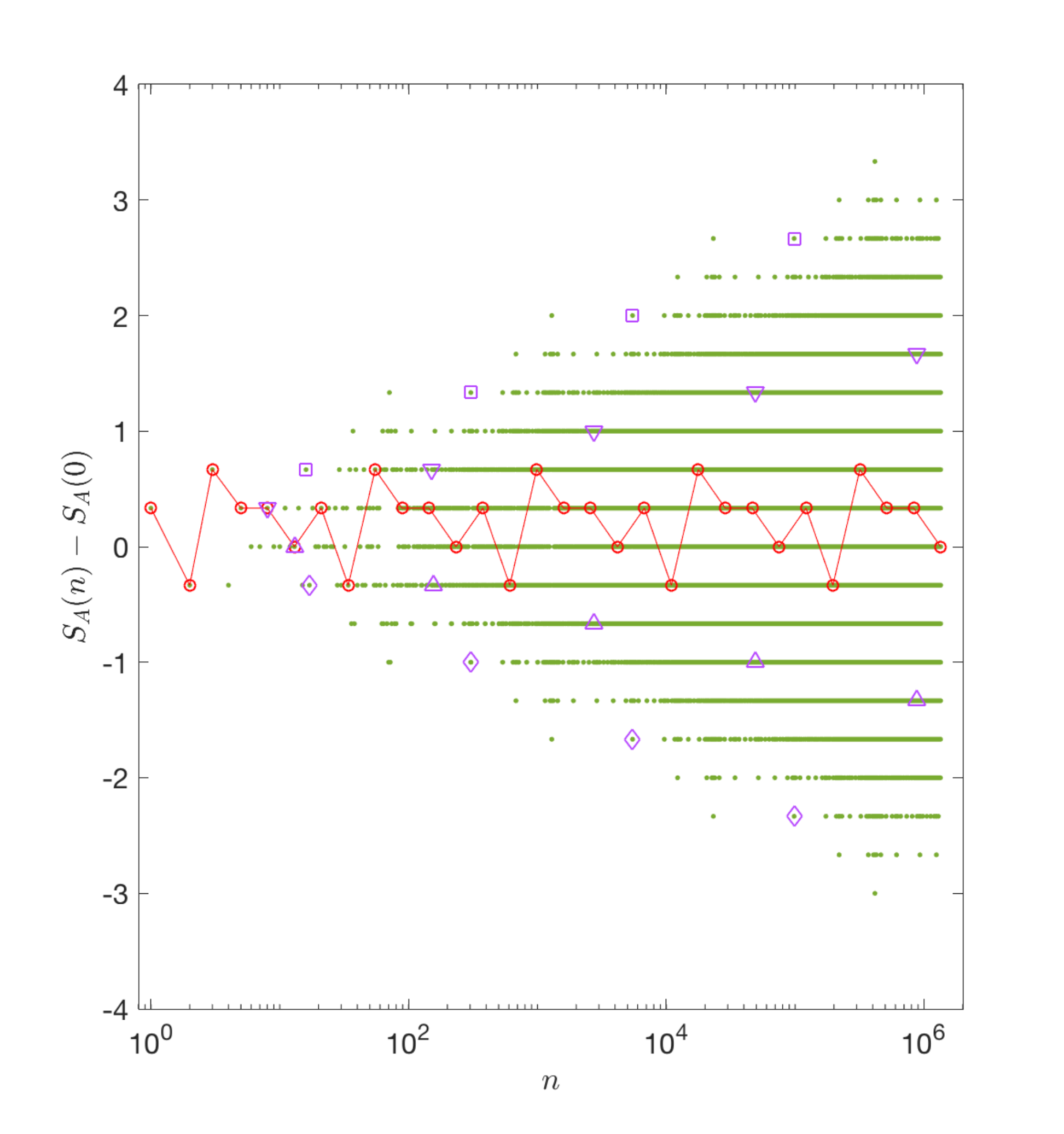}}
\subfloat[$\theta=0.2$]{\includegraphics[width=3.20in]{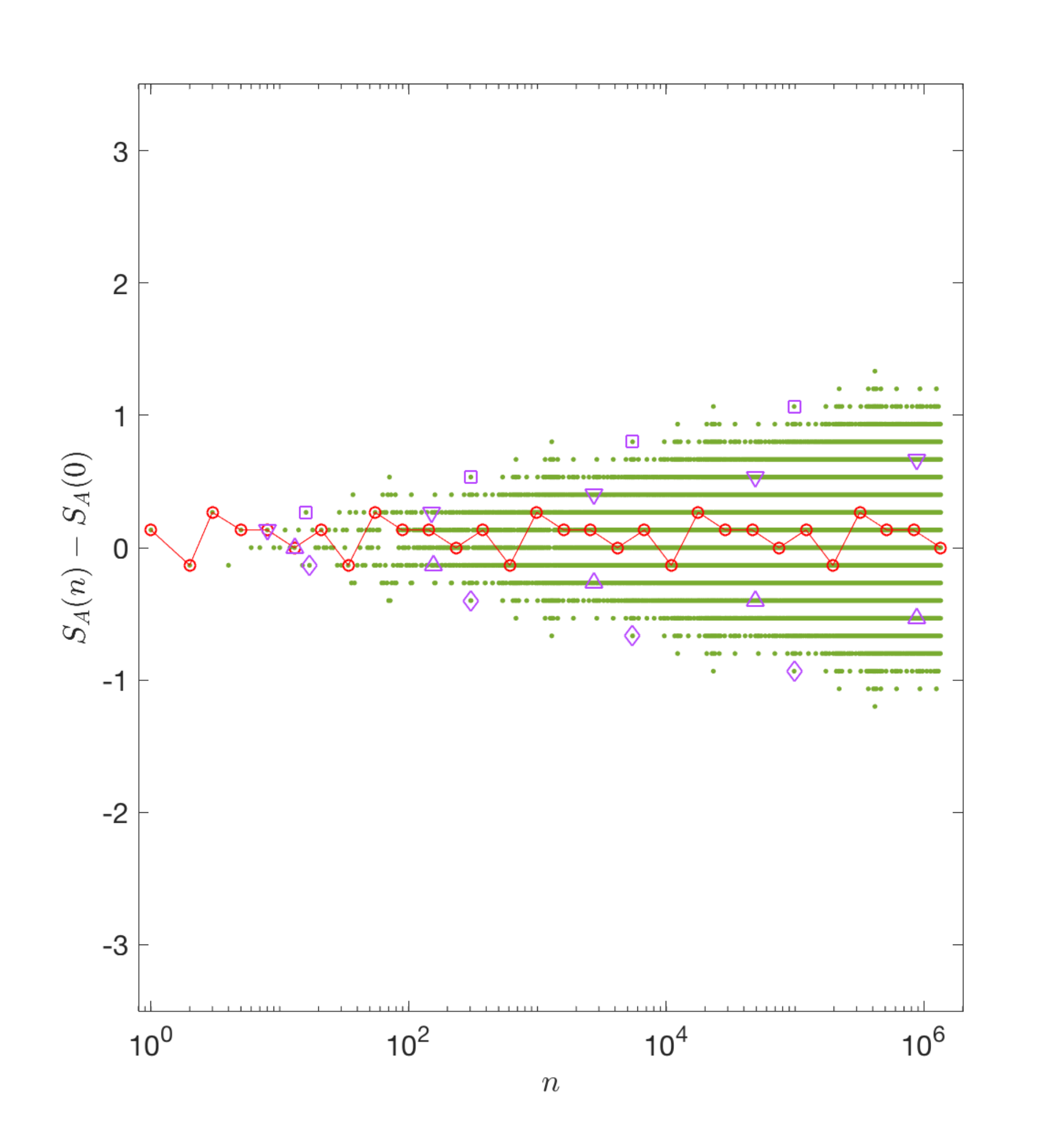}} 
\caption{
Time evolution of the entanglement entropy of $A=[0, \, L/2]$ 
at the non-heating fixed point in Eq.~\eqref{InitialConditionTheta}  
for $\theta=0.5$ (left) and $\theta=0.2$ (right).
The red solid lines correspond to the entanglement entropy 
at the Fibonacci numbers $n=F_j$, 
with the expression given in Eq.~\eqref{EE_nonHeatingFixedPoint}, where we 
choose $c=1$.
The points in rectangles, downward triangles, 
upward triangles, and diamonds correspond to the entanglement
entropy evolution at the non-Fibonacci numbers, with the expressions
given  in Eqs.\eqref{SA_Etotal_1}, \eqref{SA_Etotal_2},
\eqref{SA_Etotal_3}, and \eqref{SA_Etotal_4}, respectively.
}
\label{EEnonFib}
\end{figure}

\begin{figure}[t]
\centering
\subfloat[$\theta=0.5$]{\includegraphics[width=3.20in]{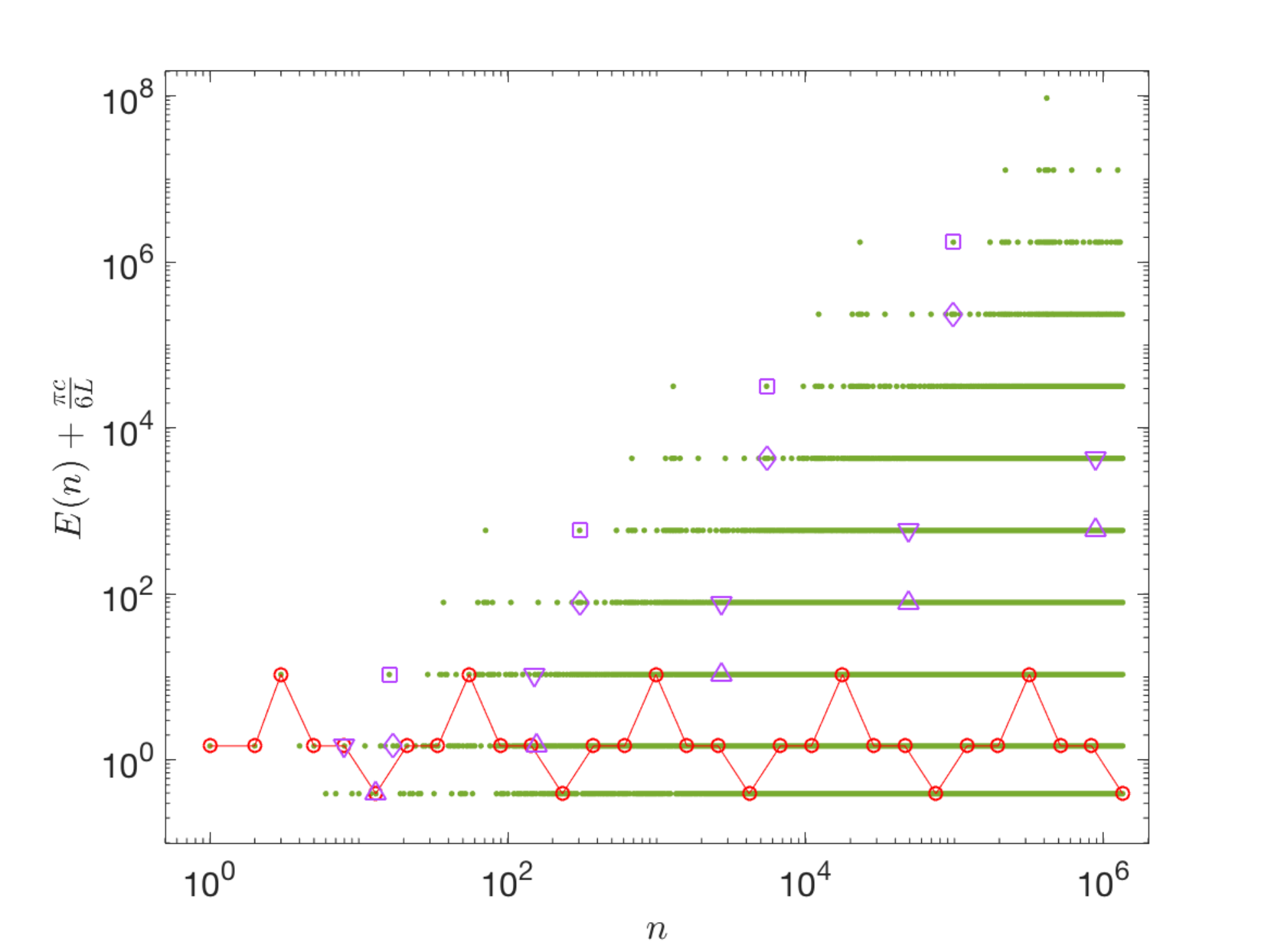}}
\subfloat[$\theta=0.2$]{\includegraphics[width=3.20in]{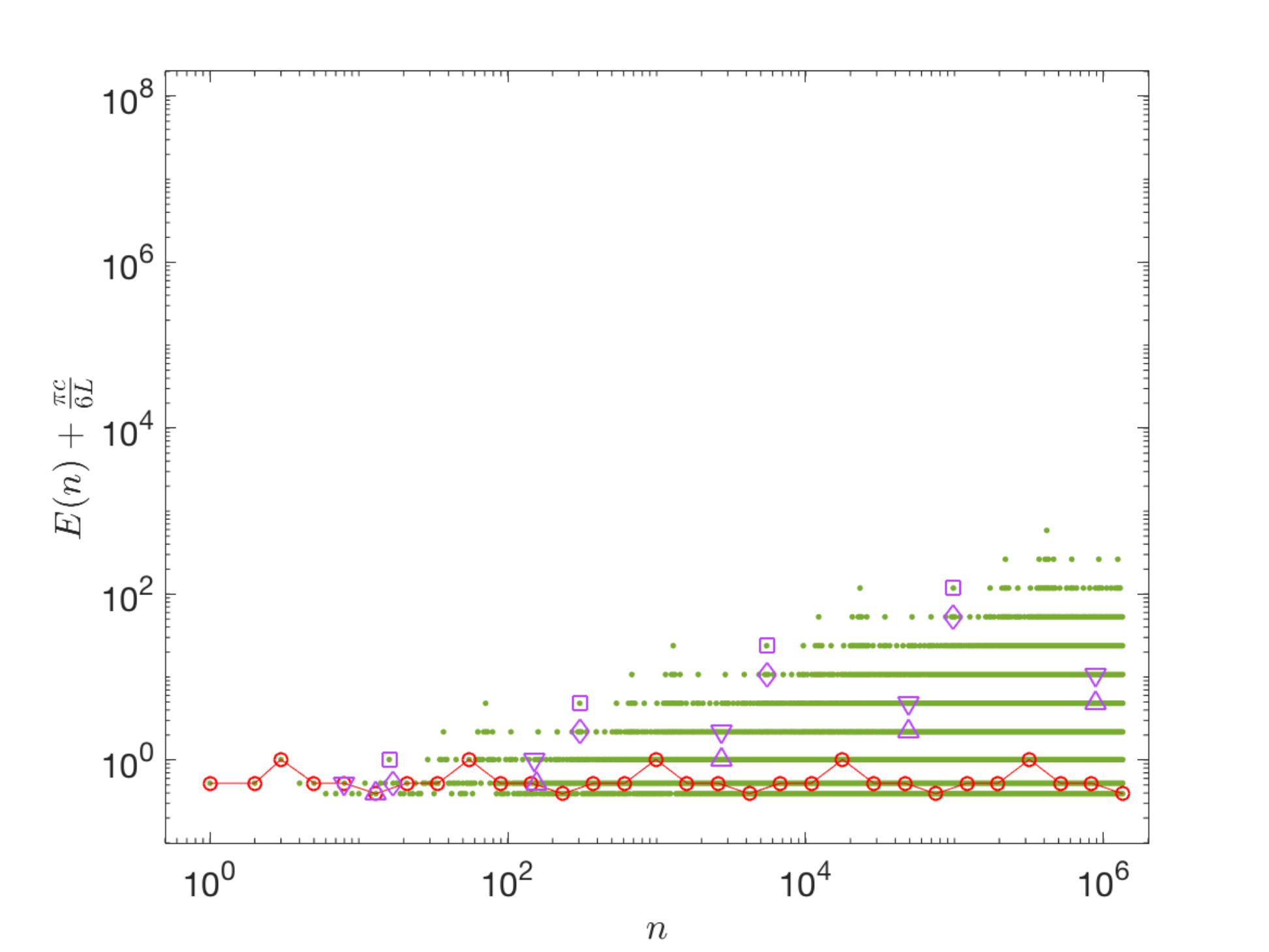}} 
\caption{
Time evolution of the total energy at the non-heating fixed point
in Eq.~\eqref{InitialConditionTheta} for
$\theta=0.5$ (left) and $\theta=0.2$ (right).
The red solid lines are the total energy at the Fibonacci numbers $n=F_j$, 
with the expression given in Eq.~\eqref{EE_nonHeatingFixedPoint}, 
where we choose $c=1$ and $L=1$.
The points in rectangles, downward triangles, 
upward triangles, and diamonds correspond to the entanglement
entropy evolution at the non-Fibonacci numbers, with the expressions
given  in Eqs.\eqref{SA_Etotal_1}, \eqref{SA_Etotal_2},
\eqref{SA_Etotal_3}, and \eqref{SA_Etotal_4}, respectively.
}
\label{EnergynonFib}
\end{figure}

Thus, the time evolution of entanglement entropy of the half-system $A=[0, L/2]$ 
and the total energy at the Fibonacci numbers are 
\be\label{EE_nonHeatingFixedPoint}
\small
S_A(F_j)-S_A(0)=\left\{
\begin{split}
&\frac{2\theta c}{3} , \quad &j=6k+1\\
-&\frac{2\theta c}{3}, \quad &j=6k+2\\
&\frac{4\theta c}{3}, \quad &j=6k+3\\
&\frac{2\theta c}{3}, \quad &j=6k+4\\
&\frac{2\theta c}{3}, \quad &j=6k+5\\
&\,\,\, 0, \quad &j=6k+6\\
\end{split}
\right.
\quad\quad
E(F_j)=\left\{
\begin{split}
&\frac{\pi c}{8L}\cdot \cosh(4\theta)-\frac{\pi c}{6L}, \quad &j=6k+1\\
&\frac{\pi c}{8L}\cdot \cosh(4\theta)-\frac{\pi c}{6L}, \quad &j=6k+2\\
&\frac{\pi c}{8L}\cdot \cosh(8\theta)-\frac{\pi c}{6L}, \quad &j=6k+3\\
&\frac{\pi c}{8L}\cdot \cosh(4\theta)-\frac{\pi c}{6L}, \quad &j=6k+4\\
&\frac{\pi c}{8L}\cdot \cosh(4\theta)-\frac{\pi c}{6L}, \quad &j=6k+5\\
&\frac{\pi c}{8L}-\frac{\pi c}{6L},\, \quad &i=6k+6\\
\end{split}
\right.
\ee
where 
$S_A(0)$ denotes the entanglement 
entropy 
in the initial state (which is the ground state of $H_{\theta=0}$
here), and
$E_{F_{6k}}=-\frac{\pi c}{24L}$ corresponds to the Casimir energy. 
Also see 
Fig.~\ref{EEnonFib} and Fig.~\ref{EnergynonFib} for examples with $c=1$ and $\theta=0.5$ and $0.2$. 

An interesting remark is that the special initial condition we choose that forms the fixed point are conformal maps that correspond to the ``reflection matrix'' (since we require the initial traces to vanish, see discussions near \eqref{M0M1Exp}). Note, the product of two distinct reflection is hyperbolic, i.e. if we drive the system with periodic driving we will end up heating the system. However, what we present just now is that if we drive it in a Fibonacci pattern, they happen to return and form a non-heating point in the quasi-periodic driving phase diagram.


\subsubsection{Entanglement and energy evolution at non-Fibonacci numbers}
\label{Sec:EEnonFib}

As shown in Fig.~\ref{EEnonFib} and Fig.~\ref{EnergynonFib}, between the Fibonacci numbers, the numerical simulation shows that entanglement and energy still increase for initial condition at fixed point. 
In the context of Fibonacci quasi-crystal, it was found that 
the wavefunction amplitude in the energy spectrum 
and some physical observables (e.g., the
resistence) have a power-law growth as a function of 
the lattice site $n$\cite{Kohmoto1987, SutherlAndresistance1987,iochum1991power,2011freefermionssd}.
In our setup, at the driving steps that are non-Fibonacci numbers,
we expect the entanglement entropy or total energy  
also grows in a certain sub-exponential way.


Now we provide analytic understanding using the property of the Fibonacci driving protocol. The idea is that, for any integer $n$ which can be written as a sum of distinct Fibonacci numbers,
\begin{equation}
    n = \sum_{j=1}^m F_{k_j} \quad \text{with} \quad k_1> k_2 > \ldots k_m
\end{equation}
the corresponding conformal transformation matrix $\Pi_n$ can be written as 
\begin{equation}
    \Pi_n = \Pi_{F_{k_1}} \cdot \Pi_{F_{k_2}} \cdots \Pi_{F_{k_m}}
\end{equation}
where each $\Pi_{F_k}$ with Fibonacci number has been obtained in Eq.~\eqref{Period6_matrix}. In particular, we can find some simple sequence
\begin{enumerate}
    \item Let 
    \begin{equation}\label{n1}
           n = F_{6m}+F_{6m-3} + F_{6m-6} +\ldots F_6+F_3
    \end{equation}
    be an integer that increase with $m$. 
    Correspondingly 
    \begin{equation}
        \Pi_n = \Pi_{6m} \cdot \Pi_{6m-3} \cdots \Pi_6 \cdot \Pi_3 = (\Pi_6 \cdot \Pi_3)^m
    \end{equation}
    Recall from \eqref{Period6_matrix}, the product $\Pi_6 \cdot \Pi_3$ and $\Pi_n$ can be evaluated explicitly
    \begin{equation}
        \Pi_6 \cdot \Pi_3 = \begin{pmatrix}
        \cosh (4\theta) & -\sinh (4\theta) \\
        -\sinh (4\theta) & \cosh(4\theta)
        \end{pmatrix} \qquad \Pi_n = \begin{pmatrix}
        \cosh (4m\theta) & -\sinh (4m\theta) \\
        -\sinh (4m\theta) & \cosh(4m\theta)
        \end{pmatrix}
    \end{equation}
    The corresponding 
    entropy for the half system $A=[0, L/2]$ and the total energy are
    \be\label{SA_Etotal_1}
    S_A(n)-S_A(0)=\frac{c}{3}\cdot 4 m\theta,\qquad
    E(n)=\frac{\pi c}{8L}\cosh(8m\theta)-\frac{\pi c}{6 L},
    \ee
where we have constrained $\theta>0$.
The plot of Eq.~\eqref{SA_Etotal_1} can be found in 
Fig.~\ref{EEnonFib} and Fig.~\ref{EnergynonFib}.
Note that $S_A(n)$ and $E(n)$ grow linearly and 
exponentially respectively as a function of $m$ for large $m$. 
However, $m$ is not the actual driving step number. We need to convert to the actual step number $n \sim e^{6m \log \varphi}$, which grows exponential with $m$ for large $m$, and $\varphi = \frac{\sqrt{5}+1}{2}$ is the golden ratio. Therefore 
\be\label{SA_Etotal_2_approx}
\begin{split}
S_A(n)-S_A(0)\simeq \frac{2\, c\cdot \theta}{9\,\log\phi}\log n,,\quad\quad
E(n)\simeq \frac{\pi c}{16 L}\cdot n^{\frac{4\theta}{3 \log\phi}}.
\end{split}
\ee
That is to say, at the non-Fibonacci numbers $n$ in Eq.~\eqref{n1}, 
the entanglement entropy $S_A(n)$ grows logarithmically in time,
and the total energy grows in a power-law in time. 
This corresponds to the feature of phase transition (or critical phase)
in the periodically driven CFT (See Table.\ref{PhaseDiagramPeriodic}).
\item 
Now we choose a difference sequence, to demonstrate that the growth rate at this fixed point depends on the sequence we pick when approaching the long time limit. Let
\begin{equation}
    n = F_{6m-1}+ F_{6m-7} + \ldots + F_5
\end{equation}
And correspondingly
\begin{equation}
    \Pi_n = \Pi_{6m-1} \cdot \Pi_{6m-7} \cdots \Pi_{5} = \left( \Pi_{F_5} \right)^m = \begin{pmatrix}
    \cosh (2m \theta) & - \sinh(2m \theta) \\
    -\sinh (2m \theta) & \cosh (2m \theta)
    \end{pmatrix}
\end{equation}
The entropy and energy formulas are
\begin{align}\label{SA_Etotal_2}
    &S_A(n)-S_A(0)=\frac{c}{3}\cdot 2 m\theta \simeq \frac{c\cdot \theta}{9\,\log\phi}\log n \\
    &E(n)=\frac{\pi c}{8L}\cosh(4m\theta)-\frac{\pi c}{6 L} \simeq \frac{\pi c}{16 L}\cdot n^{\frac{2\theta}{3 \log\phi}}.
\end{align}
\end{enumerate}

In the two examples above, the entanglement entropies all grow
with $n$. One can observe in Fig.~\ref{EEnonFib} that  
at certain points the entanglement
entropy may decrease. We will investigate these points using the following examples
\begin{enumerate}
    \item Let
    \begin{equation}
        n= F_{6m-1} + \ldots + F_{11} +F_5 + F_4
    \end{equation}
    Note the last element is important. The corresponding matrix is given as 
    \begin{equation}
        \Pi_n = (\Pi_{F+5})^m \cdot \Pi_4 = 
        i
        \begin{pmatrix}
        \cosh ((m-1) 2\theta) & \sinh ((m-1) 2\theta) \\
        - \sinh ((m-1) 2\theta) & - \cosh ((m-1) 2\theta)
        \end{pmatrix}
        \label{eqn: neg pi}
    \end{equation}
    The corresponding entropy and energy formulas are 
    \be\label{SA_Etotal_3}
        S_A(n)-S_A(0)=-\frac{c}{3}\cdot 2 (m-1)\theta, \quad\quad E(n)=\frac{\pi c}{8L}\cosh(4(m-1)\theta)-\frac{\pi c}{6 L},
    \ee
    where $\theta>0$.
    The results are similar to Eq.~\eqref{SA_Etotal_2}, with a minus sign difference in the entropy formula. In other words, we have a logarithmic decrease in the entanglement evolution and a power-law growth in the total energy evolution
    (See Fig.~\ref{EEnonFib} and Fig.~\ref{EnergynonFib}).

The entanglement decrease might look bizarre, but this could happen in a system with infinite entropy to start with, e.g. in the continuous field theories where a UV regulator is required in the entropy calculation, which itself is a manifestation of the large entanglement in the vacuum state. 

Technically, we may explain the origin of the decreasing entropy as follows: the form of the conformal transformation in \eqref{eqn: neg pi} indicates that the energy-momentum density (see \eqref{EnergyMomentumQ1}) locates exactly at $x=L/2$, which coincides with the entanglement cut we choose. 
As discussed in detail in Appendix \ref{Sec: LinearDecreaseEE}, 
in this case, the entanglement entropy will decrease in time.
Physically, it is because the degrees of freedom that carry 
the entanglement between two regions are accumulated at the 
entanglement cut. We emphasize that the points with decreasing entanglement entropy
are due to the coincidence of the energy-density peak
and the entanglement cut, and therefore are not generic. 
In general, at the non-heating fixed points, the envelopes of 
the entanglement entropy and total energy will grow logarithmically
and in a power law in time, respectively.

\item Another example with decreasing entropy we present in the Fig.~\ref{EEnonFib} and Fig.~\ref{EnergynonFib} is that
\begin{equation}
    n = F_{6m}+F_{6m-3} \ldots F_{6}+F_3+F_1
\end{equation}
with 
\begin{equation}
    \Pi_n = (\Pi_6 \cdot \Pi_3 )^m \cdot \Pi_1
\end{equation}
and the corresponding entropy and energy
\be\label{SA_Etotal_4}
S_A(n)-S_A(0)=-\frac{c}{3}\cdot (4m-2)\theta, \quad\quad E(n)=\frac{\pi c}{8L}\cosh[(8m-4)\theta]-\frac{\pi c}{6 L}.
\ee
\end{enumerate} 
Using the same procedure above, one can find many other series of 
discrete points with different growing (and decreasing) rates in the
entanglement/energy evolution in Fig.~\ref{EEnonFib} and Fig.~\ref{EnergynonFib}, these series together form the fan structure in the Figures.

\subsubsection{Comparison of CFT and lattice calculations}
\label{Sec: CompareCFTLattice_Fibonacci}

\begin{figure}[t]
\centering
\subfloat[Entanglement evolution]{\includegraphics[width=3.20in]{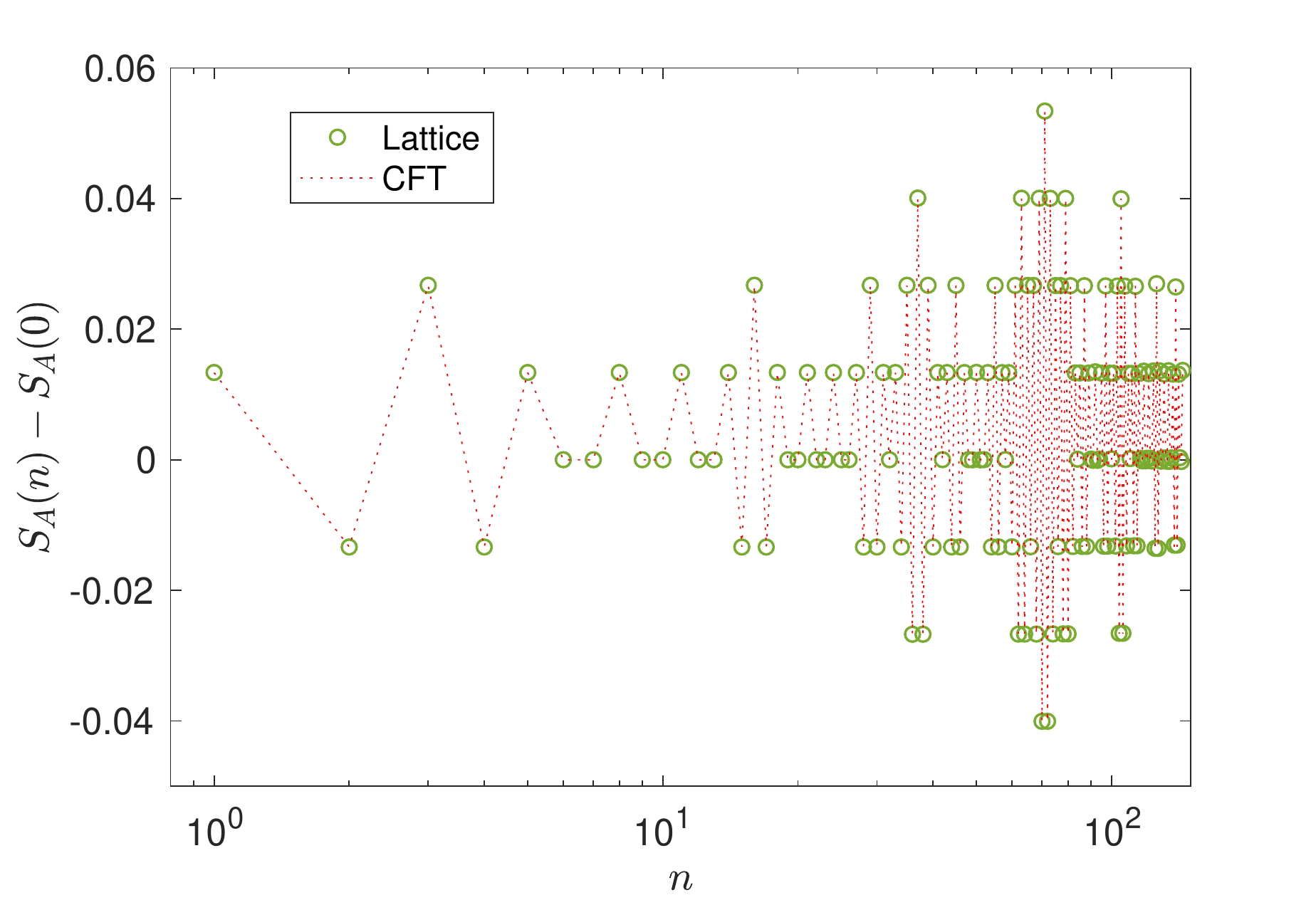}} 
\subfloat[Energy evolution]{\includegraphics[width=3.20in]{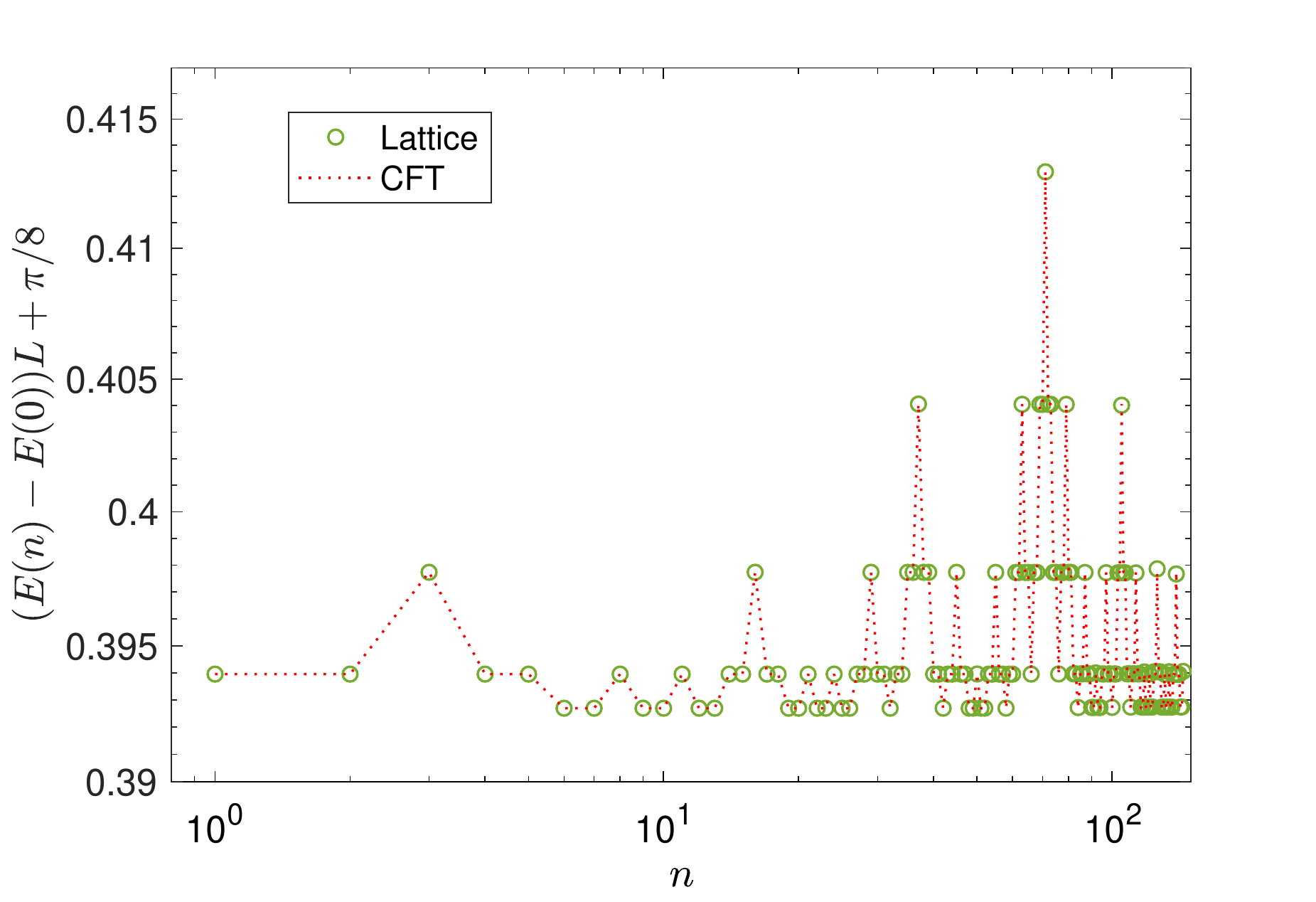}}
\caption{
Comparison of the lattice and CFT calculations at the non-heating fixed point 
 for (a) the entanglement entropy evolution and (b) the total energy evolution.
Here we choose $c=1$, $A=[0,L/2]$,
$\theta=0.02$, and 
$n_{\text{max}}=F_{11}=144$.
In the lattice calculation we consider $L=2000$.
A plot of the entanglement entropy
 for a larger $n_{\text{max}}$ can be found in Fig.~\ref{FixedPointCFTLatticeB}.
}
\label{FixedPointCFTLattice}
\end{figure}

In this part, we compare the CFT and lattice calculations 
for the time evolution of entanglement entropy/energy at 
the non-heating fixed point 
as discussed in the previous subsections. 
The lattice model we use is the same as that studied in
Sec.~\ref{Sec:LatticePeriodic}.
The two lattice Hamiltonians under consideration are 
\begin{equation}
    H_0=\frac{1}{2}\sum_{j=1}^{L-1}c_j^{\dag}c_{j+1}+h.c. \qquad H_{\theta}=\frac{1}{2}\sum_{j=1}^{L-1}f(j)c_j^{\dag}c_{j+1}+h.c
\end{equation}
where $L$ is the total length of the lattice and
$f(j)=1-\tanh(2\theta)\cdot \cos\frac{2\pi j}{L}$,
with the initial state chosen as the ground state of $H_0$.
The corresponding driving time intervals are
$T_0=L/2$ and $T_1(\theta)=
L_{\text{eff}}/2$, where $L_{\text{eff}}=L\cosh(2\theta)$.
Then we drive the system with the Fibonacci sequence as 
introduced in Sec.~\ref{Sec:FibonacciSetup}.

\begin{figure}[t]
\centering
\subfloat[CFT calculation]{\includegraphics[width=2.50in]{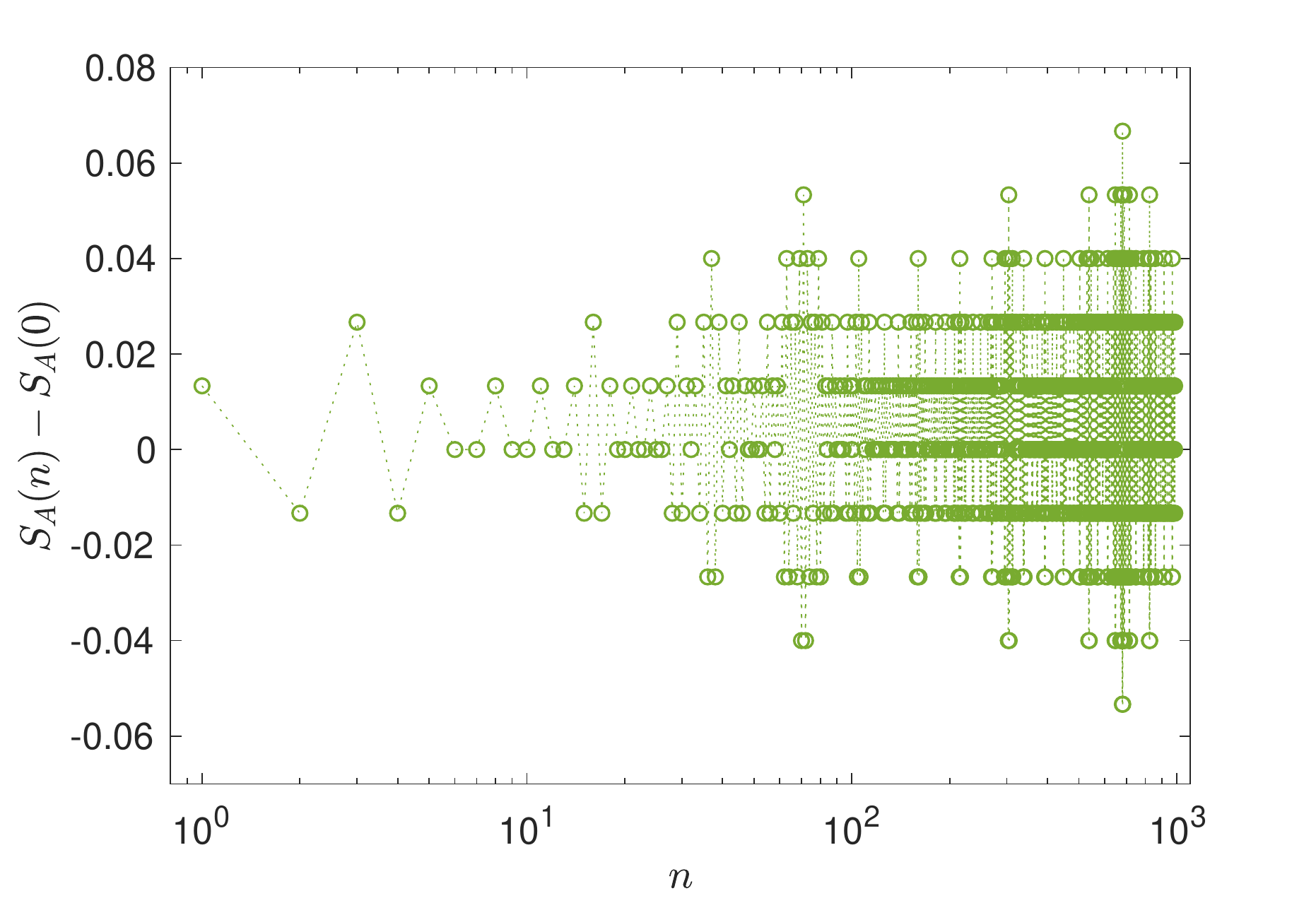}}
\subfloat[Lattice calculation with $L=500$]{\includegraphics[width=2.50in]{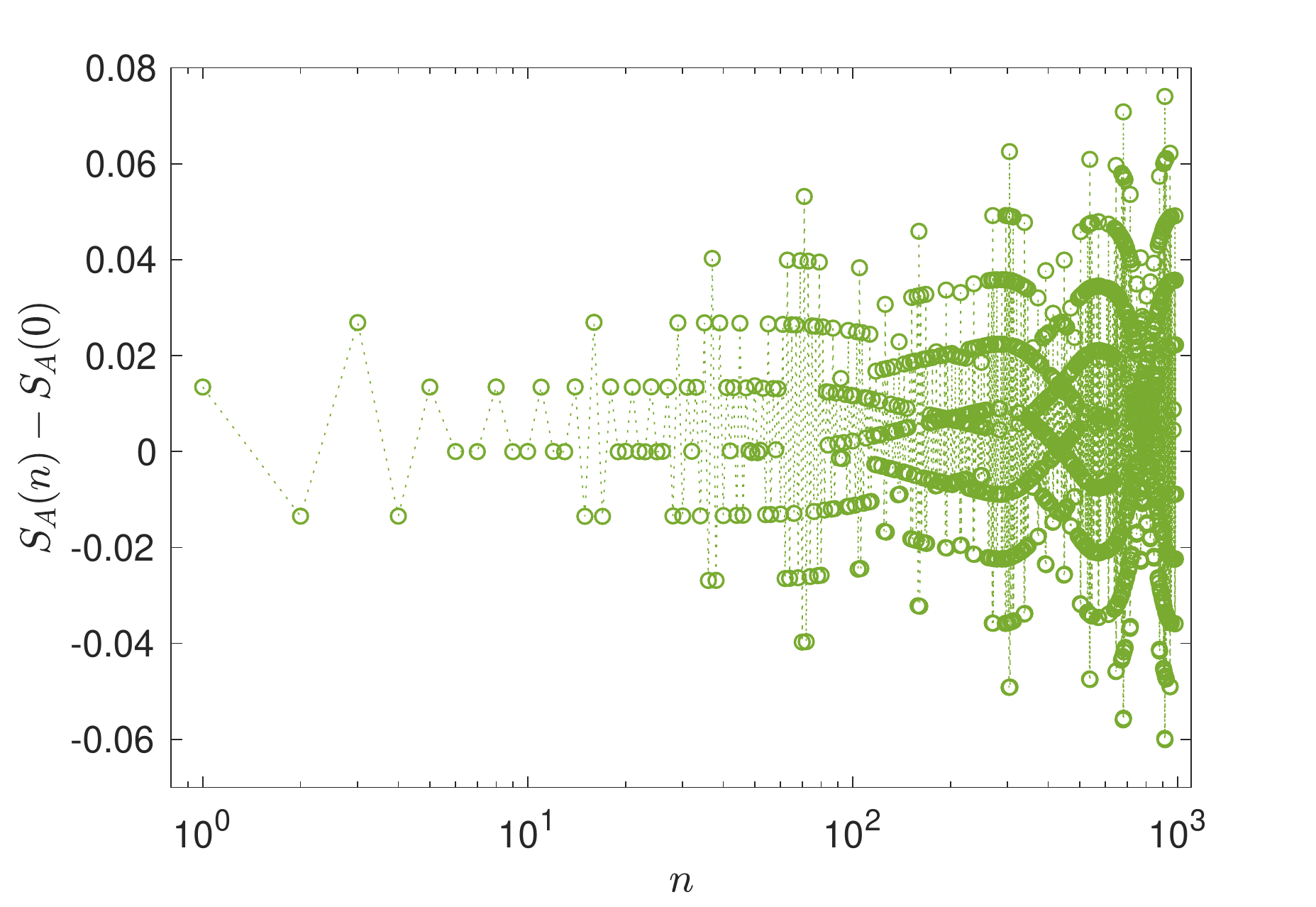}} \\
\subfloat[Lattice calculation with $L=1000$]{\includegraphics[width=2.50in]{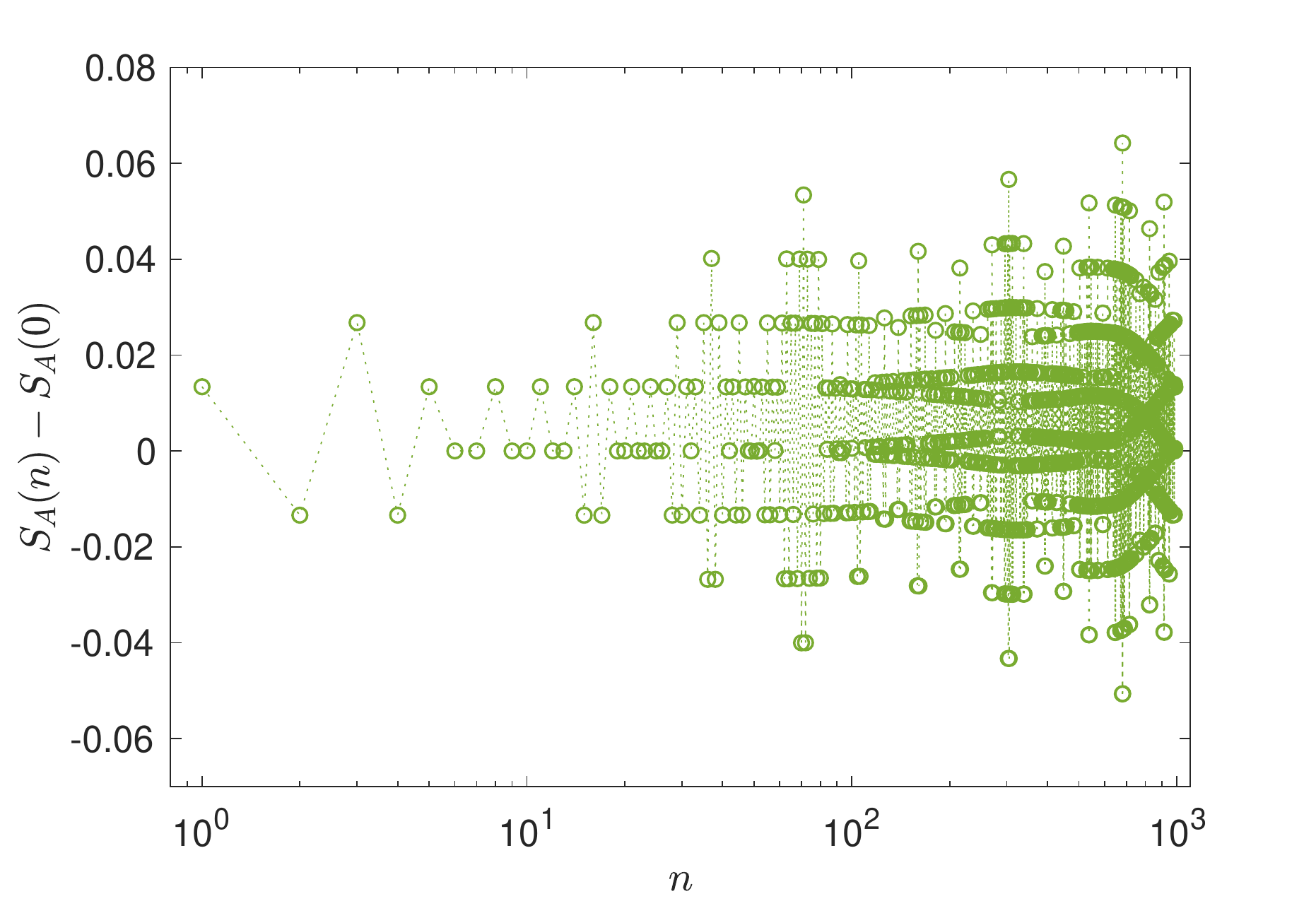}} 
\subfloat[Lattice calculation with $L=2000$]{\includegraphics[width=2.50in]{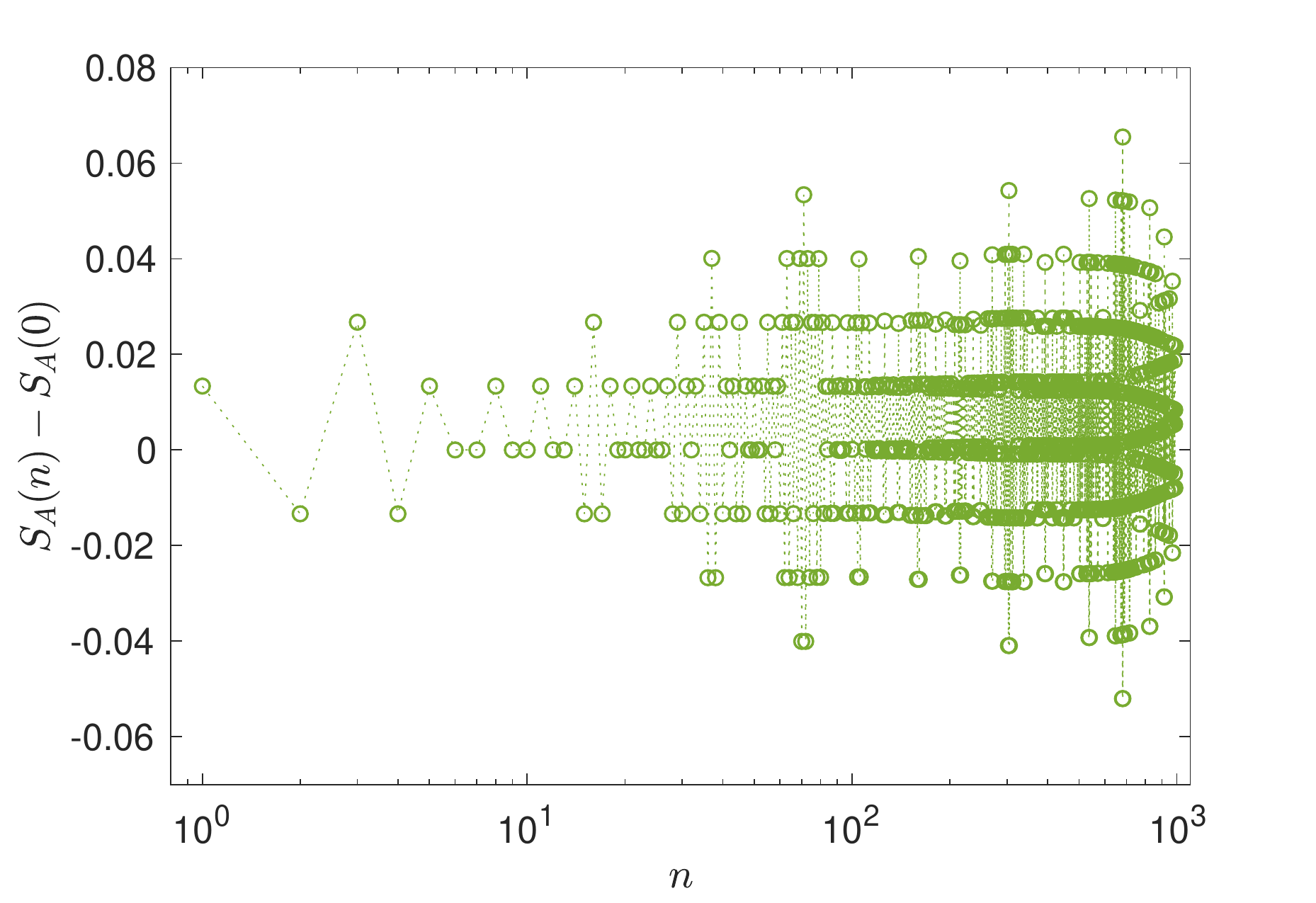}} 
\caption{
Comparison of the CFT and lattice calculations on the
entanglement entropy evolution $S_A(n)$ 
at the non-heating fixed point.
Here we choose $c=1$, $A=[0,L/2]$, $\theta=0.02$, and 
$n_{\text{max}}=F_{15}=987$.
In the lattice calculation, we consider $L=500$,  $1000$, and $2000$. 
}
\label{FixedPointCFTLatticeB}
\end{figure}

Fig.~\ref{FixedPointCFTLattice} presents the comparison of the lattice and CFT results 
on the entanglement/energy evolution. 
We find that the agreement is remarkable. 
A comparison on the entanglement entropy evolution
with larger driving steps $n$ can be found in
 Fig.~\ref{FixedPointCFTLatticeB}.
 In general, the agreement will break down for
 a large enough $n$, since more higher-energy
 modes will be involved as $n$ increases (Recall
 that the envelope of the total energy growth is power-law 
 in time at the non-heating fixed point).
 On the lattice model, the high-energy modes
 are no longer well described by a CFT, and therefore
 there must be a breakdown at certain $n$.
 \footnote{
 More precisely, let us denote $n^*$ as the driving step
 at which 
 the agreement between
 CFT and lattice calculations break down.
 From Fig.~\ref{FixedPointCFTLattice}, 
 one can observe that $n^*$ is a monotonically increasing function
 of $L$. This dependence can be understood as follows:
One may consider the wavefunction in the `Fock space' (which is
Verma module here) of a CFT of finite length $L$.
The initial state is the ground state $|0\rangle$. 
As we drive the system, higher energy modes
$|N\rangle$ ($N>1$) will be involved. 
It is noted that $N$ is independent of the 
length $L$ of the CFT. Since the energy spacing is proportional to $1/L$. one can find the energy $E_N(L)$ corresponding to $|N\rangle$ 
is higher than $E_N(L')$ if $L<L'$.
For a small $L$, $E_N(L)$ may be in the high-energy region which are no 
longer described by a CFT. However, by increasing $L$ to a large
enough $L'$, we can push $E_N(L')$ into the low-energy region
which are well described by a CFT.
This is why we have a better agreement in Fig.~\ref{FixedPointCFTLattice}
for a larger $L$.
}

\subsubsection{Exact nonheating fixed points in more general cases}
\label{Sec:NonheatingFixGeneral}

With the concrete examples illustrated in the previous discussions,
now we are ready to prove the statements as mentioned in the beginning
of Sec.~\ref{Sec: FixedPoint}, which we rewrite here:
\begin{enumerate}
    \item If both the driving Hamiltonians are chosen as elliptic types, 
one can always find \textit{exact} fixed points in the non-heating phases. 

\item At these (non-heating) fixed points, both the 
entanglement entropy and the energy evolution are of period $6$, i.e.,
$S_A(F_k)=S_A(F_{k+6})$ and $E(F_n)=E(F_{n+6})$.

\item The envelopes of the entanglement entropy and total energy
will grow logarithmically and in a power-law as a function of the 
driving steps $n$, respectively.
\end{enumerate}

\begin{defn}[Proof of claim 1.]
The non-heating fixed point has initial condition $(x_{F_2},x_{F_1},x_{F_0}) = (a,0,0)$, in terms of traces, we have
\begin{equation}
    \Tr (M_A M_B) =2a \,, \quad \Tr (M_A)=\Tr(M_B) = 0.
\end{equation}
In other words, to prove claim 1, we only need to find suitable initial conditions $T_A$ and $T_B$ for two elliptic Hamiltonian $H_A$ and $H_B$ such that the $\SU(1,1)$ matrices $M_A$ and $M_B$ are traceless, i.e. are reflection matrices (see \eqref{ReflectionMatrix}). Note the condition $\Tr (M_A M_B) =2a$ does not have a content as $a$ can be arbitrary. 

Now we explicitly find such $T_A$ and $T_B$. As discussed in Appendix \ref{Sec:OperatorEvoAppendix}, 
for a general elliptic Hamiltonian 
$H$ (see Sec.\ref{Sec:SL2deformation}) with driving interval $T$,
the corresponding M\"obius transformation is represented as follows
\be\label{GeneralEllipticMobius}
M=\left(
\begin{array}{cccc}
-\cos{\left( \frac{\calC \pi   T}{l} \right)} - i \frac{\sigma^0}{\calC} \sin{\left( \frac{\calC\pi  T}{l} \right)} &-i \frac{\sigma^+ + i\sigma^-}{\calC} \sin{\left( \frac{\calC\pi  T}{l} \right)}\\
i \frac{\sigma^+ - i\sigma^-}{\calC} \sin{\left( \frac{\calC\pi  T}{l} \right)} &-\cos{\left( \frac{\calC\pi   T}{l} \right)} + i \frac{\sigma^0}{\calC} \sin{\left( \frac{ \calC\pi T}{l} \right)}
\end{array}
\right)
\ee
where $\mathcal{C}=\sqrt{(\sigma^0)^2-(\sigma^+)^2-(\sigma^-)^2}$ and $(\sigma^0)^2-(\sigma^+)^2-(\sigma^-)^2>0$
with $\sigma^0,\, \sigma^+, \,\sigma^-\in\mathbb R$.
$l=L/q$ is the wavelength of deformation 
(See, e.g., Eq.~\eqref{Envelope_F}).
One can obtain the reflection matrix by choosing 
$T=\frac{1}{2}L_{\text{eff}}$ in Eq.~\eqref{GeneralEllipticMobius}, 
where $L_{\text{eff}}:=l/\mathcal{C}$ is the effective length.
Then Eq.~\eqref{GeneralEllipticMobius} becomes
\be\label{ReflectionM}
M=\left(
\begin{array}{cccc}
-i\frac{\sigma^{0}}{\calC} &-i\frac{\sigma^{+}+i\sigma^{-}}{\calC}\\
i\frac{\sigma^{+}- i\sigma^{-}}{\calC} &i\frac{\sigma^{0}}{\calC}
\end{array}
\right),
\ee
which is traceless obviously. 
That is to say, to arrive the 
 non-heating fixed point, we need to set $T_{A(B)}=l/2\mathcal{C}_{A(B)}$ and the corresponding reflection matrices $M_A$ and $M_B$ take the form of \eqref{ReflectionM} with subscripts $A$ and $B$. 
\end{defn}

\begin{defn}[Proof of claim 2.]
Having two reflection matrix $M_A$ and $M_B$, we immediately have the following useful property
\begin{equation}
    M_A^2 = M_B^2 = -1\,.
\end{equation}
Now let us use this to prove the claim 2. 

To verify the periodicity of entanglement entropy $S_A(F_n)$ and energy $E(F_n)$, it is sufficient to show the periodicity of the conformal transformation matrix $\Pi_{F_n}$. Let us first exam the case for $n=0\ldots 6$. Following the definition, the first 3 are
\begin{equation}
    \Pi_{F_0} = M_B \,, \quad \Pi_{F_1} = M_A \,, \quad \Pi_{F_2} = M_A M_B.
\end{equation}
Note the  $\Pi_{F_0} \neq \Pi_{F_1}$ is introduced as a convenient notation which satisfies the recurrence relation. For $n=3,4$, we can use the recurrence relation and find
\begin{equation}
    \Pi_{F_3} = \Pi_{F_2} \Pi_{F_1} = M_A M_B M_A \,, \quad \Pi_{F_4} = \Pi_{F_3} \Pi_{F_2} = M_A M_B M_A M_A M_B = M_A
 \end{equation}
where in the last equation we have used the property that $M_A^2=M_B^2=-1$. The results for $n=5,6$ are also obtained analogously 
\begin{equation}
    \Pi_{F_5}= \Pi_{F_4} \Pi_{F_3} = M_A M_A M_B M_A = - M_B M_A \,, \quad \Pi_{F_6}= \Pi_{F_5} \Pi_{F_4} = - M_B M_A M_A = M_B
\end{equation}
Note we have already verified the periodicity for $\Pi_{F_0}=\Pi_{F_6}$, further more $\Pi_{F_7}=\Pi_{F_6} \Pi_{F_5}= - M_B M_B M_A = M_A =\Pi_{F_1}$. And the recurrence relation for $\Pi_{F_n}$ only depends on the previous two elements, therefore we conclude
\begin{equation}
\wideboxed{
    \Pi_{F_{n+6}}=\Pi_{F_{n}} \,, \quad n \in \ZZ^{\geq 0} \,. 
    }
\end{equation}
The claim 2 follows immediately. 
\end{defn}

\begin{defn}[Proof of claim 3.]
First Ref.~\cite{iochum1991power,Iochum1992} generally prove that the norm of $\Pi_n$ is polynomially bounded at $\lambda_L=0$. Therefore, we only need to show there is at least a sequence of point that approaches infinity polynomially similar to what we showed in last section for a specific example. 

Note, for any two distinct reflection matrix $M_A$ and $M_B$, their product is hyperbolic. Then we can use their product to generate a sequence of points, e.g. $\Pi_{F_{6}}=M_B$ and $\Pi_{F_{7}}=M_A$  we can generate
\begin{equation}
    \Pi_{n} = \Pi_{F_{6m+1}} \Pi_{F_{6m}} \cdot \Pi_{F_{6m-5}} \Pi_{F_{6m-6}} \cdots \Pi_{F_{7}} \Pi_{F_{6}} = (M_AM_B)^m
\end{equation}
for 
\begin{equation}
    n = F_{6m+1}+F_{6m} + F_{6m-5}+ F_{6m-6} + \ldots + F_7 +F_6 \,.
\end{equation}
Since $|\Tr (M_A M_B)|>2$, we conclude that $|\Tr \Pi_n | \sim e^{m \lambda}$ at $m \rightarrow \infty$ where 
\begin{equation}
    \lambda = \log \frac{|\Tr (M_A M_B)| + \sqrt{|\Tr (M_A M_B)|-4} }{2} > 0\,.
\end{equation}
Therefore, the total energy will grow exponentially w.r.t. $m$ as if in the heating phase. However, remember the step index $n$ is also exponential in $m$ and therefore in terms of physical steps $n$, the energy grows in power law. Similarly, we have the entanglement entropy grow logarithmically.\footnote{For the subtlety about the sign in entropy growth, one can always multiply one more matrix or move the entanglement cut to ensure the entropy is growing instead of decreasing.} 
\end{defn}

\subsection{Entanglement and energy dynamics in the heating phases}

In this subsection, we present the rich features in the heating phase, including the time evolution of total energy, energy-momentum density and entanglement entropy. 

\subsubsection{Group walking and entanglement/energy evolution}
\label{Sec: GroupWalkingQuasiP}

\begin{figure}[htp]
\centering
\subfloat[Group walking of $\rho$]{\includegraphics[width=5.20in]{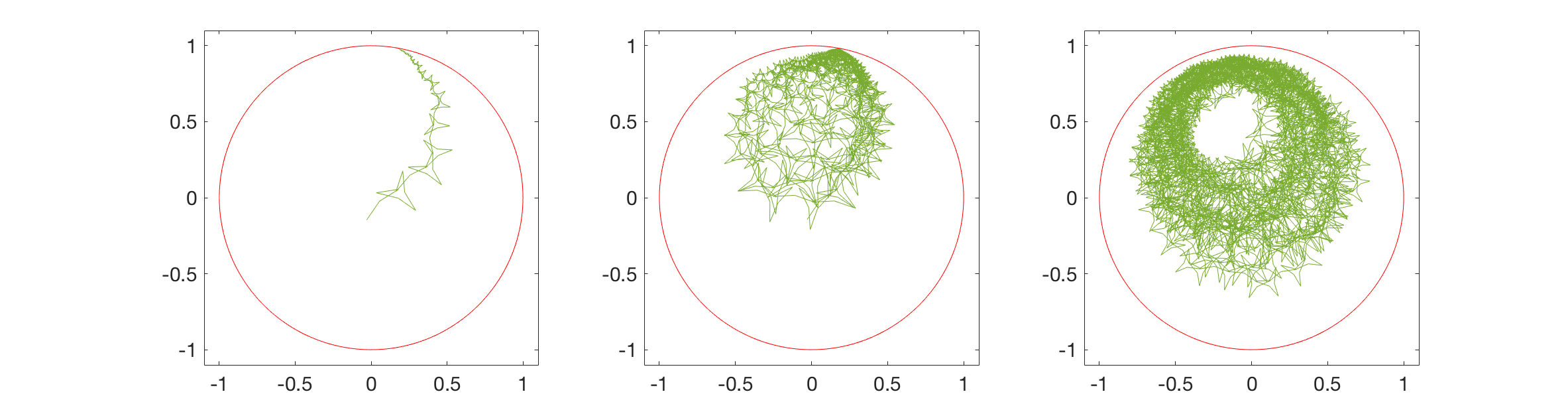}} \\
\subfloat[Group walking of $\rho\cdot \zeta$]{\includegraphics[width=5.20in]{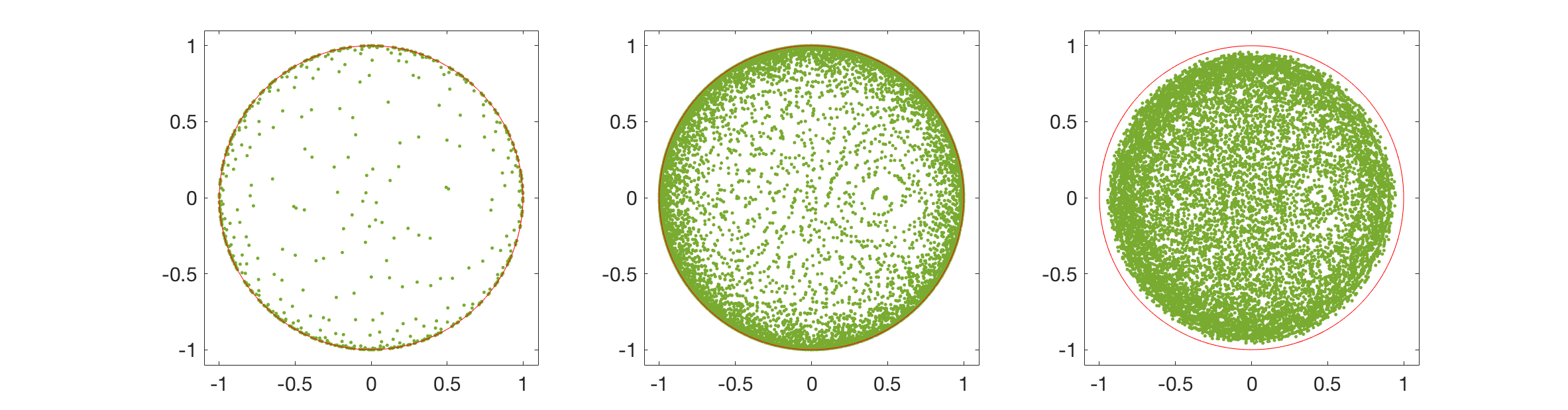}} \\
\caption{
Group walking of $\rho$ and $(\rho\,\zeta)$ in the heating
phase of a Fibonacci driven CFT. 
We consider a Fibonacci quasi-periodical driving
with $H_0$ and $H_{\theta=0.5}$.
The parameters are $T_0/L=1/2$, and (from left to right) $T_1/L_{\text{eff}}=0.041$, $0.04$, 
and $0.0401$, respectively.
The total number of driving steps is taken as $F_{20}=10946$.
}
\label{GroupWalkingFibonacci1}
\end{figure}

\begin{figure}[htp]
\centering
\subfloat[Group walking of $\rho$]{\includegraphics[width=5.10in]{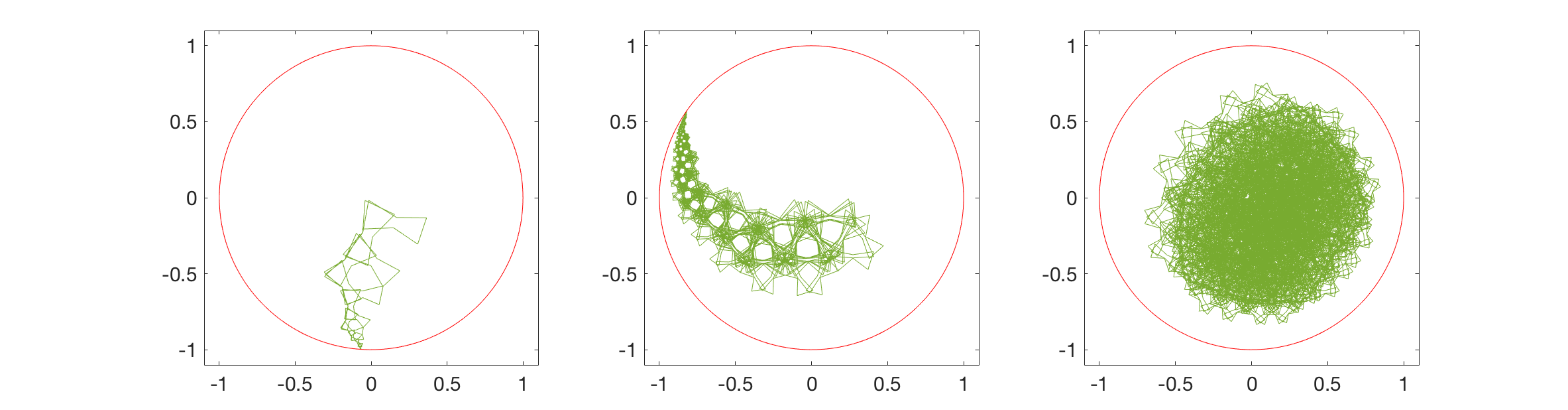}} \\
\subfloat[Group walking of $\rho\cdot\zeta$]{\includegraphics[width=5.10in]{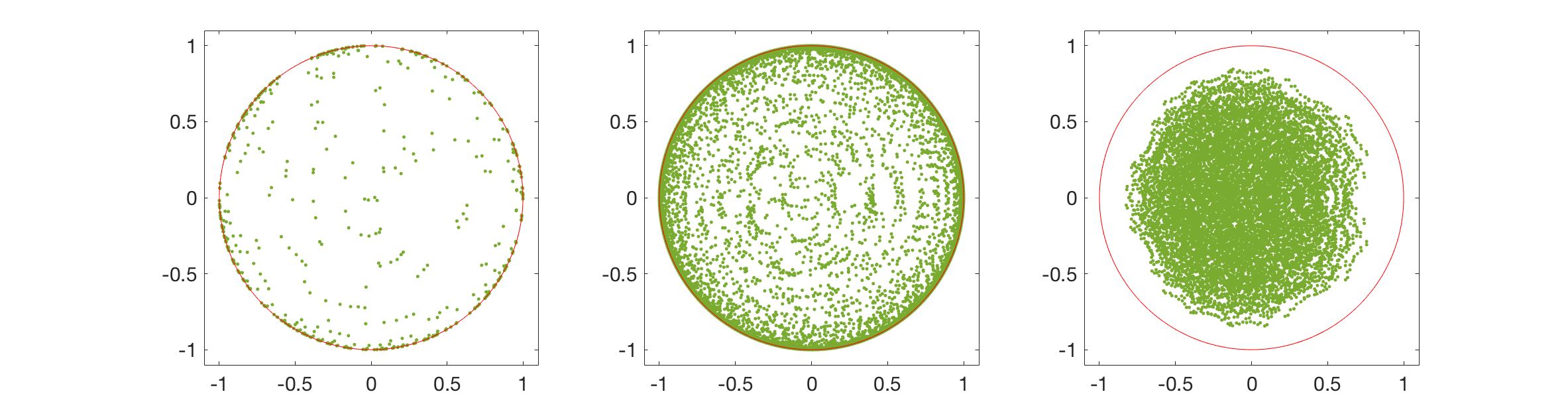}} 
\caption{
Group walking of $\rho$ and $(\rho\,\zeta)$ in the heating
phase of a Fibonacci driven CFT. 
We consider a Fibonacci quasi-periodical driving
with $H_0$ and $H_{\theta=0.5}$.
The parameters are $T_0/L=0.6$, and (from left to right) $T_1/L_{\text{eff}}=0.06$, $0.05$, 
and $0.055$, respectively.
The total number of driving steps is taken as $F_{20}=10946$.
}
\label{GroupWalkingFibonacci2}
\end{figure}

As discussed in Sec.~\ref{Sec: GroupWalking}, the group walking of $\rho$ and
$(\rho\,\zeta)$ of the $\SU(1,1)$ matrix in \eqref{M_alpha_z} reflect the time evolution 
properties of the total energy, energy-momentum density, and the entanglement entropy.
In the following, we will study the group walking of 
$\rho$ and $\rho\zeta$, as well as the energy/entanglement evolution.
The detailed discussion on energy-momentum density is left to the next subsection.

First, we consider the group walking of $\rho$ 
in $\Pi_n$ 
\be\label{M_alpha_z_quasiP}
\small
\Pi_n=\frac{1}{\sqrt{1-|\rho|^2}}
\left(
\begin{array}{cccc}
\sqrt{\zeta} &-\rho^*\frac{1}{\sqrt{\zeta}}\\
-\rho \sqrt{\zeta} &\frac{1}{\sqrt{\zeta}}\\
\end{array}
\right),\quad \rho\in \mathbb D, \,\, \zeta\in \partial \mathbb D.
\ee
As shown in Fig.~\ref{GroupWalkingFibonacci1} is a sample plot of 
the group walking of $\rho$ and $\rho\zeta$ nearby a certain point
(which is $T_0/L=1/2$ and $T_1/L_{\text{eff}}=0.04$ here, with
$L_{\text{eff}}=L\cosh(2\theta)$) in the parameter space.
See also Fig.~\ref{GroupWalkingFibonacci2} for another sample plot. 
The two driving Hamiltonians are 
$H_0$ and $H_{\theta=0.5}$, with $H_{\theta}$ given in Eq.~\eqref{H_theta_A}.
As the driving steps $n$ increase, one can find that $\rho$
will walk to a certain point $\rho_{n=\infty}$ on $\partial \mathbb D$.
This is quite an interesting feature, since $\rho$ does not
move around on $\partial \mathbb D$ in the limit $n\to \infty$.
\footnote{It is interesting that this
phenomenon also happens in the heating phases of
both the periodic and random driving CFTs, where we can prove that
$\rho_{n}$ converge to a fixed $\rho_{n=\infty}$ on $\partial \mathbb D$
(See, e.g., Eq.~\eqref{GroupWalkPeriodLimit} for the case of periodical drivings).
We suspect this is generic feature as long as the Lyapunov exponent 
is positive, and it is interesting to prove this observation, e.g., in 
the Fibonacci driven CFT here.
}

\begin{figure}[t]
\centering
\subfloat[Total energy evolution]{\includegraphics[width=5.20in]{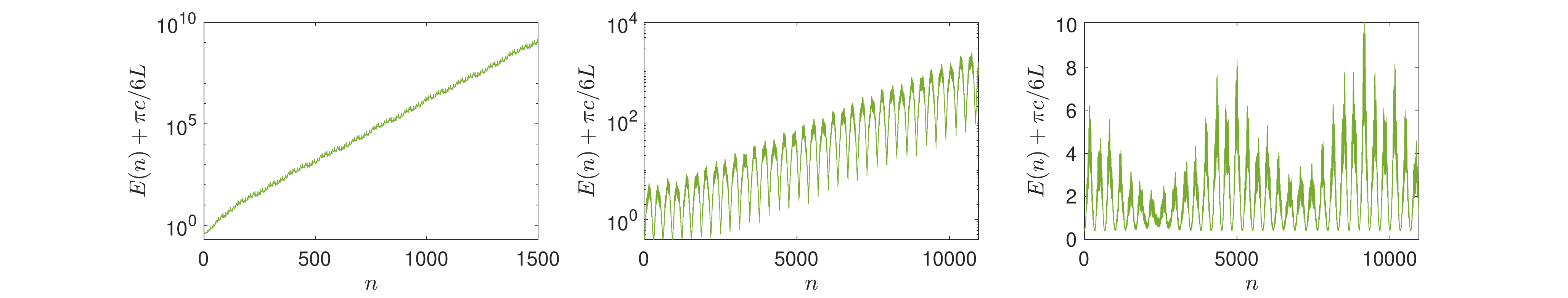}} \\
\subfloat[Entanglement entropy evolution]{\includegraphics[width=5.20in]{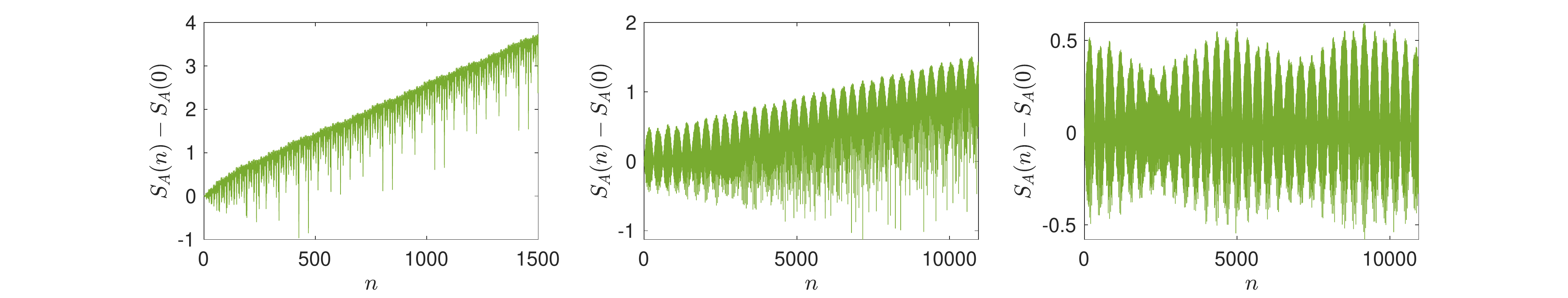}} 
\caption{
Time evolution of the total energy and the entanglement 
entropy of $A=[0,L/2]$. 
The parameters are the same as those in Fig.~\ref{GroupWalkingFibonacci1}.
Note that the total energy and entanglement entropy (in the left plot) are plotted in a small window to see the detailed oscillating structure.
}
\label{EEevolutionFibonacci1}
\end{figure}

\begin{figure}[htp]
\centering
\subfloat[Total energy evolution]{\includegraphics[width=5.10in]{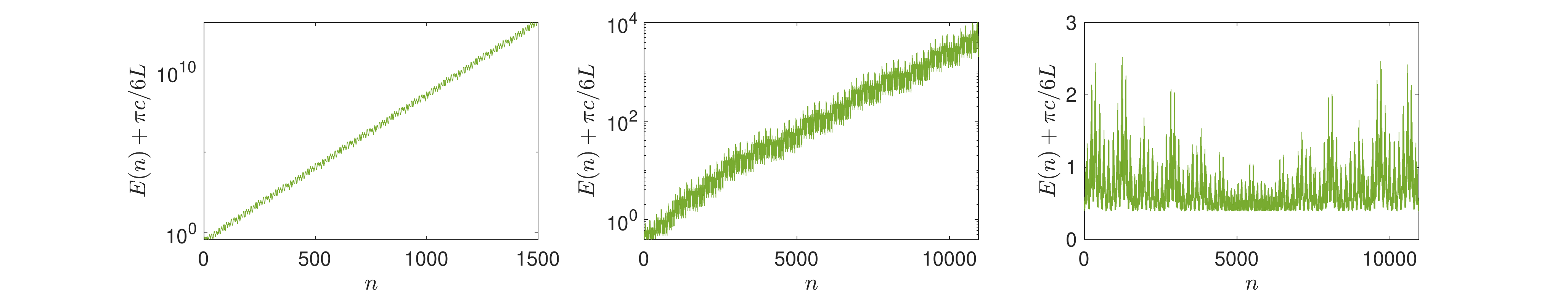}}  \\
\subfloat[Entanglement entropy evolution]{\includegraphics[width=5.10in]{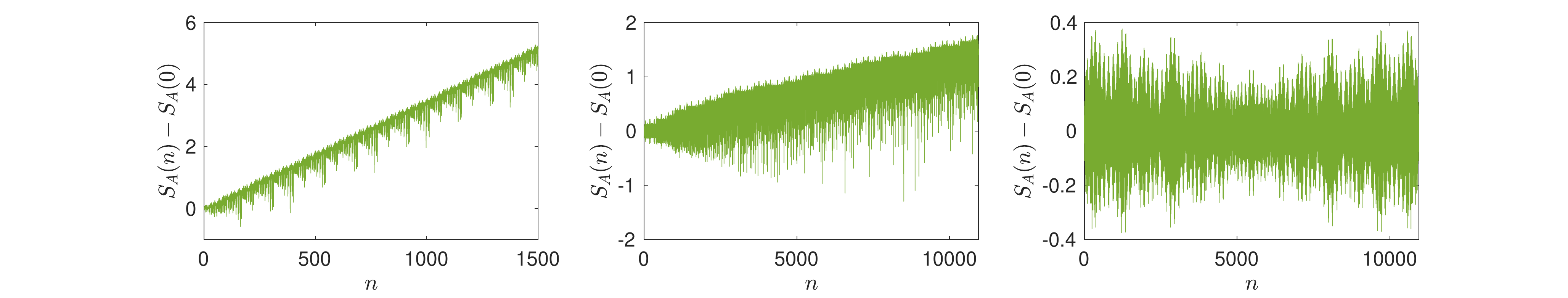}} 
\caption{
Time evolution of the entanglement entropy of $A=[0,L/2]$ and the total energy of the system.
The parameters are the same as those in Fig.~\ref{GroupWalkingFibonacci2}.
Note that the total energy and entanglement entropy (in the left plot) are plotted in a small window
to see the detailed oscillating structure.
}
\label{EEevolutionFibonacci2}
\end{figure}

As we have discussed in Sec.~\ref{Sec: GroupWalking}, as $\rho$ approaches 
the boundary $\partial\mathbb D$, both the entanglement entropy and the 
total energy of the system will grow
accordingly, as seen from Eqs.\eqref{Etotal_rho} 
and \eqref{HalfEE_general_GroupWalking}.
In Fig.~\ref{GroupWalkingFibonacci1} and Fig.~\ref{GroupWalkingFibonacci2}, 
one can find that for different $(T_0,\, T_1)$, $\rho$ approaches the 
boundary $\partial \mathbb D$ with different rates, which correspond
to different Lyapunov exponents and different growth rates
in the entanglement entropy/total energy.
This can be intuitively seen by looking at the entanglement/energy growth in
Fig.~\ref{EEevolutionFibonacci1} and Fig.~\ref{EEevolutionFibonacci2}, respectively.
Also, the patterns in $|\rho_n|$ in Fig.~\ref{GroupWalkingFibonacci1} 
and Fig.~\ref{GroupWalkingFibonacci2} result in the oscillating features in
the entanglement/energy growth in Fig.~\ref{EEevolutionFibonacci1} and \ref{EEevolutionFibonacci2}.
\footnote{It is noted that the total energy only depends on 
 $|\rho|$, as seen from Eq.\eqref{Etotal_rho}. Therefore,
the oscillating structures in $E(n)$ are totally due to $|\rho|$.
For the entanglement entropy $S_A(n)$, however, it depends on both $|\rho|$
and $\rho\zeta$, as seen from Eq.~\eqref{HalfEE_general_GroupWalking}.
Then the oscillating structures in $S_A(n)$ come from both $|\rho|$ 
and $\rho\zeta$.
}

As a remark, it is noted that for the case of $T_1/L_{\text{eff}}=0.0401$ in Figs.
\ref{GroupWalkingFibonacci1} and \ref{EEevolutionFibonacci1}, 
$\rho$ does not walk to the boundary $\partial \mathbb D$, and the total energy
seems to simply oscillate in $n$. This is because the Lyapunov exponent 
at this point is too small. One needs to take a longer time (larger $n$)
to observe the growth behavior.

Second, Let us consider the group walking of $(\rho\zeta)$.
As shown in Figs.\ref{GroupWalkingFibonacci1} and \ref{GroupWalkingFibonacci2}, 
different from the behavior of $\rho$, which walks to a fixed $\rho_{\infty}$
in the $n\to\infty$ limit, $(\rho\zeta)$ will walk around 
on the boundary $\partial \mathbb D$
even in the $n\to \infty$ limit. This will result in two interesting features:
\begin{enumerate}
    \item The locations of energy-momentum density peaks will 
    move around during the quasi-periodic driving, as seen from
    Eq.\eqref{PeakLocation}.
    
    \item More oscillating structures will be introduced in the 
    time evolution of the entanglement entropy.
    As discussed in Sec.~\ref{Sec: GroupWalking},
    the total energy only depends on $|\rho|$, but the entanglement entropy
    depends on both $|\rho|$ and $\rho\zeta$, as seen in Eq.\eqref{HalfEE_general_GroupWalking}.
    This extra oscillating structure can be found
    in Figs.\ref{EEevolutionFibonacci1} and \ref{EEevolutionFibonacci2},
    in particular for the case with large Lypunov exponents.
    
\end{enumerate}

In a short summary, there are rich patterns in the group walking of
$\rho$ and $(\rho\zeta)$ in the heating phase of a Fibonacci
quasi-periodically driven CFT.
The velocity of $\rho$ walking towards 
$\partial \mathbb D$ determines the value of Lyapunov exponents, 
and therefore the growth rate of the entanglement entropy and total 
energy. The behavior of $|\rho|$ determines the concrete features 
of the energy growth $E(n)$ through Eq.\eqref{Etotal_rho}. 
In general, one can observe various oscillating features in the 
growth of $E(n)$. Both $|\rho|$ and $(\rho\zeta)$ determine
the features of $S_A(n)$ through Eq.\eqref{HalfEE_general_GroupWalking},
where more patterns of oscillations can be observed comparing 
to $E(n)$.

\subsubsection{Locations of energy-momentum density peaks}

\begin{figure}[t]
\centering
\subfloat[Group walking of $(\rho\zeta)$]{\includegraphics[width=6.50in]{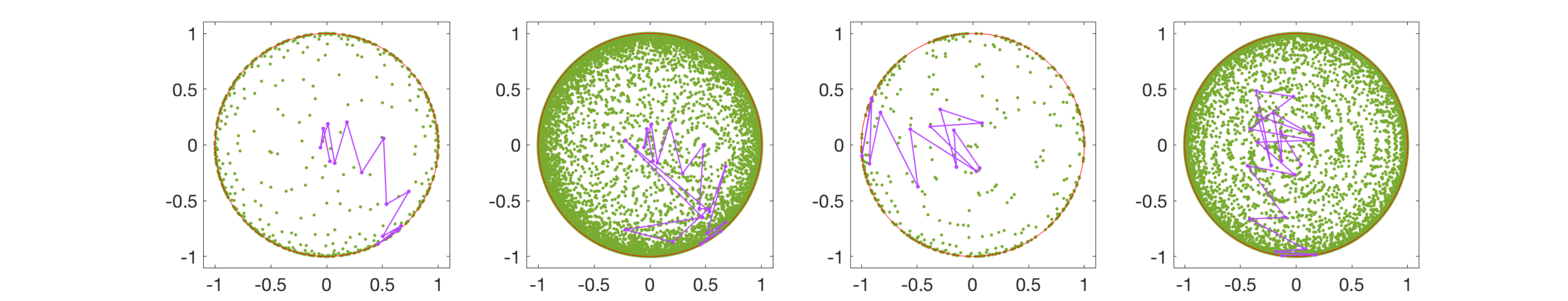}} \\
\subfloat[Arg$(\rho\zeta)/\pi$ at the Fibonacci numbers]{\includegraphics[width=6.50in]{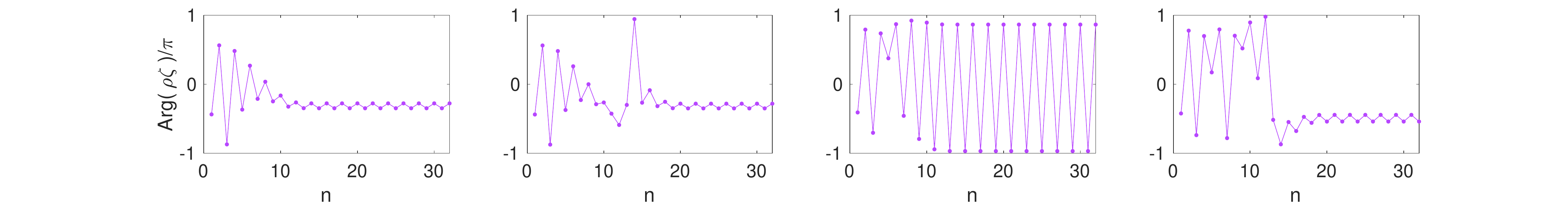}}
\caption{
(Top)
Group walking of $(\rho\zeta)$ at all numbers (green dots)
and at Fibonacci numbers (purple). 
The two driving Hamiltonians are $H_0$ and $H_{\theta=0.5}$.
The  parameters (from left to right) are
$T_0/L=1/2$ and $T_1/L_{\text{eff}}=0.041$, 
$T_0/L=1/2$ and $T_1/L_{\text{eff}}=0.04$,
$T_0/L=0.6$ and $T_1/L_{\text{eff}}=0.06$,
$T_0/L=0.6$ and $T_1/L_{\text{eff}}=0.05$,
respectively.
The first (last) two plots have the same
parameters as those in
Fig.~\ref{GroupWalkingFibonacci1} (Fig.~\ref{GroupWalkingFibonacci2}).
The total number of driving steps are 
$F_{25}=121393$.
This means there are in total 
$25$ steps of moving (purple lines) at the 
Fibonacci numbers.
(Bottom) Arg$(\rho\zeta)\in(-\pi,\pi]$ as a function of $n$, where
$n$ denotes the $n$-th Fibonacci numbers $F_n$.
The parameters are the same as the top panel, but with 
 a larger driving number $F_{32}$.
}
\label{GroupWalkingFibonacciNumber}
\end{figure}

There are several features on the distribution of energy-momentum
density we hope to point out in the heating phase of 
a Fibonacci driven CFT:

\begin{enumerate}
    \item In the heating phase, there is an array of peaks 
    of energy-momentum density distributed in the real space.
    The locations of these peaks are determined by $\Pi_n$ through Eq.~\eqref{PeakLocation}.
    In fact, as we have shown in Sec.~\ref{Sec:LyapunovHeatingPhase}, 
    a positive Lyapunov exponent $\lambda_L$ always indicates an array of peaks in the energy-momentum density $\langle T(x,n)\rangle$ in real space.
    
    \item Different from the periodically driven CFT, where the 
    peaks are located at the same positions after each period of driving (See Eq.~\eqref{PeakLocationHP}), in the quasi-periodical driving, 
    since there is 
    not a regular driving period, the locations of the 
    energy-momentum density peaks will in general move around after 
    each driving step.
    This can be seen based on the group walking of 
    $(\rho\zeta)$, which determines the locations of 
    energy-momentum density peaks [see Eq.\eqref{PeakLocation}], in Figs.\ref{GroupWalkingFibonacci1}, 
    \ref{GroupWalkingFibonacci2}, and \ref{GroupWalkingFibonacciNumber}.

    \item 
    Although the locations of the energy-momentum density peaks
    will move around, 
    in the long time driving limit ($n\gg 1$),
    we can still find some regular patterns.
    In particular, we find there is an even/odd effect in the distribution 
    of these energy-momentum density peaks when the driving steps are
    Fibonacci numbers. More concretely, let us denote $x_{\text{peak}}$ as the 
    locations of peaks of the energy-momentum density. Then we 
    observe that 
    \be\label{PeakLocation_evenodd}
    x_{\text{peak}}(F_n)=x_{\text{peak}}(F_{n+2}),
    \ee
    as shown in Fig.~\ref{GroupWalkingFibonacciNumber}.

\end{enumerate}

We find that the even/odd effect as mentioned above
is closely related to the group walking of
    $\rho$ as discussed in the previous subsection, i.e.,
    in the long time driving limit $n\to\infty$, $\rho$ will
    flow to a certain point $\rho_{\infty}$ on $\partial \mathbb D$.
By taking this as an assumption, i.e, 
$\lim_{n\to\infty} \rho_n=\rho_{\infty}\in\partial \mathbb D$, 
one can prove that there are indeed even/odd effects
in the locations of energy-momentum density peaks, as follows.

Let us denote the matrix elements of $\Pi_{F_{n-1}}$ and $\Pi_{F_{n}}$ as follows
\begin{equation}
    \Pi_{F_{n-1}} = \frac{1}{\sqrt{1-|\rho|^2}}
\begin{pmatrix}
\sqrt{\zeta} &-\rho^*\frac{1}{\sqrt{\zeta}}\\
-\rho \sqrt{\zeta} &\frac{1}{\sqrt{\zeta}}\\
\end{pmatrix} \,, \quad 
 \Pi_{F_{n}} = \frac{1}{\sqrt{1-|\rho'|^2}}
\begin{pmatrix}
\sqrt{\zeta'} &-\rho'^*\frac{1}{\sqrt{\zeta'}}\\
-\rho' \sqrt{\zeta'} &\frac{1}{\sqrt{\zeta'}}\\
\end{pmatrix}
\end{equation}
Then applying the recurrence relation, we have
\begin{equation}
\begin{aligned}
    \Pi_{F_{n+1}}= \Pi_{F_{n}} \cdot \Pi_{F_{n-1}}& = \frac{1}{\sqrt{(1-|\rho|^2)(1-|\rho'|^2)}}
    \begin{pmatrix}
   \sqrt{\zeta}\sqrt{\zeta'}+\rho(\rho')^*\frac{\sqrt{\zeta}}{\sqrt{\zeta'}}
    &
    -(\rho')^*\frac{1}{\sqrt{\zeta}\cdot \sqrt{\zeta'}}-\rho^*\frac{\sqrt{\zeta'}}{\sqrt{\zeta}}\\
-\rho'\sqrt{\zeta}\sqrt{\zeta'}-\rho\cdot\frac{\sqrt{\zeta}}{\sqrt{\zeta'}} 
&  \frac{1}{\sqrt{\zeta}\cdot \sqrt{\zeta'}}+ \rho^*\rho'\cdot\frac{\sqrt{\zeta'}}{\sqrt{\zeta}}
    \end{pmatrix} \\
    &:=
    \frac{1}{\sqrt{1-|\rho''|^2}} 
    \begin{pmatrix}
    \sqrt{\zeta''} &-\rho''^*\frac{1}{\sqrt{\zeta''}}\\
-\rho'' \sqrt{\zeta''} &\frac{1}{\sqrt{\zeta''}}\\
    \end{pmatrix}
    \end{aligned}
\end{equation}
In the heating phase, assuming $\rho \approx \rho' \rightarrow \rho_{\infty}$ at $n\rightarrow \infty$ with $|\rho_\infty|=1$, we have 
\be\label{rhozeta2}
\left\{
\begin{split}
& \rho'' \zeta''=\frac{\rho'\zeta\zeta'+\rho\zeta}{1+\rho^*\rho'\zeta'}
= \frac{\rho\zeta(1+\zeta')}{1+\zeta'}=
\rho\zeta,\\
& \rho'' = \rho' = \rho = \rho_{\infty}.
\end{split}
\right.
\ee
The first formula in Eq.~\eqref{rhozeta2} indicates that
\be
(\rho\zeta)_{F_n}=(\rho\zeta)_{F_{n+2}}, \quad n\in \mathbb Z^{>0}.
\ee
According to Eq.~\eqref{PeakLocation},
this implies that in the long time driving limit there is an 
even-odd effect in the locations of peaks of the energy momentum density
at the Fibonacci numbers (See Eq.~\eqref{PeakLocation_evenodd}).

\subsection{Other quasi-periodic driving: Aubry-Andr\'e like}
\label{Sec: Aubry Andre driving CFT}

The other well-known model for the quasi-crystal is the Aubry-Andr\'e model \cite{harper1955,aubry1980}, which describes free electron hopping on a one-dimensional lattice with the following Hamiltonian
\begin{equation*}
	H = \sum_j \left(c_j^\dag c_{j+1} + c_{j+1}^\dag c_j \right) + \lambda \sum_j \cos(2\pi \omega j + \delta) c_j^\dag c_j\,,
\end{equation*}
where $\omega$ is an irrational number that is incommensurate to the lattice periodicity and thus characterizes a quasi-periodic on-site potential. It has been used for the study of localization\cite{Soukoulis1982} and also appears in the two-dimensional integer quantum Hall effect\cite{hofstadter1976}. Given an irrational $\omega$, this model has a localization transition at $\lambda=2$ and the two phases are related by a duality transformation. This is rigorously proved in the mathematics literature \cite{damanik_2017,Soukoulis1982,simons1982} by studying the so-called almost Mathieu operator,
which is equivalent to the Aubry-Andr\'e model at the single particle level.

Motivated by the Aubry-Andr\'e model, 
here we introduce another type of quasi-periodic driving, which we call 
an Aubry-Andr\'e (quasi-periodically) driven CFT.

\subsubsection{Setup}

Let us first recall our minimal setup of the periodically driven CFT in Sec.\ref{Sec:MinimalSetup}. Our protocol for the quasi-periodic driving will be a modification to that.
In the minimal setup, the system takes the open boundary condition with $q=1$.
As depicted in \eqnref{FloquetSetup}, each cycle consists of two steps, $(H_0,T_0)$ and $(H_1,T_1)$, with $H_0$ being $H_{\theta=0}$ and $H_1$ being $H_{\theta\neq0}$. The $\SU(1,1)$ matrices associated to the unitary evolution $e^{-i H_0T_0}$ and $e^{-iH_1T_1}$ are denoted as $M_0$ and $M_1$ respectively. The formula for $M_1$ is copied below for reader's convenience. Plugging in $\theta=0$ yields $M_0$.
\begin{equation*}
    M_1 = \begin{pmatrix}
    \alpha & \beta \\ \beta^* & \alpha^*
    \end{pmatrix}\,,\quad
    \small
	\left\{
	\begin{split}
		&\alpha=\cos{\left( \frac{\pi T_{\theta}}{L_{\text{eff}}} \right)} + i\cosh(2\theta)\cdot \sin{\left( \frac{\pi T_{\theta} }{L_{\text{eff}}} \right)},\\
		&\beta=- i\sinh(2\theta)\cdot\sin{\left( \frac{\pi T_{\theta}}{L_{\text{eff}}} \right)}.
	\end{split}
	\right.
\end{equation*}
Note that $T_0$ and $T_1$ appear in $M_0$ and $M_1$ through $\cos$ and $\sin$, which implies that $T_0 \rightarrow T_0 + L$ and $T_1 \rightarrow T_1 + L_{\text{eff}}$ yield the same set of $\SU(1,1)$ matrices\footnote{Up to an overall minus sign that does not affect the dynamics because the actual M\"obius transformation acts as $\SU(1,1)/\ZZ_2$.}. 

\begin{figure}
	\centering
	\small
	\begin{tikzpicture}[baseline={(current bounding box.center)}]
	\node at (-5pt, 7pt){$H(t)$:};
	\draw [thick](20pt,0pt)--(40pt,0pt) --(40pt,20pt) -- (60pt,20pt);
	
	\draw [thick](60pt,20pt)--(60pt,0pt) --(100pt,0pt) --(100pt,20pt) -- (120pt,20pt);
	
	\draw [thick](120pt,20pt)--(120pt,0pt)--(180pt,0pt) -- (180pt,20pt) -- (200pt,20pt);
	
	\node at (10+40pt,28pt){$H_1$};
	\node at (30pt, 8pt){$H_0$};
	
	\draw[>=stealth,<->] (20pt,-5pt) --node[below]{$\omega L$} (40pt,-5pt);
	\draw[>=stealth,<->] (60pt,-5pt) --node[below]{$2\omega L$} (100pt,-5pt);
	\draw[>=stealth,<->] (120pt,-5pt) --node[below]{$3\omega L$} (180pt,-5pt);
	
	\draw [>=stealth,->] (210pt, -10pt)--(230pt,-10pt);
	\node at (235pt, -10pt){$t$};
	\end{tikzpicture}
	\caption{\label{fig:aubry Andre setup}Aubry-Andr\'e quasi-periodic driving. In the $n$-th cycle, $H_0$ is applied for time $T_0=n\omega  L$ and $H_1$ is applied for time $T_1$.}
\end{figure}
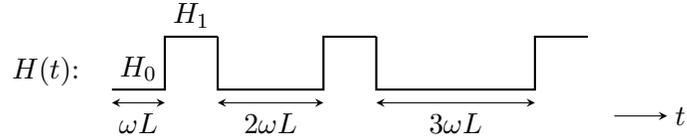

This observation leads to the following way of introducing the quasi-periodicity. We keep $T_1$ fixed and let $T_0$ have a dependence on the number of cycles. In the $n$-th cycle, $H_0$ is applied for time $T_0=n\omega L$, where $L$ is the system size. The whole sequence is depicted in \figref{fig:aubry Andre setup}. If $\omega = p/q$ is a rational number, the unitary evolution generated by $e^{-iH_0T_0}$ will repeat after every $q$ cycles, and protocol is reduced to a periodic driving with each period consisting of $q$ cycles. If $\omega$ is irrational, such periodicity disappears which gives rise to a quasi-periodic driving. One can also fix $T_0$ and vary $T_1$ instead, which does not change the result qualitatively and thus will not be discussed.

\subsubsection{Phase structure for a single irrational number}

It is subtle to directly access the dynamics with an irrational $\omega$, especially when there is no analytical tool. 
For example, to extract the Lyapunov exponent by numerically computing the matrix product can be unstable due to numerical errors. Therefore, we will track a sequence of rational numbers to approach the physics at the irrational numbers. Each rational number corresponds to a periodic driving system and we can apply the trace classifier to identify the phase structure. This is exactly the same as our discussion for the Fibonacci driven CFT in Sec.\ref{Sec:PhaseDiagramPeriodicFib}.

To illustrate the general features, let us consider $\omega = (\sqrt{5}-1)/2$ being the inverse golden ratio as a concrete example, which can be approximated by the sequence $\omega_n = F_{n-1}/F_{n}, n=1,2,3\cdots$. For a given $n$, the driving repeats after every $F_n$ cycles. We can compute the matrix that corresponds to the evolution for one period and determine the dynamics from the trace.

\begin{figure}
\centering
\subfloat[$n=3$]{
	\includegraphics[width=5cm]{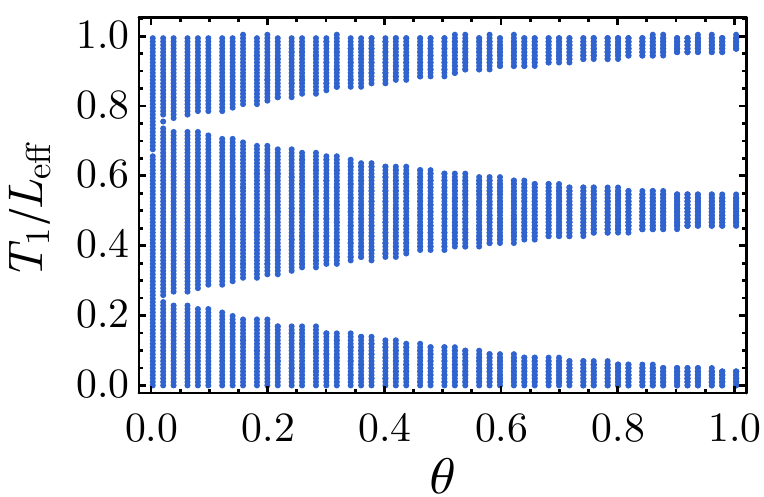}
}
\hspace{0pt}
\subfloat[$n=5$]{
	\includegraphics[width=5cm]{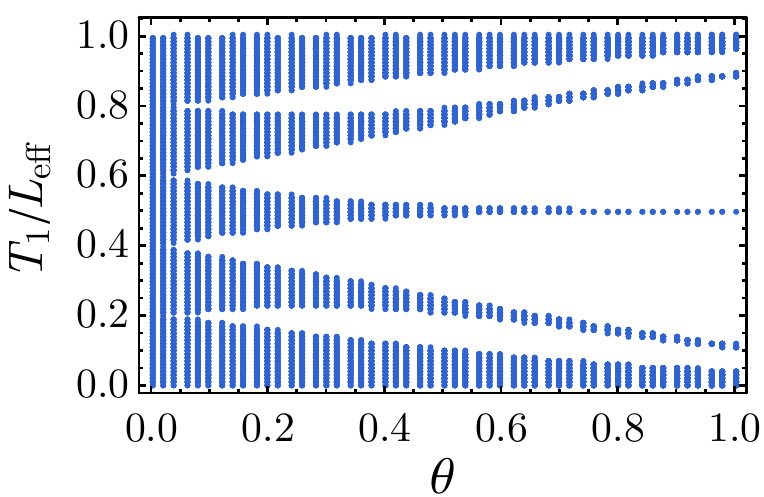}
}\\
\hspace{0pt}
\subfloat[$\theta=1/5$]{
	\includegraphics[width=5cm]{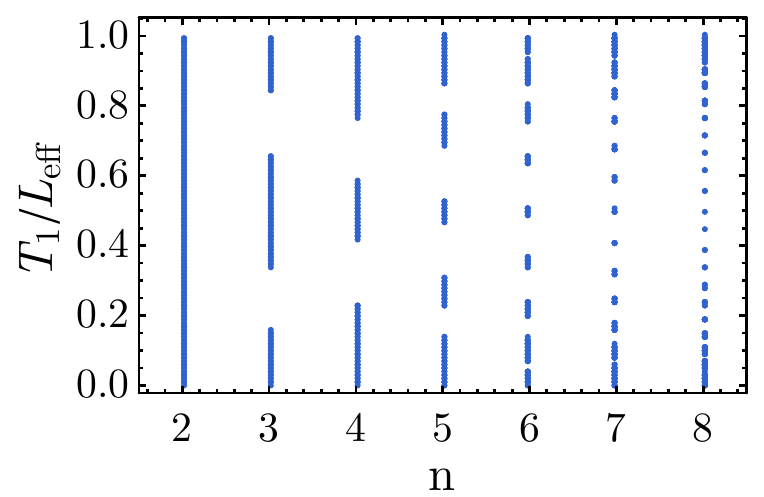}
}
\hspace{0pt}
\subfloat[$\theta=1/5,n=8$]{
	\includegraphics[width=5cm]{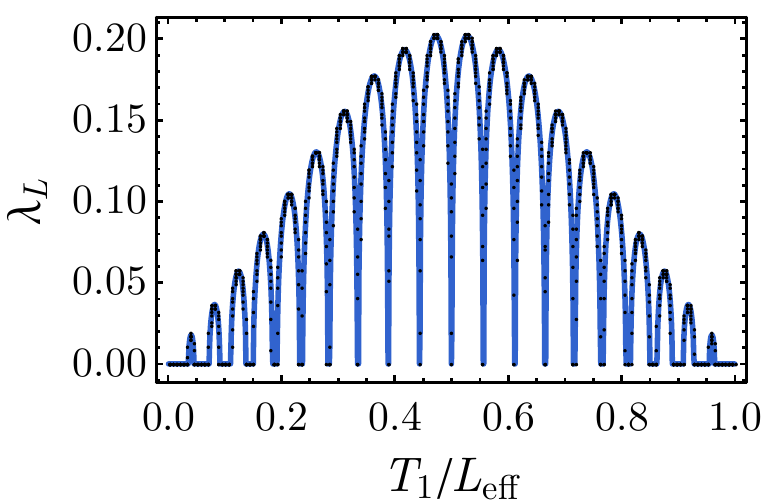}
}
\caption{\label{fig:AA golden ratio phase structure} Phase structure and Lyapunov exponents for the sequence of periodically driven CFT determined by $\omega_n=F_{n-1}/F_{n}$. In (a) and (b), the blue (white) regions are the region where the system is non-heating (heating). In (c), for a given $n$, the blue lines (blank region) correspond to the non-heating (heating) phase.}
\end{figure}

The phase structures for different values of $n$ and $\theta$ are shown in \figref{fig:AA golden ratio phase structure}, which shares a lot similarity with the Fibonacci case but also has some difference. 
Given $n$, the region for the non-heating phase gradually decreases as $\theta$ increases.
Given $\theta$, the non-heating region splits into many `bands'. Notice that $T_1=0$ and $T_1=L_{\text{eff}}$ are actually identified, the number of bands is exactly equal to $F_n$ when $F_n$ is even and $F_n-1$ when $F_n$ is odd.
If one compute the Lyapunov exponent for a fixed $n$, it shows peaks with equal spacing, as shown in \figref{fig:AA golden ratio phase structure}(d). It implies that as $\omega\rightarrow (\sqrt{5}-1)/2$, the self-similarity structure will not appear as in the Fibonacci case. The reason why the peaks have equal spacing will explained later.

\begin{figure}
	\centering
	\subfloat[$\sigma_n(\theta)$ v.s. $n$]{
		\includegraphics[width=5.5cm]{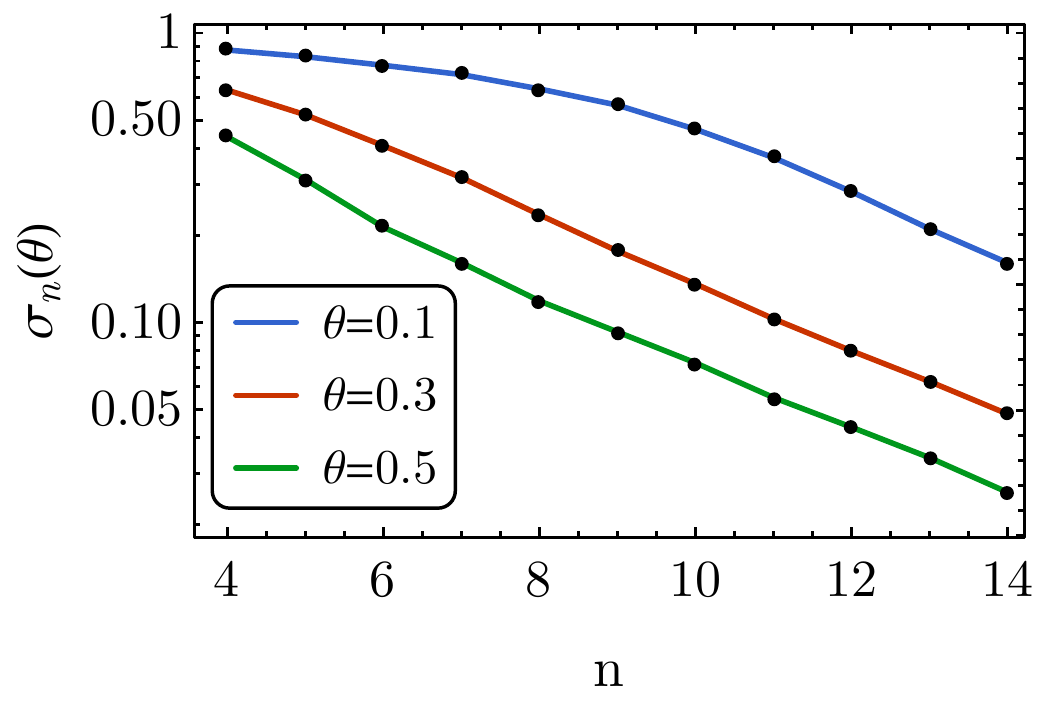}
	}
	\hspace{20pt}
	\subfloat[Decay rate $\lambda(\theta)$]{
		\includegraphics[width=5.5cm]{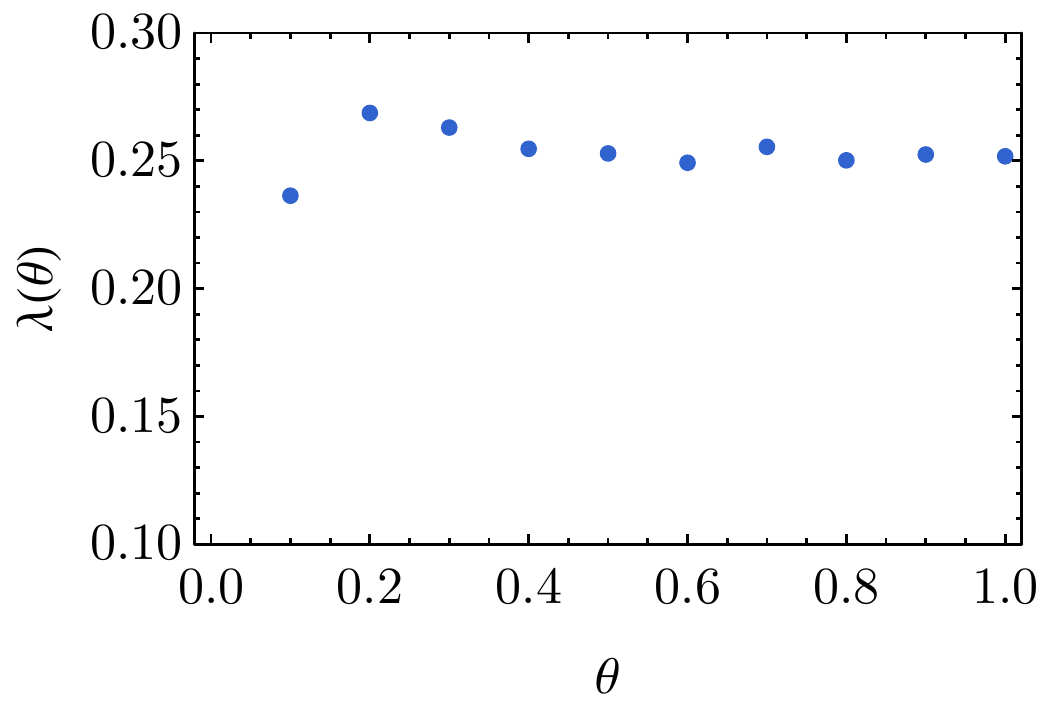}
	}
	\caption{\label{fig:AA measure of nonheating} (a) The measure of the non-heating phase $\sigma_n(\theta)$. (b) The decay rate $\lambda(\theta)$ as a function of $\theta$. For the curve with $\theta=0.1$, only the data for $n\ge 8$ are used for the fitting.}
\end{figure}

We can also check explicitly the measure of the non-heating phase as approaching the quasi-periodic limit. Following the same prescription as \eqref{Eq:Measure}, we call $\sigma_n(\theta)$ the measure of the non-heating phase, which is a function of $n$ and $\theta$. The result is shown in \figref{fig:AA measure of nonheating}(a). For a given $\theta$, the measure is exponentially decaying at large $n$, similar to what we have found for the Fibonacci case. We introduce the decaying rate $\lambda(\theta)$ as $\sigma_n(\theta) \propto e^{-\lambda(\theta) n}$, and find that it does not has a strong dependence on $\theta$.

These two features also hold for generic rational and irrational numbers and can be summarized by the following statements.
\begin{enumerate}
	\item If $\omega = p/q$ is a rational number with $p,q$ being co-prime, then the region for the non-heating phase splits into multiple bands. The number of bands depends on $q$ as
	\begin{equation}
		\text{number of bands} = \left\{
		\begin{array}{cl}
			q & q\in2\bbZ \\
			q-1 & q\in2\bbZ+1
		\end{array}
		\right.\,.
	\end{equation}
	Note the periodic boundary condition in $T_1$ when count the number of bands, namely the first band and last band (in vertical order) are considered to be the same band. 
	\item When $\omega = a_n/b_n$ approaches an irrational number as $n\rightarrow\infty$, the measure for the non-heating phase decreases exponentially with $n$. 
\end{enumerate}
The second statement is only empirical and based on the numerical observation. The first statement can be understood by a perturbative argument as follows.

\begin{defn}[Perturbative proof of the statement 1.]
Let us assume $\omega = p/q$ and analyze the matrix of the unitary evolution for one period, which consists of $q$ cycles. Notice that $L$ and $L_{\text{eff}}$ are merely the units for $T_0$ and $T_1$, we can set both of them to be $1$ without changing the physics.

We first consider the limit $\theta=0$, so that $M_0$ and $M_1$ are both diagonal matrices of pure phases. For the $k$-th cycle of the period, $T_0 = k\frac{p}{q} $ and we have
\begin{equation}
	M_{(k)}^{(0)} = M_0(T_0) M_1(T_1;\theta=0) = \begin{pmatrix}
	e^{i\pi(kp/q+T_1)} & \\ & e^{-i\pi(kp/q+T_1)} \end{pmatrix}\,.
\end{equation}
The lower index denotes the cycle and the upper index means it is the zero-th order term in the small $\theta$ expansion.
The matrix for the whole period $\Pi_q^{(0)}=M_{(1)}^{(0)} M_{(2)}^{(0)} \cdots M_{(q)}^{(0)}$ and its trace can be easily computed, and we have
\begin{equation}
	\label{eqn:AA piq}
	|\Tr \Pi_q^{(0)}| = \left\{ \begin{array}{cl} 2|\sin q\pi T_1| & q\in2\bbZ \\ 2|\cos q\pi T_1| & q\in 2\bbZ+1 \end{array}
	\right.\,,
\end{equation}
where we have used the condition that $p$ and $q$ are co-prime. A typical curve for $|\Tr \Pi_q^{(0)}|$ is shown in \figref{fig:AA piq}(a). Therefore, even without adding $\theta$, the trace can touch the critical value $|\Tr \Pi_q|=2$ at the following positions
\begin{equation}
	\begin{aligned}
	&q\in 2\bbZ :  & T_1 =& \frac{1}{q} \left(r+ \frac{1}{2}\right),\,\quad r=0,1,\cdots q-1\,; \\
	&q\in 2\bbZ+1 : & T_1 =& \frac{r}{q},\,\quad r=0,1,\cdots q-1\,.
	\end{aligned}
\end{equation}
Note that $T_1$ has a period $1$, which fixes the range of $r$.

\begin{figure}
	\centering
	\subfloat[$\theta=0$]{
		\includegraphics[width=5cm]{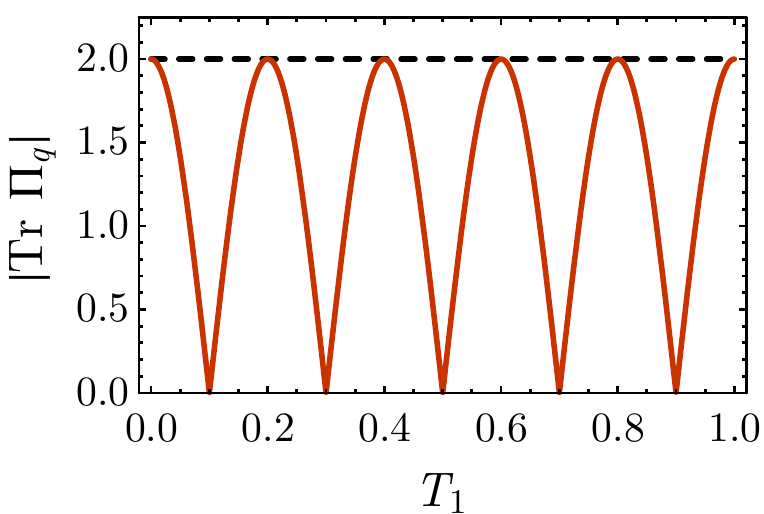}
	}
	\subfloat[$\theta=0.1$]{
		\includegraphics[width=5cm]{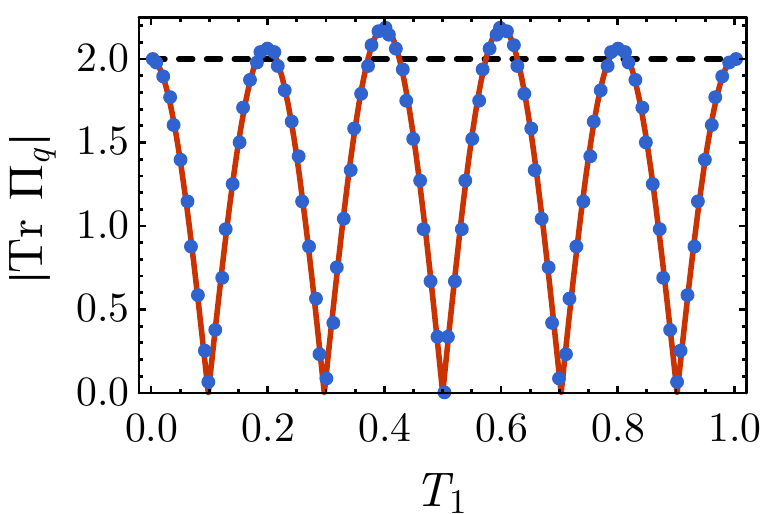}
	}
	\caption{\label{fig:AA piq} The absolute value of the trace of the matrix for one period. We choose $p=2$, $q=5$ for both plots. In (b), the blue dots are the approximated result in \eqnref{eqn:piq large q}. It matches the exact value quite well.}
\end{figure}

Then, we turn on a tiny $\theta$ and consider its contribution to $\Tr \Pi_q$ perturbatively. 
In particular, if $|\Tr\Pi_q|$ exceeds $2$ for a certain range of $T_1$, a heating phase appears there and will continue to exist for larger $\theta$. 
At small $\theta$, such thing is more likely to happen at those special positions where $|\Tr\Pi_q^{(0)}|$ already touches $2$. A numerical calculation for all small values of $q$ confirms that such thing does always happen. One typical example is shown in \figref{fig:AA piq}(b). For large $q$, we consider the following perturbative calculation. Each $M_{(k)}$ can be expanded to the second order in $\theta$
\begin{equation}
	\begin{gathered}
	M_{(k)} = M_{(k)}^{(0)} + \theta M_{(k)}^{(1)} + \theta^2 M_{(k)}^{(2)} + \calO(\theta^3) \\
	M_{(k)}^{(1)} = 2i\sin \pi T_1\begin{pmatrix} & -e^{i\pi kp/q} \\ e^{-i\pi kp/q} & \end{pmatrix}\,,\quad M_{(k)}^{(2)} = 2i\sin \pi T_1 \begin{pmatrix} e^{i\pi kp/q} & \\ & -e^{-i\pi kp/q} \end{pmatrix}
	\end{gathered}
\end{equation}
The first non-vanishing contribution comes from the second order of $\theta$, which has two terms. One is the cross term of $M_{(k)}^{(1)}$, the other only contains $M_{(k)}^{(2)}$.
In the limit of large $q$, second term dominates and we have
\begin{equation}
	\label{eqn:piq large q}
	\Tr \Pi_q \approx 2\cos\left(\pi qT_1 + \pi \frac{p(q+1)}{2}\right) - 4q\theta^2 \sin \pi T_1 \sin\left( \pi(q-1) T_1 + \pi \frac{p(q+1)}{2}\right)\,.
\end{equation}
Notably, this is a good approximation even for small $q$, the $q=5$ case is shown in \figref{fig:AA piq}(b) as an illustration. 
If $q$ is an odd number, one can check that at those special positions $T_1 = r/q$, we have
\begin{equation}
	\Tr \Pi_q \Big|_{T_1=r/q} = (-1)^{r+p(q+1)/2} \left(2 + 4\theta^2 q \left( \sin \frac{\pi r}{q} \right)^2 \right)\,,
\end{equation}
from which we can see that the value of $|\Tr \Pi_q|$ indeed exceeds $2$ except at $r=0$ ($T_1=0$). The proof for $q$ being even is similar and one can find that $|\Tr \Pi_q|$ exceeds $2$ for all $T_1 = (r+\frac{1}{2})/q$. 

So far, we have perturbatively shown that $|\Tr\Pi_q|$ will exceed $2$ at some equally spaced special positions.
Therefore, the heating phase will appear there as long as one turns on $\theta$, which accordingly opens `gaps' in the phase structure. This gives rise to the multiple bands of the non-heating phase and also explains the equally spaced peaks observed for the Lyapunov exponent. Furthermore, when $q$ is even, gaps will appear at all those special positions which leads to $q$ disconnected bands. When $q$ is odd, the gap cannot appear at $T_1=0$ which yields $q-1$ bands (after identifying $T_1/L_{\text{eff}} =0$ with $T_1/L_{\text{eff}}=1$). This completes the proof of our first statement.
\end{defn}

\subsubsection{Phase diagram and nested structure}

\begin{figure}
	\centering
	\subfloat[Full phase diagram]{
		\includegraphics[width=6.8cm]{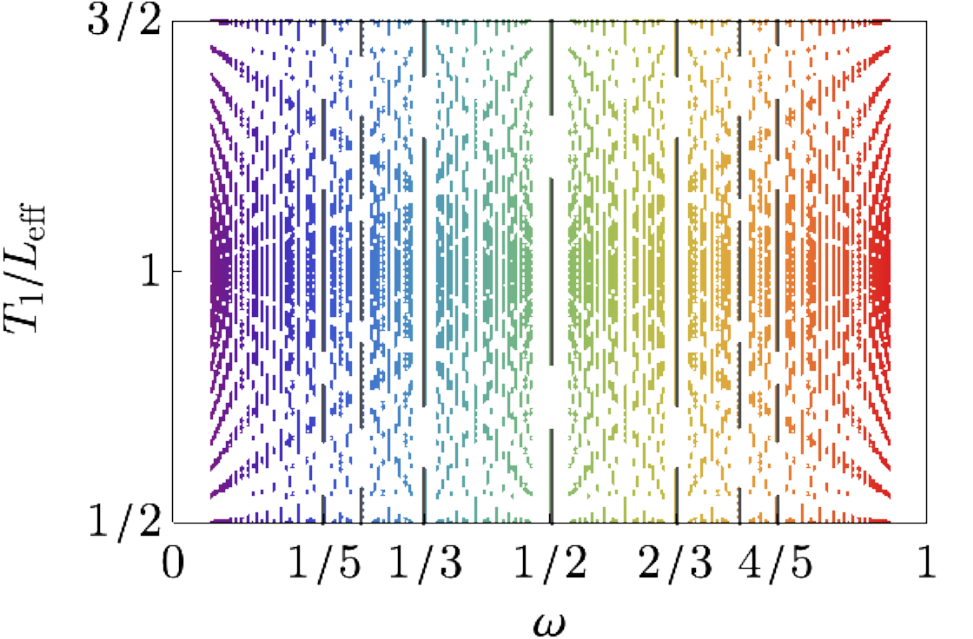}
	}
	\hspace{10pt}
	\subfloat[Subregion between $\omega=1/3$ and $\omega=1/2$]{
		\includegraphics[width=6.6cm]{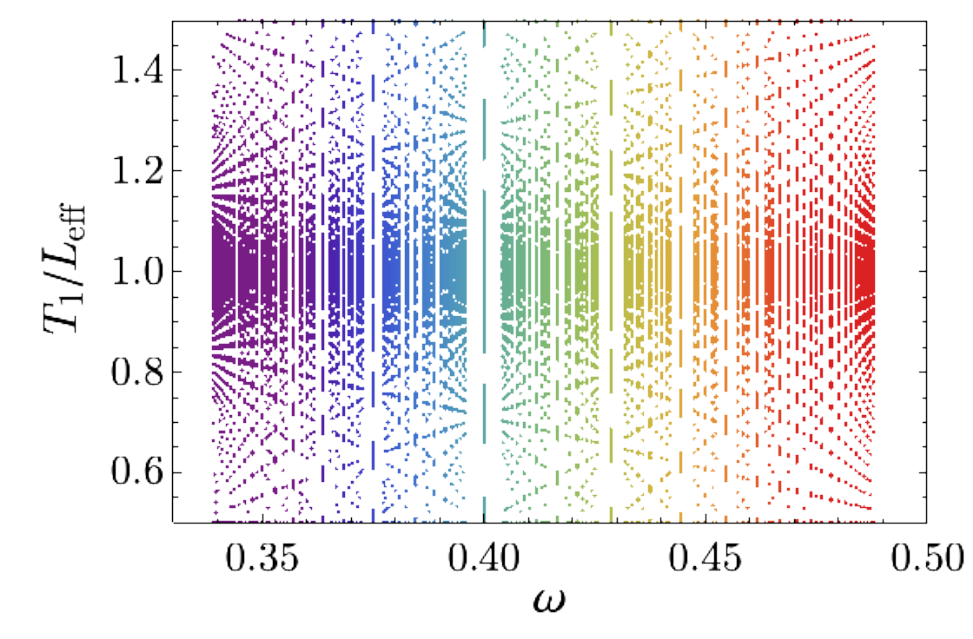}
	}
	\caption{\label{fig:AA phase diagram} The phase diagram for the Aubry-Andr\'e quasi-periodic driving CFT. The plot uses $\theta=0.2$ and includes all the rational number whose denominators are equal or less than $20$. The usage of different colors is only for the purpose of presentation and has no physical meaning. We shift the origin of $T_1/L_{\text{eff}}$ by $1/2$ when presenting the data.}	
\end{figure}

We can also study the phase diagram for generic $\omega$ including both rational and irrational numbers. The result is shown in \figref{fig:AA phase diagram}, with the colored region being the non-heating phase and blank region being the heating-phase. Here are some comments on the features of the phase diagram:
\begin{enumerate}
	\item The whole diagram is symmetric with respect to $\omega = 1/2$, which is a direct consequence of our analysis above. Namely, for a rational number $\omega=p/q$, the gap opening is independent of the numerator, which implies that the phase diagram should be invariant under $p/q\rightarrow 1-p/q$. Our following discussion will focus on the part $\omega\leq 1/2$.
	\item The diagram has (infinitely) many empty regions,  whose center is at simple rational numbers $\omega = 1/n,n\ge 2$. Some representatives are drew explicitly by the gray lines in \figref{fig:AA phase diagram}. The reason why they all sit in the relative empty region is that one has to use a rational number with large denominator to approach one of them, the non-heating bands of which are too fragmented to read by eyes.
	\item In between every two neighboring simple rationals, the structure of the subregion resembles the original phase diagram. This implies that the phase diagram has a nested structure, which is similar to the famous Hofstadter butterfly \cite{hofstadter1976} and can be understood in the following way.
\end{enumerate}

As reviewed in \appref{appendix Fib}, every rational and irrational number $\omega\in[0,1]$ can be uniquely represented by a continued fraction. Here we adopt the idea and consider a generalization
\begin{equation}
	\omega = \frac{1}{N_1\pm\frac{1}{N_2 \pm \ldots}}\,,\quad N_i=2,3,\cdots\,.
\end{equation}
We will see that it provides a useful guidance to resolve the diagram layer by layer.
Those with only $N_1$ being nonzero are dubbed as the principal series, those with nonzero $N_1,N_2$ as the first descendants and so on.

\begin{figure}
	\centering
	\subfloat[Principal series]{
		\includegraphics[width=7.2cm]{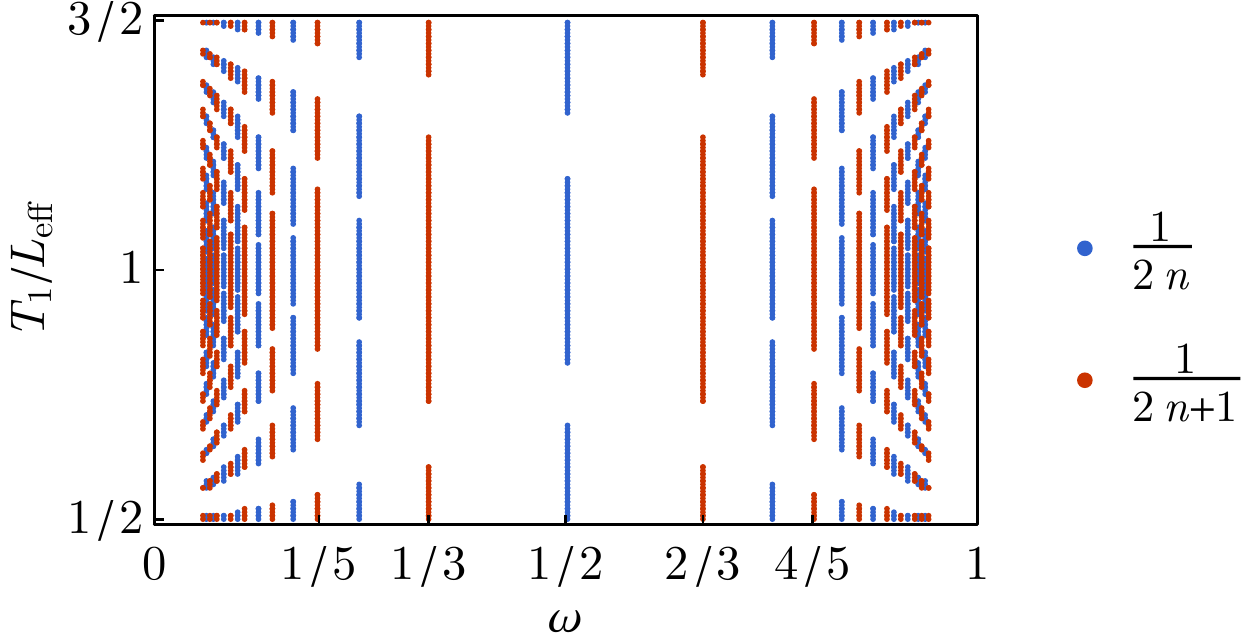}
	}
	\hspace{20pt}
	\subfloat[First descendant]{
		\includegraphics[width=7cm]{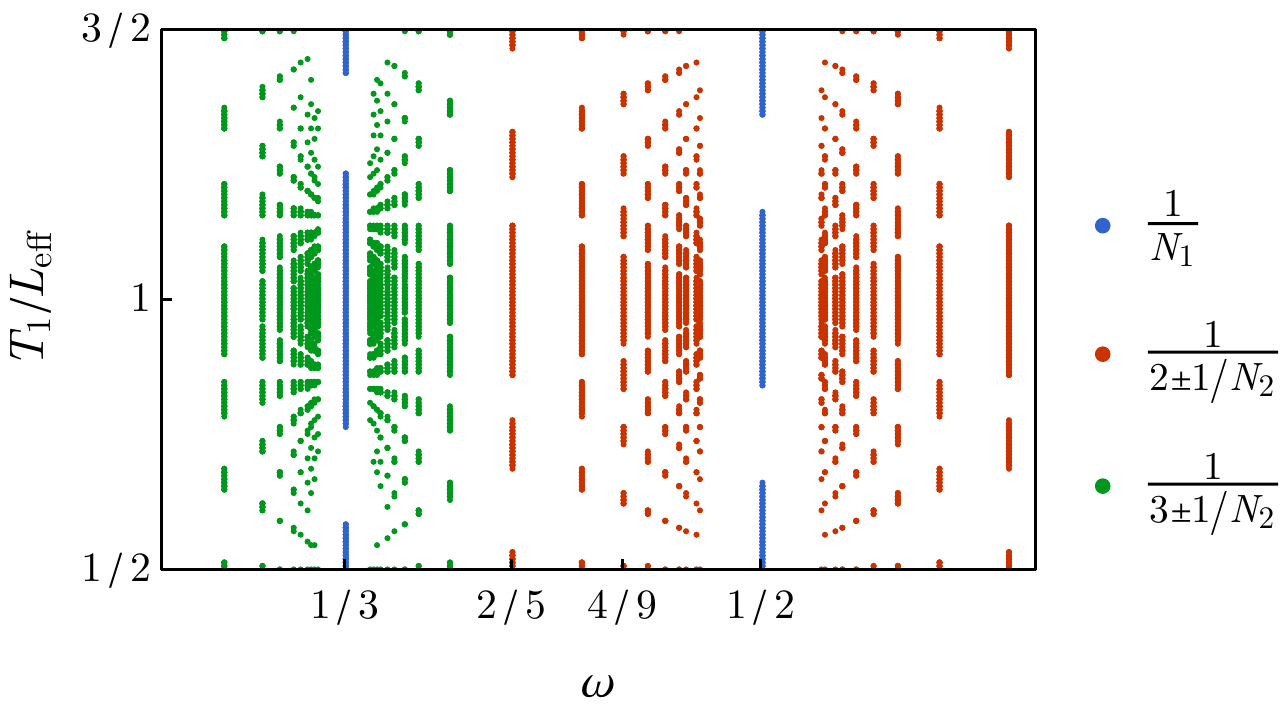}
	}
	\caption{\label{fig:AA continued fraction} The nested structure of the phase diagram for the Aubry-Andr\'e quasi-periodic driving CFT.}
\end{figure}

The nested structure of the phase diagram exactly follows such an organization:
\begin{enumerate}
    \item First, the principal series $\{1/N_1\},N_1=2,3,\cdots$ form the skeleton of the phase diagram, which is shown in \figref{fig:AA continued fraction}(a). They also sit in the relatively empty region in the full phase diagram.
    \item The first descendants $\{\frac{1}{N_1\pm \frac{1}{N_2}}\}$ fill in the blank region between the principal series and serves as the skeleton for the next descendants. For example, $\frac{1}{2 + 1/N_2}$ and $\frac{1}{3-1/N_2}$ fills the region between $1/3$ and $1/2$, as shown in \figref{fig:AA continued fraction}(b).
    \item Notice that all the descendants of $1/2$ that are smaller than $1/2$ are $\frac{1}{2+p/q},1<p<q$. Similarly all the descendants of $1/3$ that are larger than $1/3$ are $\frac{1}{3-p/q},1<p<q$, which can also be written as $\frac{1}{2+(1-p/q)}$. These two series, filling the subregion between $\omega=1/3$ and $\omega=1/2$, can be considered as the `mirror reflection' to each other with $\omega=2/5$ being the reflection center. This is similar to what is observed for the full diagram, with the difference that each pair $\{\frac{1}{2+p/q}, \frac{1}{2+(1-p/q)} \}$ do not have the same number of bands. This explains the similarity and also difference between \figref{fig:AA phase diagram}(a) and (b). The nested structure follows from continuing such kind of game.
\end{enumerate}
  
\subsubsection{Features of the group walking}

So far, the results resemble what have been found for the well 
known Aubry-Andr\'e model or almost Mathieu operator. The unique advantage of our CFT setup is that it brings physical meaning to the group walking. As discussed in \secref{Sec: GroupWalking}, the group walking of $\rho$ and $\rho\zeta$ are related to the energy, energy-momentum density and the entanglement. We will give a corresponding discussion in this section.

\begin{figure}
	\centering
	\subfloat[Group walking of $\rho$ ]{
		\includegraphics[width=6cm]{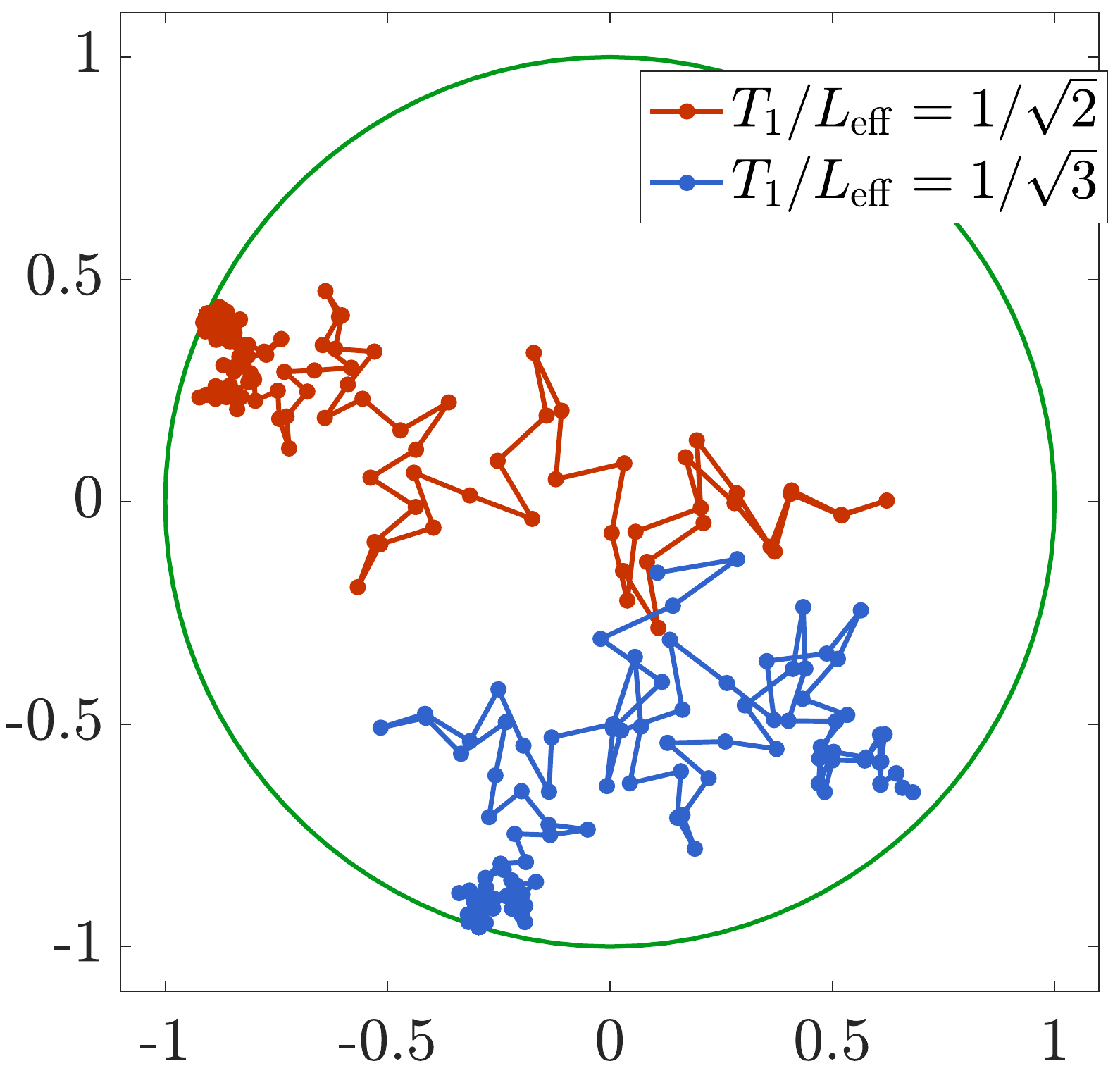}
	}
	\hspace{20pt}
	\subfloat[Group walking of $\rho\zeta$]{
		\includegraphics[width=6.6cm,height=5.8cm]{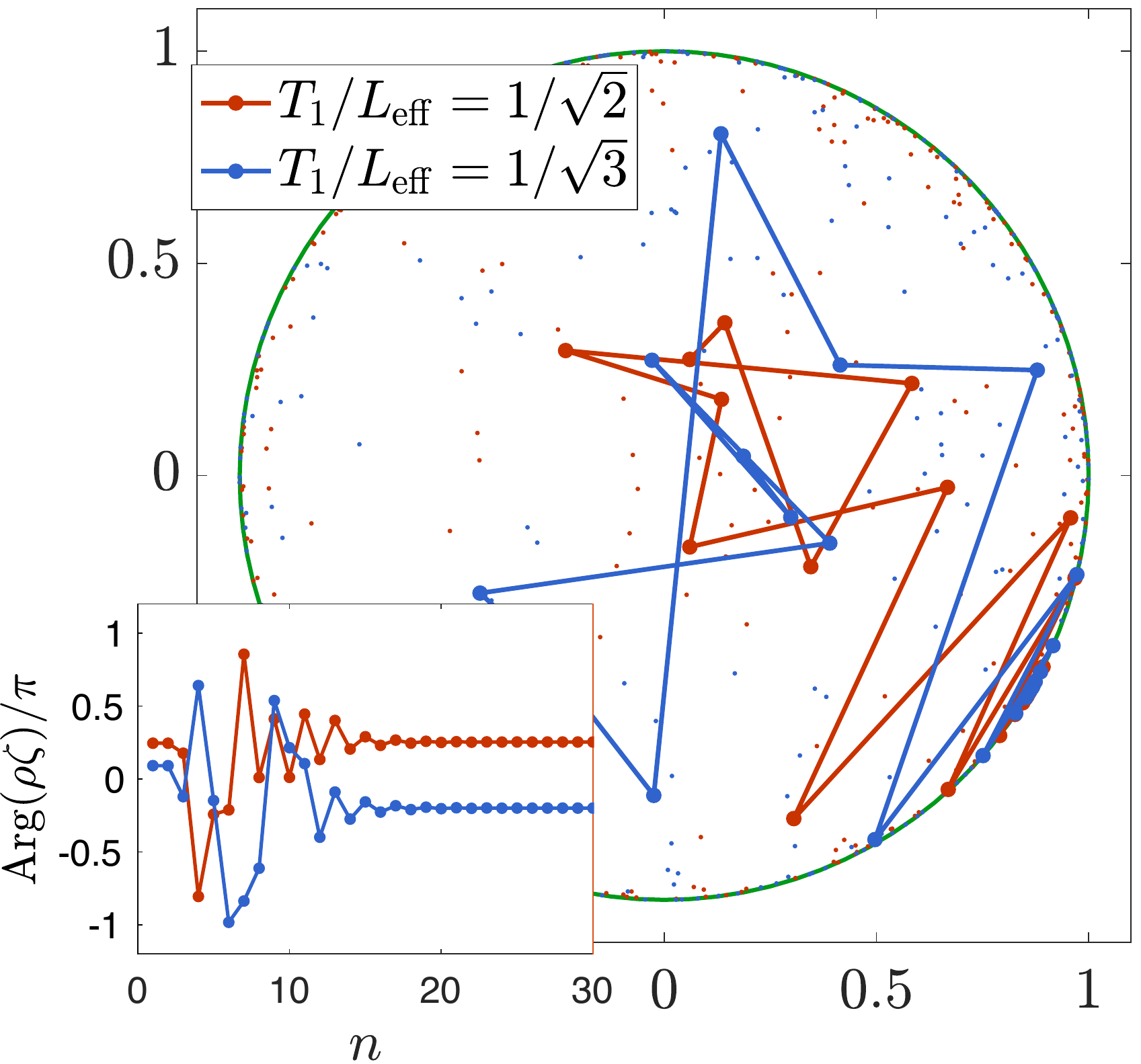}
	}
	\caption{\label{fig:AA group walking fibonacci} The group walking of $\rho$ and $\rho\zeta$. We choose $\omega$ being $(\sqrt{5}-1)/2$ and $\theta=0.1$. The choice for $T_1/L_{\text{eff}}$ is not special, one can choose any other generic values. In the numerics, we choose $\omega=F_{29}/F_{30}$ as an approximation to plot (a) and (b). In (b), the scattered dots in the circle represent $(\rho\zeta)_n$ for $n\le 500$. The inset of (b) is Arg$(\rho\zeta)/\pi$ as a function $n$, with $n$ denoting the $n$-th Fibonacci number $F_n$. }
\end{figure}

To illustrate the general feature, let us choose the inverse golden ratio $\omega=\frac{\sqrt{5}-1}{2}$ as a concrete example. The results are shown in \figref{fig:AA group walking fibonacci}. 
The behavior of $\rho$ has the same qualitative feature as reported in the Fibonacci driving case. Namely, for a generic choice of $T_1/L_{\text{eff}}$, $\rho$ will flow exponentially close to a certain point on $\partial\bbD$ in the long time limit $n\rightarrow \infty$. It follows from \eqnref{HalfEE_general_GroupWalking} and \eqnref{Etotal_rho} that the 
entanglement entropy and total energy have a linear and exponential growth respectively, which is consistent with our general claim for the heating phase.  

The behavior of $\rho\zeta$ is quite different. 
In the short time, $\rho\zeta$ seems to have a random distribution in the disk. In the late time, it flows onto $\partial \mathbb D$, which implies the formation of energy-momentum peaks. Although $(\rho\zeta)_n$ and $(\rho\zeta)_{n+1}$ does not show a strong correlation, as shown by the background scattered dots in \figref{fig:AA group walking fibonacci}(b), the sub-sequence $\{(\rho\zeta)_{k+F_n}\},n\ge 1$ for any fixed $k$ does have a definite limit as $n\rightarrow \infty$ and the detailed value of the limit depends on $k$. Physically it means that the peaks observed at the time $k+F_n$ with fixed $k$ will appear at the same position.

Such behavior of $\rho\zeta$ is generic as long as the irrational number is $\omega = (\sqrt{r^2+4}-r)/2, r=1,2,\cdots$. For example, one will observe the same feature if choosing $\omega=\sqrt{2}-1,(\sqrt{13}-4)/2$ and so on. An intuitive reason is that the $n$-th principal convergent for this type of irrational numbers can be written as $b_{n-1}/b_n$. (For the inverse golden ratio, $b_n$ is the $n$-th Fibonacci number.) Such a sequence $b_n$ provides us with a natural choice of the observation time. For a generic irrational number, its $n$-th principal convergent is $p_n/q_n$ with $\{p_n\},\{q_n\}$ being two different sequence and a ``natural choice" of the observation time becomes less clear.

\section{Discussion}
\label{Sec: Discussion}

In this paper, we propose a general framework to study the non-equilibrium dynamics of $(1+1)$D CFTs with $\SL_2$ deformation. We exam the details of the dynamical phases that emerge in the periodic and quasi-periodic driving using the tools we propose. In the sequel of this paper \cite{RandomCFT}, we will apply the framework to the random driving sequence, where the use of the Lyapunov exponent and group walking becomes a necessity rather than a convenient option. 

\bigskip

Here we highlight some of the unexpected features that we have found: 
\begin{enumerate}
    \item For the driving protocol that uses the  $\SL_2$ deformed Hamiltonian, when the total energy and entanglement start to grow, they always grow in a pattern where energy is concentrated in discrete points and form peaks; while the entanglement is shared within nearest neighbours. This phenomenon was first observed in Ref.~\cite{fan2019emergent} and is found to persist in the more general setting here.

    \item Introducing irregularity in the driving protocol usually enlarges the heating phase, which is what we have observed in the quasi-periodic driving where the non-heating regime shrinks to a set with measure zero.\footnote{For Fibonacci driving with SSD Hamiltonian, we prove it is a Cantor set by mapping to the quasi-crystal.} 
    However, for the \emph{Fibonacci} driving, the phase diagram is found to have a special non-heating fixed point, where the total energy and entanglement will return in a pattern following the Fibonacci sequence.
    More explicitly, if we observe the system only at the steps coinciding with the Fibonacci numbers, what we see is a state returning to itself with period 6. 
    
    It is worth to mention that if we pick out the two unitaries $U_A$, $U_B$ that underlies the aforementioned 
    Fibonacci driving, and 
    apply them in a periodic fashion $U_A U_B U_A U_B U_A U_B \ldots$, 
    the system will end up with a heating state.  
    This 
    is a surprise since it implies that at this special point, the ``irregularity''  actually converts the heating protocol to a non-heating one.
    The reason is that the pattern in the Fibonacci driving sequence manages to conspire with the special unitaries in a way that they happen to cancel each other and result in a return, which is explained in Sec.~\ref{Sec:NonheatingFixGeneral}.
    
    \item 
    Besides the last point, the ``order" of the quasi-periodic driving manifests itself in another way. In the heating phase of the quasi-periodic driving, ordinary stroboscopic observation of the energy peaks is featureless. However, if we consider the Fibonacci driving and observe the state at the steps coinciding with the Fibonacci numbers, the energy peaks oscillate between two fixed  positions rather than randomly distribute (which is what we expect for a general irregular driving). Similar feature is also observed in the Aubry-Andr\'e driving, where the energy peaks return to the same positions.
\end{enumerate}
\bigskip 

In the rest of this Discussion, we would like to comment on the special setting we use and some future directions. 
$\SL_2$ deformation\footnote{Or more generally ``Virasoro deformation'' which involves modulation with multiple wavelengths.} is kind of a ``shortcut'' in analyzing the driven systems, because its effect in a single driving period can be characterized by a conformal transformation without introducing external sources to the system. The simplicity of the single driving allows us to pursue the ``complexity'' in the pattern of the driving sequence as we do in this paper. For future directions 
\begin{enumerate}
    \item Within the $\SL_2$ deformation framework: so far we have been only probing the driven state by simple observables such as one-point function of energy-momentum tensor for energy distribution or two-point function of twist operator for the entanglement entropy, both of which only depend on the central charge. To explore more CFT data such as the operator content and the OPE coefficients, we need to consider more complicated observables. For example, we may consider measuring multi-point functions during the driving and ask how could a carefully designed driving protocol help us extract more CFT data. 
    
    \item Beyond the $\SL_2$ deformation framework: the deformation of 
    driving Hamiltonians considered in this work are generated by $\SL_2$ algebra.
    Most recently, the periodically driven CFTs are generalized to the 
cases where the driving Hamiltonians are deformed by arbitrary smooth
functions\cite{fan2020FloquetGeneral,lapierre2020geometric}. The underlying algebra
is the infinite dimensional Virasoro algebra. 
It is found that both the heating and non-heating phases can still be observed in general.
In particular, the phase diagrams are determined by whether there are emergent spatial
fixed points in the operator evolution. If there exist spatial fixed points, then the
driven system is in the heating phase; otherwise, the system is in the non-heating
phase.

However, it is not obvious what will happen when we perturb the driving Hamiltonian by introducing operators other than energy-momentum in the driving Hamiltonian. 
In general, if the driving Hamiltonians break the conformal symmetry, we expect that the system will finally be thermalized.
    
    Another related question is that if we treat the CFT we have as a low energy effective theory, then in the heating phase we will finally drive the system to an energy scale where we need to consider its UV completion, i.e. we need to include some irrelevant operators in the driving Hamiltonian. We could ask what will happen at that point? For example, how does the energy peak and entanglement pattern get modified? These questions are relevant in explaining the data from lattice simulation beyond the conformal regime.

    \item It is also desirable to consider the possible experimental realization of our setup. Since our driving Hamiltonians are 
    inhomogeneous in space, we expect it is natural to study the physics
    here in the cold-atom experiments, where the interactions among cold
    atoms can be optically controlled\cite{bakr2009quantum,Borish_2020}.
    In experiments, the dissipation effects caused by environments need
    to be considered\cite{breuer2002theory}. It is expected that the physics studied in this work
    can be observed if the time scale is shorter than the 
    decoherence time.
\end{enumerate}

\bigskip

\textit{Note added:} During the preparation of this manuscript, 
we learnt that the Fibonacci quasi-periodically 
driven CFT is also studied in 
\cite{ZurichQuasiPeriodic}, which
will appear on arXiv on the same day.
We thank the authors for sending us their manuscript 
before posting.

\section{Acknowledgement}

We thank for helpful discussions with Bo Han, 
Daniel Jafferis,
Eslam Khalaf, Ivar Martin, Shinsei Ryu, Hassan Shapourian, Tsukasa Tada, 
Michael Widom, Jie-Qiang Wu, Yahui Zhang and Di Zhou. 
In particular, we thank Yahui Zhang for suggestions on the study of the quasi-periodic
driving CFTs, and thank Michael Widom for pointing out Ref.~\cite{KKT1983}, 
which stimulated our interest in considering the Fibonacci 
sequence in  quasi-periodically driven CFTs. 
XW is supported by the Gordon and Betty Moore Foundations EPiQS initiative through Grant No.GBMF4303 at MIT. 
Y.G. is supported by the Gordon and Betty Moore Foundation EPiQS Initiative through Grant (GBMF-4306) and DOE grant, de-sc0019030. 
AV and RF are supported by the DARPA DRINQS program (award D18AC00033) 
and by a Simons Investigator Award.

\appendix



\section{More on time-dependent driven CFTs}
\label{Sec: TechnicalDetail}

In this appendix, we give more details on 
some formulas/results as used in the main text.

\subsection{Operator evolution with arbitrary SL$_2$ deformations}
\label{Sec:OperatorEvoAppendix}

In this appendix, we introduce the procedures of obtaining the concrete form of M\"obius transformation in Eq.~\eqref{Mobius_Transf}
in the main text. Some related details can be found in Refs.\cite{WenWu2018quench,wen2018floquet,fan2019emergent,
HanEffectiveHamiltonian,HanWenClassify}.
Let us  illustrate the calculation with a simple example, and then
give results for an arbitrary SL$_2$ deformation. 

The illustrative example we consider has the following 
deformed Hamiltonian:
\be\label{Ht_appendix}
H_{\text{deform}}=\int_0^L f(x) \, T_{00}(x) dx,\quad \text{with } f(x)=1-\tanh(2\theta)\cos\frac{2\pi q x}{L},\quad \theta>0,\,
q\in\mathbb Z,
\ee
where $T_{00}(x)$ is the Hamiltonian density with 
$T_{00}(x)=\frac{1}{2\pi}\left(T(x)+\bar{T}(x)\right)$.
For $q=1$ with open boundary conditions, 
this corresponds to the example we considered in 
Sec.\ref{Sec:MinimalSetup}, Sec.\ref{Sec: Fibonacci driving CFT} and Sec.\ref{Sec: Aubry Andre driving CFT}.
To study the M\"obius transformation in Eq.~\eqref{Mobius_Transf}, 
our derivations below apply to both periodic and open boundary conditions.
First, it is noted that $H_{\text{deform}}$ can be written in terms
of the Virasoro generators in Eq.\eqref{VirasoroGenerator} as
\be\label{HthetaVira}
H_{\text{deform}}=\frac{2\pi}{L}\left[
L_0-\tanh(2\theta)\frac{L_q+L_{-q}}{2}-\frac{c}{24}
\right]+\text{anti-chiral parts.}
\ee
As a remark, for $\theta=0$, $H_{\text{deform}}$ corresponds
to a uniform one without any deformation; for $\theta=\infty$,
$H_{\text{deform}}$ corresponds to a SSD Hamiltonian,
whose energy spectrum has been recently studied in detail in
Refs.\cite{ishibashi2015infinite,ishibashi2016dipolar,Okunishi:2016zat,Ryu1604}. 

To evaluate the correlation function, such as the simplest one $\langle \Psi(t)| \mathcal{O}(x)|\Psi(t)\rangle$, where $|\Psi(t)\rangle=e^{-iH_{\text{deform}}t}|\Psi_0\rangle$, one can study the operator evolution  $\mathcal{O}(x,t)=e^{iH_{\text{deform}}t}\mathcal{O}(x)e^{-iH_{\text{deform}}t}$, as follows.
The correlation function $\langle \Psi(t)|\mathcal{O}(x)|\Psi(t)\rangle$ can be considered as the path integral on a $w$-cylinder with the operator $\mathcal{O}$ inserted, as depicted in Fig.~\ref{ConformalMapQ}, where $w=\tau+ix$.
This cylinder can be mapped to a $q$-sheet Riemann surface with a conformal map $z=e^{\frac{2\pi q w}{L}}$ (See Fig.~\ref{ConformalMapQ}). The energy-momentum tensor transforms as
$
T_{\text{cyl}}(w)=\left(\frac{dw}{dz}\right)^{-2}
\left[ T(z)-\frac{c}{12}\{w,z\} \right]
$,
with
$
\{w,z\}=\frac{d^3w/dz^3}{dw/dz}-\frac{3}{2}\left(
\frac{d^2w/dz^2}{dw/dz}
\right)^2.
$
Then, one can find that
$
T_{\text{cyl}}(w)=\left(
\frac{2\pi z}{l}
\right)^2\left[
T(z)-\frac{c}{24}\cdot\frac{1}{z^2}
\right],
$
where we have defined $l:=L/q$.
Then the Hamiltonian in Eq.\eqref{Ht_appendix}
can be written as $H=H^{(z)}+H^{(\bar{z})}$, where
\be
\small
H^{(z)}
=\frac{2\pi}{l\cosh(2\theta)}
\oint\frac{1}{2\pi i}
\left[
\cosh(2\theta)z-\frac{\sinh(2\theta)}{2}(z^2+1)
\right]T(z)dz-\frac{\pi c}{12 l}.
\ee
A further M\"obius transformation
$\tilde{z}=\frac{-\cosh(\theta)\cdot z+\sinh(\theta)}
{\sinh(\theta)\cdot z-\cosh(\theta)}$ will transform
$H^{(z)}$ to the following simple form:
\be
\small
H^{(\tilde{z})}=\frac{2\pi}{l_{\text{eff}}}
\oint\frac{1}{2\pi i} \tilde{z}\,T(\tilde{z})\,d\tilde{z}
-\frac{\pi c}{12 l},\quad l_{\text{eff}}=l\cosh(2\theta),
\ee
and similarly for the anti-holomorphic part.
On this $\tilde{z}$ Riemann surface, the operator
evolution becomes a dilatation:
$e^{H^{(\tilde{z})}\tau}\mathcal{O}(\tilde{z},\bar{\tilde{z}})
e^{-H^{(\tilde{z})}\tau}=\lambda^h\lambda^{\bar{h}}\mathcal{O}
(\lambda \tilde{z}, \lambda\bar{\tilde{z}})$, where
$\lambda=e^{\frac{2\pi\tau}{l_{\text{eff}}}}$.
Then by mapping back to the $z$-surface, one can find 
the operator evolves as 
\be\label{OPevolutionAppendix}
e^{H^{(z)}\tau}\mathcal{O}(z,\bar{z})
e^{-H^{(z)}\tau}=
\left(\frac{\partial z'}{\partial z}\right)^h
\left(\frac{\partial \bar{z}'}{\partial \bar{z}}\right)^{\bar{h}}
\mathcal{O}(z',\bar{z}'), \quad \text{where }
z'=\frac{a z+b}{c z+d}.
\ee
By imposing the normalization condition $ad-bc=1$, and
doing an analytical continuation $\tau=it$, one has
\be\label{ZpZ_appendix}
z'=\frac{\alpha z+\beta}{\beta^* z+\alpha^*},
\ee
where $\alpha=\cos{\left( \frac{\pi t}{l_{\text{eff}}} \right)} + i\cosh(2\theta)\cdot \sin{\left( \frac{\pi t }{l_{\text{eff}}} \right)}$
and $\beta=- i\sinh(2\theta)\cdot\sin{\left( \frac{\pi t}{l_{\text{eff}}} \right)}$.
When $q=1$ such that $l=L$, we get the result 
as presented in Eq.\eqref{MobiusTheta} in the main text.

It is straightforward to generalize 
the above approach to the cases with
arbitrary SL$_2$ deformations.
A general result was recently calculated in Refs.\cite{HanEffectiveHamiltonian,HanWenClassify}.
Let us cite and briefly summarize the results here.
First, for arbitrary SL$_2$ deformations in Eq.\eqref{Envelope_F}, 
the deformed Hamiltonian can be written as 
$H_{\text{deform}}=H_{\text{deform}}^{\text{chiral}}
+H_{\text{deform}}^{\text{anti-chiral}}$, with 
$H_{\text{deform}}^{\text{chiral}}$ given in Eq.\eqref{Hdeform_Virasoro}, which we rewrite here:
\be\label{Hdeform_chiral}
H^{\text{chiral}}_{\text{deform}}=\frac{2\pi}{L}
\left(\sigma^0 L_0+\sigma^+ L_{q,+} +\sigma^-L_{q,-}
\right)-\frac{\pi c}{12L},
\ee
and similarly for the anti-chiral part. Here we have defined
$L_{q,+}=\frac{1}{2}(L_q+L_{-q})$,
and $L_{q,-}=\frac{1}{2i}(L_q-L_{-q})$.
One can further define the quadratic Casimir element: 
$
c^{(2)}:= -(\sigma^0)^2 + (\sigma^+)^2 + (\sigma^-)^2
$
\cite{ishibashi2015infinite,ishibashi2016dipolar,
HanEffectiveHamiltonian,HanWenClassify},
based on which one can classify the deformed Hamtilonians in
Eq.\eqref{Hdeform_chiral} into three types:
\be\label{HamiltonianType_appendix}
\left\{
\begin{split}
&c^{(2)}<0:\quad \text{Elliptic Hamiltonian},\\
&c^{(2)}=0:\quad \text{Parabolic Hamiltonian},\\
&c^{(2)}>0:\quad \text{Hyperbolic Hamiltonian}.\\
\end{split}
\right.
\ee
Second, we consider the operator evolution 
$e^{iH_{\text{deform}}T}\mathcal{O}(z,\bar{z})
e^{-iH_{\text{deform}}T}$ with the Hamiltonian in 
Eq.\eqref{Hdeform_chiral} for a time interval $T$.
Then one can obtain the general form of operator evolution
in Eqs.\eqref{OPevolutionAppendix} 
and \eqref{ZpZ_appendix}. 
The corresponding $SU(1,1)$ matrix
$M=\begin{pmatrix}
\alpha &\beta\\
\beta^* &\alpha^*
\end{pmatrix}$ depends on the types of Hamiltonian 
in Eq.\eqref{HamiltonianType_appendix} as
follows:\cite{HanEffectiveHamiltonian,HanWenClassify}
\be\label{MobiusThreeTypeAppendix}
\small
\left\{
\begin{split}
&\text{Elliptic:}\quad
\alpha= -\cos{\left( \frac{\calC\pi T}{l} \right)} - i \frac{\sigma^0}{\calC} \sin{\left( \frac{ \calC\pi  T}{l} \right)},\quad
\beta= -i \frac{\sigma^+ + i\sigma^-}{\calC} \sin{\left( \frac{\calC \pi T}{l} \right)},\\
&\text{Parabolic:}\quad
\alpha=-1- i  \frac{\sigma^0 \pi T}{l},\quad
\beta=-i\frac{(\sigma^+ + i\sigma^-)\pi T }{l},\\
&\text{Hyperbolic:}    \quad
\alpha = -\cosh{\left( \frac{\calC\pi  T}{l} \right)} - i\frac{\sigma^0}{\calC} \sinh{\left( \frac{\calC\pi  T}{l} \right)},\quad
\beta= -i \frac{\sigma^+ + i\sigma^-}{\calC} \sin{\left( \frac{\calC\pi  T}{l} \right)},
\end{split}
\right.
\ee
where $\calC=\sqrt{|-(\sigma^0)^2+(\sigma^+)^2+(\sigma^-)^2|}$ 
and $l=L/q$.
One can check explicitly that for elliptic, parabolic,
and hyperbolic Hamiltonians in Eqs.\eqref{Hdeform_chiral} 
and \eqref{HamiltonianType_appendix}, the corresponding 
$SU(1,1)$ matrices in Eq.\eqref{MobiusThreeTypeAppendix} 
have the properties $|\text{Tr}(M)|<2$, 
$|\text{Tr}(M)|=2$, and $|\text{Tr}(M)|>2$ respectively,
as expected.  


As a remark, the specific example in Eqs.\eqref{Ht_appendix} and \eqref{HthetaVira} is always elliptic for finite $\theta$, and parabolic for $\theta\to \infty$.

For the elliptic case in Eq.\eqref{MobiusThreeTypeAppendix}, 
by choosing $T=\frac{l}{2\calC}$, one has $M=\left(
\begin{array}{cccc}
-i\frac{\sigma^{0}}{\calC} &-i\frac{\sigma^{+}+i\sigma^{-}}{\calC}\\
i\frac{\sigma^{+}- i\sigma^{-}}{\calC} &i\frac{\sigma^{0}}{\calC}
\end{array}
\right)$.
This is a reflection matrix of the form in Eq.~\eqref{ReflectionMatrix},
with the property $\text{Tr}(M)=0$ and $M^2=(M^{-1})^2=-\mathbb I$.
Given two arbitrary relfection matrices $M_A$ and $M_B$
($M_B\neq \pm M_A$), as
discussed in Ref.~\cite{simon2005orthogonal},
there exists a $SU(1,1)$ matrix $V$ such that 
$VM_AV^{-1}=\begin{pmatrix}
-i &0\\
0 &i
\end{pmatrix}$, and 
$VM_BV^{-1}=\begin{pmatrix}
\alpha &\beta\\
\beta^* &\alpha^*
\end{pmatrix}$, where 
$|\alpha|^2-|\beta|^2=1$ and $\alpha,\beta\in\mathbb C$.
Since $M_B$ is a reflection matrix and $M_B\neq \pm M_A$, then we have
$\text{Tr}(VM_BV^{-1})=\text{Tr}(M_B)=0$, which indicates that 
$VM_BV^{-1}=\begin{pmatrix}
ia &\beta\\
\beta^* &-ia
\end{pmatrix}$, where $a\in\mathbb R$,
$a^2-|\beta|^2=1$, and $\beta\neq 0$.
Then one can check 
$\text{Tr}(M_AM_B)=\text{Tr}(VM_AV^{-1}VM_BV^{-1})
=2a$. Since $a^2=1+|\beta|^2>1$, we always have
$\big|\text{Tr}(M_AM_B)\big|>2$, i.e., 
$M_AM_B$ is a hyperbolic matrix.
These properties will be useful in the study
of the non-heating fixed point in a Fibonacci
driven CFT in Sec.\ref{Sec:NonheatingFixGeneral}.

\subsection{Time evolution of two-point correlation functions}
\label{app:two-point functions}

In this appendix, we study the time evolution of equal-time two-point correlation functions. Besides the time evolution of entanglement entropy and energy, this quantity can also be used to detect different phases of the dynamics.

We consider the two-point correlation function 
$\langle \Psi_n|\mathcal{O}(x_1)\mathcal{O}(x_2)|\Psi_n\rangle$,
where $|\Psi_n\rangle$ is the wavefunction after $n$ steps of driving and $\mathcal{O}(x)$ is a general primary field with conformal dimension $(h,\bar h)$. One can further obtain the correlation functions for descendants of $\mathcal{O}$. 
Here, $\mathcal{O}(x_i)$ is defined on the spacetime cylinder. We do the computation in the imaginary time and thus use the coordinate $w=\tau+ix$. Let us consider a conformal mapping $z=e^{\frac{2\pi q w}{L}}=e^{\frac{2\pi w}{l}}$ to map the $w$-cylinder to the $q$-sheet $z$-Riemann surface (see Fig.\ref{ConformalMapQ}), on which the operator evolution of $\calO(z_1)$ and $\calO(z_2)$ is determined by Eq.\eqref{OP_evolution} and Eqs.\eqref{ZpZ_appendix}, \eqref{MobiusThreeTypeAppendix}.
Next, we map the $q$-sheet $z$-Riemann surface to the complex $\zeta$-plane via a conformal mapping $\zeta=z^{1/q}$, and one can obtain
\be
\label{2point_correlation}
\begin{split}
&\langle \Psi_n|\calO(w_1,\bar{w}_1)
\calO(w_2,\bar{w}_2)|\Psi_n\rangle
=\prod_{i=1,2}
\left(\frac{\partial \zeta_i}{\partial w_i}\right)^h 
\prod_{i=1,2}
\left(\frac{\partial \bar{\zeta}_i}{\partial \bar{w}_{i}}\right)^{\bar{h}}
\langle 
\calO(\zeta_1,\bar{\zeta}_1)
\calO(\zeta_2,\bar{\zeta}_2)
\rangle_{\zeta}
\end{split}
\ee
where $w_j=0+ix_j$, and $(h,\bar{h})$ are the conformal dimensions of 
the operator $\calO$. More explicitly, the contribution of the holomorphic part in Eq.\eqref{2point_correlation} can be expressed in terms of the $\SU(1,1)$ matrix elements in Eq.\eqref{Accumlate_MobiusMatrix2}
as follows:
\be\label{TwistOP_chiral}
\small
\begin{split}
&\left(\frac{2\pi }{L}\right)^{2h}\cdot \frac{z_1^h}{(\beta_n^*z_1+\alpha_n^*)^{2h}}
\cdot \frac{z_2^h}{(\beta_n^*z_2+\alpha_n^*)^{2h}}\cdot
\left(\frac{\alpha_n z_1+\beta_n}{\beta_n^*z_1+\alpha_n^*}\right)^{(\frac{1}{q}-1)h}
\left(\frac{\alpha_n z_2+\beta_n}{\beta_n^*z_2+\alpha_n^*}\right)^{(\frac{1}{q}-1)h}
\\
&\cdot
\left[
\left(\frac{\alpha_n z_1+\beta}{\beta_n^*z_1+\alpha_n^*}\right)^{\frac{1}{q}}
-
\left(\frac{\alpha_n z_2+\beta}{\beta_n^*z_2+\alpha_n^*}\right)^{\frac{1}{q}}
\right]^{-2h},
\end{split}
\ee
where $z_i=e^{\frac{2\pi w_i}{l}}$.
The contribution of the anti-holomorphic part 
can be obtained by replacing $\alpha_n\to\alpha_n'$, 
$\beta_n\to \beta_n'$
and $z_i\to\bar{z}_i$ in the above equation.
Noting that $z$ lives on a $q$-sheet 
Riemann surface (see Fig.~\ref{ConformalMapQ}), 
one should be careful when evaluating Eq.\eqref{TwistOP_chiral},
by tracking if $z_i$ cross the branch cuts and move from 
one layer to another. This is subtle but important especially when the system is in a heating phase.
The relative distance between $z_1$ and $z_2$ will depend on whether there are energy-momentum density peaks between them\cite{fan2019emergent}.

As an illustration, we study the two-point correlation functions in the heating and non-heating phases 
of a periodically driven CFT, respectively.
For simplicity, we only drive the holomorphic part, and keep the 
anti-holomorphic parts untouched.
In the non-heating phase, as discussed in Sec.\ref{Sec: PeriodicDriving},
$\alpha_N$ and $\beta_N$ are periodic functions of $n$ 
[see Eq.\eqref{Psi_np}], 
and so are the correlation functions.
In the heating phase, the Lyapunov exponent is positive, i.e.,
$\lambda_L>0$. In this case, peaks of energy density will form
in the real space. In particular, when there are energy density
peaks between $x_1$ and $x_2$ (and $x_1$ and $x_2$ are not
located at the centers of the energy density peaks), one can find that 
in the long time limit $\lambda_L\cdot N\gg 1$ (recall that the total number
of driving steps is $N=n\cdot p$, where $p$ is the period of driving steps), 
\be
\frac{
\langle \Psi_n|\mathcal{O}(x_1) \mathcal{O}(x_2) |\Psi_n\rangle
}{
\langle \Psi_0|\mathcal{O}(x_1) \mathcal{O}(x_2) |\Psi_0\rangle
}\simeq e^{-2 \lambda_L\cdot h\cdot N}\cdot \left(
\frac{L}{\pi}\sin\frac{\pi(x_1-x_2)}{L}
\right)^{-2h}.
\ee
That is, the correlation function decays exponentially
as a function of the driving time.
Recently, this result is generalized in the heating phase of
more general cases where
the driving Hamiltonians are deformed by an arbitrary smooth function.
See Ref.\cite{fan2020FloquetGeneral} for more details.

\subsection{Entanglement entropy evolution}

In this appendix, we give some details on the time evolution of 
the entanglement entropy in a time-dependent driven CFT.

\subsubsection{General formula}

We give a derivation of Eq.\eqref{EE_general} in the main text.
The $m$-th Renyi entropy of $A=[x_1,\,x_2]$ can be obtained by 
studying the correlation function of twist operators:
\be
\label{RenyiEntropy_appendix}
S^{(m)}_A(n)=\frac{1}{1-m}\log\,\langle \Psi_n| \mathcal{T}_m(x_1) \bar{\mathcal{T}}_m(x_2) |\Psi_n\rangle,
\ee
where $|\Psi_n\rangle$ denotes the wavefunction after $n$
steps of drvings, and $\mathcal{T}_m$ ($\bar{\mathcal{T}}_m$) 
are primary operators with conformal dimensions $h = \bar{h} = \frac{c}{24}(m-\frac{1}{m})$.
The evalution of Eq.\eqref{RenyiEntropy_appendix} follows the previous Appendix.\ref{app:two-point functions} directly and we have
\be
\label{TwistOP_correlation}
\begin{split}
&\langle \Psi_n|\mathcal{T}_m(w_1,\bar{w}_1)
\bar{\mathcal{T}}_m(w_2,\bar{w}_2)|\Psi_n\rangle
=\prod_{i=1,2}
\left(\frac{\partial \zeta_i}{\partial w_i}\right)^h 
\prod_{i=1,2}
\left(\frac{\partial \bar{\zeta}_i}{\partial \bar{w}_{i}}\right)^{\bar{h}}
\langle 
\mathcal{T}_m(\zeta_1,\bar{\zeta}_1)
\bar{\mathcal{T}}_m(\zeta_2,\bar{\zeta}_2)
\rangle_{\zeta}
\end{split}
\ee
where $w_i=0+ix_i$ are the coordinates in the imaginary time and $z_i = e^{2\pi qw/l}$, $\zeta_j = z_j^{1/q}$.
We choose the subsystem within one deformation wavelength as 
$A=[(k-1/2)l, (k+j-1/2)l]$ or $A=[kl, (k+j)l]$ where $k,\,j\in \mathbb Z$, $j<q$, and $l=L/q$.
In this case, $z_1$ ($\bar{z}_1$) and $z_2$ ($\bar{z}_2$) always live on different layers labeled by $j$.
Let us take $A=[(k-1/2)l, (k+j-1/2)l]$ for example.
Based on Eqs.\eqref{TwistOP_chiral} and \eqref{TwistOP_correlation},
one can check explicitly that
\be
\small
\langle \Psi_n|\mathcal{T}_m(w_1,\bar{w}_1)
\bar{\mathcal{T}}_m(w_2,\bar{w}_2)|\Psi_n\rangle
=\left(\frac{2\pi}{L}\right)^{4h}
\cdot 
\frac{1}{|\alpha_n-\beta_n|^{4h}}\cdot \frac{1}{|\alpha'_n-\beta'_n|^{4h}}
\cdot\frac{1}{\left(2\sin\frac{\pi j}{q}\right)^{4h}}.
\ee
Using $h = \bar{h} = \frac{c}{24}(m-\frac{1}{m})$
and Eq.\eqref{RenyiEntropy_appendix}, we can obtain
\be\label{SAappendix01}
S_A^{(m)}(n)-S_A^{(m)}(0)=\frac{c}{6}\cdot\frac{1+m}{m}\Big(\log|\alpha_n-\beta_n|+\log|\alpha_n'-\beta_n'|
\Big),
\ee
which reduces to Eq.\eqref{EE_general} for $m\to 1$.

With the same procedure, if one chooses the subsystem as 
$A=[kl, (k+j)l]$ where $k,\,j\in\mathbb Z$ and $j<q$, 
then one can obtain 
$
S_A^{(m)}(n)-S_A^{(m)}(0)=\frac{c}{6}\cdot\frac{1+m}{m}\Big(\log|\alpha_n+\beta_n|+\log|\alpha_n'+\beta_n'|
\Big).
$
The difference between this result and Eq.\eqref{SAappendix01}
reflects the non-uniform property of the driven CFT.

\subsubsection{Linear decrease of the entanglement entropy}
\label{Sec: LinearDecreaseEE}

In this appendix, we show that if the subsystem is chosen in 
such a way that the two entanglement cuts lie on the
centers of two chiral (anti-chiral) energy-momentum density peaks,
and at the same time we keep the anti-chiral (chiral) part \textit{undriven}, 
then the entanglement entropy may decreae in time.
The choice of subsystem $A$ can be understood as follows:
\be
\label{CutCenter}
    \begin{tikzpicture}[scale=0.8, baseline={([yshift=-6pt]current bounding box.center)}]
    \draw [dotted][red][thick](-31pt,0pt)..controls (-30pt,2pt) and (-29pt,5pt)..(-28pt,40pt)..controls (-27pt,5pt) and (-26pt,2pt)..(-25pt,0pt);
    \draw [blue][thick](31pt,0pt)..controls (30pt,2pt) and (29pt,5pt)..(28pt,40pt)..controls (27pt,5pt) and (26pt,2pt)..(25pt,0pt);
    \draw [thick] (-50pt,0pt)--(50pt,0pt);
    \draw [thick] (-50pt,-5pt) -- (-50pt,5pt);

     \draw [dotted][red][xshift=100pt,thick](-31pt,0pt)..controls (-30pt,2pt) and (-29pt,5pt)..(-28pt,40pt)..controls (-27pt,5pt) and (-26pt,2pt)..(-25pt,0pt);
    \draw [blue][xshift=100pt,thick](31pt,0pt)..controls (30pt,2pt) and (29pt,5pt)..(28pt,40pt)..controls (27pt,5pt) and (26pt,2pt)..(25pt,0pt);
    \draw [xshift=100pt,thick] (-50pt,0pt)--(50pt,0pt);
    \draw [xshift=100pt,thick] (-50pt,-5pt) -- (-50pt,5pt);

    \draw [dotted][red][xshift=200pt,thick](-31pt,0pt)..controls (-30pt,2pt) and (-29pt,5pt)..(-28pt,40pt)..controls (-27pt,5pt) and (-26pt,2pt)..(-25pt,0pt);
    \draw [blue][xshift=200pt,thick](31pt,0pt)..controls (30pt,2pt) and (29pt,5pt)..(28pt,40pt)..controls (27pt,5pt) and (26pt,2pt)..(25pt,0pt);
    \draw [xshift=200pt,thick] (-50pt,0pt)--(50pt,0pt);
    \draw [xshift=200pt,thick] (-50pt,-5pt) -- (-50pt,5pt);

    \draw [dotted][red][xshift=300pt,thick](-31pt,0pt)..controls (-30pt,2pt) and (-29pt,5pt)..(-28pt,40pt)..controls (-27pt,5pt) and (-26pt,2pt)..(-25pt,0pt);
    \draw [blue][xshift=300pt,thick](31pt,0pt)..controls (30pt,2pt) and (29pt,5pt)..(28pt,40pt)..controls (27pt,5pt) and (26pt,2pt)..(25pt,0pt);
    \draw [xshift=300pt,thick] (-50pt,0pt)--(50pt,0pt);
    \draw [xshift=300pt,thick] (-50pt,-5pt) -- (-50pt,5pt);

    \draw[dashed] (28pt,0pt)--(28pt,65pt);
    \draw[dashed] (228pt,0pt)--(228pt,65pt);
    
     \node at (30pt, -10pt){$x_1$};
      \node at (230pt, -10pt){$x_2$};
      
        \draw[->,>=stealth] (128pt,50pt)--(228pt,50pt);
         \draw[->,>=stealth] (128pt,50pt)--(28pt,50pt);

 \node at (130pt, 63pt){subsystem $A$};
    
     \draw [thick] (350pt,-5pt) -- (350pt,5pt);
       \end{tikzpicture}
\ee

Let us consider the setup of periodically driven CFT in 
Sec.\ref{Sec: PeriodicDriving}, such that the locations 
of peaks are fixed in the long time driving limit.
Without loss of generality, we drive the chiral modes 
in time, but keep the anti-chiral modes undriven.
As seen from Eq.\eqref{PeakLocationHP}, the distance between 
two chiral enerngy-momentum density peaks are 
$x_{\text{peak},2}-x_{\text{peak},1}=jl$, where $j\in\mathbb Z$,
$j<q$ and $l=L/q$.
In addition, since the locations of peaks correspond to
the fixed point $\gamma_2$ (i.e., $z_{\text{peak}}=\gamma_2$, 
where we have assumed $0<\eta<1$)
in Eq.\eqref{FixedPoint}, this means 
$x_{\text{peak},1}$ and $x_{\text{peak},2}$ will not 
move around in the stroboscopic sense.
Then based on Eqs.\eqref{TwistOP_correlation} 
and \eqref{TwistOP_chiral}, one can find that
\be
\small
\langle \Psi_n|\mathcal{T}(w_1,\bar{w}_1)
\bar{\mathcal{T}}(w_2,\bar{w}_2)|\Psi_n\rangle
=\left(\frac{2\pi}{L}\right)^{4h}
\cdot 
\frac{1}{(\alpha_{np}\cdot \gamma_2+\beta_n)^{2h}
(\beta_{np}^*\cdot \gamma_2+\alpha^*_n)^{2h}
}
\cdot\frac{1}{\left(2\sin\frac{\pi j}{q}\right)^{4h}}
\cdot\frac{1}{(\gamma_2^*)^{2h}},
\ee
Now by considering the expressions of $\alpha_{np}$ and $\beta_{np}$
in Eq.\eqref{AlphaBeta}, one can find that
\be
\alpha_{np}\cdot \gamma_2+\beta_n=\eta^{\frac{n}{2}}\cdot \gamma_2,
\quad
\beta_{np}^*\cdot \gamma_2+\alpha^*_n=\eta^{\frac{n}{2}}.
\ee
and therefore $\langle \Psi_n|\mathcal{T}_m(w_1,\bar{w}_1)
\bar{\mathcal{T}}_m(w_2,\bar{w}_2)|\Psi_n\rangle=
\left(\frac{2\pi}{L}\right)^{4h}\cdot\frac{1}{\eta^{2nh}}\cdot 
\frac{1}{\left(2\sin\frac{\pi j}{q}\right)^{4h}}$, based on which
one can find that 
\be
\left\{
\begin{split}
&S^{(m)}_A(n)-S^{(m)}_A(0)=-\frac{c}{12}\cdot \frac{1+m}{m}\cdot n\cdot
\log\frac{1}{\eta},\\
&S_A(n)-S_A(0)=-\frac{c}{6}\cdot n\cdot \log\frac{1}{\eta},
\end{split}
\right.
\ee
where $0<\eta<1$.
Recalling that the total driving step number is $N=np$, we can 
write the entanglement entropy evolution in terms of the Lyapunov
exponent in Eq.\eqref{Lyapunov_Eta} as follows:
\be
S_A(n)-S_A(0)=-\frac{c}{3}\cdot \lambda_L\cdot N.
\ee
It is interesting to compare this formula with the result in 
Eq.\eqref{EEpeak1}. 
Here the linearly decreasing entanglement entropy is due
to the coincidence of the entanglement cuts with the 
centers of the chiral energy-momentum density peaks 
(while keeping the anti-chiral modes undriven).
During the driving, the degrees of freedom that 
entangle $A$ and $\bar{A}$ will flow and accumulate
at the energy-momentum density peaks, which are located
at the entanglement cut.
Intuitively, the Bell pairs that are non-local in space
now become local, which results in a decrease in 
the entanglement entropy evolution.

It is emphasized that although the entanglement entropy 
decreases in time, which is due to the choice of entanglement cuts,
the total energy of the system still grows  
in time (in the heating phase) in a periodically driven CFT.

\subsubsection{Comparison of CFT and lattice calculations}
\label{app: CompareCFTLattice_Fibonacci}

To confirm the linearly decreasing feature of the entanglement
entropy evolution in the previous
subsections, we compare the CFT and lattice calculations in this appendix.
The results here are also related to the entanglement evolution at the 
non-heating fixed point in a Fibonacci driven CFT in Sec.~\ref{Sec:EEnonFib}
(See Fig.~\ref{EEnonFib}).

As an example, let us consider the minimal setup of 
periodically driven CFTs in Sec.\ref{Sec:MinimalSetup}.
We drive the CFT periodically with $H_{\theta=0}$ and $H_{\theta}$
in Eq.\eqref{H_theta_A}, with open boundary conditions.
Now we consider two driving protocols:
\be\label{TwoProtocols}
\left\{
\begin{split}
&\text{Protocol I:}\quad\text{Driving with $H_0$ first, and then $H_1=H_{\theta}$}, 
\\
&\text{Protocol II:} \quad\text{Driving with $H_1=H_{\theta}$ first, and then $H_{0}$},
\\
\end{split}
\right.
\ee
The driving time intervals are chosen as $T_0=L/2$ for $H_0$,
and $T_1=L_{\text{eff}}/2$ for $H_1$, respectively.
The resulting M\"obius transformations in one driving period correspond 
to the following $SU(1,1)$ matrices:
\be\label{TwoProtocolsM}
\left\{
\begin{split}
&\text{Protocol I:}\quad M=M_0M_{\theta}=
\begin{pmatrix}
-\cosh(2\theta) &\sinh(2\theta)\\
\sinh(2\theta) &-\cosh(2\theta)
\end{pmatrix},
\\
&\text{Protocol II:} 
\quad M'=M_{\theta}M_{0}=
\begin{pmatrix}
-\cosh(2\theta) &-\sinh(2\theta)\\
-\sinh(2\theta) &-\cosh(2\theta)
\end{pmatrix}.
\\
\end{split}
\right.
\ee
One can find $|\text{Tr}(M)|=|\text{Tr}(M')|=|2\cosh(2\theta)|>2$,
i.e., both $M$ and $M'$ are hyperbolic. 
Therefore, in both protocols, the CFT is in a heating phase. 
The difference is that, the chiral and anti-chiral energy-momentum
density peaks are located separately at $x=0$ and $x=L$ in protocol I,
but are located at the same position $x=L/2$ in protocol II,
as pictorially shown as follows:
\be
   \begin{tikzpicture}[scale=1.0, baseline={(current bounding box.center)}]
    \draw[thick] (-60pt,0pt) -- (60pt,0pt);
   \draw[thick] (-60pt,-5pt)-- (-60pt,5pt);
    \fill[pattern = south west lines] (-70pt,-5pt) rectangle (-60pt,5pt); 
    \draw[thick] (60pt,-5pt) -- (60pt,5pt);
    \fill[pattern= south west lines] (60pt,-5pt) rectangle (70pt,5pt); 
    \draw [red][xshift=-30pt][thick](-31pt,0pt)..controls (-30pt,2pt) and (-29pt,5pt)..(-28pt,40pt)..controls (-27pt,5pt) and (-26pt,2pt)..(-25pt,0pt);
    \draw [blue][xshift=30pt][thick](31pt,0pt)..controls (30pt,2pt) and (29pt,5pt)..(28pt,40pt)..controls (27pt,5pt) and (26pt,2pt)..(25pt,0pt);
    
     \draw[thick] (0pt,-3pt)-- (0pt,3pt);
      \node at (0pt,-7pt){${x_1}$};
            \node at (0pt,-18pt){Protocol I};
   \end{tikzpicture}
   \quad\quad\quad
      \begin{tikzpicture}[scale=1.0, baseline={(current bounding box.center)}]
    \draw[thick] (-60pt,0pt) -- (60pt,0pt);
   \draw[thick] (-60pt,-5pt)-- (-60pt,5pt);
    \fill[pattern = south west lines] (-70pt,-5pt) rectangle (-60pt,5pt); 
    \draw[thick] (60pt,-5pt) -- (60pt,5pt);
    \fill[pattern= south west lines] (60pt,-5pt) rectangle (70pt,5pt); 

    \draw [blue][xshift=-28pt][thick](31pt,0pt)..controls (30pt,2pt) and (29pt,5pt)..(28pt,40pt)..controls (27pt,5pt) and (26pt,2pt)..(25pt,0pt);
        \draw [dashed][red][xshift=28pt][thick](-31pt,0pt)..controls (-30pt,2pt) and (-29pt,5pt)..(-28pt,40pt)..controls (-27pt,5pt) and (-26pt,2pt)..(-25pt,0pt);
    
     \draw[thick] (0pt,-3pt)-- (0pt,3pt);
      \node at (0pt,-7pt){${x_1}$};
            \node at (0pt,-18pt){Protocol II};
   \end{tikzpicture}
\ee
By choosing the subsystem $A=[0, \, x_1]$ with $x_1=L/2$, 
the entanglement cut will not cut any peaks in protocol I, 
but will cut both the chiral and anti-chiral peaks in 
protocol II.
Intuitively, one can unfold the CFT with open boudary conditions
to a single copy of chiral CFT with a periodic boundary condition.
After the unfolding, the entanglement cut in protocol II lies on the 
centers of two chiral energy-moemntum density peaks.
Based on our study in the previous subsection, the entanglement
entropy will decrease linearly in this case.
Next, we show this is indeed the case rigorously .

\begin{figure}[t]
\centering
\includegraphics[width=6.20in]{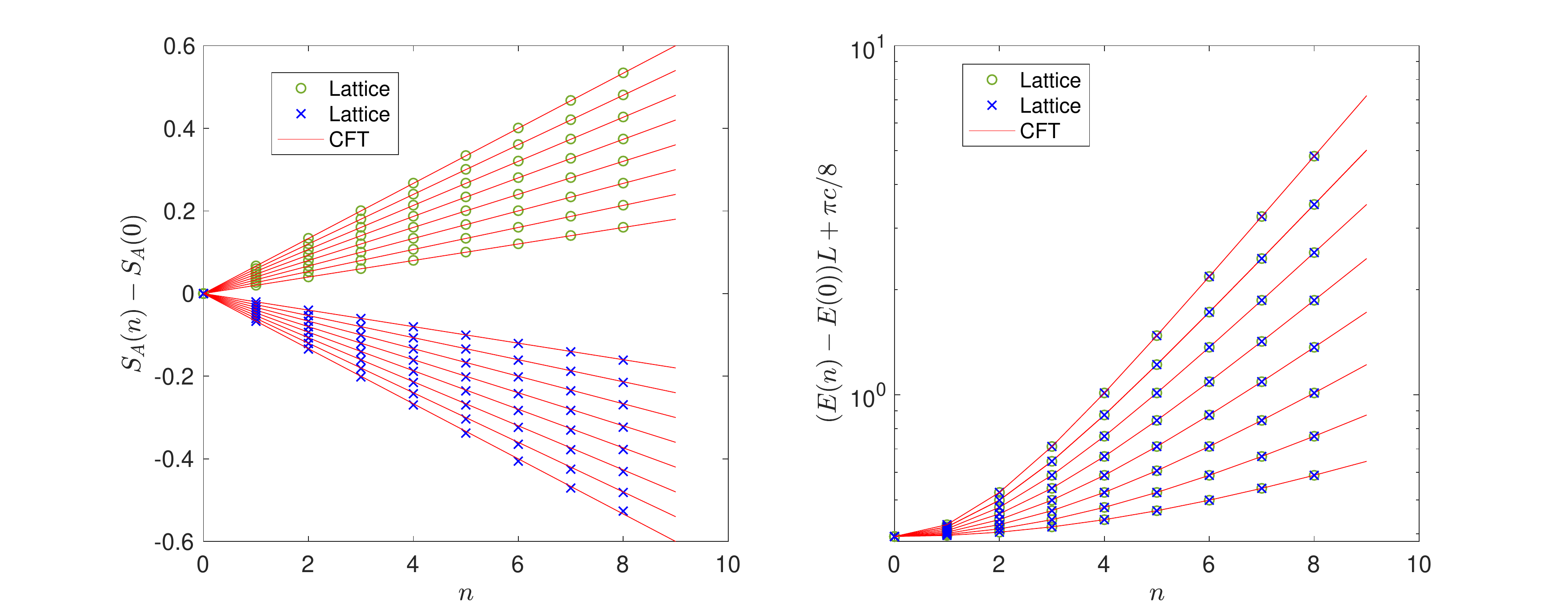}
\caption{
Comparison of the CFT and lattice calculations on the
entanglement entropy (left) and the total energy (right) evolution in 
the heating phase of a periodically driven CFT.
The numerical data in $\circ$($\times$) correspond to protocol I(II) in Eq.\eqref{TwoProtocols}.
The CFT is periodically driven with $H_0$ and $H_{\theta}$ with time intervals $T_0=L/2$ and $T_1=L_{\text{eff}}/2$, respectively.
From bottom to top (in the right plot), we choose $\theta=0.03$, $0.04$, $0.05$, $0.06$, $0.07$,
$0.08$, $0.09$, and $0.1$.
The CFT results are plotted according to Eq.~\eqref{EElinearDecrease}
and \eqref{EElinearDecreaseEnergy}.
}
\label{EELinear}
\end{figure}

Based on Eqs.\eqref{HalfEE_general} and \eqref{EnergyTotal}, 
one can obtain the entanglement entropy/total energy
evolution as follows:
\be\label{EElinearDecrease}
S_A(n)=\left\{
\begin{split}
&\frac{2n c}{3}\cdot \theta,\quad &\text{Protocol I},\\
&-\frac{2n c}{3}\cdot \theta,\quad &\text{Protocol II},
\end{split}
\right.
\ee 
where $A=[0,L/2]$.
One can find that the entanglement entropy grows (decreases) 
linearly as a function of $n$ in Protocol I (II).
On the other hand, 
the total energy of the system grows in both protocols:
\be\label{EElinearDecreaseEnergy}
E(n)-E(0)+\frac{\pi c}{8L}=\frac{\pi c}{8L}\cdot \cosh(4n\theta),\quad
\text{for both Protocols I and II}.
\ee

Now we compare the CFT and lattice calculations on the
entanglement entropy/total energy evolution.
The lattice model we consider is the same as that 
in Sec.\ref{Sec:LatticePeriodic}.
That is, 
the two lattice Hamiltonians under consideration are 
$H_0=\frac{1}{2}\sum_{i=1}^{L-1}c_i^{\dag}c_{i+1}+h.c.$,  
and
$H_{\theta}=\frac{1}{2}\sum_{i=1}^{L-1}f(i)
c_i^{\dag}c_{i+1}+h.c.$
where $L$ is the total length of the lattice and
$f(i)=1-\tanh(2\theta)\cdot \cos\frac{2\pi i}{L}$,
with the initial state chosen as the ground state of $H_0$.
The comparison is shown in Fig.\ref{EELinear}.
The agreement between CFT and lattice calculations are remarkable.

\begin{figure}[t]
\centering
\includegraphics[width=5.10in]{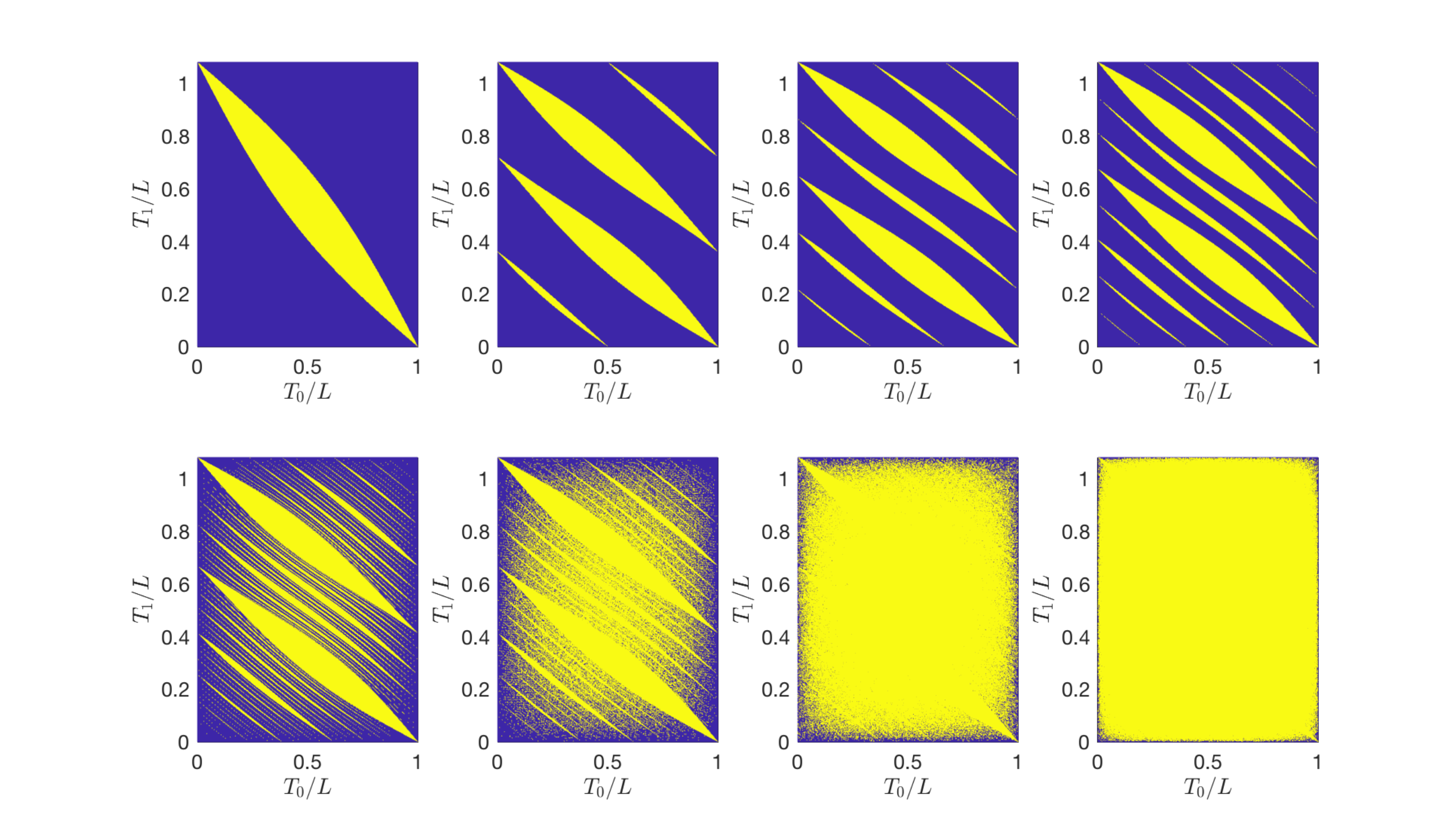}
\caption{
Phase diagrams in a periodically driven CFT with the sequence generated by finitely truncated Fibonacci bitstring, i.e. $\{X_j\}$ with 
 $\omega_n=F_{n-1}/F_n$.
 Here we choose $n=2$, $4$, $5$, $6$, $10$, $20$, $100$, and $1000$, 
 respectively. The two Hamiltonians we use are $H_0(\theta=0)$ and $H_1(\theta=0.2)$ in \eqref{H_theta_A}.
 The phase diagram is periodic in $T_0$ direction with period $L$ and in $T_1$ direction with
 period $L\cosh(2\theta)\simeq 1.08 L$.
  The blue (yellow) regions correspond to the heating (non-heating) phases.
}
\label{FibonacciPhase3}
\end{figure}

\begin{figure}[t]
\centering
\includegraphics[width=5.10in]{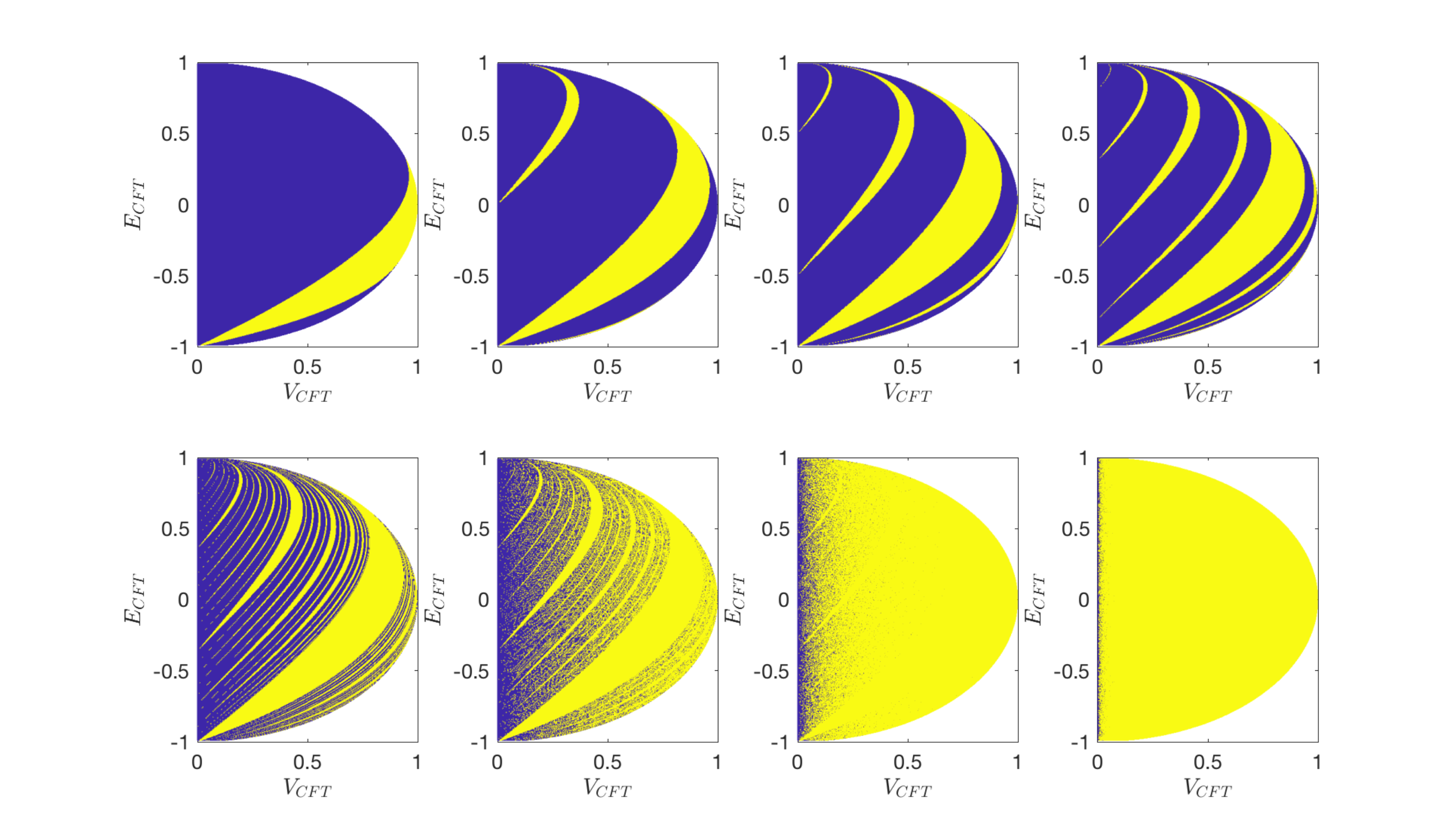}
\caption{
Phase diagrams in a periodically driven CFT with the sequence generated by finitely truncated Fibonacci bitstring, i.e. $\{X_j\}$ with 
 $\omega_n=F_{n-1}/F_n$.
 Here we choose $n=2$, $4$, $5$, $6$, $10$, $20$, $100$, and $1000$, 
 respectively. The two Hamiltonians we use are $H_0(\theta=0)$ and $H_1(\theta=0.2)$ in \eqref{H_theta_A}.
 The phase diagram is periodic in $T_0$ direction with period $L$ and in $T_1$ direction with
 period $L\cosh(2\theta)\simeq 1.08 L$.
  The blue (yellow) regions correspond to the heating (non-heating) phases.
}
\label{FibonacciPhaseEV3}
\end{figure}

\subsection{Phase diagrams from periodic to quasi-periodical driving}

In this appendix, we present one more group of results on the 
evolution of phase diagrams as we use a periodical driving to 
approach the Fibonacci quasi-periodical driving limit, 
as shown in Fig.~\ref{FibonacciPhase3} and Fig.~\ref{FibonacciPhaseEV3}.

In Fig.\ref{FibonacciPhase3}, 
we consider a periodic driving
with the sequence generated by finitely truncated Fibonacci bitstring, i.e. $\{X_j\}$ with $\omega_n=F_{n-1}/F_n$. 
The two driving Hamiltonians are
 $H_{\theta=0}$ and $H_{\theta=0.2}$ (See Sec.\ref{Sec:HthetaExample}
 for more details).
 
 In Fig.\ref{FibonacciPhaseEV3}, we replot the phase diagrams in 
 Fig.\ref{FibonacciPhase3} with the new variables as defined in 
 Eq.\eqref{EV_finiteTheta}. Based on Fig.\ref{FibonacciPhaseEV3},
 we obtain the measure of non-heating phases in Fig.\ref{RatioCFTEV}.

\section{Fibonacci bitstring/word and recurrence relation} 
\label{appendix Fib}

In this appendix, we review some basics of the Fibonacci bitstring/word (in this paper, we will call it Fibonacci bitstring instead of Fibonacci word\footnote{The term ``Fibonacci word'' was used in combintorics, whose definition is off by an overall bit flipping $0\leftrightarrow 1$ comparing to the one commonly used in the Fibonacci quasi-crystal literature. We adopt the latter convention, and rename it as Fibonacci bitstring to avoid confusion.}) for the readers' convenience. In particular, we explain the equivalence of two ways to generate the Fibonacci bitstring: (1) the quasi-periodic potential and (2) the substitution rule.  
Based on the substitution rule, we explain the recurrence relation of the traces and the constant of motion that are used in the main text.

\subsection{The substitution rule for Fibonacci bitstring} 
\begin{figure}[t]
\center
\begin{tikzpicture}[baseline={(current bounding box.center)}]
\draw[thick,->,>=stealth] (-200pt,-1pt) -- (200pt,-1pt) node[right]{$t$};
\filldraw[yellow] (-161.8pt,0pt) rectangle (-123.6pt,10pt);
\filldraw[blue] (-123.6pt,0pt) rectangle (-61.8pt,10pt);
\filldraw[yellow] (-61.8pt,0pt) rectangle (-23.6pt,10pt);
\filldraw[blue] (-23.6pt,0pt) rectangle (38.2pt,10pt);
\filldraw[yellow] (38.2pt,0pt) rectangle (77.4pt,10pt);
\filldraw[blue] (77.4pt,0pt) rectangle (138.2pt,10pt);
\filldraw[yellow] (138.2pt,0pt) rectangle (177.4pt,10pt);
\filldraw (0pt,-1pt) circle (1pt) node[below]{$0$};
\filldraw (-161.8pt,-1pt) circle (1pt) node[below]{\scriptsize $\omega^2-2$};
\filldraw (-123.6pt,-1pt) circle (1pt) node[below]{\scriptsize  $-\omega^3-1$};
\filldraw (-61.8pt,-1pt) circle (1pt) node[below]{\scriptsize $\omega^2-1$};
\filldraw (-23.6pt,-1pt) circle (1pt) node[below]{\scriptsize  $-\omega^3$};
\filldraw (38.2pt,-1pt) circle (1pt) node[below]{\scriptsize  $\omega^2$};
\filldraw (77.4pt,-1pt) circle (1pt) node[below]{\scriptsize  $-\omega^3+1$};
\filldraw (138.2pt,-1pt) circle (1pt) node[below]{\scriptsize  $\omega^2+1$};
\filldraw (177.4pt,-1pt) circle (1pt) node[below]{\scriptsize  $-\omega^3+2$};
\end{tikzpicture}
\caption{An illustration for the characteristic function $\chi(t)$ for the Fibonacci sequence. The blue regime stands for value $1$ and the  yellow for $0$.}
\label{fig: char}
\end{figure}
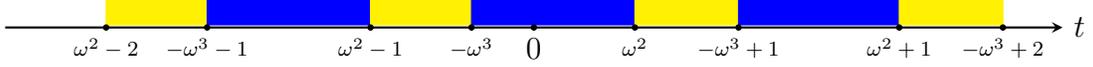

In the main text, we generate the Fibonacci quasi-periodic driving  using the following bitstring (``Fibonacci bitstring'') 
\begin{equation}
X_j=\chi((j-1)\omega)\,, \quad j=1,2,3\ldots
\label{eqn: fib potential}
\end{equation}
where $\chi(t)=\chi(t+1)$ is a
period-1 characteristic function
\begin{equation}
\chi(t)=\begin{cases}
1 & ~~ -\omega^3\leq t < \omega^2\\
0&   ~~ \omega^2\leq t < 1-\omega^3
\end{cases}
\label{eqn: char}
\end{equation}
and $\omega=\frac{\sqrt{5}-1}{2}$ is the inverse of golden ratio, see Fig.~\ref{fig: char} for an illustration of function $\chi(t)$. 
For instance,  the first a few bits of $\{X_j\}$ are given as follows,
\begin{equation}
X_{j=1,2,3\ldots} = 10110101 \ldots
\end{equation}
This definition is straightforward but not useful in our application. Instead, we will follow 
the presentation in Ref.~\cite{suto1987} to show that the above bitstring can be generated by a substitution rule. The equivalence is based on the following two properties of $\{X_j\}$
\begin{enumerate}
    \item Let us use the notation ``$\lfloor x \rfloor :=  \max\{m\in \ZZ | m \leq x\} $'' for the floor function, and we have
\begin{equation}
X_j=\lfloor (j+1) \omega \rfloor - \lfloor j \omega \rfloor  \,.
\end{equation}
To prove this statement, let us first use the relation $\omega^2+\omega-1=0$ to rewrite $\omega^2=1-\omega$ and $-\omega^3=1-2\omega$. Then according to the rule \eqref{eqn: char} we have
\begin{equation}
\begin{aligned}
X_j=1 &\Leftrightarrow \exists m \in \ZZ\,:~ m-2\omega \leq (j-1)\omega < m-\omega \\
&\Leftrightarrow \exists m \in \ZZ\,:~ j\omega< m \leq (j+1)\omega 
\end{aligned}
\end{equation}

\item Let us use $F_n$ to denote the $n$-th Fibonacci number, namely  $F_n=F_{n-1}+F_{n-2}$ with $F_0=F_1=1$, then we have
\begin{equation}
\label{eqn: 2nd prop}
X_{j+F_n}=X_j\,, \quad \text{for} ~ n\geq 2\,,~ 1 \leq j <F_n\,.
\end{equation} 
To prove this statement, it is sufficient to show that for $n\geq 2$, 
the difference of two parts in
$
X_j=\lfloor (j+1) \omega \rfloor - \lfloor j \omega \rfloor $
is unchanged while shifting the argument of the floor function $\lfloor \cdot \rfloor $ by $F_n \omega$ for $1\leq j < F_n$: we can write $F_n \omega = m + r$ where $m$ is the integer that is closest to $F_n\omega$ and $|r|={\rm dist} (F_n  \omega, \ZZ )$ denotes the distance (with sign) between $F_n \omega$ and the nearest integer $m$. Obviously, shifting by an integer will not affect the difference, so we only need to check the effect of shifting by $r$. 
From the fact that the convergent\footnote{For irrational real number $x$, we always have an infinite continued fraction representation 
\begin{equation}
x=  a_0+ \frac{1}{a_1+\frac{1}{a_2+\ldots \frac{1}{a_n+\ldots}}} \,, \qquad a_0 \in \ZZ\,, \quad a_{j\geq 1} \in \ZZ^{>0}
\end{equation}
The n-th {\bf principal convergent} is the rational number $p_n/q_n$ obtained by a truncation at $a_n$. The irrational number $\omega = \frac{\sqrt{5}-1}{2}$ has a particularly simple continued fraction representation $a_0=0$, $a_1=a_2=\ldots =1$, and its finite truncation is the ratio of two nearby Fibonacci numbers $F_{n-1}/F_n$ as one can easily check.
}
$F_{n-1}/F_n$ is a best Diophantine approximation\footnote{For a real number $x$, a rational number $p/q$ is a best approximation (of second kind) if
\begin{equation}
    \left| qx - p \right| <   \left| q'x - p' \right| \,, \quad \forall q>q'>0 \,.
\end{equation}
For the proof that the best approximations are given by the convergent, see e.g. Ref.~\cite{lang1995introduction}. 
}
 of the irrational number $\omega$, 
we have the following inequality 
\begin{equation}
|r|={\rm dist} (F_n  \omega, \ZZ ) <  {\rm dist} (j  \omega, \ZZ ) \quad \text{for} \quad 1\leq j < F_n
\end{equation}
in other words, for $j\omega$ with $1\leq j < F_n$, adding $r$ will not be able to fill the gap between $j\omega$ and a nearby integer. This statement further holds for $j=F_n$, since $\omega$ is irrational, $F_n \omega+r$ can not be an integer. To summarize, neither $\lfloor (j+1) \omega \rfloor$ nor $\lfloor j \omega \rfloor$ will change its value after a shift of $r$ for $1\leq j < F_n$ and therefore we have proved \eqref{eqn: 2nd prop}.

\end{enumerate}

The second property \eqref{eqn: 2nd prop} provides an efficient algorithm to generator the $0,1$ bitstring for $\{X_j\}$. 
For instance, let us denote the first $F_n$ bits as string $A_n B_n$ where $A_n$ stands for the first $F_{n-1}$ bits and $B_n$ stands for the next $F_{n-2}$, and they together has length $F_n=F_{n-1}+F_{n-2}$ as required, here are examples for first few $n$ 
\begin{equation}
F_3=3 : ~
\underbrace{10}_{A_3}\underbrace{1}_{B_3} \,, \qquad
F_4=5 : ~
 \underbrace{101}_{A_4}\underbrace{10}_{B_4}   \,, \qquad
 F_5=8 : ~
  \underbrace{10110}_{A_5}\underbrace{101}_{B_5} \,.
\end{equation}
Then, we group the string $A_n B_n$, rename it as $A_{n+1}=A_n B_n$ and according to \eqref{eqn: 2nd prop}, the $B_{n+1}$ is obtained by copying the first $F_{n-1}$ bits of $A_{n+1}$ which is exactly $A_n$, i.e. we have the following recurrence relation
\begin{equation}
A_{n+1}=A_n B_n \,, \quad B_{n+1}=A_{n} \,.
\label{eqn: recurrence relation}
\end{equation}
One may concern about that \eqref{eqn: 2nd prop} actually produces a  longer bitstring than the above recurrence relation since it also generates an additional segment $B_n$ after $A_{n+1}B_{n+1}$ which will overlap the first $F_{n-2}$ bits of $B_{n+2}$. One can check that the overlapping part is consistent with the rule here, because the first $F_{n-2}$ bits of $B_{n+2}$ is indeed $A_{n-1}=B_n$.

A final comment is that the above recurrence relation can be recast into a ``local'' substitution rule that is closer to the rabbit populations problem Fibonacci originally considered. Let us start with a single bit $1$, and apply the following substitution rule
\begin{equation}
1 \rightarrow 10 \,, \quad 0\rightarrow 1
\label{eqn: substitution}
\end{equation}
at each step, then we will generate the following sequence
\begin{equation}
1 \rightarrow 10 \rightarrow 101 \rightarrow 10110 \rightarrow 10110101 \rightarrow \ldots
\end{equation}
which approaches to the Fibonacci bitstring after infinite steps.

\subsection{Recurrence relation and constant of motion}

In the main text, we are interested in the product $\Pi_n$ of $n$  $\SU(1,1)$ matrices
\begin{eqnarray}
\Pi_n = M_1 \cdot M_2 \ldots M_n
\end{eqnarray}
where matrix $M_j$ depends on $X_j$ in the Fibonacci bitstring. 
The substitution rule 
\eqref{eqn: recurrence relation} directly leads to the following recurrence relation for $\Pi_n$
\begin{equation}
\Pi_{F_k}=\Pi_{F_{k-1}} \cdot \Pi_{F_{k-2}} \,, \quad \forall k\geq 3\,.
\label{eqn: M recurrence}
\end{equation}
The relation can be extended to $k=2$ by defining an auxiliary $\Pi_{F_0}=M_2$ that is distinct from $\Pi_{F_1}=M_1$, although strictly speaking $F_0=F_1=1$. 
A key observation made in Ref.~\cite{kohmoto1983} is that their traces obey the following recurrence relation 
\begin{equation}
\wideboxed{
x_{F_{k+1}} = 2 x_{F_k} x_{F_{k-1}} -x_{F_{k-2}} \,, \quad \text{where} \quad x_{F_k}=\frac{1}{2} \Tr (\Pi_{F_k}) = \frac{1}{2} \Tr (\Pi_{F_k}^{-1}) \,.
}
\label{eqn: trace recur}
\end{equation}
To derive this relation, we start with \eqref{eqn: M recurrence} and find 
\begin{equation}
\Pi_{F_{k+1}}+\Pi_{F_{k-2}}^{-1}=\Pi_{F_k}\Pi_{F_{k-1}}  +\Pi_{F_k}^{-1} \Pi_{F_{k-1}} \,.
\end{equation}
Then we insert identity $
\Pi_{F_k}+\Pi_{F_k}^{-1}= \Tr (\Pi_{F_k}) {\mathbb{I}}
$ 
for the uni-determinant $2\times 2$ matrice $\Pi_{F_k}$, 
and obtain 
\begin{equation}
\Pi_{F_{k+1}}+\Pi_{F_{k-2}}^{-1}=\Tr (\Pi_{F_k}) \Pi_{F_{k-1}} \,,
\end{equation}
whose trace gives \eqref{eqn: trace recur}. 

Using the trace relation, Ref.~\cite{kohmoto1983} further notes a constant of motion
\begin{equation}
\wideboxed{
I = - 1 + x_{F_k}^2   + x_{F_{k-1}}^2  + x_{F_{k-2}}^2  - 2 x_{F_{k}}  x_{F_{k-1}} x_{F_{k-2}} \,.
}
\end{equation}
Indeed, one can check that the change of r.h.s. under the shifting $k\rightarrow k+1$ is zero: 
\begin{equation}
\begin{aligned}
\Delta \text{r.h.s.} &= x^2_{F_{k+1}} - x_{F_{k-2}}^2 - 2 (x_{F_{k+1}}-x_{F_{k-2}} ) x_{F_k} x_{F_{k-1}} \\
&=  (x_{F_{k+1}}-x_{F_{k-2}} )  ( x_{F_{k+1}} + x_{F_{k-2}} - 2x_{F_k} x_{F_{k-1}} )=0\,.
\end{aligned}
\end{equation}

\bibliography{QuasiRef}

\end{document}